# Quantum Entanglement with Geometric Measures

by

Xuanran Zhu

A Thesis Submitted to
The Hong Kong University of Science and Technology
in Partial Fulfilment of the Requirements for
the Degree of Doctor of Philosophy
in Physics

June 2025, Hong Kong

# Quantum Entanglement with Geometric Measures

by Xuanran Zhu

Department of Physics

The Hong Kong University of Science and Technology

Abstract

Quantifying quantum entanglement is a pivotal challenge in quantum information science, particularly for high-dimensional systems, due to its computational complexity. This thesis extends the geometric measure of entanglement (GME) to introduce and investigate a suite of GME-based entanglement monotones tailored for diverse quantum contexts, including pure states, subspaces, and mixed states. These monotones are applicable to both bipartite and multipartite systems, offering a unified framework for characterizing entanglement across various scenarios. Notably, the proposed monotones are adept at identifying entanglement with varying entanglement dimensionalities, making them particularly effective for detecting high-dimensional entanglement. To support practical computation, we develop a non-convex optimization framework that yields accurate upper bounds, complemented by semidefinite programming techniques to establish robust lower bounds. Together, these approaches provide a consistent and efficient computational methodology. This work advances both the theoretical understanding and algorithmic tools for entanglement quantification, contributing to the study of complex quantum correlations in entangled systems.

# Authorization

I hereby declare that I am the sole author of the thesis.

I authorize the Hong Kong University of Science and Technology to lend this thesis to other institutions or individuals for the purpose of scholarly research.

I further authorize the Hong Kong University of Science and Technology to reproduce the thesis by photocopying or by other means, in total or in part, at the request of other institutions or individuals for the purpose of scholarly research.

Xuanran Zhu

# Quantum Entanglement with Geometric Measures

by

Xuanran Zhu

This is to certify that I have examined the above PhD thesis
and have found that it is complete and satisfactory in all respects,
and that any and all revisions required by
the thesis examination committee have been made.

---

Prof. Xi Dai, Thesis Supervisor

---

Prof. Jiannong Wang, Head of Department

Department of Physics

# ACKNOWLEDGEMENTS

First and foremost, I would like to express my deepest gratitude to my academic advisor during the first three years of my Ph.D. studies, Professor Bei Zeng. She introduced me to the fascinating world of theoretical research in quantum information and guided me toward identifying my research interests. Her insights and encouragement inspired me to explore new research methods and topics, urging me to think beyond fixed paradigms and resist the temptation to follow fleeting trends. Her perspective not only profoundly benefited my academic journey but also had a lasting impact on my personal development. Professor Zeng encouraged me to reflect on what kind of life I truly want to pursue and whether I am genuinely suited for academic research. These moments of self-reflection and introspection, I believe, are sometimes even more meaningful than simply working hard in research. As I often find in my passion for hiking, life is akin to a long journey where pauses for rest are essential for better progress ahead. I am also deeply grateful to my advisor during the final year of my Ph.D., Professor Xi Dai, for his invaluable support and assistance in helping me complete my doctoral studies.

I would also like to extend my heartfelt thanks to my undergraduate advisor, Professor Dawei Lu, who can be regarded as the person who introduced me to the quantum world. During my undergraduate years, he opened my eyes to the interdisciplinary nature of quantum mechanics and its potential to generate fascinating phenomena that deviate from classical intuition. His guidance and care for undergraduate students greatly influenced me and directly sparked my continued exploration of quantum information during my Ph.D. studies. Under his mentorship, I gained my first hands-on experience in the entire research process—from generating ideas to numerical simulations, experimental validation, manuscript writing, and finally, submitting papers. These invaluable experiences laid a solid foundation for me as a budding researcher.

Additionally, I would like to express my sincere appreciation to my collaborators and members of the research group, including but not limited to: Chao Zhang, Chenfeng Cao, Zipeng Wu, Jiahui Wu, Yichi Zhang, Zheng An, Jinfeng Zeng, Yuhan Huang, Shu Zhou, Siyuan Jin, Youning Li, Shi-Yao Hou, Yiu Tung Poon, Shilin Huang, Xinfang Nie, Yu-ang Fan, Xinyue Long, Hongfeng Liu, Keyi Huang, Cheng Xi, Liangyu Che, Yuxuan Zheng, Yufang Feng, Xiaodong Yang, Yishan Li, Yingjie Zhang, Jun Li, Qihang Liu, Tao Xin, Kai Tang, Zidong Lin, Yu Tian, Chudan Qiu, and Ying Dong. Thank you all for your support and contributions to my research. I also wish to thank the following people, including but not limited to: Ningping Cao,

Song Cheng, Liu Yang, Zhenhuan Liu, Yuguo Shao, Xiaodie Lin, Ruiqi Zhang, Zihao Li, Jie Wang, Guofeng Zhang, Xin Wang, Chengkai Zhu, Chenghong Zhu, Jianxin Chen, Qinyun Qian, and Jinguo Liu, for their valuable suggestions, academic discussions, and intellectual inspiration that sparked numerous ideas.

Finally, I would like to thank my family and friends. Without their unwavering support and understanding, I would not be who I am today. Thank you for encouraging me during my lowest moments, helping me overcome countless challenges, and reminding me during my apparent peaks to stay humble and focused on the road ahead. Life, much like the non-convex optimization problems I studied in quantum information, offers no guarantees of reaching the global optimum. All we can do is strive to improve locally, one step at a time.

# TABLE OF CONTENTS

# LIST OF FIGURES

# LIST OF TABLES

# LIST OF PUBLICATIONS

## Journal Publications

[1]**X. Zhu**, C. Zhang, Z. An, and B. Zeng, "Unified framework for calculating convex roof resource measures", npj Quantum Inf **11**, 56 (2025).

[2]J. Wu, Z. An, C. Zhang, **X. Zhu**, S. Huang, and B. Zeng, "Investigating pure state uniqueness in tomography via optimization", arXiv preprint arXiv:2501.00327 (2024).

[3]**X. Zhu**, C. Zhang, and B. Zeng, "Quantifying subspace entanglement with geometric measures", Phys. Rev. A **110**, 012452 (2024).

[4]C. Zhang*, **X. Zhu***, and B. Zeng, "Variational approach to unique determinedness in pure-state tomography", Phys. Rev. A **109**, 022425 (2024).

[5]X. Nie*, **X. Zhu***, Y.-a. Fan, X. Long, H. Liu, K. Huang, C. Xi, L. Che, Y. Zheng, Y. Feng, X. Yang, and D. Lu, "Self-consistent determination of single-impurity anderson model using hybrid quantum-classical approach on a spin quantum simulator", Phys. Rev. Lett. **133**, 140602 (2024).

[6]**X. Zhu**, C. Zhang, C. Cao, Y. Li, Y. T. Poon, and B. Zeng, "Detecting entanglement by pure bosonic extension", Phys. Rev. Res. **6**, 013249 (2024).

[7]Y. Li, C. Zhang, S.-Y. Hou, Z. Wu, **X. Zhu**, and B. Zeng, "Tapping into permutation symmetry for improved detection of k-symmetric extensions", Entropy **25**, 1425 (2023).

[8]X. Nie*, **X. Zhu***, K. Huang, K. Tang, X. Long, Z. Lin, Y. Tian, C. Qiu, C. Xi, X. Yang, J. Li, Y. Dong, T. Xin, and D. Lu, "Experimental realization of a quantum refrigerator driven by indefinite causal orders", Phys. Rev. Lett. **129**, 100603 (2022).

[9]T. Xin, Y. Li, Y.-a. Fan, **X. Zhu**, Y. Zhang, X. Nie, J. Li, Q. Liu, and D. Lu, "Quantum phases of three-dimensional chiral topological insulators on a spin quantum simulator", Phys. Rev. Lett. **125**, 090502 (2020).

* Indicates co-first authorship

# CHAPTER 1

# OVERVIEW

## 1.1 Background

Quantum mechanics, emerging in the early 20th century, transformed our understanding of matter and energy at atomic and subatomic scales. The framework was pioneered by Max Planck, who introduced the concept of quantized energy levels to address the blackbody radiation problem [1], and Albert Einstein, whose explanation of the photoelectric effect provided compelling evidence for the particle-like nature of light [2]. These groundbreaking ideas were further developed by Niels Bohr [3], Erwin Schrödinger [4], Werner Heisenberg [5], and others [6, 7], who established foundational principles such as wave-particle duality and the probabilistic interpretation of quantum phenomena. At the heart of quantum theory lies the principle of superposition, which asserts that particles can exist in multiple states simultaneously, fundamentally challenging classical deterministic views. This revolutionary shift not only redefined the foundations of physics but also set the stage for the discovery of quantum entanglement, a phenomenon that has since become a cornerstone of modern quantum science.

Quantum entanglement, first introduced by Schrödinger in 1935 [8] and explored in depth in the seminal Einstein-Podolsky-Rosen (EPR) paper [9], describes a non-classical correlation between particles, where the state of one particle is instantaneously connected to the state of another, regardless of the distance separating them. Einstein famously referred to this phenomenon as "spooky action at a distance", reflecting his skepticism. Decades later, Bell tests, beginning with Alain Aspect's pioneering experiments in the 1980s [10], provided experimental confirmation of entanglement. Today, entanglement is widely regarded as a cornerstone of quantum information science, underpinning revolutionary technologies such as quantum cryptography [11–13], quantum communication [14–16], quantum computing [17, 18] and quantum metrology [19–21]. Its profound theoretical significance and practical applications have cemented quantum entanglement as a pivotal focus of modern research.

Despite its importance, detecting and quantifying entanglement remain challenging due to the complexity of quantum systems [22]. Entanglement detection typically relies on criteria such as the positive partial transpose (PPT) test [23], which is computationally efficient but only holds as a sufficient and necessary condition for low-dimensional systems. Similarly, entanglement witnesses [24], while providing a practical and experimental approach, can only detect some entangled states and are thus limited in scope. Quantifying entanglement is even more computationally demanding, as it often requires solving non-linear optimization problems over high-dimensional state spaces. These bottlenecks have motivated the search for more efficient theoretical and computational tools to address the complexities of entanglement characterization.

Semidefinite programming (SDP), a specialized branch of convex optimization [25], has proven to be an indispensable tool in addressing entanglement-related challenges [26]. SDP optimizes linear functions over the intersection of the cone of positive semidefinite matrices and affine subspaces, making it particularly well-suited for applications in quantum information theory. For instance, SDP has been utilized to implement the symmetric extension test for entanglement detection [27, 28], design entanglement witnesses [29], and establish bounds on various entanglement measures [30, 31]. Its key advantages include polynomial-time solvability and scalability for systems of moderate dimensions. However, SDP encounters significant challenges when applied to high-dimensional systems due to the exponential growth of the optimization space, which renders computations increasingly intractable. Moreover, while SDP often provides tight bounds for certain problems, it may fail to deliver exact solutions in more complex scenarios. These limitations underscore the necessity for advancing algorithmic techniques and exploring hybrid approaches that combine SDP with complementary methods.

Non-convex optimization has emerged as a dynamic and rapidly advancing field within applied mathematics and computational science, with diverse applications spanning machine learning [32], signal processing [33], and quantum information science [34]. In contrast to convex optimization, which benefits from the guarantee of global optimality due to the convexity of the objective function or feasible region, non-convex problems are inherently more complex, often exhibiting multiple local minima and saddle points. Notable examples of non-convex optimization challenges include determining the ground state energy of large-scale quantum systems [35], training deep neural networks [36], and addressing combinatorial optimization problems [37]. Recent progress in this domain has been driven by the development of advanced techniques such as gradient-based methods [38, 39], heuristic approaches (*e.g.*, simulated annealing [40] and genetic algorithms [41]), and machine learning-inspired strategies (*e.g.*, reinforcement learning [42] and meta-optimization [43]). These innovations, coupled with significant computational advancements, have rendered non-convex optimization increasingly feasible and effective, despite its intrinsic challenges.

## 1.2 Contributions and Outline

In this thesis, we propose a unified computational framework for evaluating quantum information problems through gradient-based optimization methods. Motivated by recent advances in manifold optimization [44, 45], we begin by formulating the quantum information problems as constrained optimization problems on suitable smooth manifolds. To facilitate efficient numerical implementation, we introduce a manifold trivialization approach [46] that systematically transforms these constrained problems into unconstrained ones. This reformulation allows for the direct application of mature gradient descent techniques, significantly improving computational tractability and scalability.

A central focus of our study is the geometric measure of entanglement (GME) [47–49], which we generalize and extend to a broad class of quantum scenarios. Specifically, we construct novel GME-based entanglement monotones applicable not only to pure states but also to quantum subspaces and mixed states. Our proposed methods are valid in both bipartite and multipartite settings, offering a versatile tool for entanglement characterization across a wide range of quantum systems.

Importantly, the family of GME-based monotones developed in this thesis enables the quantification of entanglement with respect to different entanglement dimensionalities, which reflect the dimensions needed for producing the quantum correlations among the different parties. This feature is crucial for detecting and certifying high-dimensional entanglement [50–52], which is of significant interest in quantum information processing. In contrast to conventional entanglement measures, our approach provides refined insights into the structure and strength of entanglement beyond the standard dichotomy of separability versus entanglement.

To facilitate the computation of the proposed quantities, we employ the robust non-convex optimization framework described above to obtain upper bounds on the entanglement quantities of interest. Compared to convex optimization methods, particularly the semidefinite programming (SDP) approach, this framework generally requires fewer parameters, thereby significantly reducing computational time while maintaining satisfactory accuracy in practice. In addition, we demonstrate that SDP techniques can be utilized to compute rigorous lower bounds. Together, these complementary approaches establish a consistent and efficient computational paradigm for analyzing entanglement in a wide variety of quantum systems.

The structure of this thesis is as follows. Chapter 2 provides a comprehensive review of foundational concepts in quantum information theory, along with an introduction to tensor network representations. These preliminaries establish the mathematical framework necessary for formulating the quantum information problems addressed in subsequent chapters. Chapter 3 delves into the concept of quantum entanglement, covering both bipartite and multipartite systems, and examines the interconversion processes between various pure entangled states. Chapter 4 surveys prominent entanglement detection criteria and explores a range of entanglement measures used for quantification. In Chapter 5, we present an overview of semidefinite programming (SDP), a versatile tool in quantum information science, and introduce a novel optimization framework leveraging gradient descent techniques. Chapter 6 focuses on the development of a family of entanglement monotones based on the geometric measure of entanglement (GME). These monotones are applied to diverse quantum scenarios, including pure states, subspaces, and mixed states, demonstrating the efficacy of the proposed non-convex optimization framework for accurate computation. Additionally, the use of SDP to derive rigorous lower bounds is illustrated.

# CHAPTER 2

# PRELIMINARIES

## 2.1 Fundamentals of Quantum Information

Quantum information is a multidisciplinary field that integrates concepts from quantum mechanics, computer science, information theory, etc. This section introduces the foundational principles of quantum information and the associated mathematical framework of linear algebra.

### 2.1.1 Quantum states

A quantum state encapsulates the physical description of a quantum mechanical system and resides in a Hilbert space, denoted by $\mathcal{H}$. Pure states, representing fully determined quantum configurations, are described by state vectors, while mixed states, which encode partial information about a system, are represented by density operators. The introduction of Dirac's bra-ket notation [53] has significantly streamlined the mathematical formalism of quantum mechanics.

#### Pure and mixed states

A *pure state* is represented by a normalized vector in $\mathcal{H}$, expressed in Dirac notation as a ket vector $|\psi\rangle$. This vector contains all the information about the system, enabling the prediction of any measurable property. In a $d$-dimensional Hilbert space with the computational basis $\{|i\rangle\}_{i=0}^{d-1}$, a pure state is written as:

$$|\psi\rangle = \sum_{i=0}^{d-1} c_i |i\rangle, \tag{2.1}$$

where $c_i$ are complex coefficients satisfying the normalization condition $\sum_i |c_i|^2 = 1$. For instance, the state of a *qubit*, which is a two-level quantum system, is expressed as:

$$|\psi\rangle = c_0|0\rangle + c_1|1\rangle = c_0 \begin{pmatrix} 1 \\ 0 \end{pmatrix} + c_1 \begin{pmatrix} 0 \\ 1 \end{pmatrix} = \begin{pmatrix} c_0 \\ c_1 \end{pmatrix}. \tag{2.2}$$

Here, $|0\rangle$ and $|1\rangle$ are represented as unit column vectors. The corresponding bra vector $\langle\psi|$ is obtained by taking the Hermitian conjugate of the ket vector:

$$\langle\psi| = c_0^*\langle 0| + c_1^*\langle 1| = \begin{pmatrix} c_0^* & c_1^* \end{pmatrix}. \tag{2.3}$$

The *inner product* of two vectors $|\psi\rangle = \sum_{i=0}^{d-1} a_i|i\rangle$ and $|\phi\rangle = \sum_{i=0}^{d-1} b_i|i\rangle$ is defined as:

$$\langle\psi|\phi\rangle = \begin{pmatrix} a_0^* & \cdots & a_{d-1}^* \end{pmatrix} \begin{pmatrix} b_0 \\ \vdots \\ b_{d-1} \end{pmatrix} = \sum_i a_i^* b_i. \tag{2.4}$$

From the inner product, the *norm* of a vector $|\psi\rangle$ is given by $\||\psi\|| = \sqrt{\langle\psi|\psi\rangle}$, and the normalization condition for a quantum state is $\||\psi\|| = 1$. An orthonormal basis $\{|e_i\rangle\}$ satisfies the orthogonality condition:

$$\langle e_i|e_j\rangle = \delta_{ij} = \begin{cases} 1 & i = j, \\ 0 & i \neq j. \end{cases} \tag{2.5}$$

In scenarios where the system's state is not fully determined, it is described as a *mixed state*. A mixed state arises when the system can occupy one of several pure states $|\psi_i\rangle$ with probabilities $p_i$, where the set of states is denoted by $\{|\psi_i\rangle\}$. Such a state is represented by a density operator $\rho$, which is a positive semi-definite Hermitian matrix with unit trace, expressed as:

$$\rho = \sum_i p_i |\psi_i\rangle\langle\psi_i|, \tag{2.6}$$

where the probabilities $p_i$ satisfy the condition $p_i \geq 0, \sum_i p_i = 1$.

For example, the density matrix corresponding to the qubit state described in Eq. (2.2) is:

$$\begin{aligned} \rho =& |\psi\rangle\langle\psi| \\ =& c_0c_0^*|0\rangle\langle 0| + c_0c_1^*|0\rangle\langle 1| + c_1c_0^*|1\rangle\langle 0| + c_1c_1^*|1\rangle\langle 1| \\ =& c_0c_0^*\begin{pmatrix} 1 \\ 0 \end{pmatrix}\begin{pmatrix} 1 & 0 \end{pmatrix} + c_0c_1^*\begin{pmatrix} 1 \\ 0 \end{pmatrix}\begin{pmatrix} 0 & 1 \end{pmatrix} + c_1c_0^*\begin{pmatrix} 0 \\ 1 \end{pmatrix}\begin{pmatrix} 1 & 0 \end{pmatrix} + c_1c_1^*\begin{pmatrix} 0 \\ 1 \end{pmatrix}\begin{pmatrix} 0 & 1 \end{pmatrix} \\ =& \begin{pmatrix} c_0c_0^* & c_0c_1^* \\ c_1c_0^* & c_1c_1^* \end{pmatrix} \end{aligned} \tag{2.7}$$

### 2.1.2 Quantum operations

Once a quantum state is defined, various quantum operations can be applied to it. These operations include quantum-to-quantum transformations (*quantum channels*) as well as quantum-to-classical transformations (*quantum measurements*).

#### Quantum channels

The evolution of pure quantum states is governed by *unitary operators* $U$, which satisfy the condition $UU^\dagger = I$. These operators map a pure state $|\psi\rangle$ to another pure state $|\psi'\rangle = U|\psi\rangle$.

In the case of qubit systems, the Pauli matrices constitute a fundamental set of $2\times2$ complex Hermitian and unitary matrices, widely utilized in quantum information theory. These matrices are defined as:

$$\sigma_x = \begin{pmatrix} 0 & 1 \\ 1 & 0 \end{pmatrix}, \quad \sigma_y = \begin{pmatrix} 0 & -i \\ i & 0 \end{pmatrix}, \quad \sigma_z = \begin{pmatrix} 1 & 0 \\ 0 & -1 \end{pmatrix}. \tag{2.8}$$

Any single-qubit unitary transformation $U$ can be expressed in terms of the Pauli matrices $\{\sigma_x, \sigma_y, \sigma_z\}$ and the identity matrix $I$ as:

$$U = e^{i\alpha}\left(\cos\frac{\theta}{2} I - i\sin\frac{\theta}{2}(\hat{n}\cdot\vec{\sigma})\right), \tag{2.9}$$

where $\alpha$ is a global phase, $\theta$ is the rotation angle, $\hat{n} = (n_x, n_y, n_z)$ is a unit vector specifying the rotation axis, and $\vec{\sigma} = (\sigma_x, \sigma_y, \sigma_z)$ is the vector of Pauli matrices.

For mixed quantum states, the framework of *quantum channels* is employed to describe their transformations. Quantum channels are mathematically formalized as linear, completely positive, and trace-preserving (CPTP) maps. A quantum channel $\mathcal{E}$ can be expressed in terms of *Kraus operators* as:

$$\mathcal{E}(\rho) = \sum_i K_i \rho K_i^\dagger, \tag{2.10}$$

where the Kraus operators $K_i$ satisfy the completeness relation $\sum_i K_i^\dagger K_i = I$.

The Kraus representation offers a versatile framework for modeling various noise processes and interactions in quantum systems. Below, we provide examples of commonly encountered qubit channels:

**Example 1 (Amplitude Damping Channel)** This channel models energy dissipation due to spontaneous emission in a two-level quantum system (*i.e.*, a qubit). The Kraus operators for this channel, parameterized by the damping rate $\gamma$, are expressed as:

$$K_1 = \begin{pmatrix} 1 & 0 \\ 0 & \sqrt{1-\gamma} \end{pmatrix}, \quad K_2 = \begin{pmatrix} 0 & \sqrt{\gamma} \\ 0 & 0 \end{pmatrix}. \tag{2.11}$$

**Example 2 (Dephasing Channel)** This channel captures the loss of phase coherence in a quantum state without energy dissipation. It introduces random phase errors, reducing the coherence between the basis states $|0\rangle$ and $|1\rangle$. The Kraus operators for this channel are given by:

$$K_1 = \sqrt{p} I, \quad K_2 = \sqrt{1-p}\sigma_z, \tag{2.12}$$

where $p$ denotes the dephasing probability.

**Example 3 (Depolarizing Channel)** This channel describes a process where the qubit state is replaced by a completely mixed state with a certain probability. The Kraus operators for the single-qubit depolarizing channel are:

$$K_1 = \sqrt{1 - \frac{3p}{4}} I, \quad K_2 = \sqrt{\frac{p}{4}}\sigma_x, \quad K_3 = \sqrt{\frac{p}{4}}\sigma_y, \quad K_4 = \sqrt{\frac{p}{4}}\sigma_z, \tag{2.13}$$

where $p$ represents the depolarizing probability. Alternatively, the action of the channel can be expressed as $\mathcal{E}(\rho) = (1-p)\rho + p\frac{I}{2}$.

### Quantum measurements

To investigate specific physical properties of a quantum system, one typically employs observables. *Observables* are represented by Hermitian operators, which correspond to measurable quantities, with their eigenvalues denoting the possible outcomes of a quantum measurement.

An observable $O$ can be expressed through its eigenvalue decomposition as:

$$O = \sum_i \lambda_i P_i = \sum_i \lambda_i |\lambda_i\rangle\langle\lambda_i|, \tag{2.14}$$

where $\lambda_i$ are the eigenvalues, and $P_i = |\lambda_i\rangle\langle\lambda_i|$ are the orthogonal projectors onto the eigenstates of $O$.

The measurement of an observable $O$ is associated with a *projective measurement*, defined by the set of projectors $\{P_i\}$. The probability of obtaining the eigenvalue $\lambda_i$ is given by $p_i = \mathrm{Tr}[P_i\rho]$, where $\rho$ is the quantum state of the system. Following the measurement, the quantum state collapses to the post-measurement state:

$$\rho_i = \frac{1}{p_i} P_i \rho P_i. \tag{2.15}$$

The expectation value of the observable $O$ is then computed as:

$$\langle O \rangle = \sum_i p_i \lambda_i = \sum_i \lambda_i \mathrm{Tr}[P_i \rho] = \mathrm{Tr}[(\sum_i \lambda_i P_i)\rho] = \mathrm{Tr}[O\rho]. \tag{2.16}$$

Beyond projective measurements, a more general framework for quantum measurements can be established using the concept of quantum channels. This approach does not impose the requirement of orthogonality among the resulting states. Such measurements, referred to as *generalized quantum measurements*, are typically implemented by performing a projective measurement on an extended system that includes the system of interest. These measurements are characterized by a set of Kraus operators $\{E_i\}_{i=1}^n$, where the index $i \in \{1, \dots, n\}$ corresponds to the possible measurement outcomes. For an initial quantum state $\rho$, the probability of obtaining the outcome $i$ is given by:

$$p_i = \mathrm{Tr}[E_i^\dagger E_i \rho], \tag{2.17}$$

and the corresponding post-measurement state is expressed as:

$$\rho_i = \frac{1}{p_i} E_i \rho E_i^\dagger. \tag{2.18}$$

Here, the operators $M_i = E_i^\dagger E_i$ define a *positive operator-valued measure* (POVM). POVMs extend the concept of projective measurements by relaxing the requirement of orthogonality among the post-measurement states. Notably, projective measurements represent a special case of POVMs, where the Kraus operators $E_i = P_i$ are orthogonal projectors satisfying $P_i P_j = \delta_{ij} P_i$.

### 2.1.3 Composite quantum systems

Sometimes, the focus is not solely on a system as a whole, but also on its *subsystems*. This approach is particularly useful when analyzing correlations among various subsystems.

If we examine $n$ subsystems $A_i$ for $i = 1, \dots, n$, with each subsystem associated with a Hilbert space $\mathcal{H}_{A_i}$ having dimension $d_{A_i}$, then the Hilbert space representing the entire system is the *tensor product* of all the subsystem spaces:

$$\mathcal{H}_{A_1 \dots A_n} = \mathcal{H}_{A_1} \otimes \cdots \otimes \mathcal{H}_{A_n}. \tag{2.19}$$

In this combined Hilbert space, the orthonormal basis is denoted as $\{|i_1\rangle_{A_1} \otimes \cdots \otimes |i_n\rangle_{A_n}\}$, which is often simplified to the notation $|i_1 \ldots i_n\rangle$.

For example, if $\mathcal{H}$ is a two-dimensional Hilbert space with the basis $\{|0\rangle, |1\rangle\}$, then the set $\{|00\rangle, |01\rangle, |10\rangle, |11\rangle\}$ forms the basis of $\mathcal{H} \otimes \mathcal{H}$. These basis vectors are defined as follows:

$$|00\rangle = |0\rangle \otimes |0\rangle = \begin{pmatrix}1\\0\end{pmatrix} \otimes \begin{pmatrix}1\\0\end{pmatrix} = \begin{pmatrix}1\\0\\0\\0\end{pmatrix}, \quad |01\rangle = |0\rangle \otimes |1\rangle = \begin{pmatrix}1\\0\end{pmatrix} \otimes \begin{pmatrix}0\\1\end{pmatrix} = \begin{pmatrix}0\\1\\0\\0\end{pmatrix},$$
$$|10\rangle = |1\rangle \otimes |0\rangle = \begin{pmatrix}0\\1\end{pmatrix} \otimes \begin{pmatrix}1\\0\end{pmatrix} = \begin{pmatrix}0\\0\\1\\0\end{pmatrix}, \quad |11\rangle = |1\rangle \otimes |1\rangle = \begin{pmatrix}0\\1\end{pmatrix} \otimes \begin{pmatrix}0\\1\end{pmatrix} = \begin{pmatrix}0\\0\\0\\1\end{pmatrix}. \tag{2.20}$$

We can also define the tensor product of operators on the subsystems. For example, if we are given two density matrices $\rho_A$ and $\rho_B$ from the Hilbert space $\mathcal{H}_A$ and $\mathcal{H}_B$, the density matrix of two subsystems is just the tensor product of $\rho_A$ and $\rho_B$, as follows

$$\begin{aligned}\rho_{AB} = \rho_A \otimes \rho_B &= \begin{pmatrix}a_1 & a_2\\ a_3 & a_4\end{pmatrix} \otimes \begin{pmatrix}b_1 & b_2\\ b_3 & b_4\end{pmatrix}\\ &= \begin{pmatrix}a_1b_1 & a_1b_2 & a_2b_1 & a_2b_2\\ a_1b_3 & a_1b_4 & a_2b_3 & a_2b_4\\ a_3b_1 & a_3b_2 & a_4b_1 & a_4b_2\\ a_3b_3 & a_3b_4 & a_4b_3 & a_4b_4\end{pmatrix}\end{aligned} \tag{2.21}$$

In a composite quantum system, the *partial trace* over a subsystem $A_i$ of an operator $Q$ is mathematically defined as:

$$\mathrm{Tr}_{A_i}[Q] = \sum_{j=0}^{d_{A_i}-1} {}_{A_i}\langle j|Q|j\rangle_{A_i}. \tag{2.22}$$

The outcome of the partial trace is an operator acting on the Hilbert space of the remaining subsystems. This operation is particularly useful when the focus is on specific subsystems rather than the entire composite system.

For instance, consider a bipartite quantum state $\rho_{AB}$ comprising subsystems $A$ and $B$. The reduced density matrix (RDM) of subsystem $A$ is obtained by tracing out subsystem $B$:

$$\rho_A = \mathrm{Tr}_B[\rho_{AB}]. \tag{2.23}$$

The partial trace operation satisfies the following property:

$$\mathrm{Tr}_B[|a_1\rangle\langle a_2| \otimes |b_1\rangle\langle b_2|] = |a_1\rangle\langle a_2| \, \mathrm{Tr}_B[|b_1\rangle\langle b_2|],$$

where $|a_1\rangle, |a_2\rangle \in \mathcal{H}_A$ and $|b_1\rangle, |b_2\rangle \in \mathcal{H}_B$ are arbitrary vectors in the respective Hilbert spaces.

Furthermore, for a local observable $O = O_A \otimes I_B$, its expectation value can be expressed as:

$$\langle O\rangle = \mathrm{Tr}[(O_A \otimes I_B)\rho_{AB}] = \mathrm{Tr}_A[O_A(\mathrm{Tr}_B[\rho_{AB}])] = \mathrm{Tr}_A[O_A\rho_A]. \tag{2.24}$$

This result highlights that reduced density matrices (RDMs) encapsulate all the information necessary to describe the local properties of the composite quantum system.

## 2.2 Tensor Network Representation

In this section, we revisit the fundamental principles of *tensor networks* (TNs), which serve as a graphical formalism for representing tensors and their associated operations. This framework provides a versatile and intuitive approach for describing quantum systems and their underlying structures.

### 2.2.1 Graphical representation of tensors

A *tensor* is a multi-dimensional array of complex numbers, indexed with respect to a fixed standard basis. An $n$th-order tensor, characterized by its shape $(d_1, d_2, \dots, d_n)$, possesses $n$ indices, each corresponding to dimensions $d_1, d_2, \dots, d_n$.

For instance, consider the Hilbert space $\mathcal{H}$ spanned by the computational basis $\{|i\rangle\}_{i=0}^{d-1}$. A ket vector $|v\rangle \in \mathcal{H}$ is a first-order tensor, expressed in terms of its components $v_i$ as:

$$|v\rangle = \sum_i v_i |i\rangle. \tag{2.25}$$

Similarly, linear operators $U$ acting on $\mathcal{H}$, such as unitary matrices, are represented as second-

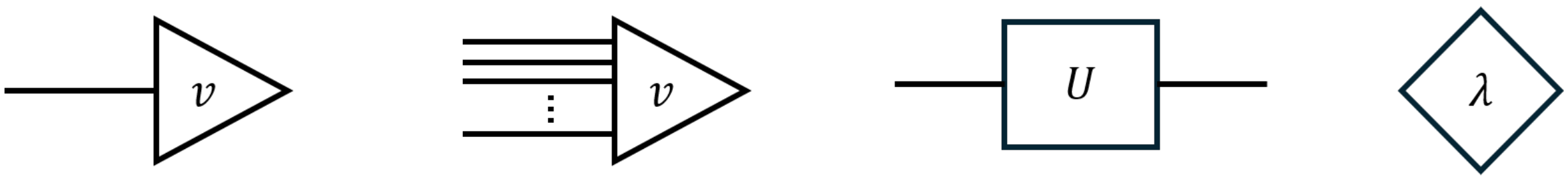

**(a)** Vector $|v\rangle \in \mathcal{H}$ **(b)** Vector $|v\rangle \in \bigotimes_{i=1}^{n} \mathcal{H}_i$ **(c)** Linear Operator $U$ **(d)** Scalar $\lambda \in \mathbb{C}$

Figure 2.1: Graphical representation of basic tensors: (a, b) vectors, (c) a linear operator, and (d) a scalar. A tensor with wires oriented in the same direction is referred to as a vector, whereas a tensor with wires oriented in two distinct directions is referred to as an operator.

order tensors with components $U_{ij}$, given by:

$$U = \sum_{i,j=0}^{d-1} U_{ij} |i\rangle\langle j|. \tag{2.26}$$

Throughout this thesis, we employ Penrose's graphical notation [54] to represent tensors. In this notation, tensors are depicted as geometric shapes: vectors are represented by triangles, linear operators by rectangles, and scalars by diamonds, as illustrated in Fig. 2.1. Each index of a tensor corresponds to an open wire in the diagram, with the number of wires indicating the tensor's order. Higher-order tensors are represented by increasing the number of wires connected to the geometric shape.

### 2.2.2 Tensor operations

Here, we summarize several fundamental mathematical operations expressed through tensor operations. The symbols around the wires in the diagrams indicate the corresponding dimensions of the indices:

**Example 4 (Tensor Reshaping)** This operation allows the manipulations like the combination of multiple indices into a single index (*merging*) or the decomposition of a single index into multiple indices (*splitting*), as illustrated below:

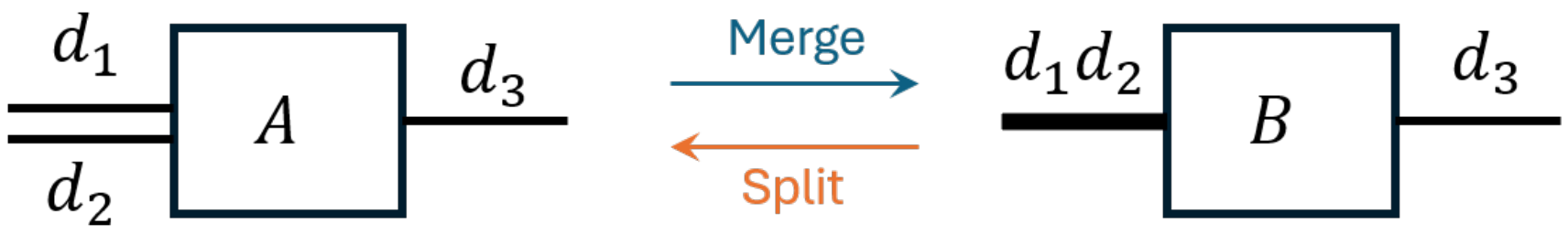

In this example, $A$ is a third-order tensor with shape $(d_1, d_2, d_3)$, and $B$ is a second-order tensor with shape $(d_1 d_2, d_3)$. Mathematically, this relationship is expressed as $A_{ijk} \leftrightarrow B_{lk} = A_{(ij)k}$.

More generally, tensor reshaping involves changing the shape of a tensor without altering its underlying data. Specifically, a $n$th-order tensor with the shape $(d_1, d_2, \ldots, d_n)$ can be reformatted as a $m$th-order tensor with the shape $(d'_1, d'_2, \ldots, d'_m)$ as long as

$$d'_1 d'_2 \cdots d'_m = d_1 d_2 \ldots d_n. \tag{2.27}$$

**Example 5 (Tensor Product)** When two or more tensors appear disconnected in the same diagram, their relationship represents a tensor product operation. This is depicted as:

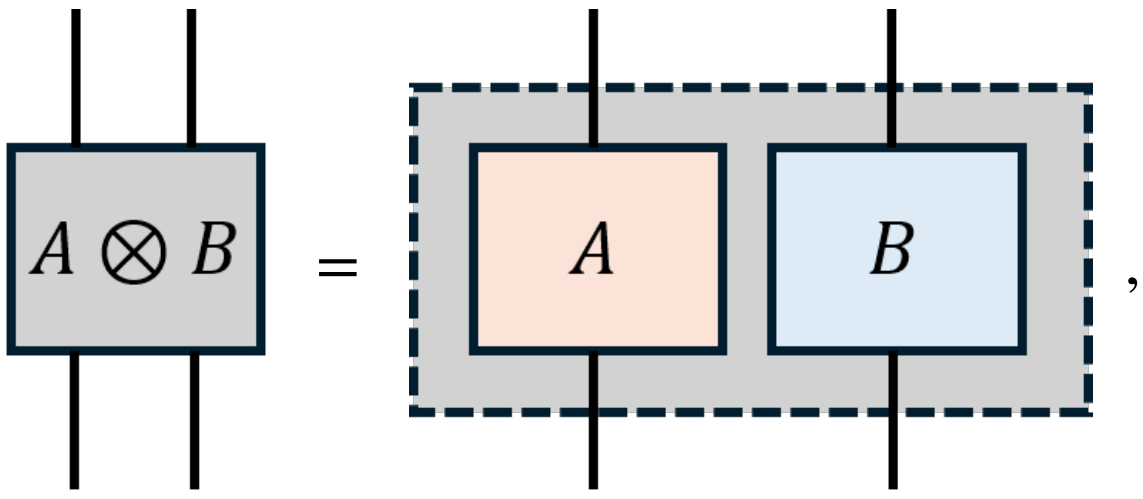

(2.28)

where the tensor product is defined as $(A \otimes B)_{ijkl} = A_{ij} B_{kl}$.

**Example 6 (Tensor Concatenation and Stacking)** Tensor concatenation combines tensors along an *existing* dimension, provided their shapes are compatible, meaning all dimensions except the concatenation axis must match. In contrast, tensor stacking aggregates tensors of identical shape along a *new* dimension, thereby increasing the order of the resulting tensor. These operations are illustrated below:

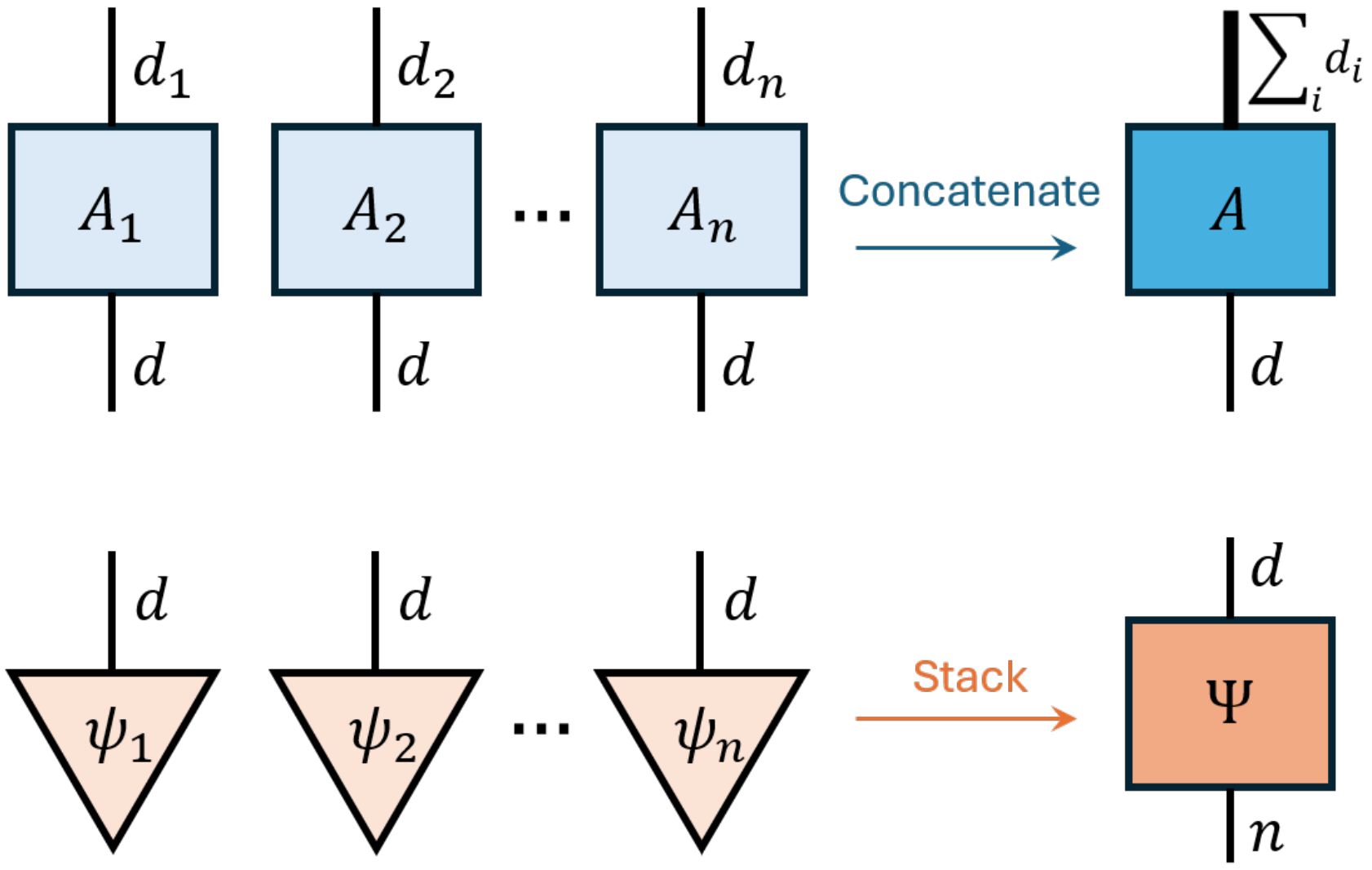

**Example 7 (Tensor Contraction)** Tensor contraction is a fundamental operation in tensor networks, involving the summation over shared indices of two or more tensors. This operation generalizes the concept of matrix multiplication to higher-dimensional tensors. Below, we illus-trate several common types of tensor contractions:

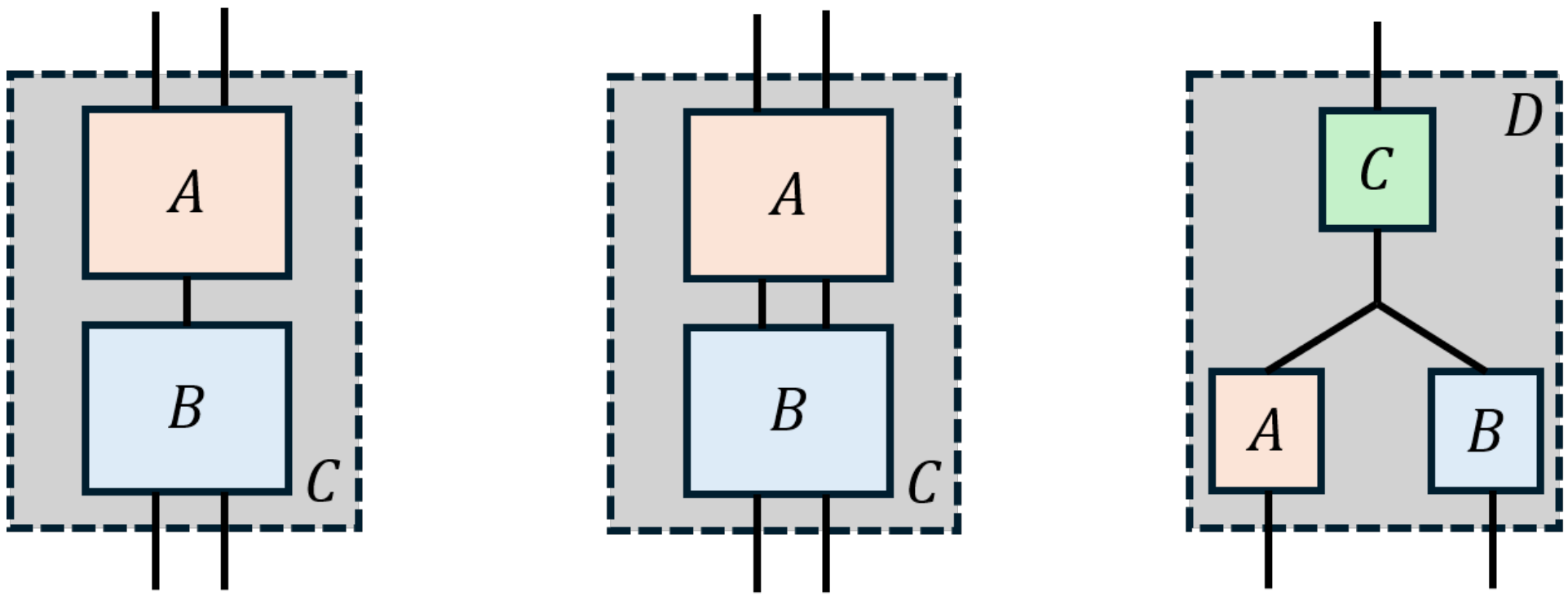

**(a)** **(b)** **(c)**

In these examples:

(a) *Single-index contraction*: Mathematically expressed as $C_{ijkl} = \sum_m A_{ijm} B_{klm}$.

(b) *Multi-index contraction*: Given by $C_{ijkl} = \sum_{m,n} A_{ijmn} B_{klmn}$.

(c) *Multi-tensor contraction*: Described as $D_{ijk} = \sum_m A_{im} B_{jm} C_{km}$.

**Example 8 (Trace and Partial Trace)** In tensor network diagrams, the trace operation is represented by connecting wires into closed loops, while the partial trace is depicted by connecting only a subset of the wires. These operations can be visualized as follows:

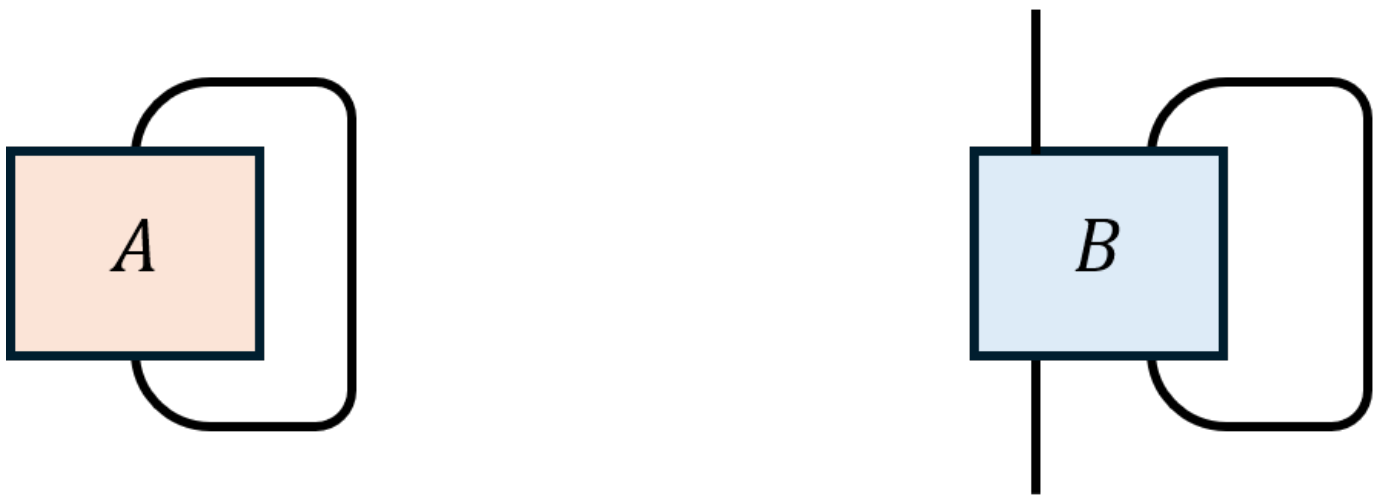

**(a)** Trace **(b)** Partial trace

In the examples above:

(a) For a matrix $A$ with shape $(d, d)$, the trace of $A$ is given by $\text{Tr}[A] = \sum_i A_{ii}$.

(b) For a tensor $B$ with shape $(d_1, d_2, d_1, d_2)$, the partial trace of $B$ over the $(d_2, d_2)$ dimensions is expressed as $(\text{Tr}_2[B])_{ij} = \sum_k B_{ikjk}$.

At the end, we summarize the common operations for scalars, vectors, and matrices using the tensor graphical representation, as illustrated in Table 2.1.

| Operation Name | Input Tensor | Formula | Output Tensor |
|---|---|---|---|
| **Scalar multiplication** | 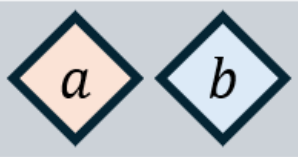  | $y = a \times b$ | 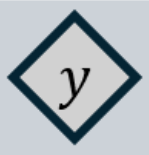  |
| **Inner product** | 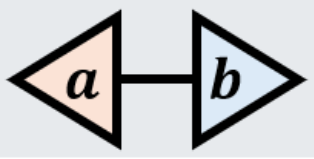  | $y = \sum_i a_i \times b_i$ | 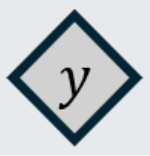  |
| **Outer product** | 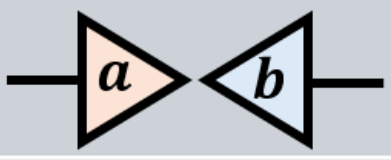  | $Y_{ij} = a_i \times b_j$ | 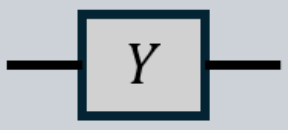  |
| **Matrix-vector product** | 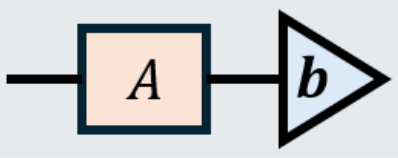  | $y_i = \sum_j A_{ij} \times b_j$ | 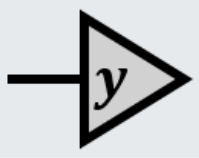  |
| **Matrix multiplication** | 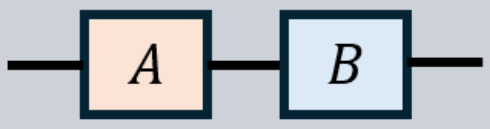  | $Y_{ij} = \sum_k A_{ik} \times B_{kj}$ | 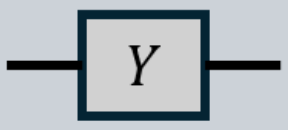  |
| **Matrix trace** | 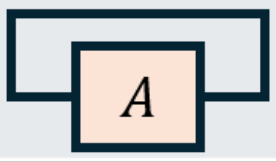  | $y = \sum_i A_{ii}$ | 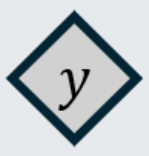  |

Table 2.1: Summary of common operations for scalars, vectors, and matrices in the tensor graphical representation.

# CHAPTER 3

# WHAT IS QUANTUM ENTANGLEMENT?

## 3.1 Bipartite Entanglement

In this section, we introduce the foundational concepts underlying bipartite entangled states. The study of bipartite entanglement serves as a cornerstone for understanding the broader framework of quantum entanglement.

### 3.1.1 Pure states

Consider a scenario involving two quantum systems, one associated with Alice and the other with Bob. The states of Alice's system are described by a Hilbert space $\mathcal{H}_A$ of dimension $d_A$, while Bob's system is characterized by a Hilbert space $\mathcal{H}_B$ of dimension $d_B$.

The composite system is represented by state vectors in the tensor product space $\mathcal{H} = \mathcal{H}_A \otimes \mathcal{H}_B$. A general pure state of this bipartite system can be written as:

$$|\psi\rangle = \sum_{i=1}^{d_A} \sum_{j=1}^{d_B} \psi_{ij} |i\rangle_A \otimes |j\rangle_B, \tag{3.1}$$

where $\{|i\rangle_A\}$ and $\{|j\rangle_B\}$ are orthonormal bases for $\mathcal{H}_A$ and $\mathcal{H}_B$, respectively, and $\psi_{ij}$ are complex coefficients satisfying the normalization condition $\sum_{i,j} |\psi_{ij}|^2 = 1$.

**Definition 1 (Entanglement for Bipartite Pure States)** A pure state $|\psi\rangle \in \mathcal{H}$ is termed *product* or *separable* if there exist states $|\phi_A\rangle \in \mathcal{H}_A$ and $|\phi_B\rangle \in \mathcal{H}_B$ such that

$$|\psi\rangle = |\phi_A\rangle \otimes |\phi_B\rangle. \tag{3.2}$$

Otherwise, $|\psi\rangle$ is classified as *entangled.*

From a physical standpoint, product states are characterized by the absence of quantum correlations between the subsystems. Specifically, if Alice measures an observable $A$ and Bob measures an observable $B$, the joint probabilities of their measurement outcomes can be expressed as the product of their individual probabilities. Consequently, the measurement outcomes on Alice's side are statistically independent of those on Bob's side.

Before delving into the concept of entangled mixed states, it is crucial to introduce a fundamental mathematical tool known as the *Schmidt decomposition*.

**Theorem 3.1.1 (Schmidt Decomposition [55])** Any bipartite pure state of the form given in Eq. (3.1) can be expressed in terms of orthonormal bases $\{|\alpha_i\rangle\}$ of $\mathcal{H}_A$ and $\{|\beta_i\rangle\}$ of $\mathcal{H}_B$ as follows:

$$|\psi\rangle = \sum_{k=1}^{r} \mu_k |\alpha_k\rangle |\beta_k\rangle, \tag{3.3}$$

where $\mu_k$ are positive real coefficients, uniquely determined by the coefficients $\psi_{ij}$ in Eq. (3.1). The integer $r \leq \min\{d_A, d_B\}$ is referred to as the *Schmidt rank* of $|\psi\rangle$, denoted by $\mathrm{SR}(|\psi\rangle)$.

This decomposition can be conveniently understood using a tensor network, as follows:

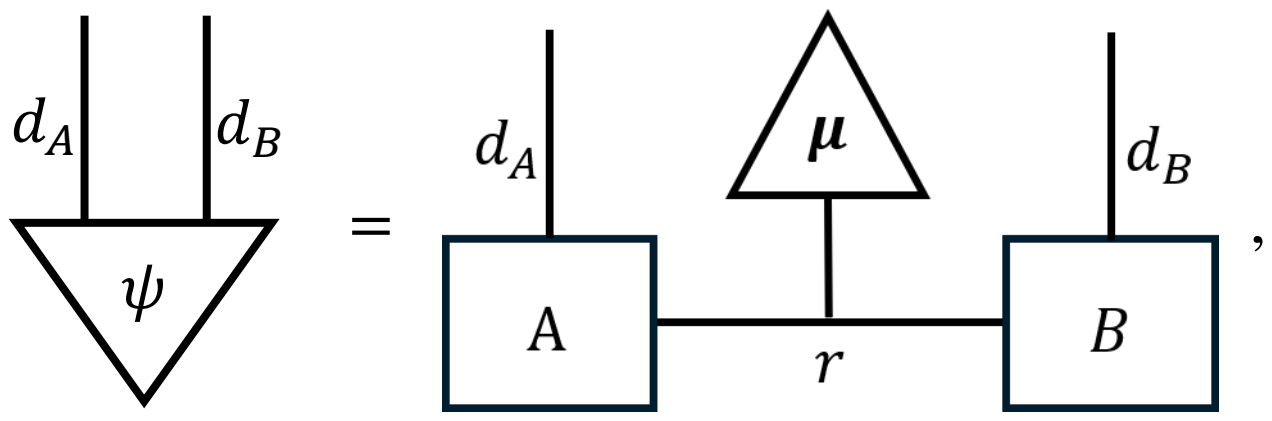

(3.4)

where the symbols $d_A$, $d_B$, and $r$ denote the dimensions of the respective indices. In this representation, vector $\boldsymbol{\mu} = (\mu_1, \mu_2, \ldots, \mu_r) \in \mathbb{R}_+^r$ satisfies $\sum_i \mu_i^2 = 1$, matrix $A$ corresponds to the set $\{|\alpha_i\rangle\}_{i=1}^r$ and matrix $B$ corresponds to $\{|\beta_i\rangle\}_{i=1}^r$, *i.e.*,

$$A = \begin{pmatrix} | & | & & | \\ |\alpha_1\rangle & |\alpha_2\rangle & \cdots & |\alpha_r\rangle \\ | & | & & | \end{pmatrix}, \quad B = \begin{pmatrix} | & | & & | \\ |\beta_1\rangle & |\beta_2\rangle & \cdots & |\beta_r\rangle \\ | & | & & | \end{pmatrix}. \tag{3.5}$$

Equivalently, *singular value decomposition* (SVD) of $\psi_{ij}$ provides the Schmidt decomposition of $|\psi\rangle$. From the definition, a separable pure state implies that the Schmidt rank is one, *i.e.*, $\boldsymbol{\mu}$ reduces to a scalar.

### 3.1.2 Mixed states

In situations where the exact state of a quantum system is not fully determined, the system is described by a mixed state $\rho$. The notions of separability and entanglement for mixed states, as introduced in Ref. [56], generalize the concepts established for pure states. The underlying principle remains consistent: a state is classified as separable if it lacks quantum correlations between its subsystems, implying that measurement outcomes on one subsystem are statistically independent of those on the other, up to some classical correlations. Conversely, the presence of such correlations indicates that the state is entangled.

**Definition 2 (Entanglement for Bipartite Mixed States)** Let $\rho$ represent the density matrix of a bipartite quantum system $AB$. The state $\rho$ is termed a *product state* if there exist density matrices $\rho_A$ and $\rho_B$ corresponding to subsystems $A$ and $B$, respectively, such that

$$\rho = \rho_A \otimes \rho_B. \tag{3.6}$$

Furthermore, $\rho$ is defined as *separable* if it can be expressed as a *convex combination* of product states. Formally, there exist a set of classical probabilities $p_i$ and product states $\rho_A^{(i)} \otimes \rho_B^{(i)}$ such that

$$\rho = \sum_i p_i \rho_A^{(i)} \otimes \rho_B^{(i)}, \tag{3.7}$$

where $p_i \geq 0$ and $\sum_i p_i = 1$. If $\rho$ cannot be expressed in this form, it is classified as entangled.

On the other hand, states of the form given in Eq. (3.7) are considered separable because they can be prepared locally by the two parties, Alice and Bob, using only classical resources. Specifically, Alice and Bob can independently generate any separable state while sharing access to a classical random number generator that outputs the index $i$ with probability $p_i$. Upon receiving the outcome $i$, they can agree to prepare the corresponding state $\rho_A^{(i)} \otimes \rho_B^{(i)}$. By following this protocol, they can construct the state $\rho = \sum_i p_i \rho_A^{(i)} \otimes \rho_B^{(i)}$ without requiring any quantum communication or non-local operations. In this framework, the correlations between the two parties are purely classical in nature. This paradigm, where local quantum operations are supplemented by classical communication, is referred to as *local operations and classical communication* (LOCC), a concept that will be explored in greater detail in subsequent sections.

Any density matrix can be expressed as a weighted sum of pure states. Consequently, a separable state can be written as a convex combination of pure product states, given by

$$\rho = \sum_i p_i |\psi_A^{(i)}\rangle\langle\psi_A^{(i)}| \otimes |\psi_B^{(i)}\rangle\langle\psi_B^{(i)}|,$$

where $|\psi_A^{(i)}\rangle \in \mathcal{H}_A$ and $|\psi_B^{(i)}\rangle \in \mathcal{H}_B$. Importantly, the sets $\{|\psi_A^{(i)}\rangle\}$ and $\{|\psi_B^{(i)}\rangle\}$ are not required to form orthogonal bases.

From a mathematical perspective, it is noteworthy that if $\rho_1$ and $\rho_2$ are separable states, then any convex combination $p_1\rho_1 + p_2\rho_2$, where $p_1, p_2 \geq 0$ and $p_1 + p_2 = 1$, also remains separable. This property establishes that the set of separable states constitutes a *convex set*. Furthermore, pure product states represent the *extreme points* of this convex set. Consequently, the set of separable states can be described as the *convex hull* of pure product states. The number of pure product states required to represent any separable state is constrained by Carathéodory's theorem, which is formally stated as follows:

**Theorem 3.1.2 (Carathéodory's theorem [57])** Any separable state $\rho \in \mathcal{H}_A \otimes \mathcal{H}_B$ can be expressed as a convex combination of at most $d_A^2 d_B^2$ pure product states, where $d_A$ and $d_B$ denote the dimensions of the Hilbert spaces $\mathcal{H}_A$ and $\mathcal{H}_B$, respectively.

This constraint stems from the convexity property of separable states, which guarantees that any element within a $d$-dimensional convex set can be expressed as a convex combination of no more than $d + 1$ extreme points of the set. In the specific case of two-qubit systems, this bound can be further refined. Notably, any separable state in such systems can be represented as a convex combination of at most four pure product states [58, 59].

The notion of Schmidt rank, originally defined for pure bipartite states, can be extended to bipartite mixed states, as introduced in [60]:

**Definition 3 (Schmidt Number of Bipartite Mixed State)** The Schmidt number of a bipartite density matrix $\rho$, denoted as $\mathrm{SN}(\rho)$, is defined as follows: (i) There does not exist any pure-state decomposition of $\rho$, *i.e.*, $\rho = \sum_i p_i |\psi_i\rangle\langle\psi_i|$ with $\{p_i, |\psi_i\rangle\}$, such that all pure state in the decomposition have Schmidt rank less than $k$. (ii) There exists at least one decomposition of $\rho$ where all pure states $|\psi_i\rangle$ in the decomposition have Schmidt rank no larger than $k$. Formally, $\mathrm{SN}(\rho) = k$ if and only if both conditions are satisfied.

This definition can be expressed mathematically as:

$$\begin{aligned} &\mathrm{SN}(\rho) = \min k, \\ s.t. \quad &\rho = \sum_i p_i |\psi_i\rangle\langle\psi_i|, \sum_i p_i = 1, \\ &\mathrm{SR}(|\psi_i\rangle) \leq k, 0 \leq p_i \leq 1, \quad \forall i. \end{aligned} \tag{3.8}$$

Let $S_k$ represent the set of all density matrices on $\mathcal{H}_A \otimes \mathcal{H}_B$ with Schmidt number at most $k$. This set forms a convex and compact subset within the space of all density matrices, exhibiting a hierarchical structure such that $S_{k-1} \subset S_k$. Notably, the set of separable states corresponds to $S_1$, as pure product states inherently possess a Schmidt rank of one.

Figure 3.1 provides a schematic visualization of the $d \otimes d$ quantum state space, where $d \otimes d$ indicates that the dimensions of $\mathcal{H}_A$ and $\mathcal{H}_B$ are both $d$. The figure highlights several prominent examples of pure states, including a product state as defined in Eq. (3.2), a *Bell state* $\frac{1}{\sqrt{2}}(|00\rangle +$

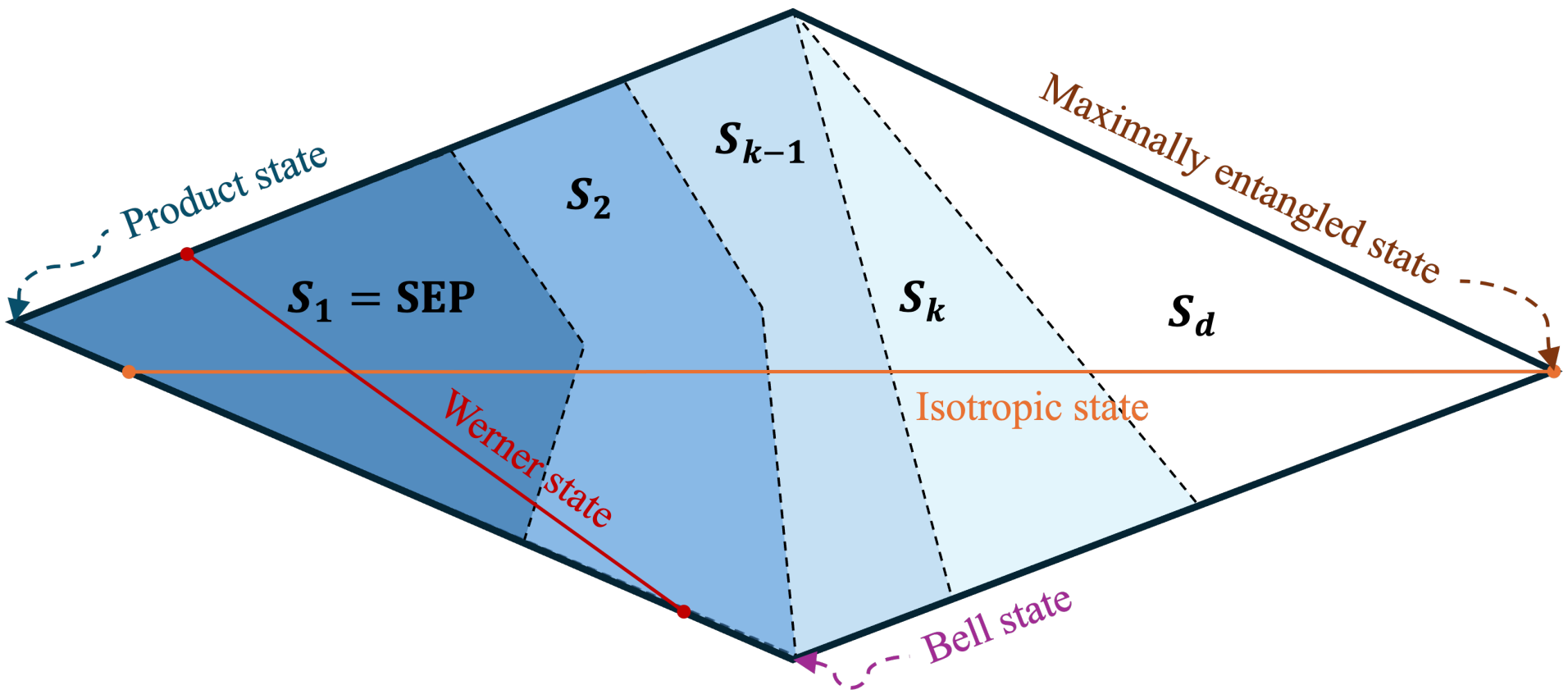

Figure 3.1: Illustration of the $d \otimes d$ ($d > 2$) quantum state space, partitioned according to Schmidt numbers. The set $S_k$ comprises states with Schmidt number SN $\leq k$, forming a convex subset of the quantum state space and satisfying the hierarchy $S_1 \subset S_2 \subset \cdots \subset S_d$. The set of separable states is denoted by SEP. The figure also highlights the placement of notable quantum states within the state space, including product states, Bell states, maximally entangled states, Werner states, and isotropic states.

$|11\rangle$), and a *maximally entangled state*, expressed as:

$$|\Psi_d^+\rangle = \frac{1}{\sqrt{d}} \sum_{i=0}^{d-1} |i\rangle|i\rangle. \tag{3.9}$$

As for mixed states, the figure also includes two lines representing the *isotropic states* and the *Werner states*, both of which are characterized by a single parameter. Isotropic states, parameterized by $F \in [0, 1]$, are defined as follows:

$$\rho_I(F) = \frac{1-F}{d^2-1}(I_{d^2} - |\Psi_d^+\rangle\langle\Psi_d^+|) + F|\Psi_d^+\rangle\langle\Psi_d^+|, \tag{3.10}$$

where $I_{d^2}$ denotes the $d^2 \times d^2$ identity matrix, and $|\Psi_d^+\rangle$ represents the maximally entangled state as defined in Eq. (3.9). These states are a mixture of the maximally entangled state and the maximally mixed state. As the parameter $F$ varies, the Schmidt number of isotropic states transitions from 1 to $d$. Consequently, these states are depicted as a line spanning the entire quantum state space.

Similarly, Werner states, another well-known one-parameter family of $d \otimes d$ states, are parameterized by $\alpha \in [-1, 1]$ and are expressed as:

$$\rho_W(\alpha) = \frac{1}{d^2 - d\alpha} I_{d^2} - \frac{\alpha}{d^2 - d\alpha} \sum_{ij} |i, j\rangle\langle j, i|, \tag{3.11}$$

where $|i, j\rangle$ denotes the computational basis states. It has been shown that the Schmidt number of Werner states is at most 2 [56, 61]. As a result, these states are represented as a line confined within the subset $S_2$ of the quantum state space.

## 3.2 Multipartite Entanglement

This section explores the intricate structure of entanglement in multipartite systems, emphasizing its significantly greater complexity and richness compared to the bipartite case.

### 3.2.1 Pure states

Multipartite entanglement in pure states can be broadly classified into distinct categories:

**Definition 4 (Entanglement for Multipartite Pure States)** A multipartite pure state $|\psi\rangle \in \mathcal{H} = \mathcal{H}_1 \otimes \mathcal{H}_2 \otimes \cdots \otimes \mathcal{H}_n$ is said to be *fully product* or *separable* if it can be expressed as a tensor product of states $|\phi^{(i)}\rangle \in \mathcal{H}_i$, such that

$$|\psi\rangle = |\phi^{(1)}\rangle \otimes |\phi^{(2)}\rangle \otimes \cdots \otimes |\phi^{(n)}\rangle. \tag{3.12}$$

If this condition is not satisfied, the state $|\psi\rangle$ is classified as *completely entangled*.

A state is termed *biproduct* or *biseparable* if there exists a *bipartition* $K|K^c$, where $K$ is a subset of $\{1, 2, \ldots, n\}$ and $K^c$ denotes its complement, such that the state is separable with respect to this bipartition. If no such bipartition exists, the state $|\psi\rangle$ is considered *genuinely entangled*.

In the specific case of three-qubit pure states, two paradigmatic examples of genuinely entangled states are the Greenberger-Horne-Zeilinger (GHZ) state [62, 63] and the W state [64, 65], which are defined as follows:

$$\begin{aligned} |\mathrm{GHZ}\rangle &= \frac{1}{\sqrt{2}}(|000\rangle + |111\rangle), \\ |\mathrm{W}\rangle &= \frac{1}{\sqrt{3}}(|100\rangle + |010\rangle + |001\rangle). \end{aligned} \tag{3.13}$$

From a physical perspective, biseparable states can be prepared using only local interactions within specific subsystems, without requiring global operations among all parties. In contrast, the creation of genuinely entangled states necessitates non-local interactions that involve all subsystems, reflecting their inherently global nature.

A natural question arises: given two three-qubit states, $|\phi\rangle$ and $|\psi\rangle$, is it possible to transform a single copy of $|\phi\rangle$ into $|\psi\rangle$ using local operations and classical communication (LOCC) with a nonzero probability? Such transformations are referred to as *stochastic local operations and classical communication* (SLOCC).

Remarkably, it has been shown that genuinely entangled three-qubit states fall into two inequivalent SLOCC classes, represented by the GHZ and W states, respectively. These two classes are fundamentally distinct under SLOCC transformations, meaning that no SLOCC protocol can convert one into the other [65].

Furthermore, the GHZ and W states exhibit strikingly different properties. For instance, the GHZ state maximally violates several well-known Bell inequalities [66], highlighting its strong nonlocal correlations. In contrast, the W state demonstrates superior robustness against particle loss. Specifically, if one qubit is lost from the GHZ state, the resulting reduced density matrix $\rho_{AB} = \mathrm{Tr}_C[|\mathrm{GHZ}\rangle\langle\mathrm{GHZ}|]$ becomes separable, indicating the complete loss of entanglement. Conversely, for the W state, the reduced density matrix $\rho_{AB} = \mathrm{Tr}_C[|\mathrm{W}\rangle\langle\mathrm{W}|]$ retains its entanglement, underscoring its resilience in the face of particle loss.

Unlike bipartite pure states, where the Schmidt decomposition provides a straightforward method to determine the Schmidt rank, extending this concept to the multipartite case [67, 68] is significantly more challenging. This complexity arises because multipartite states are represented by tensors rather than matrices. In particular, the *tensor rank*, defined as the minimum number of product terms required for decomposition, is notoriously difficult to compute. Its computational complexity is believed to be at least NP-complete [69, 70]. Formally, the tensor rank is defined as follows:

**Definition 5 (Tensor Rank of Multipartite Pure State)** For a multipartite pure state $|\psi\rangle \in \mathcal{H} = \mathcal{H}_1 \otimes \mathcal{H}_2 \otimes \cdots \otimes \mathcal{H}_n$, the *tensor rank* is the smallest integer $r$ such that there exist states $|\phi_i^{(k)}\rangle \in \mathcal{H}_k$ for $1 \leq i \leq r$ and $1 \leq k \leq n$ satisfying

$$|\psi\rangle = \sum_{i=1}^{r} \mu_i |\phi_i^{(1)}\rangle \otimes |\phi_i^{(2)}\rangle \otimes \cdots \otimes |\phi_i^{(n)}\rangle. \tag{3.14}$$

Here, the coefficients $\mu_i$ are positive, and the product terms need not be orthogonal. The tensor rank of a pure state $|\psi\rangle$ is denoted as $\mathrm{TR}(|\psi\rangle)$.

In addition, the concept of *border rank* has been introduced [71–73]. This notion accounts for the fact that certain tensors can be approximated arbitrarily closely using tensors with smaller tensor ranks—an effect that does not occur with matrices.

A well-known example is the W state, as defined in Eq. (3.13), which has a tensor rank of 3

but can be approximated to any desired precision by states with a minimum tensor rank of 2, as shown below:

$$|\mathrm{W}\rangle = \frac{1}{\sqrt{3}}\left(\lim_{t\to\infty}\frac{1}{t}\left(|0\rangle + t|1\rangle\right)^{\otimes 3} - |0\rangle^{\otimes 3}\right). \tag{3.15}$$

For any $t \neq 0$, the tensor on the right-hand side has a rank of 2, and in the limit $t \to 0$, it converges to the W state. In this sense, we say the border rank of W state is 2. For convenience, the border rank of a pure state $|\psi\rangle$ is denoted as $\mathrm{BR}(|\psi\rangle)$.

From a practical standpoint, border rank may be more relevant for characterizing multipartite entanglement, compared to the tensor rank. This is because a multipartite pure state with a border rank $r$ can always be approximated to any desired precision using states with tensor rank $r$, even if its actual tensor rank is higher. Mathematically, it can be interpreted that the border rank is in a sense the *closure* of tensor rank, characterizing the *asymptotic complexity* of multipartite entanglement.

### 3.2.2 Mixed states

The classification of mixed states builds upon the framework established for bipartite entanglement, utilizing convex combinations as described below:

**Definition 6 (Entanglement for Multipartite Mixed States)** Let $\rho$ be a density matrix defined on the composite Hilbert space $\mathcal{H} = \mathcal{H}_1 \otimes \mathcal{H}_2 \otimes \cdots \otimes \mathcal{H}_n$. The state $\rho$ is termed *fully product* if it can be expressed as a tensor product of density matrices $\rho^{(i)} \in \mathcal{H}_i$, such that

$$\rho = \rho^{(1)} \otimes \rho^{(2)} \otimes \cdots \otimes \rho^{(n)}. \tag{3.16}$$

A state $\rho$ is classified as *fully separable* if it can be written as a convex combination of fully product pure states. Specifically, there exist classical probabilities $p_i$ and fully product pure states $|\phi_i^{\mathrm{fs}}\rangle \in \mathcal{H}$ such that

$$\rho = \sum_i p_i |\phi_i^{\mathrm{fs}}\rangle\langle\phi_i^{\mathrm{fs}}|. \tag{3.17}$$

On the other hand, a state $\rho$ that is not fully separable is referred to as *completely entangled.* A state is said to be *biseparable* if it can be expressed as a convex combination of biseparable pure states, *i.e.*,

$$\rho = \sum_i p_i |\phi_i^{\mathrm{bi}}\rangle\langle\phi_i^{\mathrm{bi}}|, \tag{3.18}$$

where the biseparable states $|\phi_i^{\mathrm{bi}}\rangle$ may correspond to different bipartitions of the subsystems. If a state cannot be decomposed in this manner, it is classified as *genuinely entangled.*

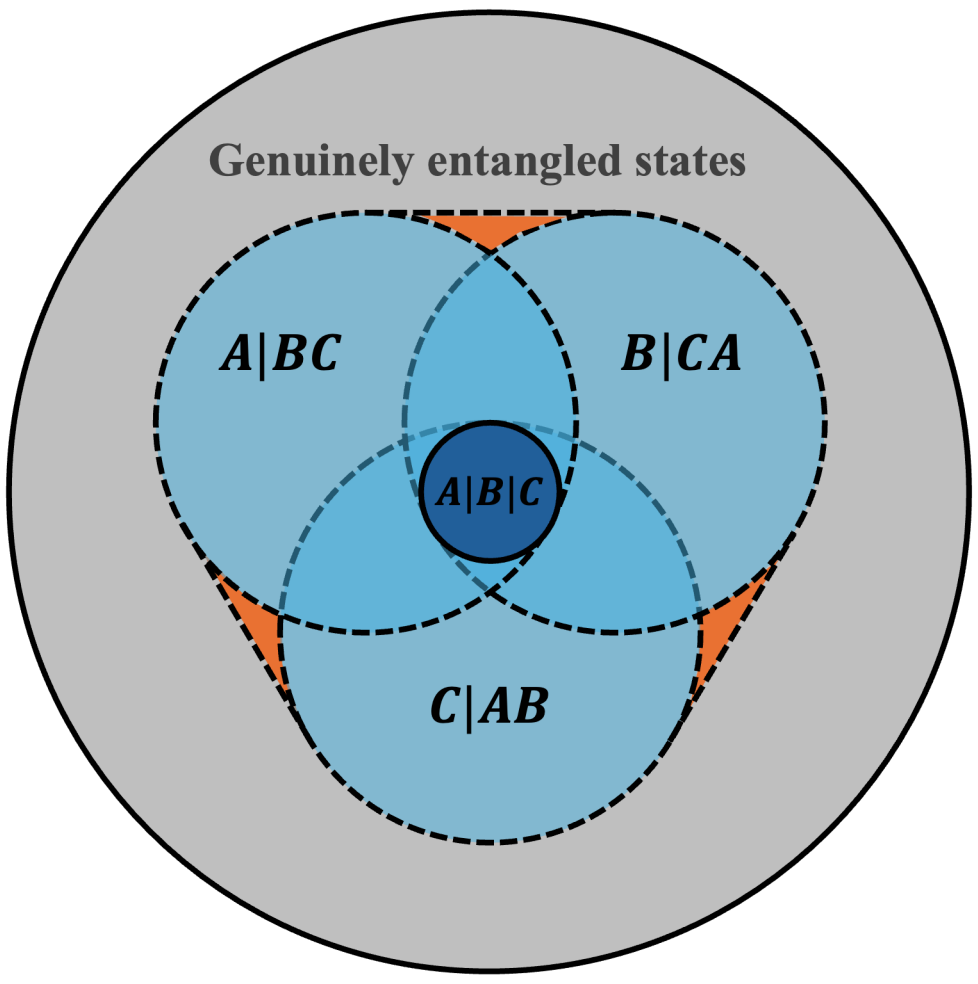

Figure 3.2: The entanglement structure of tripartite mixed states $\rho_{ABC}$. The dark blue region represents the set of fully separable states, which is entirely contained within the light blue region corresponding to biseparable states with fixed bipartitions. The orange region depicts biseparable states that lack a fixed bipartition, while the gray area outside the convex hull signifies genuinely entangled states.

In Fig. 3.2, this classification is illustrated using a schematic representation of nested convex sets. The inclusion and exclusion relations among the different sets are clearly depicted. Surprisingly, as shown in the figure, there exist states that are biseparable with respect to each fixed bipartition but are not fully separable [74, 75], which is impossible for multipartite pure states [76].

## 3.3 Entanglement Transformation

Although local operations and classical communication (LOCC) cannot create entanglement from product states, they enable the manipulation of one entangled state into another. In the following subsections, we explore the transformation of bipartite entangled states, as well as the protocols for entanglement distillation and formation.

### 3.3.1 Deterministic and probabilistic transformations

Local operations and classical communication (LOCC) refers to a framework in quantum information theory where multiple parties, each holding part of a quantum system, are restricted to performing local quantum operations on their respective subsystems and coordinating their actions via classical communication. This paradigm is widely used to study tasks such as quantum state transformation, entanglement manipulation, and quantum communication under the constraint that no direct quantum communication between the parties is allowed. LOCC protocols can involve one-way communication, where information flows in a single direction, or two-way communication, where parties exchange information back and forth, enabling more complex and cooperative transformations of quantum systems.

The outcome of an LOCC operation on a state $\rho$ typically consists of a classical measurement result $i$ and the corresponding post-measurement state $\sigma_i$. Formally, an LOCC operation can be described as a *quantum instrument* $\{\mathcal{F}_i\}$, where $\mathcal{F}_i$ are completely positive maps satisfying $\mathcal{F} = \sum_i \mathcal{F}_i$, which constitutes a quantum channel. The post-measurement state is given by

$$\sigma_i = \frac{\mathcal{F}_i(\rho)}{p_i}, \tag{3.19}$$

where $p_i = \mathrm{Tr}[\mathcal{F}_i(\rho)]$ represents the probability of obtaining the outcome $i$. An LOCC operation is said to transform $\rho$ into $\sigma$ *deterministically* if $\sigma_i = \sigma$ for all $i$.

The success probability of transforming $\rho$ into $\sigma$ via an LOCC operation is defined as the total probability of the outcomes $i$ that yield $\sigma$, and is expressed as:

$$p_{\mathrm{succ}}(\rho \to \sigma | \{\mathcal{F}_i\}) = \sum_{i:\sigma_i=\sigma} p_i. \tag{3.20}$$

**Definition 7 (Optimal Success Probability of Entanglement Transformation)** For two bipartite quantum states $\rho$ and $\sigma$ in the system $AB$, the optimal success probability $p_{\mathrm{succ}}(\rho \to \sigma)$ of transforming $\rho$ into $\sigma$ is the maximum success probability over all LOCC instruments, *i.e.*,

$$p_{\mathrm{succ}}(\rho \to \sigma) = \max_{\{\mathcal{F}_i\} \in \mathrm{LOCC}(A:B)} p_{\mathrm{succ}}(\rho \to \sigma | \{\mathcal{F}_i\}). \tag{3.21}$$

For bipartite pure states, the most general LOCC operations can be effectively reduced to one-way communication [77]. This simplification arises from the Schmidt decomposition, which provides a symmetric representation (up to local unitaries) under the exchange of subsystems $A$ and $B$. Consequently, for pure states, any LOCC operation can be implemented equivalently by a single measurement on one party, followed by local unitary operations on both parties conditioned on the measurement outcome.

Consider the following example of a bipartite pure state shared between Alice and Bob [78]:

$$|\psi\rangle = \frac{1}{\sqrt{2}}(|01\rangle + |10\rangle). \tag{3.22}$$

The objective is to transform this state into the target state:

$$|\phi\rangle = \cos\theta|01\rangle + \sin\theta|10\rangle. \tag{3.23}$$

This transformation can be implemented as follows. Alice begins by performing a generalized measurement on her subsystem, characterized by the following Kraus operators:

$$\begin{aligned} E_1 &= \begin{pmatrix} \cos\theta & 0 \\ 0 & \sin\theta \end{pmatrix}, \\ E_2 &= \begin{pmatrix} \sin\theta & 0 \\ 0 & \cos\theta \end{pmatrix}. \end{aligned} \tag{3.24}$$

The post-measurement states resulting from this operation are:

$$\begin{aligned} |\psi_1\rangle &= \sqrt{2}(E_1 \otimes I)\frac{1}{\sqrt{2}}(|01\rangle + |10\rangle) = \cos\theta|01\rangle + \sin\theta|10\rangle, \\ |\psi_2\rangle &= \sqrt{2}(E_2 \otimes I)\frac{1}{\sqrt{2}}(|01\rangle + |10\rangle) = \sin\theta|01\rangle + \cos\theta|10\rangle. \end{aligned} \tag{3.25}$$

If the measurement outcome corresponds to the first state, $|\psi_1\rangle$, the transformation is complete. However, if the outcome corresponds to the second state, $|\psi_2\rangle$, Alice applies a unitary operation on her subsystem:

$$U = \begin{pmatrix} 0 & 1 \\ 1 & 0 \end{pmatrix}. \tag{3.26}$$

This operation transforms the global state into $\sin\theta|11\rangle + \cos\theta|00\rangle$. Subsequently, Alice communicates the measurement outcome to Bob, who applies the same unitary operation $U$ on his subsystem. This ensures that the desired target state $|\phi\rangle$ is achieved with certainty.

Nielsen subsequently established a rigorous framework for determining the feasibility of such transformation protocols for arbitrary bipartite pure states [79]. This framework, which provides both necessary and sufficient conditions, is based on the Schmidt coefficients of the initial and target states and leverages the mathematical concept of *majorization* [80]. Below, we briefly introduce the notion of majorization:

**Definition 8 (Majorization)** Let $\boldsymbol{x}, \boldsymbol{y} \in \mathbb{R}^n$ be two vectors. We say that $\boldsymbol{x}$ *weakly majorizes* $\boldsymbol{y}$, denoted as $\boldsymbol{x} \succ^{w} \boldsymbol{y}$, if the following inequality holds:

$$\sum_{i=1}^{k} x_i^{\downarrow} \geq \sum_{i=1}^{k} y_i^{\downarrow}, \tag{3.27}$$

for all $k = 1, \ldots, n$, where $x_i^{\downarrow}$ represents the $i$-th largest entry of $\boldsymbol{x}$. If, in addition, $\boldsymbol{x}$ and $\boldsymbol{y}$ satisfy $\sum_{i=1}^{n} x_i = \sum_{i=1}^{n} y_i$, then $\boldsymbol{x}$ is said to *majorize* $\boldsymbol{y}$, denoted as $\boldsymbol{x} \succ \boldsymbol{y}$.

For a bipartite pure state $|\psi\rangle$ expressed in its Schmidt decomposition, as given in Eq. (3.3), the reduced density matrix for subsystem $A$ can be computed as:

$$\rho_A = \mathrm{Tr}_B[|\psi\rangle\langle\psi|] = \sum_{i=1}^{r} \mu_i^2 |\alpha_i\rangle\langle\alpha_i| = \sum_{i=1}^{r} \lambda_i |\alpha_i\rangle\langle\alpha_i|, \tag{3.28}$$

where the eigenvalues of $\rho_A$, denoted as $\lambda_i$, correspond to the squares of the Schmidt coefficients. The feasibility of deterministic transformations between pure states is determined by these eigenvalues, as formalized in the following theorem:

**Theorem 3.3.1 (Nielsen's Theorem [79])** A bipartite pure state $|\psi\rangle$ can be deterministically transformed into another bipartite pure state $|\phi\rangle$ via LOCC if and only if

$$\boldsymbol{\lambda}(\psi_A) \prec \boldsymbol{\lambda}(\phi_A), \tag{3.29}$$

where $\boldsymbol{\lambda}(\psi_A)$ is the vector of eigenvalues of the reduced density matrix $\mathrm{Tr}_B[|\psi\rangle\langle\psi|]$ for subsystem $A$.

Nielsen's theorem provides a clear criterion for determining whether a deterministic transformation is possible. Furthermore, it implies the existence of *incomparable* states, where neither $|\psi\rangle$ nor $|\phi\rangle$ can be transformed into the other via LOCC. Building on Nielsen's theorem, Vidal extended the framework by introducing the concept of *probabilistic* transformations:

**Theorem 3.3.2 (Vidal's Theorem [81])** In the $d_A \otimes d_B$ system, a bipartite pure state $|\psi\rangle$ can be transformed into another bipartite pure state $|\phi\rangle$ with probability $p$ by LOCC if and only if

$$\boldsymbol{\lambda}(\psi_A) \prec^{w} p\boldsymbol{\lambda}(\phi_A), \tag{3.30}$$

where $\boldsymbol{\lambda}(\psi_A)$ represents the vector of eigenvalues of the reduced density matrix $\mathrm{Tr}_B[|\psi\rangle\langle\psi|]$ on subsystem A, and the symbol $\prec^{w}$ denotes weak majorization. The optimal success probability

is given by

$$p_{\text{succ}}(|\psi\rangle \to |\phi\rangle) = \min_{1\leq k\leq r} \left\{ \frac{\sum_{i=k}^{d} \lambda_i^{\downarrow}(\psi_A)}{\sum_{i=k}^{d} \lambda_i^{\downarrow}(\phi_A)} \right\}, \tag{3.31}$$

where $\lambda_i^{\downarrow}(\psi_A)$ denotes the $i$-th largest eigenvalue of the reduced density matrix of $|\psi\rangle$, $r$ is the Schmidt rank of $|\phi\rangle$ and $d = \min\{d_A, d_B\}$.

Consider the following example of bipartite pure states:

$$\begin{aligned} |\psi\rangle &= \sqrt{\frac{2}{5}}|00\rangle + \sqrt{\frac{2}{5}}|11\rangle + \sqrt{\frac{1}{10}}|22\rangle + \sqrt{\frac{1}{10}}|33\rangle, \\ |\phi\rangle &= \sqrt{\frac{1}{2}}|00\rangle + \sqrt{\frac{1}{4}}|11\rangle + \sqrt{\frac{1}{4}}|22\rangle, \end{aligned} \tag{3.32}$$

with the eigenvalue vectors $\boldsymbol{\lambda}(\psi_A) = (\frac{2}{5}, \frac{2}{5}, \frac{1}{10}, \frac{1}{10})$ and $\boldsymbol{\lambda}(\phi_A) = (\frac{1}{2}, \frac{1}{4}, \frac{1}{4}, 0)$. According to Nielsen's theorem, the condition $\boldsymbol{\lambda}(\psi_A) \nprec \boldsymbol{\lambda}(\phi_A)$ implies that a deterministic transformation from $|\psi\rangle$ to $|\phi\rangle$ via LOCC is not feasible. However, Vidal's theorem permits a probabilistic transformation with an optimal success probability of $\frac{4}{5}$. Remarkably, it has been shown that the introduction of a *catalyst state* can enable an exact transformation in such scenarios [82].

In the catalytic LOCC framework, Alice and Bob are allowed to utilize a preshared auxiliary bipartite pure state, referred to as the catalyst. The transformation must ensure that the catalyst state is returned to its original form after the process. For the states defined in Eq. (3.32), a suitable $2 \otimes 2$ catalyst state is given by:

$$|\omega\rangle = \sqrt{\frac{3}{5}}|00\rangle + \sqrt{\frac{2}{5}}|11\rangle. \tag{3.33}$$

By applying Nielsen's theorem to the composite transformation $|\psi\rangle \otimes |\omega\rangle \to |\phi\rangle \otimes |\omega\rangle$, it can be verified that the transformation is now achievable deterministically. This example highlights the utility of catalyst states in enhancing the success probability of entanglement transformations, even enabling a transition from $p < 1$ to $p = 1$.

### 3.3.2 Entanglement distillation and formation

Entanglement distillation and entanglement formation are two fundamental processes in quantum information theory that highlight the operational significance of entanglement as a resource. Entanglement distillation refers to the process of extracting high-quality Bell states from a larger number of partially entangled states using local operations and classical communication (LOCC), and it is essential for enabling tasks such as quantum communication [83] and quan-tum key distribution [84]. On the other hand, entanglement formation quantifies the minimal amount of entanglement required to create a given quantum state using LOCC, reflecting the cost of preparing entangled states. Together, these concepts provide an operational framework for understanding how entanglement can be both consumed and refined, playing a key role in the characterization of quantum resources and their applications in quantum technologies, as depicted in Fig. 3.3.

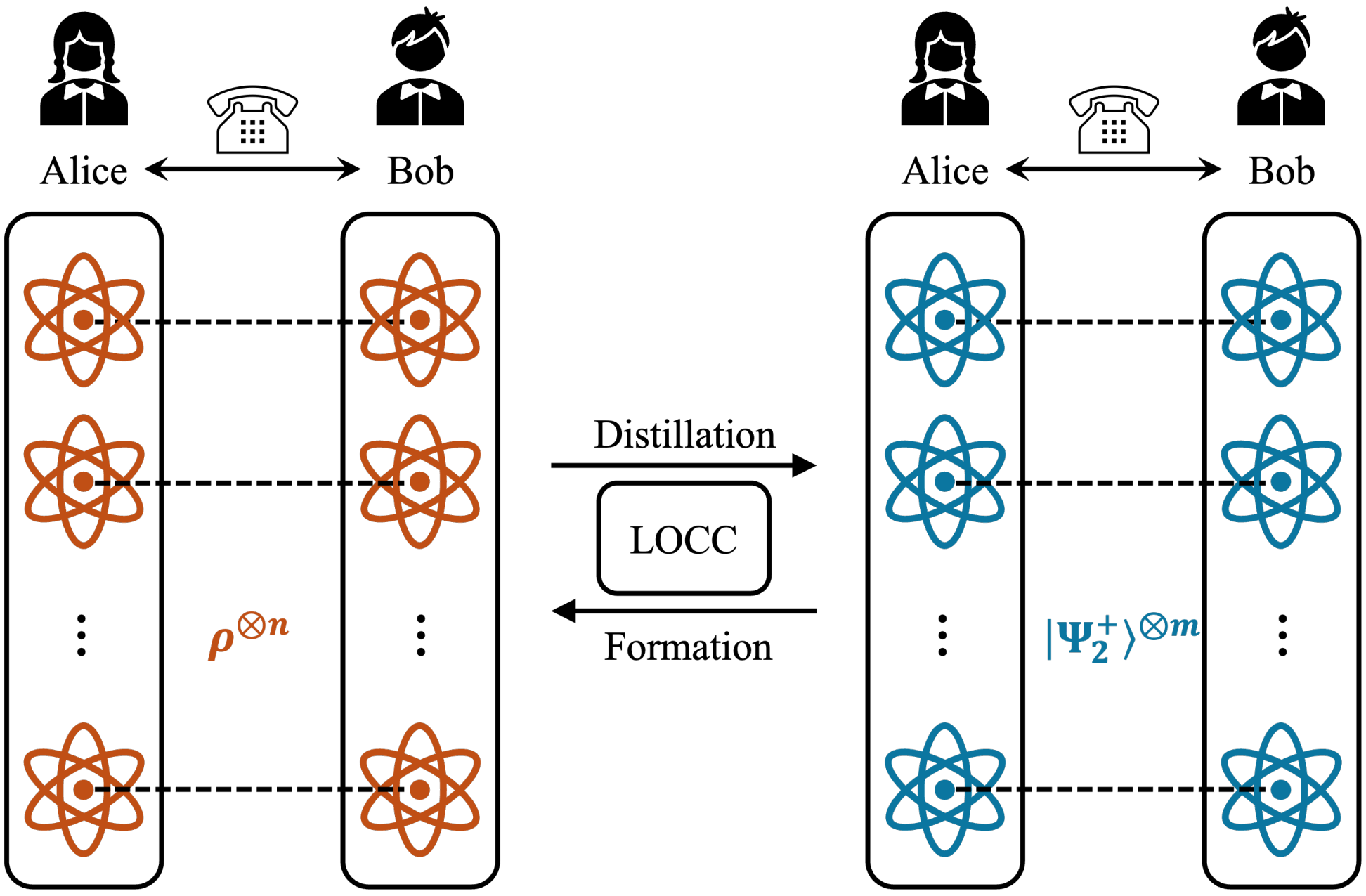

Figure 3.3: Illustration of entanglement distillation and formation protocols. In the distillation process, Alice and Bob utilize LOCC to convert $n$ copies of a noisy quantum state $\rho$ into $m$ copies of maximally entangled Bell states. Conversely, the formation process involves reconstructing $n$ copies of $\rho$ from $m$ Bell states.

In the protocol of *entanglement distillation*, Alice and Bob share $n \gg 1$ copies of a bipartite quantum state $\rho$ and aim to convert them into $m$ copies of Bell state $|\Psi_2^+\rangle$ using LOCC operations [85]. The asymptotic rate at which Bell states can be extracted from $\rho$, known as the *distillable entanglement*, is formally defined as:

$$E_D(\rho) = \sup \left\{ \frac{m}{n} : \lim_{n\to\infty} \inf_{\mathcal{E}\in \mathrm{LOCC(A:B)}} \left\| \mathcal{E}\left[\rho^{\otimes n}\right] - (|\Psi_2^+\rangle\langle\Psi_2^+|)^{\otimes m} \right\|_1 = 0 \right\}, \tag{3.34}$$

where the trace norm $\|\rho - \sigma\|_1 = \mathrm{Tr}[\sqrt{(\rho-\sigma)(\rho-\sigma)^\dagger}]$ quantifies the difference between two quantum states.

Conversely, the *entanglement formation* protocol addresses the reverse process, wherein the goal is to determine the optimal rate at which $n$ copies of $\rho$ can be synthesized from $m$ copies of the Bell state $|\Psi_2^+\rangle$ via LOCC [86]. The corresponding asymptotic rate, referred to as the *entanglement cost*, is defined as:

$$E_C(\rho) = \inf\left\{ \frac{m}{n} : \lim_{n\to\infty} \inf_{\mathcal{E}\in\mathrm{LOCC(A:B)}} \left\| \mathcal{E}\left[ (|\Psi_2^+\rangle\langle\Psi_2^+|)^{\otimes m} \right] - \rho^{\otimes n} \right\|_1 = 0 \right\}. \tag{3.35}$$

In general, the inequality $E_D(\rho) \neq E_C(\rho)$ underscores the fundamental *irreversibility* of entanglement transformations, even in the asymptotic limit. However, for bipartite pure states $|\psi\rangle$, these two quantities are equal and coincide with the *entanglement entropy* [85], defined as:

$$H(\boldsymbol{\lambda}(\psi_A)) = -\sum_i \lambda_i \log_2 \lambda_i, \tag{3.36}$$

where $\boldsymbol{\lambda}(\psi_A)$ denotes the eigenvalue spectrum of the reduced density matrix of subsystem $A$.

This inherent irreversibility gives rise to a fascinating phenomenon known as *bound entanglement* [87]. For certain entangled states $\rho$, entanglement is required for their preparation, yet no entanglement can be distilled from them, *i.e.*, $E_C(\rho) > E_D(\rho) = 0$. Such states are referred to as *bound entangled states*, in contrast to *free entangled states*, which are distillable. Determining whether a given state is distillable remains an open and challenging problem. In the following chapter, we will explore intriguing connections between distillability and various separability criteria, shedding light on this fundamental question.

At first glance, bound entangled states may appear to have limited practical utility, as they cannot be directly employed for tasks such as quantum key distribution, which typically relies on the availability of Bell states [88]. However, it has been demonstrated that certain bound entangled states can indeed enable secure quantum key distribution [89, 90], highlighting the distinction between entanglement distillation and secure key distillation. Following this breakthrough, numerous applications of bound entanglement have been proposed, including its roles in quantum steering [91] and nonlocality [92].

Another common misconception is that bound entangled states are weakly entangled. However, families of bound entangled states have been identified with logarithmically increasing Schmidt numbers relative to the local dimensions [93], as well as states exhibiting a linear scaling of Schmidt numbers [94, 95]. These findings reveal that *high-dimensional entanglement* can also manifest within bound entangled states.

# CHAPTER 4

# HOW TO DETECT AND QUANTIFY ENTANGLEMENT?

## 4.1 Separability Criteria

A fundamental challenge in quantum information theory is determining whether a given quantum state is separable or entangled, a problem known as the *separability problem* [96]. This problem has been shown to be NP-hard [22], meaning that no deterministic polynomial-time algorithm is currently known to solve it. Separability criteria are typically formulated based on specific properties that all separable states must satisfy. If a quantum state fails to meet such a criterion, it can be conclusively identified as entangled. However, satisfying the criterion does not guarantee separability, as these criteria provide only *necessary but not sufficient* conditions for detecting entanglement.

### 4.1.1 Positive partial transpose

One of the earliest and most widely recognized separability criteria was introduced by Peres [23], based on the concept of partial transposition. Consider a bipartite quantum state $\rho$ in the Hilbert space $\mathcal{H}_A \otimes \mathcal{H}_B$, expressed as:

$$\rho = \sum_{i,j}^{d_A} \sum_{k,l}^{d_B} \rho_{ij,kl} |i\rangle\langle j| \otimes |k\rangle\langle l|. \tag{4.1}$$

The *partial transpose* of $\rho$ with respect to subsystem A, denoted as $\rho^{\mathrm{T}_A}$, is defined by swapping the indices of subsystem A:

$$\rho^{\mathrm{T}_A} = \sum_{i,j}^{d_A} \sum_{k,l}^{d_B} \rho_{ji,kl} |i\rangle\langle j| \otimes |k\rangle\langle l|. \tag{4.2}$$

For a separable state $\rho = \sum_i p_i |\psi_A^{(i)}\rangle\langle\psi_A^{(i)}| \otimes |\psi_B^{(i)}\rangle\langle\psi_B^{(i)}|$, the partial transpose yields:

$$\begin{aligned} \rho^{\mathrm{T}_A} &= \sum_i p_i (|\psi_A^{(i)}\rangle\langle\psi_A^{(i)}|)^{\mathrm{T}} \otimes |\psi_B^{(i)}\rangle\langle\psi_B^{(i)}| \\ &= \sum_i p_i |\psi_A^{*(i)}\rangle\langle\psi_A^{*(i)}| \otimes |\psi_B^{(i)}\rangle\langle\psi_B^{(i)}|, \end{aligned} \tag{4.3}$$

where the hermiticity of pure states is used, *i.e.*, $(|\psi\rangle\langle\psi|)^{\mathrm{T}} = |\psi^*\rangle\langle\psi^*|$. Here, $|\psi^*\rangle$ denotes the complex conjugate of $|\psi\rangle$. Since $|\psi_A^{*(i)}\rangle$ remains a valid pure state in $\mathcal{H}_A$, the partial transpose $\rho^{\mathrm{T}_A}$ is positive semidefinite, ensuring that it represents a valid quantum state. The same reasoning applies to the partial transpose with respect to subsystem B, $\rho^{\mathrm{T}_B}$.

For clarity, a density matrix $\rho$ is said to have a *positive partial transpose* (PPT) if its partial transpose is positive semidefinite, meaning it has no negative eigenvalues. A state satisfying this property is referred to as a PPT state, while a state that does not satisfy it is termed non-PPT (NPT). This observation leads to the following well-established criterion:

**Theorem 4.1.1 (PPT Criterion [23])** If a bipartite state $\rho$ is NPT, then it is entangled.

This theorem provides a straightforward and effective method for detecting entanglement by analyzing the eigenvalue spectrum of the partially transposed density matrix. However, a natural question arises: is the PPT criterion also sufficient to guarantee separability? It has been shown that this is true only in specific low-dimensional cases:

**Theorem 4.1.2 (Horodecki's Theorem [97, 98])** For a quantum state $\rho$ in a $2 \otimes 2$ or $2 \otimes 3$ system, the PPT criterion is both necessary and sufficient for separability, *i.e.*, $\rho^{\mathrm{T}_A} \succeq 0$ implies that $\rho$ is separable. In higher-dimensional systems, this sufficiency no longer holds.

Interestingly, the PPT criterion also plays a pivotal role in the study of bound entanglement, as it remains the only known sufficient condition for undistillability:

**Theorem 4.1.3 (PPT Criterion for Undistillability [87])** If a bipartite state is PPT, then it is undistillable.

This criterion underscores the significance of PPT entangled states in the context of bound entanglement. The first example of a PPT entangled state was introduced in [98], with additional examples provided in [75, 99]. To illustrate, consider the construction of a PPT entangled state based on an example from [75]. In a $3 \otimes 3$ quantum system, the following product states can be

defined:

$$
\begin{aligned}
|\psi_1\rangle &= \tfrac{1}{\sqrt{2}}|0\rangle(|0\rangle - |1\rangle), & |\psi_2\rangle &= \tfrac{1}{\sqrt{2}}(|0\rangle - |1\rangle)|2\rangle, \\
|\psi_3\rangle &= \tfrac{1}{\sqrt{2}}|2\rangle(|1\rangle - |2\rangle), & |\psi_4\rangle &= \tfrac{1}{\sqrt{2}}(|1\rangle - |2\rangle)|0\rangle, \\
|\psi_5\rangle &= \tfrac{1}{3}(|0\rangle + |1\rangle + |2\rangle)(|0\rangle + |1\rangle + |2\rangle).
\end{aligned} \tag{4.4}
$$

These five product states form an *unextendible product basis* (UPB), meaning they are mutually orthogonal and no additional product state exists that is orthogonal to all of them. Using this UPB, a mixed state can be constructed as:

$$
\rho = \frac{1}{4}\left( I - \sum_{i=1}^{5} |\psi_i\rangle\langle\psi_i| \right), \tag{4.5}
$$

where $\mathcal{P}_\perp = I - \sum_{i=1}^{5} |\psi_i\rangle\langle\psi_i|$ acts as the projector onto the subspace orthogonal to $\{|\psi_i\rangle\}$. Since no product states exist in the range of $\rho$, the range criterion (see Sec. 4.1.3) implies that $\rho$ is entangled. Furthermore, as $\rho^{\mathrm{T}_A} = \rho$, the state is PPT, confirming that it is a bound entangled state.

An intriguing question is whether NPT bound entangled states exist. While some studies suggest their existence [100–102], the issue remains unresolved. If such states do exist, a remarkable phenomenon emerges: for a hypothetical NPT bound entangled state $\rho$ and another bound entangled state $\sigma$, their joint state $\rho \otimes \sigma$ may no longer exhibit bound entanglement [103, 104]. This phenomenon, known as *superactivation*, has also been observed in quantum communication, where two quantum channels, each incapable of transmitting quantum information individually, can jointly enable reliable information transmission [105].

### 4.1.2 Reduction criterion

Another separability criterion, closely related to the PPT criterion (Theorem 4.1.1), is the reduction criterion:

**Theorem 4.1.4 (Reduction Criterion [106])** A bipartite quantum state $\rho_{AB}$ in a $2 \otimes 2$ or $2 \otimes 3$ system is separable *if and only if*

$$
\rho_A \otimes I_B - \rho_{AB} \geq 0, \tag{4.6}
$$

where $\rho_A$ denotes the reduced density matrix of $\rho$ on subsystem A. For systems of higher dimensions, this criterion serves only as a necessary condition for separability.

The proof of this theorem will be presented in Sec. 4.1.4. Notably, the reduction criterion is not stronger than the PPT criterion. However, as we will observe, it can be extended to detect high-dimensional entanglement, specifically entangled states with large Schmidt numbers. Additionally, the reduction criterion offers a sufficient condition for distillability:

**Theorem 4.1.5 (Reduction Criterion for Distillability [106])** If a bipartite quantum state violates the reduction criterion, then the state is distillable.

### 4.1.3 Range criterion

For a separable state $\rho = \sum_i p_i |\psi_A^{(i)}\rangle\langle\psi_A^{(i)}| \otimes |\psi_B^{(i)}\rangle\langle\psi_B^{(i)}|$, the set of product states $\{|\psi_A^{(i)}\rangle|\psi_B^{(i)}\rangle\}$ spans the range of $\rho$, while the set $\{|\psi_A^{*(i)}\rangle|\psi_B^{(i)}\rangle\}$ spans the range of its partial transpose $\rho^{\mathrm{T}_A}$. This observation leads to the following theorem:

**Theorem 4.1.6 (Range Criterion [98])** A bipartite quantum state $\rho$ is entangled if its range cannot be spanned by a set of product states $\{|\psi_A^{(i)}\rangle|\psi_B^{(i)}\rangle\}$ or the range of its partial transpose $\rho^{\mathrm{T}_A}$ cannot be spanned by $\{|\psi_A^{*(i)}\rangle|\psi_B^{(i)}\rangle\}$.

While the range criterion can, in certain cases, be more powerful than the PPT criterion, it does not provide an efficient method for identifying the vectors $\{|\psi_A^{(i)}\rangle|\psi_B^{(i)}\rangle\}$. Furthermore, it is not applicable to noisy states, as such states are typically full rank, allowing for the construction of a product state basis in their range.

### 4.1.4 Positive but not completely positive maps

The PPT criterion (Theorem 4.1.1) and the reduction criterion (Theorem 4.1.4) are specific instances of a broader framework of separability criteria, collectively referred to as *positive but not completely positive maps*. To elucidate this concept, we first define positive and completely positive maps as follows:

**Definition 9 (Positive and Completely Positive Maps)** A linear map $\Lambda : \mathcal{A}_1 \to \mathcal{A}_2$, which acts on operators in the space $\mathcal{A}_1$ and maps them to the space $\mathcal{A}_2$, is termed *positive* if it preserves positivity, *i.e.*,

$$\Lambda(A) \geq 0 \quad \forall A \geq 0. \tag{4.7}$$

Furthermore, a positive map $\Lambda$ is called *completely positive* if its extension

$$\Lambda \otimes I_d : \mathcal{A}_1 \otimes \mathcal{M}_d \to \mathcal{A}_2 \otimes \mathcal{M}_d \tag{4.8}$$

remains positive for any dimension $d$, where $\mathcal{M}_d$ denotes the space of $d \times d$ matrices, and $I_d$ is the identity map on $\mathcal{M}_d$.

Positive maps provide a foundation for formulating a *necessary and sufficient* condition for detecting entanglement. Consider a separable state $\rho = \sum_i p_i |\psi_A^{(i)}\rangle\langle\psi_A^{(i)}| \otimes |\psi_B^{(i)}\rangle\langle\psi_B^{(i)}|$. Applying the map $I \otimes \Lambda$ to $\rho$ yields:

$$(I \otimes \Lambda)\rho = \sum_i p_i |\psi_A^{(i)}\rangle\langle\psi_A^{(i)}| \otimes \Lambda(|\psi_B^{(i)}\rangle\langle\psi_B^{(i)}|), \tag{4.9}$$

where $\Lambda$ is a positive map. Since $\Lambda(|\psi_B^{(i)}\rangle\langle\psi_B^{(i)}|)$ is positive, it follows that $(I \otimes \Lambda)\rho$ is also positive. This observation establishes that positive maps are a necessary condition for separability. Moreover, the following theorem demonstrates their sufficiency:

**Theorem 4.1.7 (Positive Map Criterion [97])** A bipartite quantum state $\rho$ is separable *if and only if*

$$I \otimes \Lambda(\rho) \succeq 0, \tag{4.10}$$

for all positive maps $\Lambda$.

Thus, the separability problem is intrinsically linked to the classification of all positive maps, a longstanding open problem in mathematics [107, 108]. Notable examples of positive but not completely positive maps include the partial transpose and the reduction map. The reduction map is defined as:

$$R(X) = \mathrm{Tr}[X] \cdot I - X, \tag{4.11}$$

which directly leads to the reduction criterion $I \otimes R(\rho) = \rho_A \otimes I - \rho \succeq 0$.

It has been established that the reduction map belongs to the class of *decomposable maps*, meaning it can be expressed as:

$$R = P_1 + P_2 \circ T, \tag{4.12}$$

where $P_1$ and $P_2$ are completely positive maps, and $T$ denotes the transpose operation. This decomposition implies that $I \otimes R$ cannot detect entanglement unless $I \otimes T$ (the partial transpose map) detects it. Consequently, the reduction criterion is strictly weaker than the PPT criterion.

An intermediate concept bridging positive and completely positive maps is the notion of a $k$-positive map:

**Definition 10 ($k$-Positive Map)** A linear map $\Lambda : \mathcal{A}_1 \to \mathcal{A}_2$, which acts on operators in the space $\mathcal{A}_1$ and maps them to the space $\mathcal{A}_2$, is termed *k-positive* if its extension

$$\Lambda \otimes I_k : \mathcal{A}_1 \otimes \mathcal{M}_k \to \mathcal{A}_2 \otimes \mathcal{M}_k \tag{4.13}$$

remains positive, where $\mathcal{M}_k$ denotes the space of $k \times k$ matrices, and $I_k$ is the identity map on $\mathcal{M}_k$.

The concept of $k$-positive maps is intricately linked to quantum entanglement through the Schmidt number (see Definition 3), as formalized in the following theorem:

**Theorem 4.1.8 ($k$-Positive Map Criterion [60])** A bipartite quantum state $\rho$ has a Schmidt number $\mathrm{SN}(\rho) \leq k$ *if and only if*

$$(I \otimes \Lambda_k)(\rho) \geq 0, \tag{4.14}$$

for all $k$-positive maps $\Lambda_k$.

A notable family of $k$-positive maps can be constructed as a generalization of the reduction map [see Eq. (4.11)], as described below:

**Theorem 4.1.9 (Generalized Reduction Map [60])** A family of positive maps is defined as

$$R_p(X) = \mathrm{Tr}[X] \cdot I - pX, \tag{4.15}$$

which is $k$-positive but not $(k+1)$-positive for $\frac{1}{k+1} < p \leq \frac{1}{k}$.

### 4.1.5 Entanglement witnesses

Let us now consider a fundamentally different approach to separability criteria. The previously discussed methods rely on having complete knowledge of the density matrix of the quantum state and involve performing specific operations on it. However, in practical scenarios, we may lack precise information about the state or wish to detect entanglement experimentally. A more practical and physically motivated approach involves using measurable observables, known as *entanglement witnesses* [24, 109, 110]:

**Definition 11 (Entanglement Witness)** An observable $W$ is called an *entanglement witness* if it satisfies the following conditions:

$$\begin{aligned} &\mathrm{Tr}[W\rho] \geq 0 \quad \text{for all separable states } \rho, \\ &\mathrm{Tr}[W\rho^\star] < 0 \quad \text{for at least one entangled state } \rho^\star. \end{aligned} \tag{4.16}$$

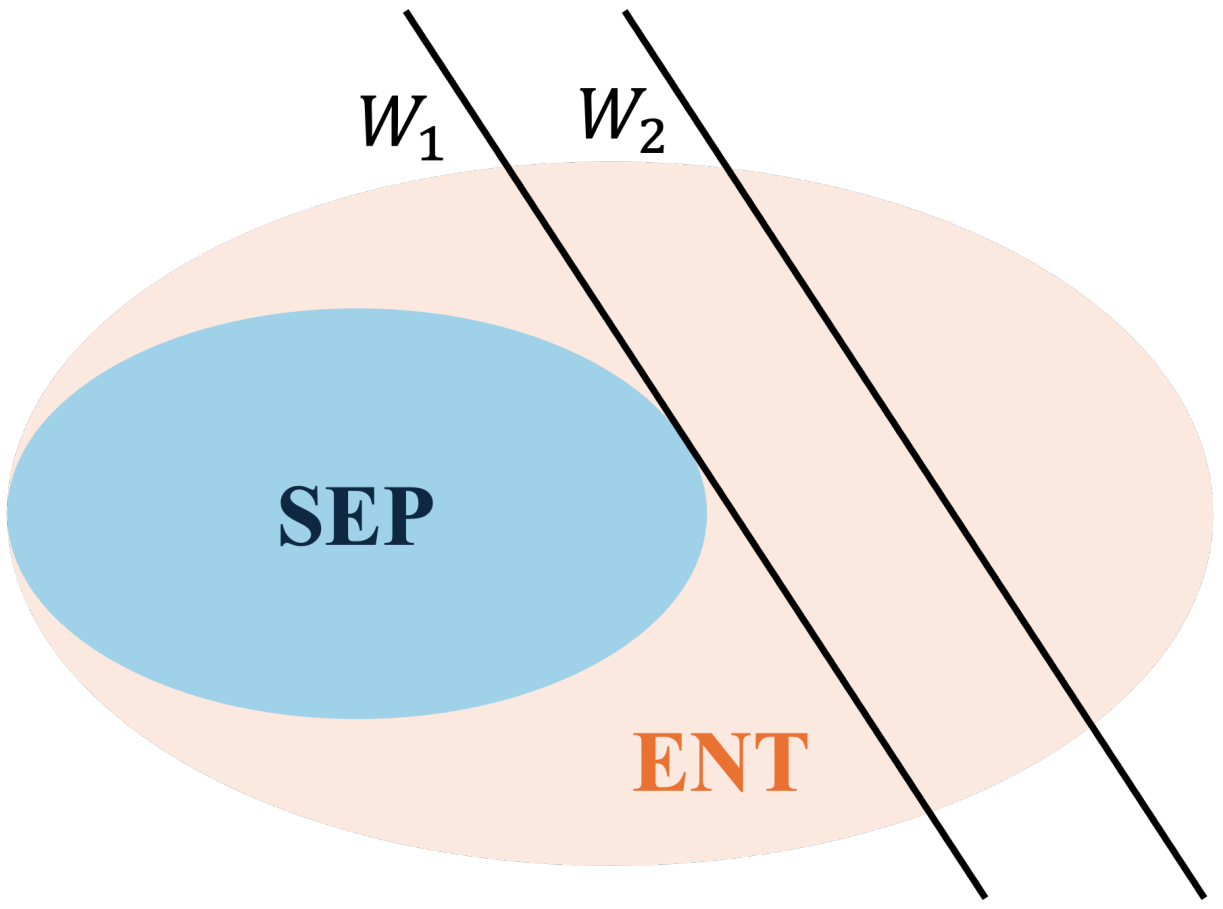

Figure 4.1: Geometric representation of entanglement witnesses. The quantum state space is divided into two regions by the hyperplanes defined by the witnesses $W_1$ and $W_2$. The hyperplanes correspond to $\mathrm{Tr}[W_1\rho] = 0$ and $\mathrm{Tr}[W_2\rho] = 0$. The region labeled SEP represents the set of separable states, while ENT denotes the set of entangled states. Witness $W_1$ is more effective than $W_2$ as it detects a larger subset of entangled states.

Thus, if the measurement yields $\mathrm{Tr}[W\rho^\star] < 0$, one can conclusively determine that $\rho^\star$ is entangled. In this case, we say that the state $\rho^\star$ is detected by the witness $W$.

Entanglement witnesses have a clear geometric interpretation: since the expectation value of an observable depends linearly on the quantum state, the set of states satisfying $\mathrm{Tr}[W\rho] = 0$ forms a *hyperplane* in the space of all quantum states. This hyperplane divides the state space into two regions. The set of separable states lies entirely on the side where $\mathrm{Tr}[W\rho] \geq 0$, while the other region contains the entangled states that can be detected by $W$. This geometric perspective is illustrated in Fig. 4.1. Using this representation, we can better understand the following result:

**Theorem 4.1.10 (Hahn-Banach Separation Theorem [111])** Let $S$ be a convex, compact set, and let $a \notin S$. Then there exists a hyperplane that separates $a$ from the set $S$.

From this theorem, it follows that for any entangled state, there always exists a hyperplane that separates it from the set of separable states. This leads to the following important result:

**Theorem 4.1.11 (Completeness of Entanglement Witnesses)** For every entangled state $\rho$, there exists an entanglement witness $W$ capable of detecting it.

While this theorem guarantees the existence of an entanglement witness for any given entangled state, the practical challenge lies in constructing such witnesses. In the following, we will explore several examples to illustrate the principles and methodologies for constructing entanglement witnesses.

The first example involves detecting an NPT state $\sigma$ in system AB. If we know that $\sigma^{\mathrm{T}_A}$ has a negative eigenvalue $\lambda_- < 0$ with a corresponding eigenvector $|\lambda_-\rangle$, the witness can be constructed in the following form:

$$W = (|\lambda_-\rangle\langle\lambda_-|)^{T_A}, \tag{4.17}$$

which satisfies:

$$\mathrm{Tr}[W\sigma] = \mathrm{Tr}[(|\lambda_-\rangle\langle\lambda_-|)^{T_A}\sigma] = \mathrm{Tr}[(|\lambda_-\rangle\langle\lambda_-|)\sigma^{T_A}] = \lambda_- < 0, \tag{4.18}$$

and

$$\mathrm{Tr}[W\rho] = \mathrm{Tr}[(|\lambda_-\rangle\langle\lambda_-|)\rho^{T_A}] \geq 0, \tag{4.19}$$

for all separable states $\rho$.

More generally, if an entangled state $\sigma$ can be detected by a positive map $\Lambda$, then $I \otimes \Lambda(\sigma)$ has a negative eigenvalue $\lambda_- < 0$ with a corresponding eigenvector $|\lambda_-\rangle$. In this case, we can construct a witness as:

$$W = I \otimes \Lambda^\dagger(|\lambda_-\rangle\langle\lambda_-|), \tag{4.20}$$

where $\Lambda^\dagger$ is the *adjoint map* of $\Lambda$, defined by the relation:

$$\mathrm{Tr}[\Lambda^\dagger(X) \cdot Y] = \mathrm{Tr}[X \cdot \Lambda(Y)], \tag{4.21}$$

for all operators $X, Y$.

Similar to Eq. (4.12), witness operators can be classified into two categories: *decomposable witnesses* and *indecomposable witnesses*. Decomposable witnesses detect only NPT entangled states, while indecomposable witnesses can detect bound entangled states. In [112], a family of indecomposable witnesses was constructed using the unextendible product basis.

Another important case is the optimization of witnesses, which provides an intuitive approach to constructing *optimal witnesses*. Starting from an entangled pure state $|\psi\rangle$, we can define a witness as:

$$W = \alpha I - |\psi\rangle\langle\psi|. \tag{4.22}$$

The underlying principle is intuitive: states in close proximity to an entangled state must also exhibit entanglement. The term $\mathrm{Tr}[\rho|\psi\rangle\langle\psi|] = \langle\psi|\rho|\psi\rangle$ quantifies the fidelity between the state $|\psi\rangle$ and $\rho$. If this fidelity surpasses a certain threshold $\alpha$, indicating that $\rho$ is sufficiently close to $|\psi\rangle$, the expectation value $\mathrm{Tr}[W\rho] < 0$ confirms that $\rho$ is entangled.

The critical task lies in determining the appropriate threshold $\alpha$ to ensure that the witness operator $W$ remains positive for all separable states. This condition can be expressed as:

$$\alpha \geq \max_{\rho \in \mathrm{SEP}} \mathrm{Tr}[\rho|\psi\rangle\langle\psi|] = \max_{|\phi\rangle = |a\rangle \otimes |b\rangle} |\langle\psi|\phi\rangle|^2. \tag{4.23}$$

The second equality follows from the fact that a linear function achieves its maximum over a convex set at one of its extreme points. Furthermore, this maximum value corresponds to the square of the largest Schmidt coefficient of $|\psi\rangle$. When this bound is saturated, the resulting witness is referred to as *optimal*, as it cannot be refined further to detect additional entangled states.

This optimization process can also be interpreted geometrically, as depicted in Fig. 4.1. Starting with a generic witness $W_2$ characterized by a higher threshold, the optimization ultimately converges to an optimal witness $W_1$ with the minimal threshold, thereby enhancing its effectiveness in detecting entanglement.

## 4.2 Entanglement Measures

In this section, we introduce the fundamental concepts of entanglement measures. After outlining the requirements for entanglement measures, we present various useful entanglement measures.

### 4.2.1 Requirements for entanglement measures

Once an entangled state has been successfully identified, the next logical step is to address the question: how can its entanglement be *quantified*? The characterization of entanglement depends on the specific quantum information tasks under consideration, such as the asymptotic costs associated with entanglement distillation and formation (see Sec. 3.3.2). Nevertheless, an *axiomatic* framework provides a unified approach to defining general criteria for entanglement quantification [86, 113]. Any function that satisfies these criteria is referred to as an *entanglement measure*. While the precise definitions of entanglement measures may vary depending on the context, the following fundamental requirements are widely accepted:

**Definition 12 (Entanglement Measure)** An entanglement measure is a function $E$ that assigns a real number to a quantum state, satisfying the following properties:

- (*Non-negativity*) For any quantum state $\rho$, $E(\rho) \geq 0$, with equality, $E(\rho) = 0$, *if and only if* $\rho$ is separable.

- (*Monotonicity*) The measure $E$ must not increase under local operations and classical communication (LOCC). Specifically, for any LOCC operation $\Lambda$ and quantum state $\rho$, it holds that $E(\Lambda(\rho)) \leq E(\rho)$.

Meanwhile, the *entanglement monotone* is another key quantity commonly used for entanglement quantification. It remains non-increasing under LOCC but is not guaranteed to be non-zero for all entangled states [114].

In addition to the two fundamental properties outlined above, several supplementary criteria can be advantageous for the study of entanglement, but are not necessary.

One such criterion is *invariance under local unitary transformations*, which requires that an entanglement measure remains unchanged when local unitary operations are applied to the quantum state. Mathematically, this is expressed as:

$$E(\rho) = E(U_A \otimes U_B \rho U_A^\dagger \otimes U_B^\dagger), \tag{4.24}$$

indicating that the measure is independent of the choice of local basis.

A stricter version of the monotonicity condition is often considered, which demands that the entanglement measure should not increase on average under local operations and classical communication (LOCC). Specifically, if an LOCC operation transforms $\rho$ into a set of states $\{\rho_i\}$ with corresponding probabilities $\{p_i\}$, then the following inequality must hold:

$$\sum_i p_i E(\rho_i) \leq E(\rho). \tag{4.25}$$

Another desirable property is *convexity*, which ensures that the entanglement measure does not increase under probabilistic mixing of quantum states. Formally, for a mixture of states $\{\rho_i\}$ with probabilities $\{p_i\}$, the measure satisfies:

$$E\left(\sum_i p_i \rho_i\right) \leq \sum_i p_i E(\rho_i). \tag{4.26}$$

This property reflects the physical intuition that losing information about the components of a quantum ensemble should not lead to an increase in entanglement.

For scenarios involving multiple copies of a quantum state, it is often expected that the entanglement measure scales linearly, *i.e.*, $E(\rho^{\otimes n}) = nE(\rho)$. A stronger version of this property, known as *additivity*, requires that the measure satisfies:

$$E(\rho_1 \otimes \rho_2) = E(\rho_1) + E(\rho_2), \tag{4.27}$$

ensuring that the entanglement of a composite system is simply the sum of the entanglements of its individuals.

### 4.2.2 Convex roof entanglement measures

A widely adopted method for defining entanglement measures is the *convex roof construction* [115]. This approach begins by specifying the entanglement measure for pure states, denoted as $E(|\psi\rangle)$, and subsequently extends it to mixed states through the following expression:

$$E(\rho) = \min_{\{p_i, |\psi_i\rangle\}} \sum_i p_i E(|\psi_i\rangle), \tag{4.28}$$

where the minimization is performed over all possible pure-state decompositions of $\rho$, such that $\rho = \sum_i p_i |\psi_i\rangle\langle\psi_i|$, with $p_i \geq 0$ and $\sum_i p_i = 1$. These measures are referred to as *convex roof entanglement measures*, as depicted in Fig. 4.2.

The convex roof construction offers several advantages. Notably, the resulting entanglement measure is inherently convex and inherits desirable properties from the pure-state measure $E(|\psi\rangle)$, which is often easier to analyze. Furthermore, a general proof of the monotonicity of all convex roof entanglement measures under local operations and classical communication (LOCC) was established in [114]. However, the optimization required in this construction is computationally challenging, and analytical solutions are typically available only for specific cases.

Two of the earliest and most well-established entanglement measures introduced within this framework are the *entanglement of formation* (EoF) [86, 116, 117] and *concurrence* [58, 118–121]. These measures have proven instrumental in characterizing quantum phase transitions in many-body systems [122–124] and have even been experimentally accessible [125].

Consider a bipartite quantum state $\rho \in \mathcal{H}_A \otimes \mathcal{H}_B$. The *entanglement of formation* (EoF) for $\rho$ is defined as:

$$E_f(\rho) = \min_{\{p_i, |\psi_i\rangle\}} \sum_i p_i E_f(|\psi_i\rangle), \tag{4.29}$$

where the minimization is performed over all possible pure-state decompositions of $\rho$, such that $\rho = \sum_i p_i |\psi_i\rangle\langle\psi_i|$.

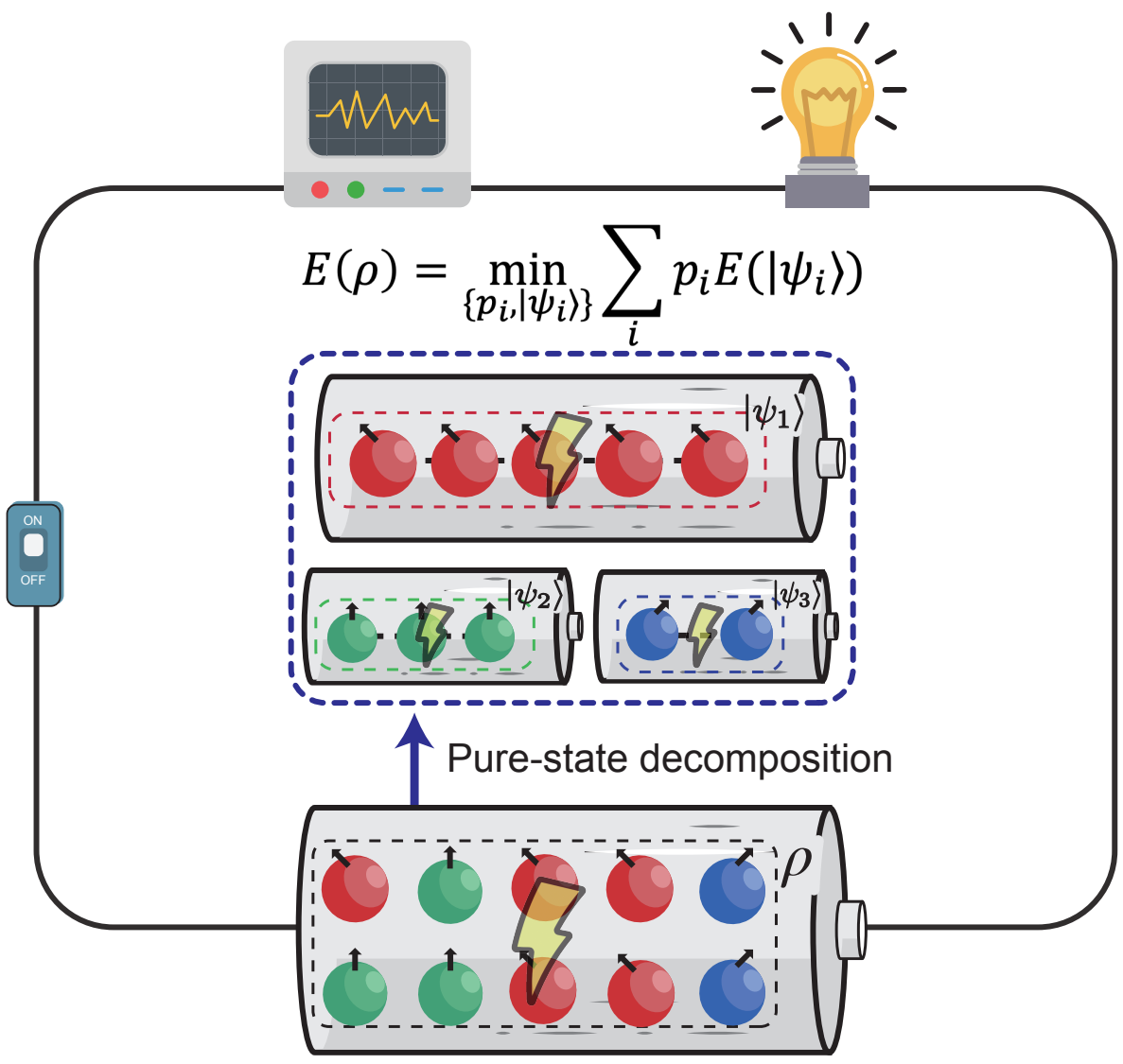

Figure 4.2: The convex roof entanglement measure $E(\rho)$ is calculated by minimizing a weighted sum of the entanglement measure for pure states $E(|\psi_i\rangle)$ over all possible pure-state decompositions $\rho = \sum_i p_i |\psi_i\rangle\langle\psi_i|$.

For a pure state $|\psi_{AB}\rangle$, the EoF, denoted as $E_f(|\psi_{AB}\rangle)$, is equivalent to the *von Neumann entropy* of the reduced density matrix of $|\psi_{AB}\rangle$. This entropy, often referred to as the *entanglement entropy*, serves as the quantum analog of the classical Shannon entropy [126] and is given by:

$$S(\rho_A) = -\mathrm{Tr}[\rho_A \log_2 \rho_A], \tag{4.30}$$

where $\rho_A = \mathrm{Tr}_B[|\psi_{AB}\rangle\langle\psi_{AB}|]$ represents the reduced density matrix of subsystem $A$. If the eigenvalues of $\rho_A$ are denoted by $\{\lambda_i\}$, the entropy can be expressed as:

$$S(\rho_A) = H(\{\lambda_i\}) = -\sum_i \lambda_i \log_2 \lambda_i, \tag{4.31}$$

which corresponds to the Shannon entropy of the eigenvalue distribution $\{\lambda_i\}$.

It is evident that $E_f(|\psi\rangle)$ vanishes only for product states. Therefore, via the convex roof construction, $E_f(\rho) = 0$ if and only if $\rho$ is separable, satisfying the non-negativity condition of entanglement measures. The monotonicity of EoF under LOCC was also proved in [86], along with other desirable properties such as convexity and asymptotic continuity [96].

Although EoF is additive for pure states, it has been shown to be *subadditive* for mixed states [96], meaning:

$$E_f(\rho \otimes \sigma) \leq E_f(\rho) + E_f(\sigma). \tag{4.32}$$

*Concurrence*, initially introduced by Hill and Wootters as an entanglement measure for two-qubit systems [118], was later generalized to bipartite quantum systems of arbitrary dimensions [121]. For a bipartite pure state $|\psi_{AB}\rangle$ in a $d_A \otimes d_B$ system, the concurrence is defined in terms of the purity of the reduced density matrix:

$$C(|\psi_{AB}\rangle) = \sqrt{2(1 - \mathrm{Tr}[\rho_A^2])} = \sqrt{2(1 - \mathrm{Tr}[\rho_B^2])}. \tag{4.33}$$

Similarly, for mixed states, the concurrence is extended using the convex roof construction.

A closely related entanglement measure is the *linear entropy of entanglement* (LEE) [127, 128], which is defined for pure bipartite states as:

$$E_l(|\psi_{AB}\rangle) = 1 - \mathrm{Tr}[\rho_A^2] = \frac{C(|\psi_{AB}\rangle)^2}{2}. \tag{4.34}$$

In the special case of two qubits, the entanglement of formation (EoF) is a monotonically increasing function of the concurrence, expressed as [58]:

$$E_f(\rho) = H_2\left(\frac{1 + \sqrt{1 - C^2(\rho)}}{2}\right), \tag{4.35}$$

where $H_2(x)$ denotes the binary entropy function:

$$H_2(x) = -x \log_2 x - (1 - x) \log_2(1 - x). \tag{4.36}$$

Fortunately, for a two-qubit density matrix $\rho$, a *spin-flipped state* can be defined as:

$$\tilde{\rho} = (\sigma_y \otimes \sigma_y)\rho^*(\sigma_y \otimes \sigma_y), \tag{4.37}$$

where $*$ represents the complex conjugate in the computational basis $\{|00\rangle, |01\rangle, |10\rangle, |11\rangle\}$. So that the concurrence of $\rho$ can now be obtained analytically as:

$$C(\rho) = \max\{0, \mu_1 - \mu_2 - \mu_3 - \mu_4\}, \tag{4.38}$$

where $\{\mu_1, \mu_2, \mu_3, \mu_4\}$ are the square roots of the eigenvalues of the matrix $\rho\tilde{\rho}$, arranged in descending order.

As a result, for two-qubit systems, the entanglement of formation can be determined analytically using the above formula, thereby circumventing the need for numerical optimization in Eq. (4.29).

For some special states with certain symmetry, the entanglement of formation (EoF) can also be obtained analytically. Two prominent examples are the isotropic states, defined in Eq. (3.10), and the Werner states, defined in Eq. (3.11).

Isotropic states are invariant under the transformation $U \otimes U^*$, for any unitary matrix $U$. For a $d \otimes d$ isotropic state, the EoF is given as follows [129]:

$$E_f\left(\rho_I(F)\right) = \begin{cases} 0, & F \in \left[0, \frac{1}{d}\right], \\ R_{1,d-1}(F), & F \in \left(\frac{1}{d}, \frac{4(d-1)}{d^2}\right), \\ \frac{d\log_2(d-1)}{d-2}(F-1) + \log_2 d, & F \in \left[\frac{4(d-1)}{d^2}, 1\right], \end{cases} \tag{4.39}$$

where

$$R_{1,d-1}(F) = H_2(\gamma(F)) + [1-\gamma(F)]\log_2(d-1), \tag{4.40}$$

with

$$\gamma(F) = \frac{1}{d}[\sqrt{F} + \sqrt{(d-1)(1-F)}]^2. \tag{4.41}$$

Werner states, on the other hand, are invariant under the transformation $U \otimes U$, for any unitary matrix $U$. For Werner states in a $d \otimes d$ system, the EoF can be derived as follows [130]:

$$E_f\left(\rho_W(\alpha)\right) = \begin{cases} 0, & \alpha \in \left[-1, \frac{1}{d}\right], \\ H_2\left(\frac{1}{2}\left(1 - \sqrt{1 - \left(\frac{d\alpha - 1}{\alpha - d}\right)^2}\right)\right), & \alpha \in \left(\frac{1}{d}, 1\right]. \end{cases} \tag{4.42}$$

Here, $H_2(x)$ is the binary entropy function defined in Eq. (4.36).

Another notable category of measures is the so-called *geometric measure of entanglement* (GME) [47–49], which quantifies the geometric "distance" between a given pure state and the set of product states. This type of measure is applicable not only to bipartite systems but can also be extended to the multipartite setting.

For an $n$-partite pure state $|\psi\rangle$, the GME for pure states is defined as

$$E_G(|\psi\rangle) = 1 - \max_{|\phi\rangle} |\langle\phi|\psi\rangle|^2 = \min_{|\phi\rangle}(1 - |\langle\phi|\psi\rangle|^2), \tag{4.43}$$

where $|\phi\rangle = \otimes_{i=1}^n |\phi^{(i)}\rangle$ represents an arbitrary fully product state. Using the convex roof construction, its extension to mixed states is expressed as:

$$E_G(\rho) = \min_{\{p_i, |\psi_i\rangle\}} \sum_i p_i E_G(|\psi_i\rangle). \tag{4.44}$$

Now, consider the bipartite case. For bipartite pure states, the term $\max_{|\phi\rangle} |\langle\phi|\psi\rangle|$ corresponds to the largest Schmidt coefficient of the state $|\psi\rangle$. Furthermore, in two-qubit cases, it is linked to the concurrence, $C(\rho)$, through the relation:

$$E_G(\rho) = \frac{1}{2}\left[1 - \sqrt{1 - C(\rho)^2}\right]. \tag{4.45}$$

For $d \otimes d$ isotropic states, as defined in Eq. (3.10), the analytical expression for the GME is given by [49]:

$$E_G\left(\rho_I(F)\right) = \begin{cases} 0, & F \in \left[0, \frac{1}{d}\right], \\ 1 - \frac{1}{d}\left[\sqrt{F} + \sqrt{(1-F)(d-1)}\right]^2, & F \in \left(\frac{1}{d}, 1\right]. \end{cases} \tag{4.46}$$

On the other hand, for $d \otimes d$ Werner states, as defined in Eq. (3.11), the GME is given by [49]:

$$E_G\left(\rho_W(\alpha)\right) = \begin{cases} 0, & \alpha \in \left[-1, \frac{1}{d}\right], \\ \frac{1}{2}\left(1 - \sqrt{1 - \left(\frac{d\alpha-1}{\alpha-d}\right)^2}\right), & \alpha \in \left(\frac{1}{d}, 1\right]. \end{cases} \tag{4.47}$$

For multipartite states, let us consider some examples of symmetric states. In an $n$-qubit system, permutation-symmetric states are often expressed using the *Dicke basis* $\{|D_n^k\rangle\}_{k=0}^n$, where the Dicke state is defined as:

$$\left|D_n^k\right\rangle = \binom{n}{k}^{-1/2} \sum_{\text{perm}} \underbrace{|0\rangle|0\rangle\cdots|0\rangle}_{n-k}\underbrace{|1\rangle|1\rangle\cdots|1\rangle}_{k}, \tag{4.48}$$

representing the equal superposition of all permutations of computational basis states with $n-k$ qubits in state $|0\rangle$ and $k$ qubits in state $|1\rangle$. It has been shown that the closest product state to a Dicke state takes the form [131]:

$$|\phi\rangle = \left(\sqrt{\frac{n-k}{n}}|0\rangle + \sqrt{\frac{k}{n}}|1\rangle\right)^{\otimes n}, \tag{4.49}$$

*i.e.*, a tensor product of $n$ identical single-qubit states. Consequently, the geometric measure of entanglement for the Dicke state is given by:

$$E_G(|D_n^k\rangle) = 1 - \binom{n}{k}\left(\frac{k}{n}\right)^k\left(\frac{n-k}{n}\right)^{n-k}. \tag{4.50}$$

For the special case of three qubits, notable examples of permutation-symmetric states include:

$$
\begin{aligned}
|\mathrm{GHZ}\rangle &= \frac{1}{\sqrt{2}}(|000\rangle + |111\rangle),\\
|\mathrm{W}\rangle &= \frac{1}{\sqrt{3}}(|100\rangle + |010\rangle + |001\rangle),\\
|\tilde{\mathrm{W}}\rangle &= \frac{1}{\sqrt{3}}(|110\rangle + |101\rangle + |011\rangle).
\end{aligned} \tag{4.51}
$$

Using a similar approach, the maximal overlaps $\max_{|\phi\rangle} |\langle\phi|\psi\rangle|$, denoted as $\Lambda_{\max}$, for these states are determined to be:

$$
\Lambda_{\max}(|\mathrm{GHZ}\rangle) = \frac{\sqrt{2}}{2}, \quad \Lambda_{\max}(|\mathrm{W}\rangle) = \Lambda_{\max}(|\tilde{\mathrm{W}}\rangle) = \frac{2}{3}. \tag{4.52}
$$

More generally, the following theorem provides insight into the geometric measure of entanglement for all permutation-symmetric states:

**Theorem 4.2.1 (GME for Permutation-Symmetric States [132])** For any multipartite pure state with permutation symmetry, the closest product state under GME, as defined in Eq. (4.43), is also symmetric, *i.e.*, it is a tensor product of identical single-party states.

Finally, it is worth emphasizing the natural connection between the geometric measure of entanglement and the entanglement witness defined in Eq. (4.22). For an entanglement witness $W = \alpha I - |\psi\rangle\langle\psi|$, the following relationship holds:

$$
\alpha \geq \max_{|\phi\rangle} |\langle\phi|\psi\rangle|^2 = \Lambda^2_{\max}(|\psi\rangle). \tag{4.53}
$$

Thus, the optimal entanglement witness for a given entangled state $|\psi\rangle$ can be constructed as:

$$
W = \Lambda^2_{\max}(|\psi\rangle)I - |\psi\rangle\langle\psi|. \tag{4.54}
$$

For example, for the GHZ state, the optimal witness is:

$$
W_{\mathrm{GHZ}} = \frac{1}{2}I - |\mathrm{GHZ}\rangle\langle\mathrm{GHZ}|. \tag{4.55}
$$

Similarly, for the W and inverted-W states, the optimal witnesses are:

$$
W_{\mathrm{W}} = \frac{4}{9}I - |\mathrm{W}\rangle\langle\mathrm{W}|, \quad W_{\tilde{\mathrm{W}}} = \frac{4}{9}I - |\tilde{\mathrm{W}}\rangle\langle\tilde{\mathrm{W}}|. \tag{4.56}
$$

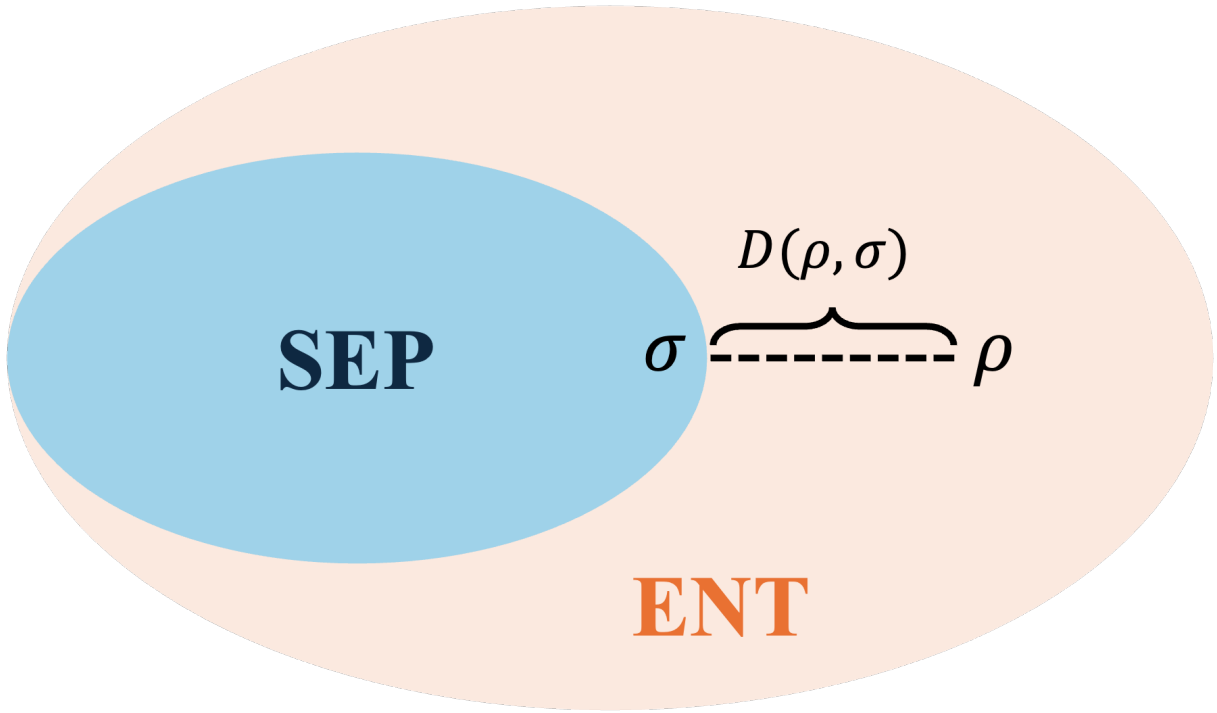

Figure 4.3: Visualization of a distance-based entanglement measure $E(\rho)$, which quantifies the entanglement of a quantum state $\rho$ as the minimum distance to the set of separable states $\sigma$, using an appropriate distance metric $D(\rho, \sigma)$.

### 4.2.3 Distance-based entanglement measures

A prominent class of entanglement measures is grounded in the intuitive notion that the closer a quantum state is to the set of separable states, the less entangled it is. These measures, referred to as *distance-based entanglement measures*, are formally defined as:

$$E(\rho) = \min_{\sigma \in \mathrm{SEP}} D(\rho, \sigma), \tag{4.57}$$

where SEP denotes the set of separable states, and $D(\rho, \sigma)$ quantifies the "distance" between the quantum states $\rho$ and $\sigma$, as illustrated in Fig. 4.3. It is worth noting that $D$ may not necessarily satisfy all the mathematical properties of a strict distance metric. To ensure monotonicity under local operations and classical communication (LOCC), it is typically required that:

$$D(\rho, \sigma) \geq D(\Lambda(\rho), \Lambda(\sigma)), \tag{4.58}$$

for any LOCC operation $\Lambda$ acting on the quantum states $\rho$ and $\sigma$.

Interestingly, the geometric measure of entanglement (GME), introduced via the convex roof construction and defined in Eq. (4.44), can also be interpreted as a distance-based measure. This equivalence is formalized in the following theorem:

**Theorem 4.2.2 (Distance-based Representation of GME [132])** For a multipartite mixed state $\rho$ in the Hilbert space $\mathcal{H} = \otimes_{i=1}^{n} \mathcal{H}_i$, the geometric measure of entanglement can be expressed as:

$$E_G(\rho) = \min_{\sigma \in \mathrm{SEP}} D(\rho, \sigma), \tag{4.59}$$

where $D(\rho, \sigma) = 1 - F(\rho, \sigma)$ is a distance measure derived from Uhlmann's fidelity $F(\rho, \sigma) =$

$\|\sqrt{\rho}\sqrt{\sigma}\|_1^2 = (\mathrm{Tr}[\sqrt{\sqrt{\rho}\sigma\sqrt{\rho}}])^2$, and SEP represents the set of fully separable states.

Another widely studied distance-based entanglement measure is the *relative entropy of entanglement* (REE). The classical relative entropy, which quantifies the divergence between two probability distributions $\{p_i\}$ and $\{q_i\}$, is defined as [133]:

$$S(\{p_i\}\|\{q_i\}) = \sum_i p_i \log_2 \frac{p_i}{q_i}. \tag{4.60}$$

This concept is extended to the quantum domain [134], where the relative entropy between two quantum states $\rho$ and $\sigma$ is given by:

$$S(\rho\|\sigma) = \mathrm{Tr}[\rho(\log_2 \rho - \log_2 \sigma)]. \tag{4.61}$$

Building on this, the REE was introduced as an entanglement measure in [113, 135], and it is defined as:

$$E_R(\rho) = \min_{\sigma\in\mathrm{SEP}} S(\rho\|\sigma) = \min_{\sigma\in\mathrm{SEP}} \mathrm{Tr}[\rho(\log_2 \rho - \log_2 \sigma)]. \tag{4.62}$$

For the distillable entanglement [see Eq. (3.34)] and the entanglement cost [see Eq. (3.35)], the relative entropy of entanglement provides an intermediate relationship between them, as stated in the following theorem:

**Theorem 4.2.3 (Lower and Upper Bounds for REE [135, 136])** The relative entropy of entanglement $E_R(\rho)$ is bounded above by the entanglement cost $E_C(\rho)$ and below by the distillable entanglement $E_D(\rho)$, *i.e.*,

$$E_D(\rho) \leq E_R(\rho) \leq E_C(\rho). \tag{4.63}$$

Analytical results for the REE are also available for isotropic states and Werner states. For a $d \otimes d$ isotropic state $\rho_I(F)$ defined in Eq. (3.10), the REE is given by [137]:

$$E_R\left(\rho_I(F)\right) = \begin{cases} 0, & F \in \left[0, \frac{1}{d}\right], \\ S\left(\rho_I(F)\middle\|\rho_I\left(\frac{1}{d}\right)\right), & F \in \left(\frac{1}{d}, 1\right]. \end{cases} \tag{4.64}$$

For a $d \otimes d$ Werner state $\rho_W(\alpha)$ defined in Eq. (3.11), the REE is given by [130]:

$$E_R\left(\rho_W(\alpha)\right) = \begin{cases} 0, & \alpha \in \left[-1, \frac{1}{d}\right], \\ S\left(\rho_W(\alpha)\middle\|\rho_W\left(\frac{1}{d}\right)\right), & \alpha \in \left(\frac{1}{d}, 1\right]. \end{cases} \tag{4.65}$$

### 4.2.4 Operational entanglement measures

Two prominent operational entanglement measures, introduced in Sec. 3.3.2, are the distillable entanglement $E_D$ and the entanglement cost $E_C$. These measures are formally defined as follows:

$$
\begin{aligned}
E_D(\rho) &= \sup\left\{ \frac{m}{n} : \lim_{n\to\infty} \inf_{\mathcal{E}\in \mathrm{LOCC(A:B)}} \left\| \mathcal{E}\left[\rho^{\otimes n}\right] - (|\Psi_2^+\rangle\langle\Psi_2^+|)^{\otimes m} \right\|_1 = 0 \right\}, \\
E_C(\rho) &= \sup\left\{ \frac{m}{n} : \lim_{n\to\infty} \inf_{\mathcal{E}\in \mathrm{LOCC(A:B)}} \left\| \mathcal{E}\left[|\Psi_2^+\rangle\langle\Psi_2^+|^{\otimes m}\right] - \rho^{\otimes n} \right\|_1 = 0 \right\},
\end{aligned} \tag{4.66}
$$

where these expressions describe the *asymptotic transformation* between multiple copies of the given quantum state $\rho$ and multiple copies of the Bell state $|\Psi_2^+\rangle = \frac{1}{\sqrt{2}}(|00\rangle + |11\rangle)$.

The properties of these measures are noteworthy. For a bipartite pure state $|\psi\rangle$, the distillable entanglement and the entanglement cost coincide and are equal to the entanglement of formation (EoF) of $|\psi\rangle$, *i.e.*,

$$E_C(|\psi\rangle) = E_D(|\psi\rangle) = E_f(|\psi\rangle), \tag{4.67}$$

indicating that the conversion between the Bell state and the given pure state is *reversible* in the asymptotic limit. In this context, the EoF provides a *unique* measure for characterizing the transformation rate.

In contrast, for a bipartite mixed state $\rho$, the relationship $E_C(\rho) \geq E_D(\rho)$ generally holds. Furthermore, it has been established that the entanglement cost is equivalent to the *regularization* of the entanglement of formation [138]:

$$E_C(\rho) = \lim_{n\to\infty} \frac{1}{n} E_f(\rho^{\otimes n}). \tag{4.68}$$

Since the entanglement of formation is additive for pure states, the entanglement cost reduces to the EoF in the case of pure states. However, for mixed states, the subadditivity of the EoF renders the regularization process highly non-trivial and computationally demanding.

# CHAPTER 5

# CONVEX AND NON-CONVEX OPTIMIZATION

## 5.1 Convex Optimization

Optimization is a fundamental area in mathematics and computer science, playing a critical role in many quantum information problems. One of the most famous and widely used optimization paradigms is *convex optimization* [25]. Below, we provide a brief introduction to some basic concepts in convex optimization. Then, we focus on a specific class of convex optimization methods, called *semidefinite programming*, along with their applications in quantum information problems.

### 5.1.1 Introduction

We start by introducing the $k$-dimensional *simplex*, which is defined as follows:

$$\Delta^k = \left\{ \boldsymbol{x} \in \mathbb{R}^{k+1} \,:\, \|\boldsymbol{x}\|_1 = \sum_i x_i = 1,\, 0 \le x_i \le 1 \;\forall i = 0, \dots, k \right\}. \tag{5.1}$$

A *convex combination* of a set of points refers to a linear combination of those points, where the coefficients form a vector belonging to the simplex. Using this notion, we can formally define a *convex set* as follows:

**Definition 13 (Convex Set)** A subset $C \subseteq \mathbb{R}^n$ is said to be a *convex set* if every convex combination of points in $C$ also belongs to $C$. Equivalently, for any $x, y \in C$ and $\lambda \in [0, 1]$, the point $\lambda x + (1 - \lambda) y$ must also lie in $C$.

Some common examples of convex sets in quantum information theory are:

- The set of probability vectors, *i.e.*, the probability simplex.
- The set of density matrices.
- The set of separable states.
- The set of quantum states with positive partial transpose.
- The set of quantum channels.

The notions of *convexity* and *concavity* for a mathematical function $f$ are formally defined as follows:

**Definition 14 (Convex and Concave Functions)** A function $f : C \to \mathbb{R}$ is said to be *convex* if, for all $x, y \in C$ and $\lambda \in [0, 1]$, the following inequality holds:

$$f(\lambda x + (1-\lambda)y) \leq \lambda f(x) + (1-\lambda) f(y). \tag{5.2}$$

Conversely, $f$ is termed *concave* if $-f$ is convex.

Several important examples of convex functions in quantum information theory include:

- The von Neumann entropy, $S(\rho) = -\mathrm{Tr}[\rho \log_2 \rho]$.
- The quantum relative entropy, $S(\rho\|\sigma) = \mathrm{Tr}[\rho(\log_2 \rho - \log_2 \sigma)]$, which is convex in either $\rho$ or $\sigma$.
- The trace norm, $\|\rho\|_1 = \mathrm{Tr}[|\rho|]$.
- The expectation value of an observable, $\mathrm{Tr}[O\rho]$, with respect to the quantum state $\rho$.

A general convex optimization problem is formulated as the minimization of a convex function $f(x)$ over a convex set $C$:

$$\min_{x \in C} f(x), \tag{5.3}$$

where $f$ is referred to as the *objective function*, and $C$ is known as the *feasible set*. If the feasible set $C$ is empty, the problem is classified as *infeasible*.

A key property of convex optimization is that any *local minimum* is also a *global minimum*. This ensures that improving the solution locally will eventually lead to the global minimum, provided it exists. However, this property does not guarantee the efficiency of specific algorithms in achieving the global minimum.

One of the most well-known methods for solving convex optimization problems is the *interior point method* [25, 139, 140]. The general convex optimization problem can be reformulated in the following way:

$$\begin{aligned} \min_{\boldsymbol{x} \in \mathbb{R}^n} \quad & f(\boldsymbol{x}) \\ \text{subject to} \quad & g_i(\boldsymbol{x}) \leq 0, \quad i = 1, \ldots, m, \\ & h_i(\boldsymbol{x}) = 0, \quad i = 1, \ldots, p, \end{aligned} \tag{5.4}$$

where $f, g_1, \ldots, g_m : \mathbb{R}^n \to \mathbb{R}$ are convex functions, and $h_1, \ldots, h_p$ are affine transformations, *i.e.*, $h_i(\boldsymbol{x}) = \boldsymbol{a}_i \cdot \boldsymbol{x} - b_i$ for vector $\boldsymbol{a_i}$ and scalar $b_i$. The central idea of the interior point method is to work within the interior of the feasible region, $\mathbf{int}\mathcal{X} = \{\boldsymbol{x} : g_i(\boldsymbol{x}) < 0, h_i(\boldsymbol{x}) = 0 \; \forall i\}$, and incorporate a *barrier function* $\phi$ that satisfies the following properties [141]:

- $\phi$ is smooth and *strongly convex*, meaning there exists a constant $\nu > 0$ such that $\nabla^2\phi(\boldsymbol{x}) \succeq \nu\boldsymbol{I}$ for all $\boldsymbol{x} \in \mathcal{X}$.
- $\phi(\boldsymbol{x}_k) \to \infty$ for any sequence $\{\boldsymbol{x}_k\}$ in $\mathbf{int}\mathcal{X}$ approaching the boundary $\partial\mathcal{X}$.

For all $t \geq 0$, the *equality-constrained* penalized problem, denoted by $P(t)$, is given by:

$$\begin{aligned} \min_{\boldsymbol{x}} \quad & tf(\boldsymbol{x}) + \phi(\boldsymbol{x}) \\ \text{subject to} \quad & h_i(\boldsymbol{x}) = 0, \quad i = 1, \dots, p. \end{aligned} \tag{5.5}$$

This problem has a unique solution, denoted by $\boldsymbol{x}^*(t)$, which satisfies $\boldsymbol{x}^*(t) \in \mathbf{int}\mathcal{X}$. The set of solutions $\{\boldsymbol{x}^*(t) : t \geq 0\}$ is known as the *central path*. Under mild regularity conditions, the central path converges to an optimal solution of the original problem as $t \to \infty$.

A common choice for the barrier function is the *log barrier* function, which is defined as:

$$\phi(\boldsymbol{x}) = -\sum_{i=1}^{m} \log\big(-g_i(\boldsymbol{x})\big). \tag{5.6}$$

The domain of the log barrier function is the set of interior points of the feasible region. The gradient and Hessian of the log barrier function are given by:

$$\begin{aligned} \nabla\phi(\boldsymbol{x}) &= -\sum_{i=1}^{m} \frac{1}{g_i(\boldsymbol{x})}\nabla g_i(\boldsymbol{x}), \\ \nabla^2\phi(\boldsymbol{x}) &= \sum_{i=1}^{m} \frac{1}{g_i(\boldsymbol{x})^2}\nabla g_i(\boldsymbol{x})\nabla g_i(\boldsymbol{x})^T - \sum_{i=1}^{m} \frac{1}{g_i(\boldsymbol{x})}\nabla^2 g_i(\boldsymbol{x}). \end{aligned} \tag{5.7}$$

With the above components, a *path-following* algorithm typically operates as follows, as illustrated in Fig. 5.1: Given the current iteration $t_k > 0, \boldsymbol{x}^{(k)} \in \mathbf{int}\mathcal{X}$ and $\boldsymbol{x}^{(k)}$ is close to $\boldsymbol{x}^*(t_k)$, the procedure is:

- Update the current value of $t_k$ to a new value $t_{k+1} > t_k$;
- Apply an equality-constrained minimization algorithm to solve the problem $P(t_{k+1})$, using $\boldsymbol{x}^{(k)}$ as the initial guess, to obtain the new point $\boldsymbol{x}^{(k+1)}$ close to $\boldsymbol{x}^*(t_{k+1})$.

Finally, we summarize the advantages of path-following algorithms in the interior point method. First, mature numerical techniques, such as Newton's method, can be employed to solve the associated equality-constrained problems efficiently. Second, for certain classes of barrier functions, it can be proven that the path-following scheme produces an $\epsilon$-suboptimal solution in *polynomial* time relative to the problem size and $\log(1/\epsilon)$. Third, the interior point method serves both theoretical and practical purposes effectively.

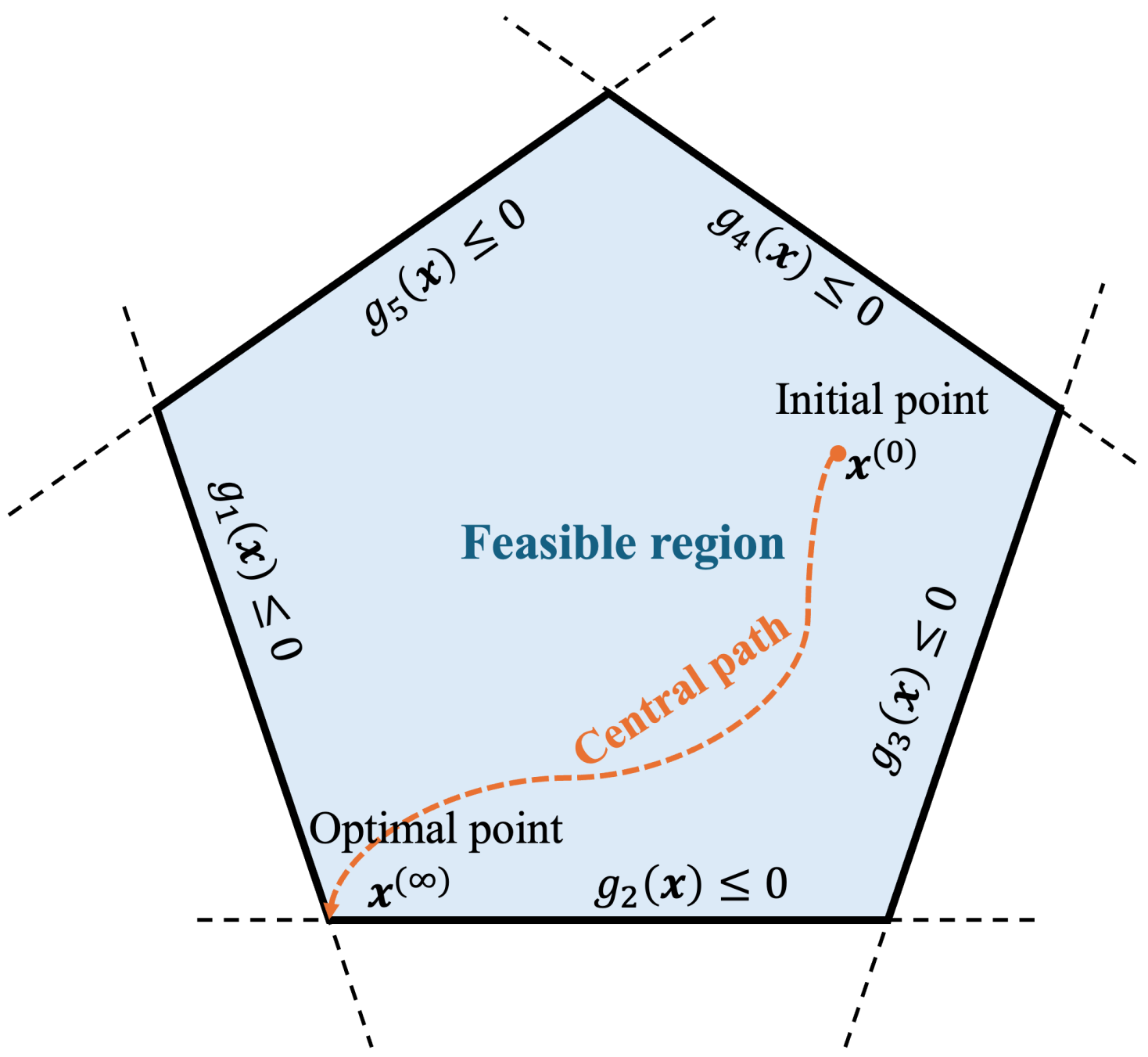

Figure 5.1: Visualization of the interior point method. The algorithm typically starts with an initial interior point $\boldsymbol{x}^{(0)}$ and iteratively solves equality-constrained minimization subproblems to generate new feasible points. These points progressively converge toward the optimal solution $\boldsymbol{x}^{(\infty)}$ of the original convex optimization problem. The trajectory of these feasible points defines the central path, which leads to the optimal solution.

### 5.1.2 Semidefinite programming and its applications

*Semidefinite programming* (SDP) represents a relatively recent and rapidly evolving subfield of convex optimization. Over the past few decades, SDPs have gained prominence as a powerful and versatile tool for advancing research in quantum information theory [26, 142].

An SDP is formulated as a constrained optimization problem involving an *operator variable* $X$, which is a Hermitian operator acting on a finite-dimensional complex vector space, satisfying $X^\dagger = X$. The objective function is a real-valued linear function of $X$, typically expressed as $\mathrm{Tr}[AX]$, where $A$ is a Hermitian operator. The constraints are composed of linear equalities and inequalities, given by $\Phi_i(X) = B_i$ for $i = 1, \ldots, m$ and $\Gamma_j(X) \leq C_j$ for $j = 1, \ldots, n$, where $B_i$ and $C_j$ are Hermitian operators, and $\Phi_i(\cdot)$ and $\Gamma_j(\cdot)$ are *hermiticity-preserving* linear maps.

The general form of an SDP can be expressed as:

$$\begin{aligned} \max_X \quad & \mathrm{Tr}[AX], \\ \text{subject to} \quad & \Phi_i(X) = B_i, \quad i = 1, \ldots, m, \\ & \Gamma_j(X) \preceq C_j, \quad j = 1, \ldots, n. \end{aligned} \tag{5.8}$$

Although the problem is presented here as a maximization, it is straightforward to reformulate a minimization problem into this framework by introducing a negative sign in the objective function. Consequently, throughout this discussion, we do not strictly differentiate between minimization and maximization problems, as they are interchangeable within this context.

One possible application of semidefinite programming is solving the *fidelity estimation problem* during state preparation. Let us start with a simple scenario where the target state is a pure state $|\psi\rangle$. Due to imperfect preparation, we may obtain an *unknown* mixed state $\rho$, along with some measurement results $\{m_i\}$. The goal is to estimate the fidelity between the target state $|\psi\rangle$ and the unknown state $\rho$, which is given by:

$$F(\rho, |\psi\rangle) = \langle\psi|\rho|\psi\rangle = \mathrm{Tr}[|\psi\rangle\langle\psi|\rho]. \tag{5.9}$$

The *maximum fidelity* can then be computed via an SDP:

$$\begin{aligned} \max_\rho \quad & \mathrm{Tr}[|\psi\rangle\langle\psi|\rho] \\ \text{subject to} \quad & \mathrm{Tr}[M_i\rho] = m_i, \quad \text{for } i = 1, \ldots, N, \\ & \mathrm{Tr}[\rho] = 1, \\ & \rho \succeq 0, \end{aligned} \tag{5.10}$$

where $M_i$ is the measurement observable corresponding to the result $m_i$. In other words, this SDP allows us to determine the *best-case* fidelity, *i.e.*, the largest fidelity among all possible states that are consistent with the observed data.

On the other hand, we can also consider the *minimization problem* by replacing max with min. In this case, the solution provides the *worst-case* fidelity, *i.e.*, the smallest fidelity among all possible states compatible with the observed data.

However, in more general cases, if we aim to estimate the fidelity with respect to a mixed state, we can only solve for the best-case scenario. The key distinction is that the fidelity with respect to a pure state is a linear function of $\rho$, which means it is both convex and concave. This linearity enables efficient computation of both the best-case and worst-case fidelities via SDP. In contrast, fidelity to a mixed state is not linear, complicating the analysis.

Before we consider the SDP for a mixed target state, we need to address the general fidelity function, which is defined as:

$$F(\rho, \sigma) = \|\sqrt{\rho}\sqrt{\sigma}\|_1^2 = \left(\mathrm{Tr}[\sqrt{\sqrt{\sigma}\rho\sqrt{\sigma}}]\right)^2, \tag{5.11}$$

where $F(\rho, \sigma)$ is a non-linear function of both $\rho$ and $\sigma$. This non-linearity means that we cannot directly use it as the objective function in an SDP.

Despite being non-linear, the fidelity is a concave function [78], satisfying:

$$F(p\rho_1 + (1-p)\rho_2, \sigma) \geq pF(\rho_1, \sigma) + (1-p)F(\rho_2, \sigma), \tag{5.12}$$

for $p \in [0, 1]$, and similarly for $\sigma$.

Interestingly, it can be shown that, even though it is non-linear, the fidelity can be expressed as an SDP by introducing auxiliary variables [143]. Specifically, we can write:

$$\begin{aligned} \sqrt{F(\rho, \sigma)} = \max_{Y,Z} \quad & \mathrm{Tr}[Y] \\ \text{subject to} \quad & \begin{pmatrix} \rho & Y + iZ \\ Y - iZ & \sigma \end{pmatrix} \geq 0. \end{aligned} \tag{5.13}$$

Here, the variables $Y$ and $Z$ are arranged into a block matrix to make the expression more concise. The inequality constraint in this formulation is equivalent to the standard form in Eq. (5.8), *i.e.*,

$$\Gamma(Y, Z) = |0\rangle\langle 0| \otimes \rho + |0\rangle\langle 1| \otimes (Y + iZ) + |1\rangle\langle 0| \otimes (Y - iZ) + |1\rangle\langle 1| \otimes \sigma, \tag{5.14}$$

which is indeed a hermiticity-preserving linear map, as required in the SDP formulation.

Returning to the problem of estimating the largest fidelity to a mixed target state, we can now use Eq. (5.13) to reformulate the problem as an SDP. The following SDP computes the best-case square-root fidelity between any state $\rho$ consistent with the experimental data $\{m_i\}$ and a target

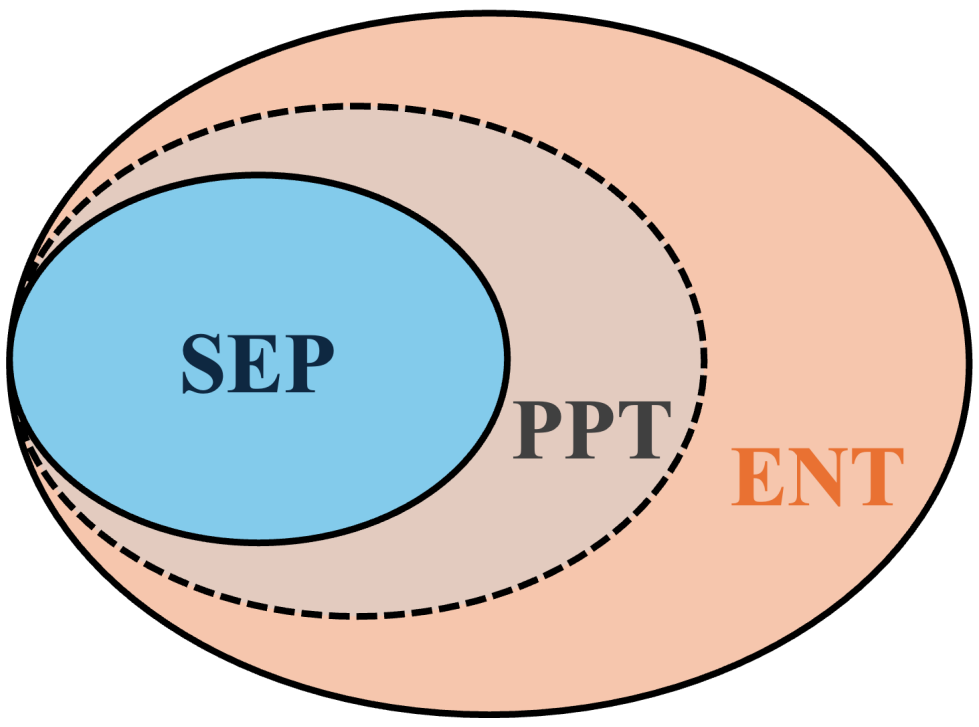

Figure 5.2: Illustration of the PPT relaxation for bipartite separable states. Here, SEP represents the set of separable states, PPT denotes the set of states with a positive partial transpose, and ENT corresponds to the set of entangled states. The intersection of PPT and ENT highlights a subset of bound entangled states.

state $\sigma$:

$$\begin{aligned} \max_{Y,Z,\rho} \quad & \mathrm{Tr}[Y] \\ \text{subject to} \quad & \mathrm{Tr}[M_i\rho] = m_i, \quad \text{for } i = 1,\dots,N, \\ & \mathrm{Tr}[\rho] = 1, \\ & \rho \succeq 0, \\ & \begin{pmatrix} \rho & Y + iZ \\ Y - iZ & \sigma \end{pmatrix} \succeq 0. \end{aligned} \tag{5.15}$$

Another significant application of SDPs is in calculating entanglement measures. For example, in Sec. 4.2.3, we noted that the geometric measure of entanglement for a quantum state $\rho$ can be expressed using the fidelity function as follows:

$$E_G(\rho) = \min_{\sigma\in\mathrm{SEP}} (1 - F(\rho,\sigma)) = 1 - \max_{\sigma\in\mathrm{SEP}} F(\rho,\sigma) = 1 - \left(\max_{\sigma\in\mathrm{SEP}} \sqrt{F(\rho,\sigma)}\right)^2, \tag{5.16}$$

where the computation of the square root of fidelity can be formulated as an SDP, as shown in Eq. (5.13).

Although the set of separable states (SEP) is convex, we cannot describe it using a finite number of hermiticity-preserving maps. As discussed in Sec. 4.1.4, this difficulty arises because the separability problem is equivalent to classifying all positive maps, a problem that is widely believed to be NP-hard.

In such cases, it is useful to find a *relaxation* of the problem, which is easier to solve and provides meaningful bounds. This is exactly the situation with the separability problem. Specifically, we can approximate the set of separable states (SEP) using the set of states with positive partial transpose (PPT), which forms an outer approximation of SEP, as depicted in Fig. 5.2. In other words, SEP is a convex subset of PPT, and PPT itself is also convex.

To make the problem computationally tractable, we can relax the original optimization problem by replacing the set SEP (separable states) with the set PPT (states with positive partial transpose). This relaxation allows the problem to be solved as an SDP:

$$\begin{aligned} \max_{Y,Z,\sigma} \quad & \mathrm{Tr}[Y] \\ \text{subject to} \quad & \mathrm{Tr}[\sigma] = 1, \\ & \sigma^{\mathrm{T}_A} \geq 0, \quad \sigma \geq 0, \\ & \begin{pmatrix} \rho & Y + iZ \\ Y - iZ & \sigma \end{pmatrix} \geq 0, \end{aligned} \tag{5.17}$$

where $\sigma^{\mathrm{T}_A}$ denotes the partial transpose of $\sigma$ with respect to subsystem $A$.

This relaxed optimization problem provides an *upper bound* for $\max_{\sigma \in \mathrm{SEP}} \sqrt{F(\rho,\sigma)}$. Consequently, since the geometric measure of entanglement is defined as:

$$E_G(\rho) = 1 - \left( \max_{\sigma \in \mathrm{SEP}} \sqrt{F(\rho,\sigma)} \right)^2, \tag{5.18}$$

the solution to the relaxed problem allows us to compute a *lower bound* for the geometric measure of entanglement $E_G(\rho)$.

### 5.1.3 Duality of semidefinite programming

For every *primal* SDP, as expressed in Eq. (5.8), there exists a corresponding *dual* SDP formulation. The optimal value of the dual SDP serves as an upper bound for the optimal value of the primal SDP. Moreover, under standard regularity conditions—which are typically satisfied by most problems encountered in quantum information theory—the optimal values of the primal and dual SDPs coincide.

Consequently, the duality framework not only simplifies the analysis of complex optimization problems but also enables efficient numerical algorithms for solving them, as many SDP solvers rely on dual formulations to compute solutions.

To derive the dual form, we first associate a Lagrange multiplier, also referred to as a *dual variable*, $Y_i$ for $i = 1, \dots, m$ to each equality constraint, and $Z_j$ for $j = 1, \dots, n$ to each inequality constraint. These dual variables are Hermitian operators. This leads to the following Lagrangian

function for the SDP:

$$\begin{aligned}\mathcal{L} &= \mathrm{Tr}[AX] + \sum_{i=1}^{m} \mathrm{Tr}[Y_i(B_i - \Phi_i(X))] + \sum_{j=1}^{n} \mathrm{Tr}[Z_j(C_j - \Gamma_j(X))], \\ &= \mathrm{Tr}\left[X\left(A - \sum_{i=1}^{m} \Phi_i^\dagger(Y_i) - \sum_{j=1}^{n} \Gamma_j^\dagger(Z_j)\right)\right] + \sum_{i=1}^{m} \mathrm{Tr}[Y_i B_i] + \sum_{j=1}^{n} \mathrm{Tr}[Z_j C_j],\end{aligned} \tag{5.19}$$

where the maps $\Phi_i^\dagger(\cdot)$ and $\Gamma_j^\dagger(\cdot)$ are the adjoint maps of $\Phi_i(\cdot)$ and $\Gamma_j(\cdot)$, respectively.

The Lagrangian is constructed by adding to the objective function two sets of terms derived from the constraints, along with the dual variables. For convenience, we refer to variables that satisfy the constraints as *feasible variables*, denoted by $X \in \mathcal{F}$. The optimal value of the primal SDP can then be expressed as:

$$\alpha = \max_{X \in \mathcal{F}} \mathrm{Tr}[AX]. \tag{5.20}$$

We now impose that the dual variable $Z_j$ must be positive semidefinite, *i.e.*, $Z_j \geq 0$. The reason for this constraint is that the Lagrangian will always be greater than or equal to the value of the objective function for any feasible variable $X$:

$$\mathcal{L} \geq \mathrm{Tr}[AX] \quad \text{for all } X \in \mathcal{F} \text{ whenever } Z_j \geq 0 \ \forall j, \tag{5.21}$$

since $C_j - \Gamma_j(X) \geq 0$ and, for any pair of positive semidefinite operators $A$ and $B$, it holds that $\mathrm{Tr}[AB] \geq 0$.

The second constraint we impose on the dual variables is:

$$A - \sum_{i=1}^{m} \Phi_i^\dagger(Y_i) - \sum_{j=1}^{n} \Gamma_j^\dagger(Z_j) = 0, \tag{5.22}$$

such that the Lagrangian $\mathcal{L}$ becomes independent of the primal variable $X$ [see the second line of Eq. (5.19)]. Consequently, the value of the Lagrangian, when evaluated for a feasible variable $X$ and for dual variables that satisfy the above constraints, reduces to:

$$\mathcal{L} = \sum_{i=1}^{m} \mathrm{Tr}[Y_i B_i] + \sum_{j=1}^{n} \mathrm{Tr}[Z_j C_j]. \tag{5.23}$$

Combining these constraints, we arrive at the dual optimization problem for Eq. (5.8):

$$
\begin{aligned}
\min_{Y_i, Z_j} \quad & \sum_{i=1}^{m} \mathrm{Tr}[Y_i B_i] + \sum_{j=1}^{n} \mathrm{Tr}[Z_j C_j], \\
\text{subject to} \quad & A - \sum_{i=1}^{m} \Phi_i^{\dagger}(Y_i) - \sum_{j=1}^{n} \Gamma_j^{\dagger}(Z_j) = 0, \\
& Z_j \succeq 0, \quad \text{for } j = 1, \dots, n.
\end{aligned}
\tag{5.24}
$$

As anticipated, the dual problem is itself a semidefinite program (SDP). Specifically, the objective function is a linear combination of the dual variables $Y_i$ and $Z_j$. The *dual feasible set*, denoted as $\tilde{\mathcal{F}}$, consists of all dual variables $(Y_1, \dots, Y_m, Z_1, \dots, Z_n)$ that satisfy the associated constraints. The optimal value of the dual SDP, represented by $\beta$, is given by:

$$
\beta = \min_{(Y_1,\dots,Y_m,Z_1,\dots,Z_n)\in\tilde{\mathcal{F}}} \sum_{i=1}^{m} \mathrm{Tr}[Y_i B_i] + \sum_{j=1}^{n} \mathrm{Tr}[Z_j C_j]. \tag{5.25}
$$

To illustrate these concepts, consider the problem of determining the maximum eigenvalue of a Hermitian operator $H$. This task can be formulated as the following optimization problem:

$$
\begin{aligned}
\max_{\rho} \quad & \mathrm{Tr}[H\rho], \\
\text{subject to} \quad & \mathrm{Tr}[\rho] = 1, \\
& \rho \succeq 0.
\end{aligned}
\tag{5.26}
$$

The first constraint, $\mathrm{Tr}[\rho] = 1$, introduces a scalar dual variable $y$, while the second constraint, $\rho \succeq 0$, introduces a dual variable $Z$. The corresponding Lagrangian is expressed as:

$$
\begin{aligned}
\mathcal{L} &= \mathrm{Tr}[H\rho] + y(1 - \mathrm{Tr}[\rho]) + \mathrm{Tr}[Z\rho], \\
&= \mathrm{Tr}[(H - yI + Z)\rho] + y.
\end{aligned}
\tag{5.27}
$$

To ensure that the Lagrangian remains bounded below, we impose the condition $Z \succeq 0$. Furthermore, setting $H - yI + Z = 0$ ensures that the Lagrangian $\mathcal{L}$ becomes independent of the primal variable $\rho$. Under these conditions, the Lagrangian simplifies to $\mathcal{L} = y$, leading to

the following dual SDP:

$$\begin{aligned}
\min_{y,Z} \quad & y, \\
\text{subject to} \quad & yI - Z = H, \\
& Z \succeq 0.
\end{aligned} \tag{5.28}$$

On closer inspection, we observe that the objective function does not depend on the dual variable $Z$. In this context, $Z$ serves as a *slack variable*. Specifically, we can combine the two constraints into a single inequality, $Z = yI - H \succeq 0$, allowing the dual SDP to be simplified as:

$$\begin{aligned}
\min_{y} \quad & y, \\
\text{subject to} \quad & yI \succeq H.
\end{aligned} \tag{5.29}$$

We now return to the important properties of *weak and strong duality*. Weak duality, which always holds, states that the optimal value of the dual SDP is an upper bound on the optimal value of the primal SDP, *i.e.*, $\alpha \leq \beta$.

Next, let us assume that the primal SDP satisfies two natural properties:

- The feasible set $\mathcal{F}$ is non-empty, *i.e.*, $\mathcal{F} \neq \emptyset$, and furthermore, there exists a *strictly feasible* solution. That is, if we replace all of the inequality constraints with *strict inequality* constraints,

$$\Gamma_j(X) \prec C_j \quad \text{for } j = 1, \ldots, n, \tag{5.30}$$

  while maintaining all of the equality constraints $\Phi_i(X) = B_i$ for $i = 1, \ldots, m$, then it is still possible to find a feasible $X$. Such an $X$ is called an *interior point* of the feasible set $\mathcal{F}$. This property can be interpreted as a type of robustness of the solution, meaning that small perturbations will not lead to infeasibility.
- The optimal value is finite, *i.e.*, $\alpha < \infty$. In other words, the optimization is not unbounded and has a non-trivial maximum value.

If the above two conditions hold, then strong duality exists, and the optimal values of the primal and dual SDPs coincide, *i.e.*, $\alpha = \beta$.

**Theorem 5.1.1 (Strong Duality of SDP [26])** When the primal SDP is *strictly feasible and*

*bounded*, strong duality holds. Specifically, for the following optimizations:

$$\begin{aligned} \alpha &= \max_{X \in \mathcal{F}} \operatorname{Tr}[AX], \\ \beta &= \min_{(Y_1,\ldots,Y_m,Z_1,\ldots,Z_n)\in\tilde{\mathcal{F}}} \sum_{i=1}^{m} \operatorname{Tr}[Y_i B_i] + \sum_{j=1}^{n} \operatorname{Tr}[Z_j C_j], \end{aligned} \tag{5.31}$$

we have $\alpha = \beta$.

This theorem is the main justification for referring to Eq. (5.24) as the dual SDP. When strong duality holds, the dual SDP can be considered a reformulation of the same optimization problem, and in many cases, it provides valuable insights into how to approach or interpret the problem.

## 5.2 Non-Convex Optimization

While convex optimization has proven effective in solving numerous problems in quantum information theory, its limitations are also evident. Firstly, many problems cannot be formulated as semidefinite programmings (SDPs) because they are fundamentally *non-convex*. For instance, when pure states are considered as the feasible set, it becomes clear that they form a non-convex set. Secondly, even if a problem meets the criteria for convex optimization, it may still be challenging to optimize efficiently due to the complexity of describing certain objects with convex constraints. An example of this is the separability problem discussed in Sec. 5.1.2.

The inherently non-convex nature of many quantum information problems highlights the need for a non-convex optimization framework. Interestingly, non-convex techniques can often provide a direct and efficient approach to addressing both convex and non-convex problems.

### 5.2.1 Gradient descent strategy

The *gradient descent strategy* [144] is a widely used method for finding the minimum of a differentiable, convex, or non-convex multivariate function $\mathcal{L}$, commonly referred to as the *cost function* or *loss function* (also known as the *objective function*). It is one of the most popular algorithms for solving *unconstrained* optimization problems, particularly in the field of machine learning [32]. The gradient descent method seeks to generate a sequence of parameters:

$$\boldsymbol{x}_1, \boldsymbol{x}_2, \ldots, \boldsymbol{x}_T, \tag{5.32}$$

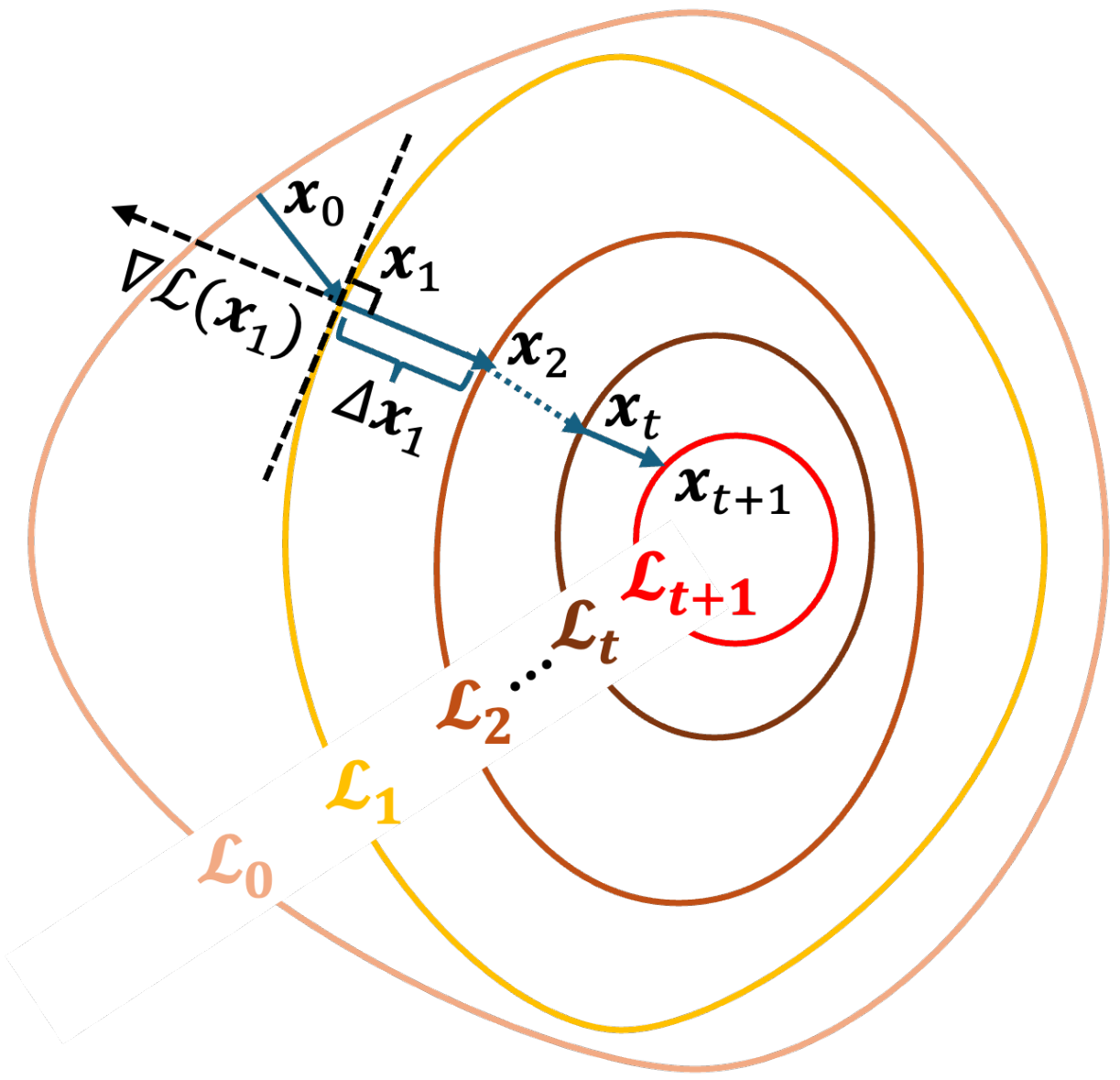

Figure 5.3: Geometric interpretation of gradient descent. Beginning from an arbitrary initial point $\boldsymbol{x}_0$, the algorithm iteratively updates the point by moving in the direction opposite to the gradient, which is perpendicular to the current level curves. A *level curve* represents a set of points $\boldsymbol{x}$ where the value of $\mathcal{L}(\boldsymbol{x})$ remains constant. This direction is mathematically proven to correspond to the steepest descent within the vicinity of the current point.

such that as $T \to \infty$, the objective function $\mathcal{L}(\boldsymbol{x}_T)$ approaches its optimal minimum value. At each iteration $t$, the parameters are updated by applying a step $\Delta\boldsymbol{x}_t$, as follows:

$$\boldsymbol{x}_{t+1} = \boldsymbol{x}_t + \Delta\boldsymbol{x}_t. \tag{5.33}$$

The most straightforward update strategy, known as the *steepest-descent method*, involves moving in the direction opposite to the gradient. This direction corresponds to the *steepest* descent, as gradients are orthogonal to *level curves* (or *level surfaces*), as illustrated in Fig. 5.3. Mathematically, this means we can set $\Delta\boldsymbol{x}_t = -\eta_t \nabla\mathcal{L}(\boldsymbol{x}_t)$, leading to the update rule:

$$\boldsymbol{x}_{t+1} = \boldsymbol{x}_t - \eta_t \nabla\mathcal{L}(\boldsymbol{x}_t), \tag{5.34}$$

where $\eta_t$ is the *step size* (or *learning rate*) at iteration $t$, and the algorithm typically begins with a random initial point $\boldsymbol{x}_0$. Using the first-order approximation, we can prove that $\mathcal{L}(\boldsymbol{x}_{t+1}) \leq \mathcal{L}(\boldsymbol{x}_t)$ for an appropriately chosen step size $\eta_t$:

$$\mathcal{L}(\boldsymbol{x}_{t+1}) = \mathcal{L}(\boldsymbol{x}_t - \eta_t \nabla\mathcal{L}(\boldsymbol{x}_t)) \approx \mathcal{L}(\boldsymbol{x}_t) - \eta_t (\nabla\mathcal{L}(\boldsymbol{x}_t))^T \nabla\mathcal{L}(\boldsymbol{x}_t) \leq \mathcal{L}(\boldsymbol{x}_t). \tag{5.35}$$

Because this method is guaranteed by the first-order approximation, it is also referred to as a *first-order iterative algorithm*. By following this iterative scheme, we ultimately converge to a *local minimum*.

A critical aspect of the gradient descent algorithm is determining the step size for each iteration. A straightforward approach involves using a constant step size, $\eta_t = \eta$, across all iterations. However, selecting an appropriate step size is challenging: a step size that is too small leads to slow convergence, while an excessively large step size may result in overshooting the minimum or even divergence.

To address the limitations of the steepest-descent method, various advanced optimization techniques have been developed to enhance the performance of gradient descent in diverse scenarios. These include gradient descent with momentum [145], which accelerates convergence by incorporating past gradients; Newton's method [146], which leverages second-order information (*i.e.*, Hessian matrix) to adaptively adjust the update direction and step size; and the limited-memory Broyden-Fletcher-Goldfarb-Shanno (L-BFGS) method [38], which approximates the Hessian matrix to efficiently handle large-scale optimization problems.

Gradient descent with momentum is an optimization algorithm designed to enhance both the convergence speed and stability of standard gradient descent. It is particularly effective in scenarios where gradients exhibit oscillatory behavior, such as in high-curvature regions, or when progress is slow due to small gradient magnitudes.

The concept of momentum is inspired by physics, where a moving object tends to maintain its velocity unless acted upon by an external force. Similarly, in optimization, momentum builds "velocity" in a consistent direction, reducing oscillations and smoothing the optimization trajectory for greater efficiency. Mathematically, the momentum update rule is:

$$\begin{cases} \boldsymbol{v}_{t+1} = \beta \boldsymbol{v}_t + \eta \nabla \mathcal{L}(\boldsymbol{x}_t), \\ \boldsymbol{x}_{t+1} = \boldsymbol{x}_t - \boldsymbol{v}_{t+1}, \end{cases} \tag{5.36}$$

where the algorithm maintains a "velocity" vector $\boldsymbol{v}_t$, which accumulates a weighted sum of past gradients. The momentum coefficient $\beta$ determines the contribution of prior gradients to the current "velocity" $\boldsymbol{v}_{t+1}$. As a result, the momentum term accelerates convergence in shallow regions of the optimization landscape while dampening oscillations in high-curvature regions.

Newton's method is an advanced optimization technique that extends gradient descent by incorporating *second-order* information about the objective function. While first-order iterative algorithms rely on the gradient (the first derivative) to update parameters, Newton's method leverages the *Hessian matrix* (a matrix of second-order partial derivatives) to dynamically adjust both the update direction and step size. This makes Newton's method particularly effective at minimizing objective functions with complex curvature, as it accounts for the local geometry of the function.

To derive Newton's method, we consider the second-order Taylor series expansion of the objective function $\mathcal{L}(\boldsymbol{x})$ around the current parameter $\boldsymbol{x}_t$:

$$\mathcal{L}(\boldsymbol{x}) \approx \mathcal{L}(\boldsymbol{x}_t) + \nabla\mathcal{L}(\boldsymbol{x}_t)^T(\boldsymbol{x} - \boldsymbol{x}_t) + \frac{1}{2}(\boldsymbol{x} - \boldsymbol{x}_t)^T H_{\mathcal{L}}(\boldsymbol{x}_t)(\boldsymbol{x} - \boldsymbol{x}_t), \tag{5.37}$$

where $\nabla\mathcal{L}(\boldsymbol{x}_t)$ is the gradient at the current parameter $\boldsymbol{x}_t$, and $H_{\mathcal{L}}(\boldsymbol{x}_t)$ is the Hessian matrix at $\boldsymbol{x}_t$. The Hessian matrix is defined as:

$$H_{\mathcal{L}} = \begin{pmatrix} \frac{\partial^2\mathcal{L}}{\partial x_1^2} & \frac{\partial^2\mathcal{L}}{\partial x_1 \partial x_2} & \cdots & \frac{\partial^2\mathcal{L}}{\partial x_1 \partial x_n} \\ \frac{\partial^2\mathcal{L}}{\partial x_2 \partial x_1} & \frac{\partial^2\mathcal{L}}{\partial x_2^2} & \cdots & \frac{\partial^2\mathcal{L}}{\partial x_2 \partial x_n} \\ \vdots & \vdots & \ddots & \vdots \\ \frac{\partial^2\mathcal{L}}{\partial x_n \partial x_1} & \frac{\partial^2\mathcal{L}}{\partial x_n \partial x_2} & \cdots & \frac{\partial^2\mathcal{L}}{\partial x_n^2} \end{pmatrix}, \tag{5.38}$$

where the element in the $i$th row and $j$th column is given by $(H_{\mathcal{L}})_{i,j} = \frac{\partial^2\mathcal{L}}{\partial x_i \partial x_j}$.

To minimize the second-order approximation in Eq. (5.37), we set the gradient of the approximation to zero. This yields the update rule for the next parameter:

$$\boldsymbol{x}_{t+1} = \boldsymbol{x}_t - H_{\mathcal{L}}(\boldsymbol{x}_t)^{-1}\nabla\mathcal{L}(\boldsymbol{x}_t), \tag{5.39}$$

where $H_{\mathcal{L}}(\boldsymbol{x}_t)^{-1}$ is the inverse of the Hessian matrix.

Although Newton's method performs effectively in certain scenarios, it encounters several challenges. First, the computation of the Hessian matrix and its inverse is computationally demanding, particularly for high-dimensional problems. For $n$ parameters, the cost of inverting the Hessian matrix is $\mathcal{O}(n^3)$ [147], which becomes infeasible for large-scale optimization tasks. Second, if the Hessian matrix is not positive definite, the update direction may fail to lead to a minimum. Third, Newton's method exhibits local convergence, meaning it is effective only when the initial point lies sufficiently close to the minimum. A poorly selected starting point can result in slow or suboptimal convergence.

The L-BFGS algorithm is a popular optimization method tailored for large-scale problems. It extends the BFGS method, a quasi-Newton approach that approximates the inverse of the Hessian matrix rather than computing it explicitly. As a result, its computational complexity is reduced to $\mathcal{O}(n^2)$, compared to the $\mathcal{O}(n^3)$ complexity of Newton's method. However, the origi-nal BFGS algorithm requires storing dense matrices, which becomes impractical for extremely large problems. To overcome the memory and computational drawbacks of BFGS, the L-BFGS method was introduced. L-BFGS avoids storing the full Hessian matrix and instead leverages a *limited amount of memory* to efficiently approximate the Hessian.

The properties of gradient descent depend on the characteristics of the objective function as well as the specific variant of gradient descent employed. The assumptions made about the objective function influence the convergence rate and other provable properties. For instance, if the objective function is assumed to be strongly convex and Lipschitz smooth, gradient descent with a fixed step size is guaranteed to converge linearly [25]. Weaker assumptions, such as convexity without strong convexity or smoothness, typically result in slower convergence rates, such as sublinear convergence, and may require more sophisticated step size selection strategies.

It is also important to note that, unlike convex optimization problems, where global optimality is guaranteed under certain conditions, the gradient descent strategy in non-convex optimization does not ensure a globally optimal solution. However, it does guarantee *local optimality* under appropriate assumptions about the problem structure and gradient behavior.

### 5.2.2 Automatic differentiation technique

While numerous optimizers have been proposed for gradient-based algorithms, a critical challenge remains: how to compute gradients both efficiently and accurately.

Methods for computing gradients in computer programs can be categorized into four main approaches:

- *Manual differentiation*, where gradients are calculated and coded by hand.
- *Numerical differentiation*, which uses finite-difference approximations.
- *Symbolic differentiation*, performed through expression manipulations in computer algebra systems such as Mathematica.
- *Automatic differentiation*, also referred to as algorithmic differentiation.

Historically, researchers have invested considerable effort in manually deriving analytical gradients. However, manual differentiation is not only labor-intensive but also highly susceptible to errors. Among the alternatives, numerical differentiation is straightforward to implement but often suffers from inaccuracies due to round-off and truncation errors [148]. Furthermore, it scales poorly for gradient calculations, making it unsuitable for large-scale problems. Symbolic differentiation addresses some of the limitations of manual and numerical methods but often produces overly complex and unwieldy expressions, a phenomenon known as "expression swell" [149]. Additionally, both manual and symbolic approaches require objective functions to be expressed in closed forms, significantly restricting their flexibility and applicability.

In this section, we focus on the fourth powerful technique, automatic differentiation (AD) [150], which combines the efficiency of numerical differentiation with the precision of symbolic differentiation. AD is implemented in many deep learning frameworks, such as PyTorch [151] and TensorFlow [152].

Automatic differentiation (AD) fundamentally represents the objective function as a *computational graph*, which consists of a sequence of *elementary operations* (*e.g.*, addition, subtraction, multiplication, division) and *elementary functions* (*e.g.*, exponential, logarithmic, trigonometric functions). By systematically applying the *chain rule* to these operations, AD efficiently computes partial derivatives of any order, requiring only a small constant overhead in arithmetic operations compared to the original program. These derivatives are then used to compute the gradient of the objective function with respect to its variables in an efficient manner.

AD operates in two primary modes: *forward mode* [153] and *reverse mode* [154]. To illustrate, consider the example function:

$$y = f(x_1, x_2) = \ln x_1 + x_1 x_2 - \sin\big(x_2\big). \tag{5.40}$$

This function can be decomposed into a sequence of elementary operations and functions, as depicted in Fig. 5.4. The computational graph provides a structured representation of the dependencies between variables. Starting from the inputs $x_1$ and $x_2$, one can traverse the graph to compute the final output, $y = v_5$. The key question then arises: how can the gradient

$$\nabla f(x_1, x_2) = \left(\frac{\partial y}{\partial x_1}, \frac{\partial y}{\partial x_2}\right) \tag{5.41}$$

from the computational graph? Assume we aim to calculate the partial derivative $\frac{\partial y}{\partial x_1}$. Using the chain rule, we perform forward derivative calculations with the intermediate outcomes $\{v_i, \frac{\partial v_i}{\partial x_1}\}$

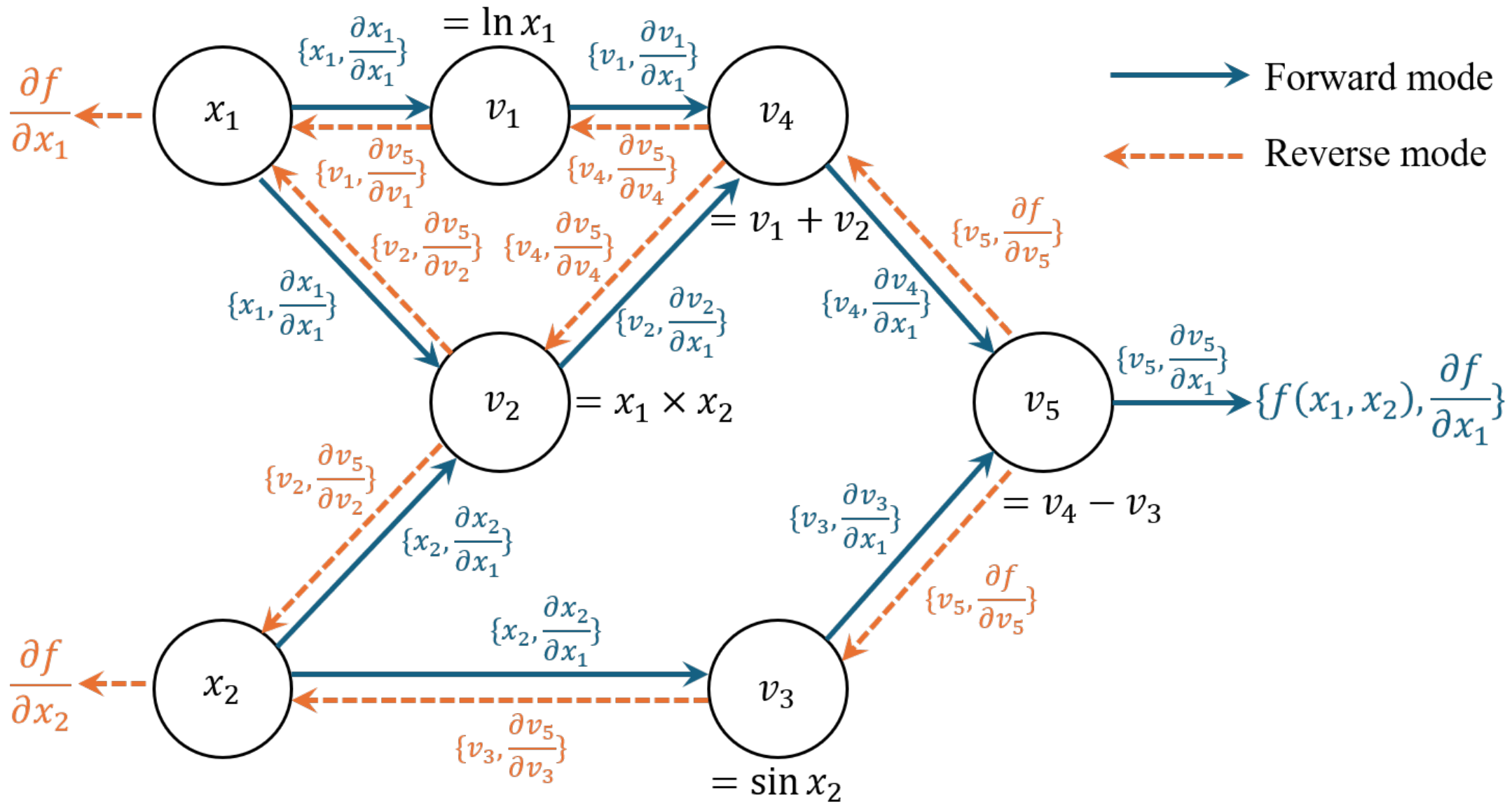

Figure 5.4: The computational graph of the function $f(x_1, x_2) = \ln x_1 + x_1 x_2 - \sin(x_2)$. $v_1, v_2, \ldots, v_5$ are used to record the intermediate variables. The solid blue lines represent the forward mode, while the dashed orange lines denote the reverse mode. In the forward mode, we use the $\{v_i, \frac{\partial v_i}{\partial x_j}\}$ to obtain the partial derivative $\frac{\partial f}{\partial x_j}$. In contrast, $\{v_i, \frac{\partial f}{\partial v_i}\}$ is used in the reverse mode to obtain the $(\frac{\partial f}{\partial x_1}, \frac{\partial f}{\partial x_2})$ in a single backward pass.

until we obtain $\frac{\partial y}{\partial x_1}$, denoted by $\dot{y}$. Mathematically, this process is written as:

| **Variable** | **Expression** |
|---|---|
| $\dot{x}_1$ | $\frac{\partial x_1}{\partial x_1} = 1$ |
| $\dot{x}_2$ | $\frac{\partial x_2}{\partial x_1} = 0$ |
| $\dot{v}_1$ | $\dot{x}_1 / x_1$ |
| $\dot{v}_2$ | $\dot{x}_1 \cdot x_2 + \dot{x}_2 \cdot x_1$ |
| $\dot{v}_3$ | $\dot{x}_2 \cdot \cos(x_2)$ |
| $\dot{v}_4$ | $\dot{v}_1 + \dot{v}_2$ |
| $\dot{v}_5$ | $\dot{v}_4 - \dot{v}_3$ |
| $\dot{y}$ | $\dot{v}_5$ |

In the forward mode, we compute $\dot{v} = \frac{\partial v}{\partial x_1}$, focusing on the derivatives of all variables with respect to a single input variable. Conversely, the reverse mode considers $\bar{v} = \frac{\partial y}{\partial v_i}$ for all intermediate variables $v_i$, *i.e.*, the partial derivatives of the output with respect to all variables.

This process can be expressed as:

$$
\begin{aligned}
\bar{v}_5 &= \bar{y} = \frac{\partial y}{\partial y} = 1, \\
\bar{v}_4 &= \bar{v}_5 \times \frac{\partial v_5}{\partial v_4} = \bar{v}_5 \times 1, \\
\bar{v}_3 &= \bar{v}_5 \times \frac{\partial v_5}{\partial v_3} = \bar{v}_5 \times (-1), \\
\bar{v}_2 &= \bar{v}_4 \times \frac{\partial v_4}{\partial v_2} = \bar{v}_4 \times 1, \\
\bar{v}_1 &= \bar{v}_4 \times \frac{\partial v_4}{\partial v_1} = \bar{v}_4 \times 1, \\
\bar{x}_1 &= \bar{v}_2 \times \frac{\partial v_2}{\partial x_1} + \bar{v}_1 \times \frac{\partial v_1}{\partial x_1} = \bar{v}_2 \times x_2 + \frac{\bar{v}_1}{x_1}, \\
\bar{x}_2 &= \bar{v}_3 \times \frac{\partial v_3}{\partial x_2} + \bar{v}_2 \times \frac{\partial v_2}{\partial x_2} = \bar{v}_3 \times \cos x_2 + \bar{v}_2 \times x_1.
\end{aligned}
\tag{5.42}
$$

The reverse mode, commonly referred to as *backpropagation*, is particularly advantageous for functions with a high-dimensional input space and a single scalar output. It efficiently computes the partial derivatives with respect to all input variables in a single backward pass. Figure 5.4 illustrates the distinction between forward and reverse modes within the computational graph, highlighting the efficiency of the reverse mode in scenarios involving multiple input variables.

The automatic differentiation technique is critically important for modern gradient-based algorithms because it enables the efficient, accurate, and scalable computation of derivatives, which form the backbone of these methods. Gradient-based algorithms depend on the gradients of objective functions to determine both the direction and magnitude of parameter updates. AD ensures that these gradients are calculated precisely and efficiently, regardless of the complexity of the function being optimized.

### 5.2.3 Manifold optimization framework

While advanced gradient-based algorithms and automatic differentiation techniques provide powerful tools for optimization, they cannot be directly applied to quantum information problems due to the inherent constraints of quantum objects.

Interestingly, many quantum objects can be characterized as residing on specific *manifolds* with well-defined geometric structures. In mathematical terms, a manifold is a topological space that locally resembles Euclidean space but may exhibit more intricate global geometry. More formally, an $n$-dimensional manifold, or $n$-*manifold*, is a topological space in which every point has a neighborhood that is *homeomorphic* to an *open* subset of $n$-dimensional Euclidean space. Examples of one-dimensional manifolds include lines and circles, while two-dimensional manifolds include planes, spheres, and torus.

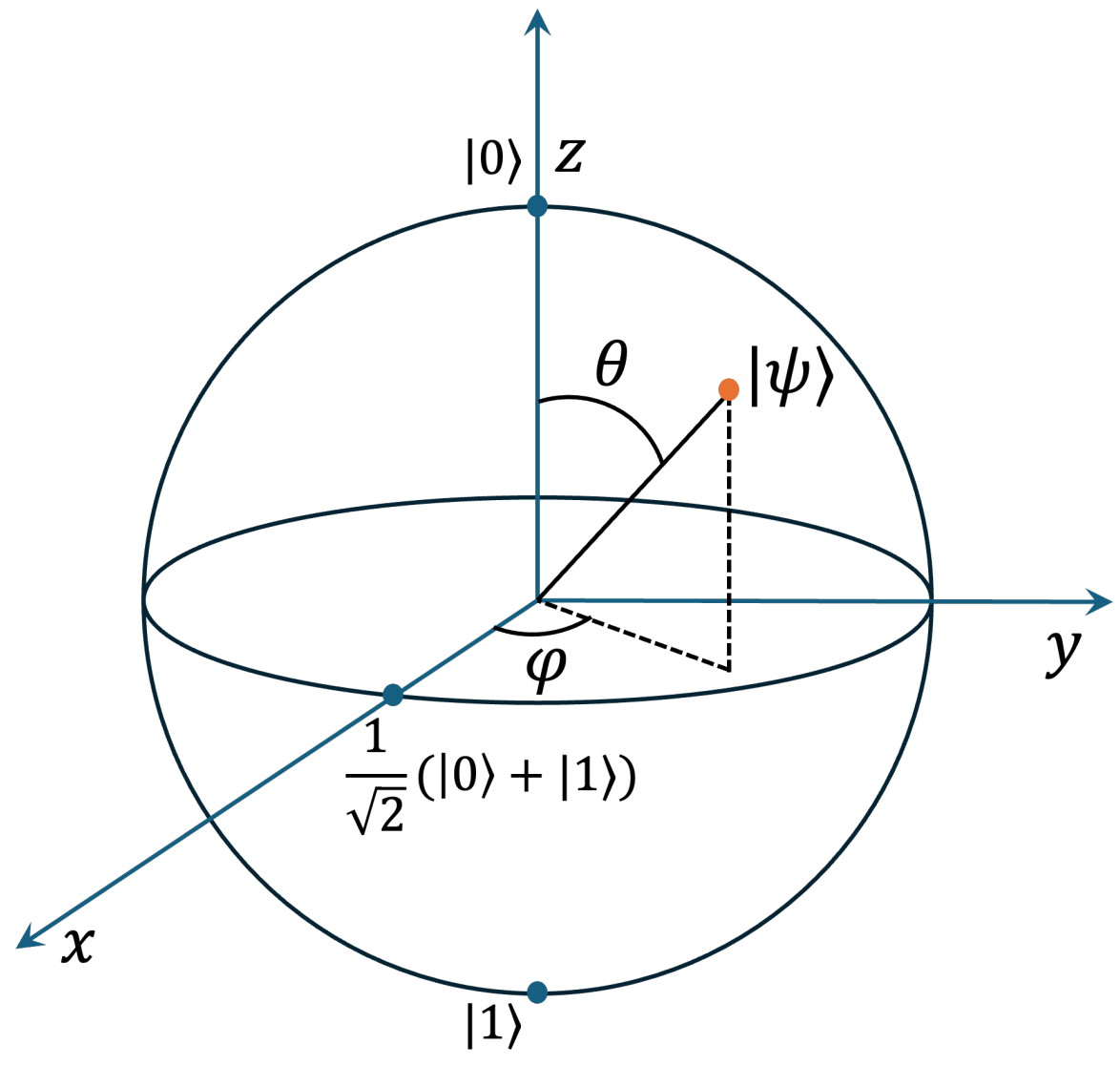

Figure 5.5: The Bloch sphere representing single-qubit pure states. Points on the surface correspond to pure states, parameterized by the Bloch vector $(\theta, \varphi)$. The sphere is a mathematical manifold.

A typical example in quantum information is the pure state of a single qubit, which can be expressed as:

$$|\psi\rangle = \cos\frac{\theta}{2}|0\rangle + e^{i\varphi}\sin\frac{\theta}{2}|1\rangle, \tag{5.43}$$

where $0 \leq \theta \leq \pi$ and $0 \leq \varphi < 2\pi$. Here, $|0\rangle$ and $|1\rangle$ denote the computational basis states of the qubit. The parameters $(\theta, \varphi)$ provide a geometric representation of the pure state space of qubits, commonly referred to as the *Bloch vector*. The set of all possible pure qubit states forms a sphere known as the *Bloch sphere*, as depicted in Fig. 5.5. Mathematically, the Bloch sphere corresponds to a sphere manifold embedded in $\mathbb{R}^3$.

More generally, for a $d$-dimensional pure state, the state can be expressed as:

$$|\psi\rangle = \sum_{k=0}^{d-1}(x_{2k} + ix_{2k+1})|k\rangle, \tag{5.44}$$

where $x_0, x_1, \ldots, x_{2d-1}$ are real numbers satisfying the normalization condition $\|\boldsymbol{x}\|_2 = 1$. This implies that a $d$-dimensional pure state can be represented as a $2d$-dimensional real vector of unit norm. The set of such vectors forms a higher-dimensional sphere, which is also a manifold:

**Definition 15 ($n$-Sphere Manifold)** The $n$-sphere manifold $\mathbb{S}^n$ is the set of all unit-norm vectors in $\mathbb{R}^{n+1}$, defined as:

$$\mathbb{S}^n = \left\{ \boldsymbol{x} \in \mathbb{R}^{n+1} \,:\, \|\boldsymbol{x}\|_2 = \sqrt{\sum_i x_i^2} = 1 \right\}. \tag{5.45}$$

Clearly, the Bloch sphere is a special case of the $n$-sphere manifold with $n = 2$, while general $d$-dimensional pure states can be described by a $(2d-1)$-sphere manifold. However, it is worth noting that Eq. (5.44) does not account for the global phase effect, meaning that one parameter is redundant in representing these states.

Another widely used manifold is the so-called *probability manifold*, which represents the interior points inside a simplex:

**Definition 16 (Probability Manifold)** The probability manifold is the set of all probability vectors with positive entries that add up to one, *i.e.*,

$$\Delta_+^n = \left\{ \boldsymbol{p} \in \mathbb{R}^{n+1} \,:\, p_i > 0 \;\forall i, \|\boldsymbol{p}\|_1 = \sum_i p_i = 1 \right\}. \tag{5.46}$$

The probability manifold provides a natural framework for representing probability distributions arising from quantum systems and measurements. Furthermore, below we summarize several common manifolds and their corresponding quantum objects that frequently appear in quantum information problems:

| **Quantum Object** | **Manifold** | **Definition** |
|---|---|---|
| Probability | Positive number $\mathbb{R}_+$ | $\{p \in \mathbb{R} : p > 0\}$ |
| Probability vector | Probability manifold $\Delta_+^n$ | $\{\boldsymbol{p} \in \mathbb{R}^{n+1} : p_i > 0 \;\forall i, \|\boldsymbol{p}\|_1 = 1\}$ |
| Pure state | Sphere manifold $\mathbb{S}^n$ | $\{\boldsymbol{x} \in \mathbb{R}^{n+1} : \|\boldsymbol{x}\|_2 = 1\}$ |
| Mixed state | Full-rank density matrix $\mathbb{D}(n)$ | $\{\rho \in \mathbb{C}^{n\times n} : \rho = \rho^\dagger, \mathrm{Tr}[\rho] = 1, \rho \succ 0\}$ |
| Observable | Hermitian matrix $\mathbb{H}(n)$ | $\{H \in \mathbb{C}^{n\times n} : H = H^\dagger\}$ |
| Quantum circuit | Unitary matrix $\mathbb{U}(n)$ | $\{U \in \mathbb{C}^{n\times n} : U^\dagger U = I\}$ |
| Quantum channel | Stiefel manifold $\mathbb{St}(n,p)$ | $\{V \in \mathbb{C}^{n\times p} : V^\dagger V = I\}$ |

Table 5.1: Common quantum objects and their corresponding manifolds with definitions. Here, the superscripts represent the dimensions of manifolds, and $(*)$ or $(*, *)$ indicate the dimensions of the space that the manifold embeds.

Different manifolds can also be combined to form new manifolds. Specifically, the *Cartesian product* provides a way to combine two manifolds to create a new manifold that incorporates both in a natural way:

**Theorem 5.2.1 (Product Manifold [44])** Let $\mathcal{M}$ and $\mathcal{N}$ be two smooth manifolds of dimension $m$ and $n$, respectively. The Cartesian product $\mathcal{M} \times \mathcal{N}$ is also a manifold with dimension $m + n$.

For instance, if you combine a line $\mathbb{R}$ and a circle $\mathbb{S}^1$, the result is an infinite cylinder. Each point on the cylinder corresponds to a pair $(x, \theta)$, where $x$ is a point on the line and $\theta$ represents a point on the circle. Similarly, combining two circles $\mathbb{S}^1 \times \mathbb{S}^1$ results in a torus.

The product manifold plays an important role in describing quantum objects. For example, a fully product state $|\psi_1\rangle \otimes |\psi_2\rangle \otimes \cdots \otimes |\psi_n\rangle$ in the composite Hilbert space $\mathcal{H} = \mathcal{H}_1 \otimes \mathcal{H}_2 \otimes \cdots \otimes \mathcal{H}_n$, with dimensions $d_1, d_2, \dots, d_n$, can be described by the product manifold of sphere manifolds:

$$\mathbb{S}^{2d_1-1} \times \mathbb{S}^{2d_2-1} \times \cdots \times \mathbb{S}^{2d_n-1}. \tag{5.47}$$

In general, for any quantum object that a single manifold cannot characterize, we can decompose it into basic objects with known manifolds and combine them using the product manifold formalism, similar to the idea of *modularization*. However, from the resulting product manifold, it is often unclear how these components are mathematically combined to describe the desired quantum object. Fortunately, tensor network representations provide a clear and convenient framework to reveal the relationships between different manifolds, which can be represented as tensors.

For example, consider the separability problem, which is notoriously difficult due to the complexity of characterizing separable states. In the context of convex optimization, the PPT relaxation is often used to approximate separable states by embedding them into a larger convex set. Within the manifold framework, however, there is a more intuitive approach based on the definition of separable states. Specifically, a bipartite separable state in $\mathcal{H}_A \otimes \mathcal{H}_B$ can be expressed as:

$$\rho = \sum_{i=1}^{n} p_i (|\psi_A^{(i)}\rangle \otimes |\psi_B^{(i)}\rangle)(\langle\psi_A^{(i)}| \otimes \langle\psi_B^{(i)}|), \tag{5.48}$$

where $n$ is bounded by Carathéodory's theorem (see Theorem 3.1.2), *i.e.*, $n \leq (d_A d_B)^2$.

Clearly, for any product state $|\psi_A^{(i)}\rangle \otimes |\psi_B^{(i)}\rangle$, it can be characterized by the product manifold $\mathbb{S}^{2d_A-1} \times \mathbb{S}^{2d_B-1}$. For the probability vector $\boldsymbol{p} = (p_1, p_2, \dots, p_n)$, the corresponding manifold is $\Delta_+^{n-1}$. Thus, in principle, the set of separable states can be described using these manifolds.

Specifically, we first combine two pure states $|\psi_A^{(i)}\rangle$ and $|\psi_B^{(i)}\rangle$ to form a product pure state

$|\psi_A^{(i)}\rangle \otimes |\psi_B^{(i)}\rangle$ using the tensor product and index merging:

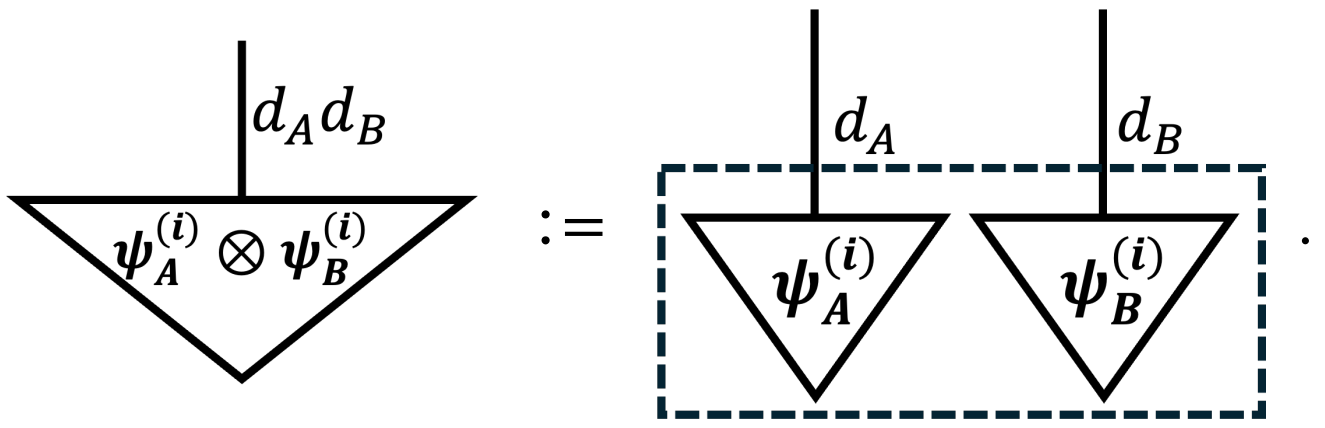

(5.49)

Next, we stack $n$ separate product pure states to form a higher-order tensor, denoted by $\Psi_A \otimes \Psi_B$:

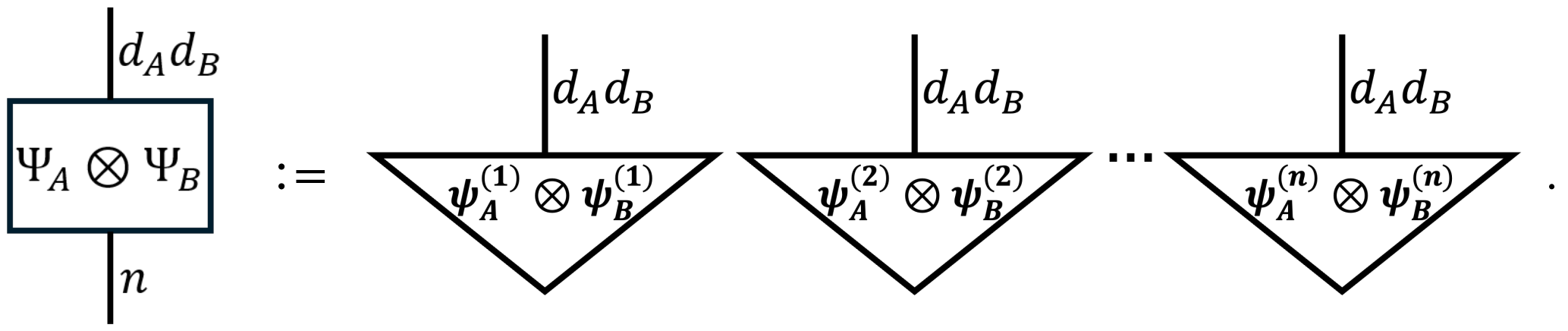

(5.50)

Finally, for the set of separable states, we can characterize them as follows:

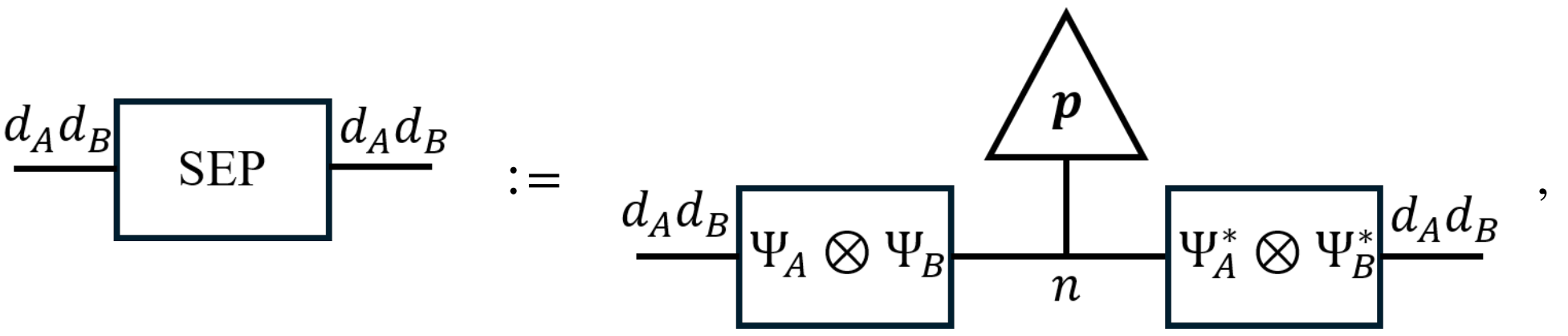

(5.51)

where $\boldsymbol{p}$ is a vector tensor that represents the probabilities associated with each pure product state, and $\Psi_A^* \otimes \Psi_B^*$ indicates a conjugation operation on $\Psi_A \otimes \Psi_B$. With a sufficiently large $n$, this characterization becomes satisfactory.

With the description of quantum objects in the manifold scenario, numerous quantum information problems can be cast into the so-called *manifold optimization framework*. In this framework, the problem is formalized as an optimization problem constrained on some manifold $\mathcal{M}$ with certain properties, *i.e.*,

$$\min_{x \in \mathcal{M}} f(x). \tag{5.52}$$

However, optimization on manifolds remains both theoretically and practically challenging due to the inherent complexity of the objects involved and the nonlinear structure of the constraints.

A pivotal advancement in this domain was achieved by Edelman, Arias, and Smith, who developed efficient algorithms for optimization on manifolds, with a particular focus on eigenvalue problems and matrix factorizations [155]. Their pioneering work laid the foundation for practical methods tailored to manifold-constrained optimization. Subsequently, the influential book by Absil, Mahony, and Sepulchre [156] established a comprehensive framework for *Riemannian optimization*, introducing key concepts such as Riemannian gradients, Hessians, and geodesics. This formalization of differential geometry in optimization significantly enhanced both convergence rates and computational efficiency.

To integrate the strengths of gradient descent and automatic differentiation within the manifold optimization framework, we propose a simplified yet effective approach for solving manifold-constrained optimization problems. This method diverges from traditional Riemannian geometry-based algorithms, prioritizing ease of implementation while achieving competitive performance.

### 5.2.4 Manifold trivialization approach

Trivialization offers a parametrization strategy that converts manifold-constrained problems into unconstrained ones [46]. Specifically, this method relies on a subjective map $\phi$ that maps Euclidean space $\mathbb{R}^n$ onto the target manifold $\mathcal{M}$, ensuring that

$$\min_{x\in\mathcal{M}} f(x) = \min_{\boldsymbol{\theta}\in\mathbb{R}^n} f(\phi(\boldsymbol{\theta})). \tag{5.53}$$

This process, referred to as *trivialization*, defines $\phi$ as the trivialization map. Consequently, trivialization transforms constrained optimization on a nontrivial manifold into an unconstrained problem in trivial Euclidean space, enabling the direct application of advanced optimizers or gradient-based algorithms. The dimension of the associated Euclidean space corresponds to the number of real parameters required to parametrize the target manifold.

The complexity of the resulting optimization problem largely depends on the properties of the trivialization map $\phi$. If $\phi$ is a diffeomorphism—meaning it is bijective, continuous, differentiable, and has a continuously differentiable inverse—it guarantees that no additional saddle points or local minima are introduced during trivialization. Therefore, the effectiveness of manifold optimization is strongly influenced by the quality of the trivialization.

A well-known example is the *parameterized quantum circuit* (PQC), which represents a quantum circuit designed with adjustable parameters. These parameters allow the modification of the behavior of quantum gates within the circuit, enabling it to execute specific computations or optimizations. PQCs play a foundational role in hybrid quantum-classical algorithms and are widely used in quantum machine learning [157] and variational quantum algorithms [158]. Mathematically, an $n$-qubit quantum circuit corresponds to a unitary matrix $U$, which lies on the unitary matrix manifold $\mathbb{U}(2^n)$. A PQC realizes the desired quantum circuit through a parameterized unitary matrix $U(\boldsymbol{\theta})$ with tunable parameters $\boldsymbol{\theta}$, as illustrated by the 3-qubit example in

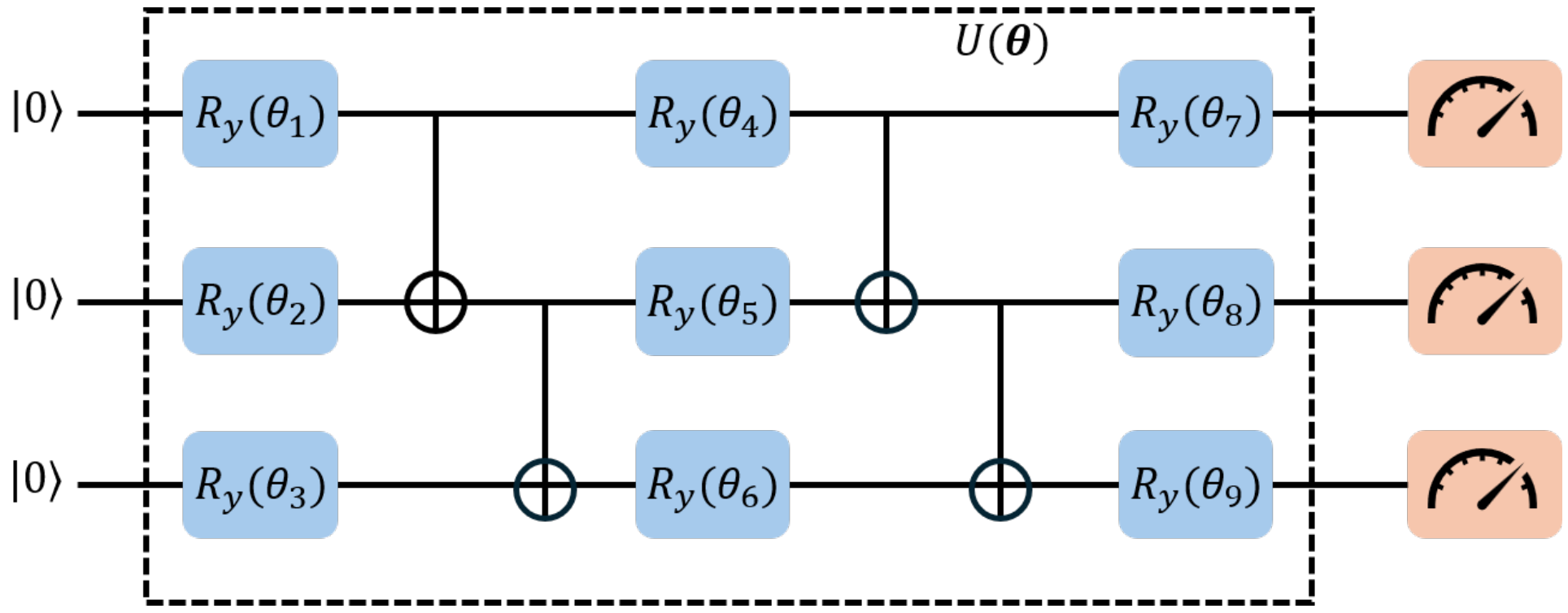

Figure 5.6: Schematic diagram of parameterized quantum circuit $U(\boldsymbol{\theta})$. Single-qubit gate $R_y(\theta) = e^{-i\frac{\theta}{2}\sigma_y}$ and two-qubit gate controlled-NOT (CNOT) are used to construct the circuit.

Fig. 5.6. Therefore, a PQC can be viewed as a specific instance of trivialization on the unitary matrix manifold.

Let us start with the simplest manifold, which consists solely of positive numbers. For this manifold, a common trivialization map is the SoftPlus function [159]:

**Theorem 5.2.2 (Trivialization on Positive Numbers)** The following map $\phi$ provides a trivialization on the set of positive numbers $\mathbb{R}_+$:

$$\text{SoftPlus}(\theta) = \ln\left(1 + e^{\theta}\right) : \mathbb{R} \to \mathbb{R}_+. \tag{5.54}$$

The SoftPlus function is a smooth approximation of the rectified linear unit (ReLU) [160] activation function and is widely utilized in machine learning and deep learning. Its differentiability everywhere makes it more mathematically convenient than ReLU. An alternative and simpler approach is the exponential function $\phi(\theta) = e^{\theta}$; however, this may result in the issue of *gradient explosion*.

Using the trivialization of positive numbers, we can derive a trivialization for the probability manifold by applying a *normalization* operation under the $\ell^1$-norm:

**Theorem 5.2.3 (Trivialization on Probability Manifold)** Let the trivialization map of positive numbers $\mathbb{R}_+$ be denoted as $g$. The following map $\phi$ provides a trivialization on the probability manifold $\Delta_+^n$:

$$\begin{aligned} \phi(\boldsymbol{\theta}) &= (p_1, p_2, \cdots, p_{n+1}) : \mathbb{R}^{n+1} \to \Delta_+^n, \\ p_i &= \frac{g(\theta_i)}{\|g(\boldsymbol{\theta})\|_1} = \frac{g(\theta_i)}{\sum_j g(\theta_j)}. \end{aligned} \tag{5.55}$$

If we choose the exponential map as the trivialization for positive numbers, the resulting trivialization map for the probability manifold is the well-known SoftMax function [161]:

$$\text{SoftMax}(\theta_i) = \frac{e^{\theta_i}}{\sum_j e^{\theta_j}}, \tag{5.56}$$

which is widely used in classification problems [162], where it interprets raw model outputs as probabilities over different classes.

For the sphere manifold $\mathbb{S}^n$, the trivialization process is analogous to that of the probability manifold. In this case, the sphere manifold is obtained through a normalization operation under the $\ell^2$-norm:

**Theorem 5.2.4 (Trivialization on $n$-Sphere Manifold)** The following map $\phi$ provides a trivialization on $\mathbb{S}^n$:

$$\phi(\boldsymbol{\theta}) = \frac{\boldsymbol{\theta}}{\|\boldsymbol{\theta}\|_2} = \frac{\boldsymbol{\theta}}{\sqrt{\sum_i \theta_i^2}} \,:\, \mathbb{R}^{n+1}\backslash\{\mathbf{0}\} \to \mathbb{S}^n. \tag{5.57}$$

It is important to note that in this trivialization, the domain is not exactly $\mathbb{R}^{n+1}$. However, this limitation is acceptable because the defects in the Euclidean space do not form a region with "volume". Consequently, the probability of encountering one of these ill-defined regions is zero when using gradient-based algorithms with random initializations. For simplicity, these measure-zero defects are often ignored, and the domain is conveniently represented as $\mathbb{R}^{n+1}$.

Finally, let us consider the domain of matrix manifolds. First, we introduce the Hermitian matrix manifold $\mathbb{H}(n)$, which represents the set of quantum observables. A common trivialization for Hermitian matrices is the *hermitianization* process:

**Theorem 5.2.5 (Trivialization on Hermitian Matrix)** The trivialization $\phi$ for Hermitian matrices is performed in two steps. First, $n^2$ real parameters are used to define a real square matrix, denoted by $g$. Second, this real square matrix is transformed into a Hermitian complex matrix via hermitianization, denoted by $h$:

$$\begin{aligned}
g(\boldsymbol{\theta}) &= \begin{pmatrix} \theta_1 & \theta_{n+1} & \cdots & \theta_{n^2-n+1} \\ \theta_2 & \theta_{n+2} & \cdots & \theta_{n^2-n+2} \\ \vdots & \vdots & \vdots & \vdots \\ \theta_n & \theta_{2n} & \cdots & \theta_{n^2} \end{pmatrix} : \mathbb{R}^{n^2} \to \mathbb{R}^{n\times n}, \\
h(A) &= (A + A^T) + i(A - A^T) \,:\, \mathbb{R}^{n\times n} \to \mathbb{H}(n), \\
\phi(\boldsymbol{\theta}) &= h(g(\boldsymbol{\theta})) \,:\, \mathbb{R}^{n^2} \to \mathbb{H}(n).
\end{aligned} \tag{5.58}$$

From the trivialization map of the Hermitian matrix, we can derive a trivialization for the unitary matrix manifold using the exponential map:

**Theorem 5.2.6 (Trivialization on Unitary Matrix)** Let $g$ be any trivialization map of the Hermitian matrix $\mathbb{H}(n)$. The following map $\phi$ provides a trivialization of the unitary matrix manifold $\mathbb{U}(n)$:

$$\phi(\boldsymbol{\theta}) = \exp(i \cdot g(\boldsymbol{\theta})) \; : \; \mathbb{R}^{n^2} \to \mathbb{U}(n). \tag{5.59}$$

This trivialization map is more general than the concept of PQC illustrated in Fig. 5.6, as it is not restricted to $n$-qubit quantum circuits but applies to arbitrary $d$-dimensional quantum systems. In this context, the Hermitian matrix corresponds to a more physical quantity called the *Hamiltonian*, which governs the quantum evolution of a given quantum system.

In this section, we have introduced several common trivialization maps for different manifolds, including the probability manifold, the sphere manifold, the Hermitian matrix, and the unitary matrix. For certain manifolds, the trivialization method is not unique and can be compared in different scenarios. However, in this thesis, we will not delve deeply into the mathematical details of these trivializations. Instead, we will primarily focus on the numerical performance of different trivialization maps.

### 5.2.5 Summary and discussion

In summary, we have developed a unified framework for addressing quantum information problems using a non-convex optimization approach, which distinguishes itself significantly from convex optimization methods.

As illustrated in Fig. 5.7, the process begins with identifying the appropriate manifold description for the given quantum information problem. This step is essential to reformulate the problem as a manifold optimization problem. Next, leveraging the idea of trivialization on the target manifold, we transform the original constrained optimization problem into an unconstrained one in Euclidean space, which serves as our parameter space. Following this, we select a suitable classical optimizer, with the objective function embedded into a computational graph implemented using the PyTorch deep learning framework. This enables automatic gradient computation in each iteration via gradient backpropagation. Among available optimizers, L-BFGS is typically chosen as the default due to its strong performance in optimization problems without randomness. Finally, the optimal solution obtained from the optimization can deepen our understanding of the original quantum information problem.

Compared to convex optimization methods, particularly the SDP approach, our framework generally requires far fewer parameters for solving the same quantum information problem. This reduction in parameters significantly decreases computational time while maintaining satisfactory accuracy in practice. However, as a non-convex method, our approach does not guarantee convergence to the global minimum, instead providing upper bounds in the worst-case scenario. To ensure reliable results, the objective function is minimized multiple times with randomly initialized parameters, treating the number of trials as a hyperparameter in the optimization process. Furthermore, the fundamental manifolds introduced in this framework are versatile and can be applied across various quantum information problems, demonstrating the universality of this approach in different scenarios.

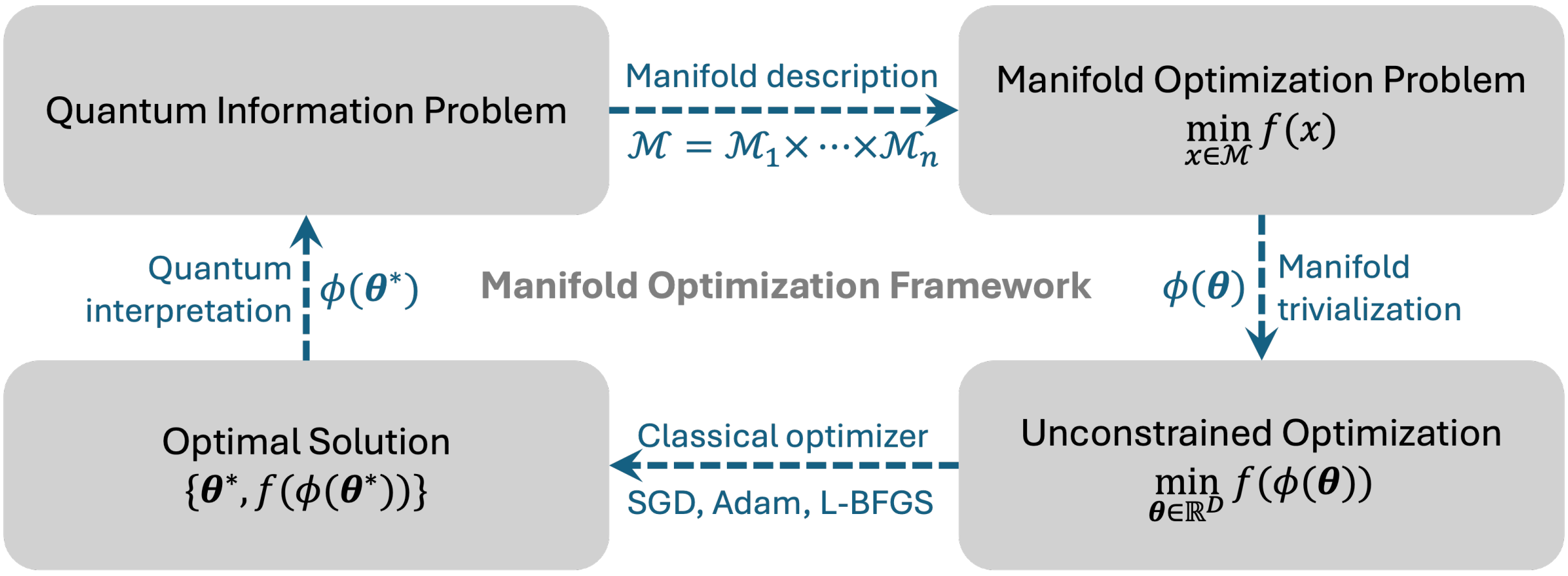

Figure 5.7: Flowchart of the manifold optimization framework for solving quantum information problems.

# CHAPTER 6

# QUANTIFYING ENTANGLEMENT WITH GEOMETRIC MEASURES

The geometric measure of entanglement (GME) [47–49, 163], as introduced in Sec. 4.2.2, quantifies the geometric "distance" between a given pure state $|\psi\rangle$ and the set of fully product states. It is defined as follows:

$$E_G(|\psi\rangle) = 1 - \max_{|\phi_{\text{prod}}\rangle} |\langle\phi_{\text{prod}}|\psi\rangle|^2. \tag{6.1}$$

In this chapter, we extend the concept of the geometric measure of entanglement (GME) to a broader range of scenarios, encompassing pure states, quantum subspaces, and mixed states in both bipartite and multipartite quantum systems. We develop a unified and rigorous framework, built upon a series of GME-based *entanglement monotones*, to analyze entanglement across these domains. This framework not only facilitates the detection and quantification of entanglement but also enables the characterization of entanglement through distinct *entanglement dimensionalities*, which correspond to the minimum dimensions required to reproduce the quantum correlations intrinsic to the entangled states (*asymptotically*). For example, in bipartite systems, entanglement dimensionalities are typically quantified by the Schmidt rank for pure states and the Schmidt number for mixed states. This capability is particularly important for the detection and certification of high-dimensional entanglement [50–52], which plays a pivotal role in advancing quantum information processing.

## 6.1 Pure State Entanglement

### 6.1.1 Bipartite pure states

Let us begin with the simplest case, *i.e.*, , bipartite pure states. In this scenario, the entanglement dimensionality is uniquely characterized by the Schmidt rank. A natural extension of GME is to quantify the geometric "distance" between the given state and the set of states with bounded Schmidt rank:

**Definition 17 (*k*-GME for Bipartite Pure States)** Consider a bipartite pure state $|\psi\rangle$ in the Hilbert space $\mathcal{H}_A \otimes \mathcal{H}_B$. The geometric measure of $k$-bounded Schmidt rank is defined as

$$E_G^{(k)}(|\psi\rangle) = 1 - \max_{\mathrm{SR}(|\phi\rangle)<k} |\langle\phi|\psi\rangle|^2, \tag{6.2}$$

where $\mathrm{SR}(|\phi\rangle)$ represents the Schmidt rank of $|\phi\rangle$.

In the special case of $k = 2$, this definition reduces to the geometric measure of entanglement (GME), $E_G$, where the maximum is taken over all product states, *i.e.*, states with Schmidt rank one. For a pure state $|\psi\rangle$ in the Schmidt decomposition form:

$$|\psi\rangle = \sum_i \mu_i |\alpha_i\rangle|\beta_i\rangle, \tag{6.3}$$

where $\mu_i$'s are the Schmidt coefficients of $|\psi\rangle$ arranged in *non-increasing* order. The $k$-GME of $|\psi\rangle$ can be expressed with the Schmidt coefficients, as follows:

$$E_G^{(k)}(|\psi\rangle) = 1 - \sum_{i=1}^{k-1} \mu_i^2, \tag{6.4}$$

Alternatively, it can also be expressed as:

$$E_G^{(k)}(|\psi\rangle) = \sum_{i=k}^{\min\{d_A,d_B\}} \lambda_i, \tag{6.5}$$

where $\lambda(\psi_A) = (\lambda_1, \lambda_2, \cdots)$ are the eigenvalues of the reduced density matrix of $|\psi\rangle$ on subsystem $A$, arranged in non-increasing order. For convenience, we define $E_G^{(1)}(|\psi\rangle) = \sum_i \lambda_i = 1$. It is evident that $\mathrm{SR}(|\psi\rangle) = k$ if and only if $E_G^{(k)}(|\psi\rangle) > 0$ and $E_G^{(k+1)}(|\psi\rangle) = 0$.

There is a significant connection between $k$-GME and the entanglement transformation introduced in Sec. 3.3. In the bipartite setting, this relation shows that the transformation between pure states is uniquely determined by $k$-GME. Specifically, Vidal's theorem for probabilistic entanglement transformation can be stated as follows:

**Theorem 6.1.1 (Probabilistic Entanglement Transformation)** For $d_A \otimes d_B$ pure bipartite states $|\psi\rangle$ and $|\phi\rangle$, the state $|\psi\rangle$ can be transformed into $|\phi\rangle$ via LOCC with the optimal success probability given by:

$$p_{\mathrm{succ}}(|\psi\rangle \to |\phi\rangle) = \min_{1\le k\le r} \frac{E_G^{(k)}(|\psi\rangle)}{E_G^{(k)}(|\phi\rangle)}, \tag{6.6}$$

where $r$ is the Schmidt rank of $|\phi\rangle$, and $E_G^{(k)}$ is defined in Eq. (6.5).

From this theorem, it is evident that a necessary and sufficient condition for an entanglement transformation with non-zero probability is that $E_G^{(k)}(|\psi\rangle) > 0$ for all $k = 1, \dots, \mathrm{SR}(|\phi\rangle)$. Furthermore, if for any $k$, $E_G^{(k)}(|\psi\rangle) \geq E_G^{(k)}(|\phi\rangle)$ holds, then $|\psi\rangle$ can be deterministically transformed into $|\phi\rangle$ via LOCC. This result is known as Nielsen's theorem [79].

An interesting application of this theorem is the distillation of an $m$-dimensional maximally entangled state $|\Psi_m^+\rangle$ from a given pure state $|\psi\rangle$ [77]:

**Theorem 6.1.2 (Distillation of Maximally Entangled State)** A $d \otimes d$ pure bipartite state $|\psi\rangle$ can be converted into an $m$-dimensional maximally entangled state $|\Psi_m^+\rangle = \frac{1}{\sqrt{m}} \sum_{i=0}^{m-1} |ii\rangle$ ($m \leq d$) via LOCC with the optimal success probability:

$$p_{\mathrm{succ}}(|\psi\rangle \to |\Psi_m^+\rangle) = \min_{1 \leq n \leq m} B_n^m, \tag{6.7}$$

where $B_n^m = \frac{m}{n} E_G^{(m-n+1)}(|\psi\rangle)$.

Interestingly, it can be shown that if $B_{n+1}^m > B_n^m$, then $B_{n+2}^m > B_{n+1}^m$. In other words, once $B_n^m$ starts to increase, it will continue increasing. Using this property, we can prove that for $m = d$:

$$p_{\mathrm{succ}} = B_1^d = d E_G^{(d)}(|\psi\rangle) = d \lambda_d(\psi_A), \tag{6.8}$$

where $\lambda_d(\psi_A)$ is the smallest eigenvalue of the reduced density matrix of $|\psi\rangle$ on subsystem $A$.

In the following, we consider the $k$-GME for $d \otimes d$ bipartite random states. For pure states, a widely-used, unitarily invariant distribution is the *Haar measure* [164, 165]. Its randomness ensures unbiased, uniform sampling over the state space, making it a useful tool for both theoretical analysis and practical applications.

From the formula in Eq. (6.5), we see that the distribution of $k$-GME for $d \otimes d$ random pure states is determined by the eigenvalues $\lambda_i$'s. The joint distribution of these eigenvalues under the Haar measure is given as follows [166–168]:

$$P(\lambda_1, \lambda_2, \dots, \lambda_d) = N \delta\left(\sum_i \lambda_i - 1\right) \prod_{i<j} |\lambda_i - \lambda_j|^2, \tag{6.9}$$

where $N$ is the normalization constant, and the eigenvalues $1 \geq \lambda_1 \geq \lambda_2 \geq \cdots \geq \lambda_d \geq 0$ are arranged in non-increasing order. In Ref. [169], the marginal distribution for each eigenvalue is also derived as follows:

**Theorem 6.1.3 (Marginal Eigenvalue Distribution for Haar Random States [169])** For the non-negative eigenvalues $\lambda_i$'s satisfying the joint distribution given in Eq. (6.9), the marginal distribution of each eigenvalue (in *non-increasing* order) is:

$$P(\lambda_i = x) = N \sum_{j=i}^{d} (-1)^{i+j} \binom{j-1}{i-1} A_j^{(d,d)}(x) \Theta(1 - jx), \tag{6.10}$$

where $A_j^{(d,d)}(x)$ is a polynomial in $x$ of order $d^2 - 1$ (specified in [169]), $\Theta$ denotes the unit step function, and $N$ is the normalization constant.

As we can see, the marginal distribution of the $i$-th largest eigenvalue is generally expressed as the summation of a series of polynomials with cutoffs. For the smallest eigenvalue $\lambda_d$, we have the simplest form: $P(\lambda_d = x) = d(d^2 - 1)(1 - dx)^{d^2-2}\Theta(1 - dx)$ [170]. This provides an explicit expression for the distribution of $E_G^{(d)}$:

**Theorem 6.1.4** For $d \otimes d$ random pure states sampled from the Haar measure, the geometric measure of $d$-bounded Schmidt rank follows the following probability distribution function (P.D.F.):

$$P(E_G^{(d)} = x) = d(d^2 - 1)(1 - dx)^{d^2-2}\Theta(1 - dx), \tag{6.11}$$

with the expectation value $\langle E_G^{(d)} \rangle = \frac{1}{d^3}$, the variance $\sigma^2 = \frac{2}{d^3(d^2+1)} - \frac{1}{d^6}$, and the cumulative distribution function (C.D.F.):

$$\Pr[E_G^{(d)} \geq x] = (1 - dx)^{d^2-1}. \tag{6.12}$$

Similarly, we can also compute $E_G^{(2)} = 1 - \lambda_1$ from the distribution of the largest eigenvalue $\lambda_1$. However, for arbitrary $d$, the distribution of $\lambda_1$ is significantly more complicated and lengthy, and the exact form of the expectation value $\langle \lambda_1 \rangle$ with respect to $d$ remains unknown. Asymptotically, it has been shown that $\langle \lambda_1 \rangle$ scales as $\frac{4}{d}$ [171, 172]. Consequently, while the average value of $E_G^{(2)}$ is indeed close to the maximal value $1 - \frac{1}{d}$ as the dimension increases, the probability of finding a highest-dimensional entangled state ($E_G^{(d)} > 0$) is exponentially small [172]. This result highlights the scarcity of high-dimensional entangled states from Haar random sampling. In other words, for random pure states generated from the Haar measure, most of them tend to exhibit low-dimensional entanglement, such as Bell states. For other geometric measures, $E_G^{(k)} = \sum_{i=k}^{d} \lambda_i$ ($k = 3, \ldots, d - 1$), the analysis becomes significantly more complex and remains unclear.

Let us discuss the $4 \otimes 4$ bipartite system as an example, we have the analytical expressions for each coefficient $A_j^{(d,d)}(x)$:

$$\begin{aligned}
A_4^{(4,4)}(x) &= 60(1-4x)^{14}, \\
A_3^{(4,4)}(x) &= 60(1-3x)^8(3-96x+1308x^2-6128x^3+29818x^4-70160x^5+67812x^6), \\
A_2^{(4,4)}(x) &= 30(1-2x)^6(6-264x+5208x^2-45920x^3+229936x^4 \\
&\quad -859040x^5+2706592x^6-5570528x^7+5517256x^8), \\
A_1^{(4,4)}(x) &= 60(1-x)^8(1-48x+1044x^2-9904x^3+44934x^4-94128x^5+73116x^6),
\end{aligned} \tag{6.13}$$

which leads to the analytical expression for each eigenvalue distribution in non-increasing order:

$$\begin{aligned}
P(\lambda_1 = x) &= -A_4^{(4,4)}(x)\Theta(1-4x) + A_3^{(4,4)}(x)\Theta(1-3x) \\
&\qquad - A_2^{(4,4)}(x)\Theta(1-2x) + A_1^{(4,4)}(x)\Theta(1-x), \\
P(\lambda_2 = x) &= 3A_4^{(4,4)}(x)\Theta(1-4x) - 2A_3^{(4,4)}(x)\Theta(1-3x) + A_2^{(4,4)}(x)\Theta(1-2x), \\
P(\lambda_3 = x) &= -3A_4^{(4,4)}(x)\Theta(1-4x) + A_3^{(4,4)}(x)\Theta(1-3x), \\
P(\lambda_4 = x) &= A_4^{(4,4)}(x)\Theta(1-4x).
\end{aligned} \tag{6.14}$$

Thus, from the probability distribution of largest eigenvalue $\lambda_1$ and smallest eigenvalue $\lambda_4$, we can obtain the distribution of $E_G^{(4)}$ and $E_G^{(2)}$, as follows

$$\begin{aligned}
P(E_G^{(4)} = x) &= A_4^{(4,4)}(x)\Theta(1-4x) = 60(1-4x)^{14}\Theta(1-4x), \\
P(E_G^{(2)} = x) &= P(\lambda_1 = 1-x) \\
&= -A_4^{(4,4)}(1-x)\Theta(4x-3) + A_3^{(4,4)}(1-x)\Theta(3x-2) \\
&\qquad - A_2^{(4,4)}(1-x)\Theta(2x-1) + A_1^{(4,4)}(1-x)\Theta(x).
\end{aligned} \tag{6.15}$$

We conducted numerical experiments to verify our analytical results, as shown in Fig. 6.1. Specifically, we randomly generated 500,000 pure states in a $4 \otimes 4$ bipartite system from the Haar measure and computed their $k$-GME values for $k = 2, 3, 4$. The statistical results were then used to calculate the numerical probability distribution function (P.D.F.) for each $k$. The numerical results are consistent with our analytical findings, revealing a generic entanglement property for random pure states.

According to Theorem 6.1.2, we can also study the optimal success probability distribution for distilling an $m$-dimensional maximally entangled state $|\Psi_m^+\rangle$ from a random pure state in a

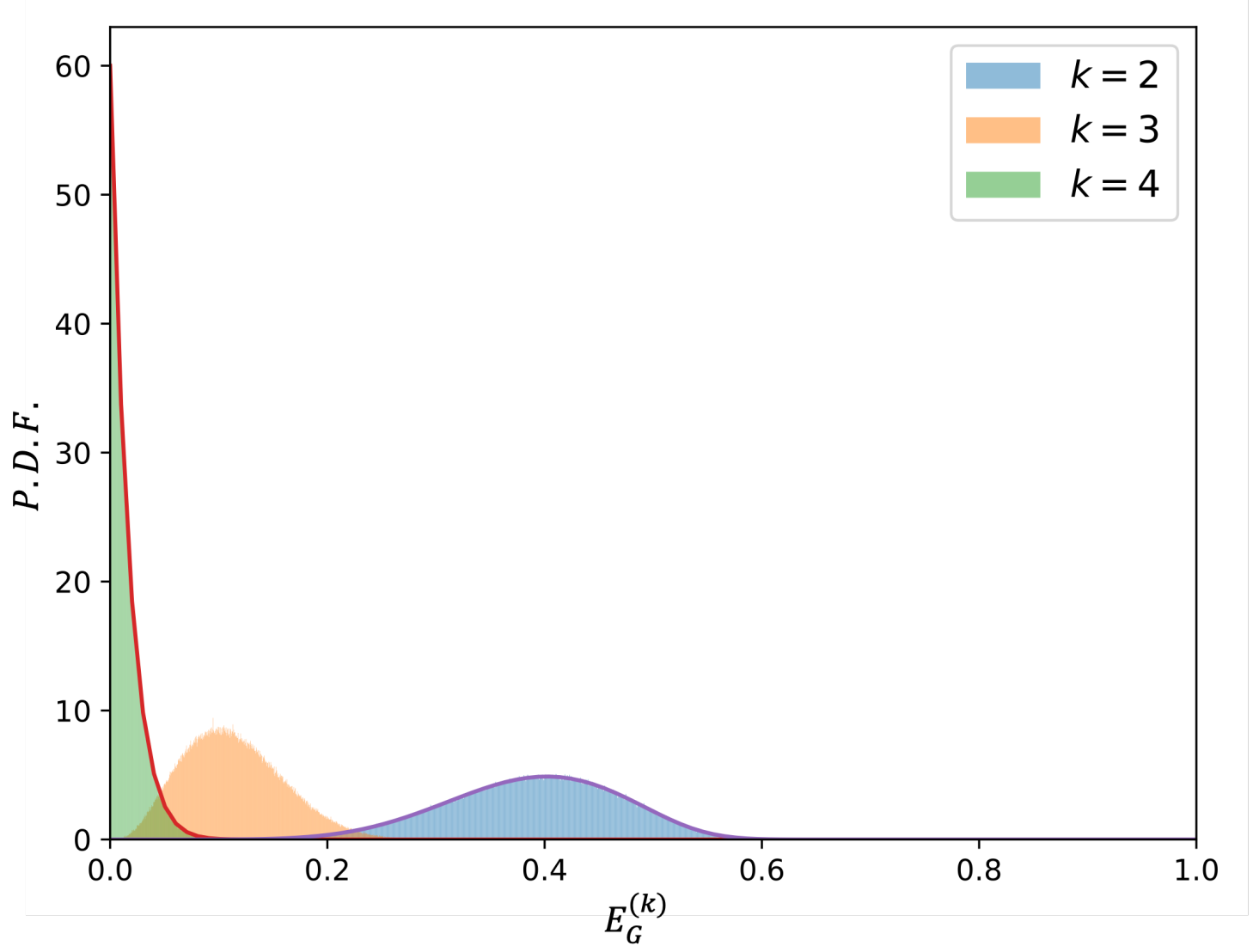

Figure 6.1: $k$-GME distribution for Haar random pure states in a $4 \otimes 4$ system. The green, orange, and blue areas represent the numerical probability distribution function (P.D.F.) of $E_G^{(k)}$ for $k = 2, 3, 4$, respectively. For comparison, the red curve denotes the analytical results for $E_G^{(4)}$, while the purple curve represents the analytical results for $E_G^{(2)}$.

$d \otimes d$ system. The probability distribution function satisfies:

$$P(p_{\text{succ}}^{(m)} = x) = P\left(\min_{1 \leq n \leq m} \frac{m}{n} E_G^{(m-n+1)} = x\right), \tag{6.16}$$

where $p_{\text{succ}}^{(m)} = p_{\text{succ}}(|\psi\rangle \to |\Psi_m^+\rangle)$ is a random variable describing the success probability for a random pure state $|\psi\rangle$. The distribution of $p_{\text{succ}}^{(m)}$ is *collectively* determined by the distributions of $E_G^{(1)} = 1, E_G^{(2)}, \ldots, E_G^{(m)}$ for random pure states.

Taking the case $m = 2$ as an example, we can compute:

$$P(p_{\text{succ}}^{(2)} = x) = P\left(\min\{1, 2E_G^{(2)}\} = x\right), \tag{6.17}$$

such that:

$$\begin{aligned} P(p_{\text{succ}}^{(2)} = x) =& P\left(E_G^{(2)} = \frac{x}{2}\right) \Pr\left[E_G^{(2)} < \frac{1}{2}\right] + \delta(x-1) \Pr\left[E_G^{(2)} \geq \frac{1}{2}\right] \\ =& P\left(E_G^{(2)} = \frac{x}{2}\right) \int_0^{\frac{1}{2}} P(E_G^{(2)} = x) dx + \delta(x-1) \int_{\frac{1}{2}}^1 P(E_G^{(2)} = x) dx, \end{aligned} \tag{6.18}$$

where $\delta(x-1)$ indicates that the two-dimensional maximally entangled state $|\Psi_2^+\rangle$ can be distilled *deterministically* from some random pure states.

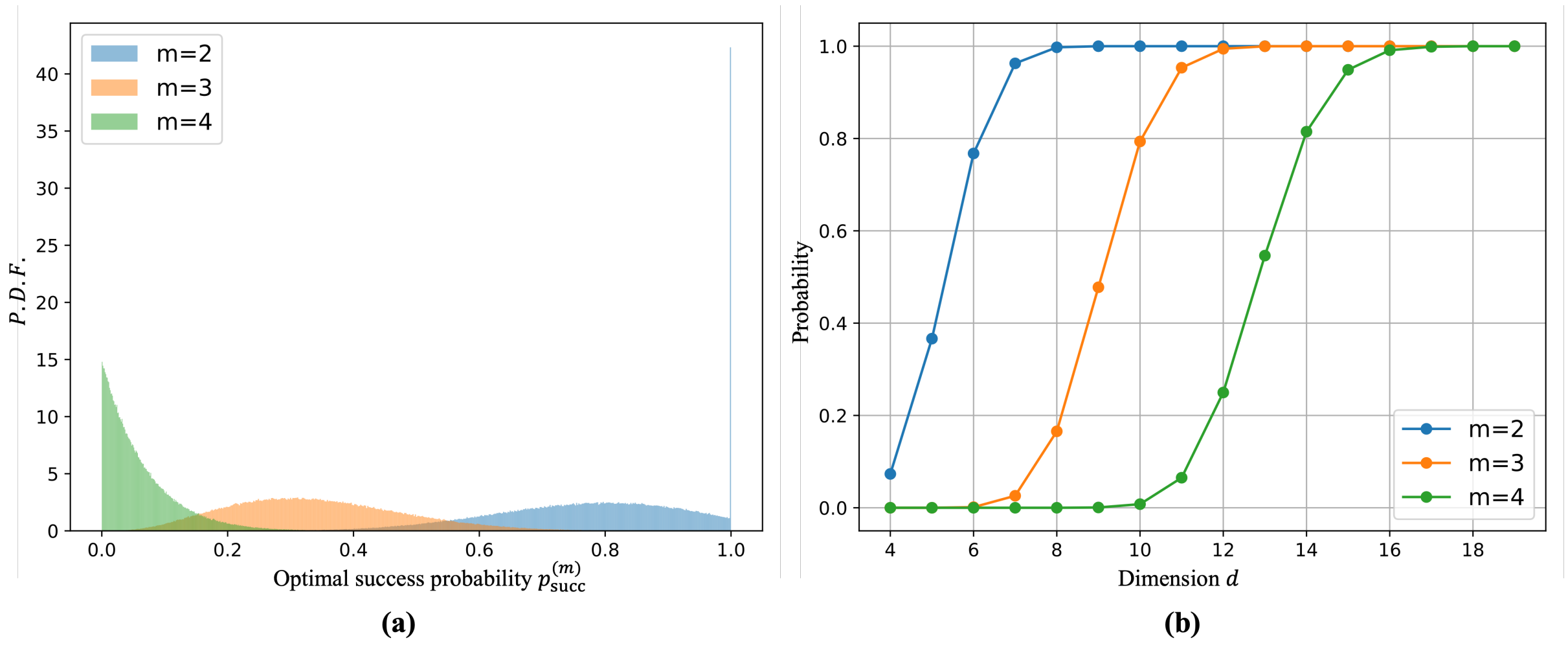

Figure 6.2: Entanglement distillation analysis for random pure states: (a) Optimal success probability distribution for distilling $m$-dimensional maximally entangled states from $4 \otimes 4$ Haar random pure states. (b) Probability of having a deterministic distillation for $m$-dimensional maximally entangled states from $d \otimes d$ Haar random pure states.

For the special case $m = d$, according to Eq. (6.8), we have:

$$P(p_{\text{succ}}^{(d)} = x) = (d^2 - 1)(1 - x)^{d^2-2}, \tag{6.19}$$

which is *solely* determined by the highest-dimensional entanglement monotone $E_G^{(d)}$ of random pure states. This indicates that the general distillation of $|\Psi_d^+\rangle$ from $d \otimes d$ random states is *probabilistic*. Additionally, the corresponding probability cannot be increased by using any *catalyst state* $|\varphi\rangle$ [82], *i.e.*,

$$p_{\text{succ}}(|\psi\rangle \to |\Psi_d^+\rangle) \geq p_{\text{succ}}(|\psi\rangle|\varphi\rangle \to |\Psi_d^+\rangle|\varphi\rangle). \tag{6.20}$$

In Fig. 6.2(a), we generated $500,000$ random pure states in a $4 \otimes 4$ bipartite system from the Haar measure and computed their optimal success probability for distilling $m$-dimensional maximally entangled states ($m = 2, 3, 4$). The statistical results were then used to calculate the corresponding probability distribution function (P.D.F.). As expected, the distillation of the lowest-dimensional maximally entangled states can be deterministic for some random states, which forms a peak in the distribution. However, the probability distribution tends to decay to near zero as we aim to distill higher-dimensional entanglement.

In Fig. 6.2(b), we calculated the probability of achieving deterministic distillation for $m$-dimensional maximally entangled states in different $d \otimes d$ systems. Unsurprisingly, if we aim to distill higher-dimensional entanglement with high probabilities, one sufficient condition is to increase the dimensions of the quantum states we have.

### 6.1.2 Multipartite pure states

The study of multipartite pure states presents a significantly more complex and nuanced challenge compared to their bipartite counterparts. While bipartite pure states can be represented as matrices, multipartite pure states are described using *tensors* [53]. Although the Schmidt decomposition provides an efficient method for determining the Schmidt rank in bipartite systems, calculating the tensor rank for multipartite systems is considerably more difficult. Specifically, the tensor rank—defined as the minimum number of product terms required for decomposition—is widely recognized as computationally intractable. Indeed, determining the tensor rank is believed to be at least NP-complete [69, 70].

In this context, the concept of border rank serves as a critical refinement [71, 72]. Border rank accounts for the fact that certain tensors can be approximated arbitrarily closely by tensors with lower ranks, a property not observed in matrices. For instance, the W state in a three-qubit system has a tensor rank of 3 but can be approximated to any desired precision by states with a tensor rank of 2 [173]. Consequently, the border rank of the W state is defined as 2. From a practical perspective, border rank often provides a more accessible characterization of entanglement dimensionality in multipartite systems, as a state with border rank $r$ can always be approximated to any desired accuracy by states with a tensor rank of $r$, even if its actual tensor rank is higher.

The investigation of tensor and border ranks is not only crucial for advancing quantum information science—particularly in applications requiring high-dimensional entanglement [50–52, 174]—but also holds significant implications in other fields, such as algebraic complexity theory [175–177].

A natural idea to extend the concept of GME in the multipartite scenario involves measuring the geometric “distance” between the given state and the set of states with a bounded tensor rank:

**Definition 18 ($k$-GME for Multipartite Pure States)** Consider a multipartite pure state $|\psi\rangle$ in the Hilbert space $\mathcal{H}_1 \otimes \mathcal{H}_2 \otimes \cdots \otimes \mathcal{H}_n$. The geometric measure of $k$-bounded tensor rank is defined as

$$E_G^{(k)}(|\psi\rangle) = 1 - \max_{\mathrm{TR}(|\phi\rangle)<k} |\langle\phi|\psi\rangle|^2, \tag{6.21}$$

where $\mathrm{TR}(|\phi\rangle)$ represents the tensor rank of $|\phi\rangle$.

This definition is consistent with the bipartite cases since the tensor rank is equivalent to the Schmidt rank for bipartite pure states. Furthermore, from the definition of border rank, we know that the border rank of a pure state $|\psi\rangle$, denoted by $\mathrm{BR}(|\psi\rangle) = k$ if and only if $E_G^{(k)}(|\psi\rangle) > 0$ and $E_G^{(k+1)}(|\psi\rangle) = 0$.

The gap between these two different ranks appears due to the non-existence of a *best rank-r approximation* for an arbitrary tensor [178], implying $\mathrm{TR}(|\psi\rangle) \geq \mathrm{BR}(|\psi\rangle)$. However, there exist two notable exceptions where $\mathrm{TR}(|\psi\rangle) = \mathrm{BR}(|\psi\rangle)$: the cases $n = 2$ (bipartite system) and $k = 2$ (approximation by fully product states). The former implies $\mathrm{TR}(|\psi\rangle) = \mathrm{BR}(|\psi\rangle)$ for a bipartite pure state $|\psi\rangle$ while the latter means that $\mathrm{BR}(|\psi\rangle) = 1$ is equivalent to $\mathrm{TR}(|\psi\rangle) = 1$.

In the following, we introduce a numerical approach to compute the $k$-GME of a given multipartite state $|\psi\rangle$. A critical aspect of this method is the characterization of the set of states with bounded tensor rank. By definition, a multipartite state $|\phi\rangle$ with $\mathrm{TN}(|\phi\rangle) \leq r$ can be expressed as

$$|\phi\rangle = \sum_{i=1}^{r} \mu_i |\phi_i^{(1)}\rangle \otimes |\phi_i^{(2)}\rangle \otimes \cdots \otimes |\phi_i^{(n)}\rangle, \tag{6.22}$$

where $|\phi_i^{(j)}\rangle$ is a pure state in the $j$th Hilbert space with dimension $d_j$, and $\mu_i$ is a positive coefficient subject to a normalization condition. Notably, orthogonality among the different $|\phi_i^{(j)}\rangle$ is not required, allowing the product terms to cancel each other out, which may reduce the tensor rank.

The key step in this approach is to consider an *unnormalized* pure state $|\tilde{\phi}\rangle$, expressed as:

$$|\tilde{\phi}\rangle = \sum_{i=1}^{r} \tilde{\mu}_i |\phi_i^{(1)}\rangle \otimes |\phi_i^{(2)}\rangle \otimes \cdots \otimes |\phi_i^{(n)}\rangle, \tag{6.23}$$

where the coefficients $\tilde{\mu}_i$ are positive numbers without any normalization constraint. For this state, each component can be described by the corresponding manifold with suitable trivialization maps, as introduced in Sec. 5.2. To parameterize the positive coefficients $\mu_i$'s, we employ the SoftPlus map. For each individual pure state $|\phi_i^{(j)}\rangle$, which lies on the sphere manifold $\mathbb{S}^{2d_j-1}$, we use a normalization operation (see Theorem 5.2.4). As a result, $|\tilde{\phi}\rangle$ can now be parameterized by some real parameters $\boldsymbol{\theta}$ without any constraint.

This process allows us to reformulate the original computation as an unconstrained optimization problem in Euclidean space, expressed as:

$$E_G^{(k)}(|\psi\rangle) = 1 - \max_{\boldsymbol{\theta} \in \mathbb{R}^D} \frac{|\langle \tilde{\phi}(\boldsymbol{\theta}) | \psi \rangle|^2}{\langle \tilde{\phi}(\boldsymbol{\theta}) | \tilde{\phi}(\boldsymbol{\theta}) \rangle}, \tag{6.24}$$

where the dimension of the Euclidean space is $D = (2\sum_j d_j + 1)(k-1)$. Furthermore, we can also represent $|\tilde{\phi}\rangle$ in the tensor network, as follows:

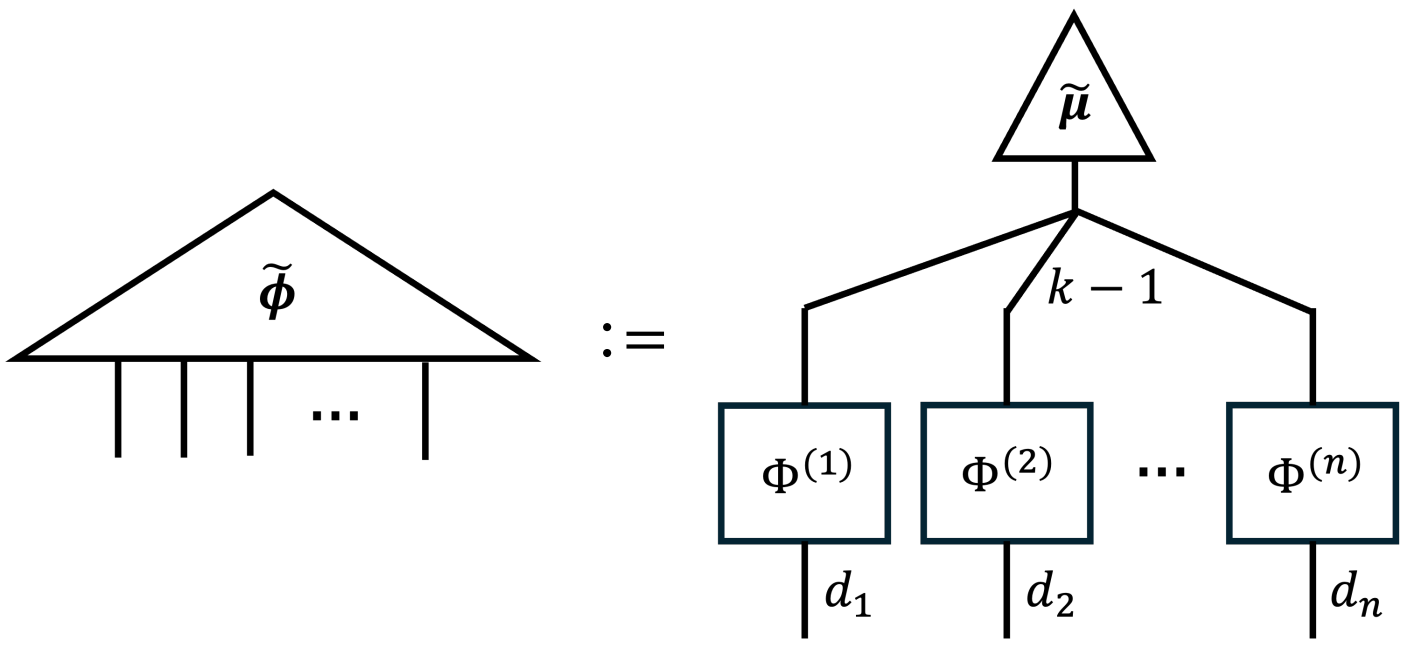

(6.25)

where the matrix $\Phi^{(j)}$ denotes the set $\{|\phi_i^{(j)}\rangle\}_{i=1}^{k-1}$ and $\tilde{\boldsymbol{\mu}} = (\tilde{\mu}_1, \tilde{\mu}_2, \ldots, \tilde{\mu}_{k-1})$ is a vector with positive entries. So that Eq. (6.24) can also be reformed as:

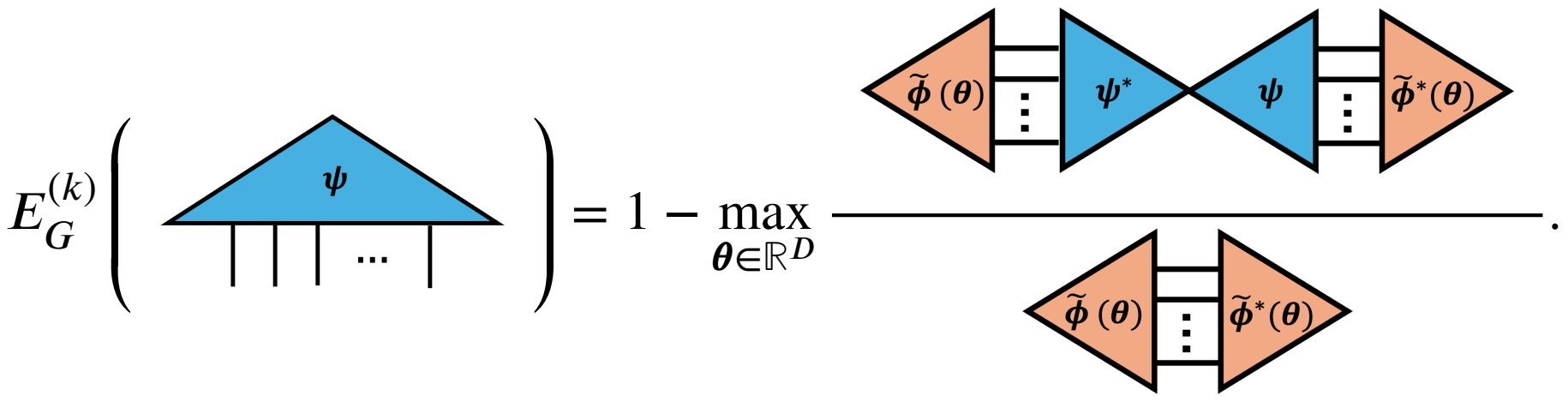

(6.26)

As the first example, the basis of permutation symmetric states in $n$-qubit system are given by the Dicke states, which are mathematically expressed as the sum of all permutations of computational basis states with $n-m$ qubits being $|0\rangle$ and $m$ being $|1\rangle$:

$$\left|D_n^m\right\rangle = \begin{pmatrix} n \\ m \end{pmatrix}^{-1/2} \sum_{\text{perm}} \underbrace{|0\rangle|0\rangle\cdots|0\rangle}_{n-m} \underbrace{|1\rangle|1\rangle\cdots|1\rangle}_{m}, \tag{6.27}$$

with $0 \le m \le n$. It is proved that the closest fully product state for computing $E_G^{(2)}$ [49, 131] has the form

$$|\phi\rangle = \left(\sqrt{\frac{n-m}{n}}|0\rangle + \sqrt{\frac{m}{n}}|1\rangle\right)^{\otimes n}, \tag{6.28}$$

*i.e.*, a tensor product of $n$ identical single qubit states. From that, the geometric measure of entanglement is found to be

$$E_G^{(2)}(|D_n^m\rangle) = 1 - \binom{n}{m}\left(\frac{m}{n}\right)^m\left(\frac{n-m}{n}\right)^{n-m}. \tag{6.29}$$

Due to the symmetry of Dicke states, we can just discuss the cases where $m \le \left\lfloor \frac{n}{2} \right\rfloor$. It has been proved that for a Dicke state $|D_n^m\rangle$, its border rank $\mathrm{BR}(|D_n^m\rangle) = m+1$ [179], which means there should be a transition between zero and nonzero for $E_G^{(m+2)}$ and $E_G^{(m+1)}$.

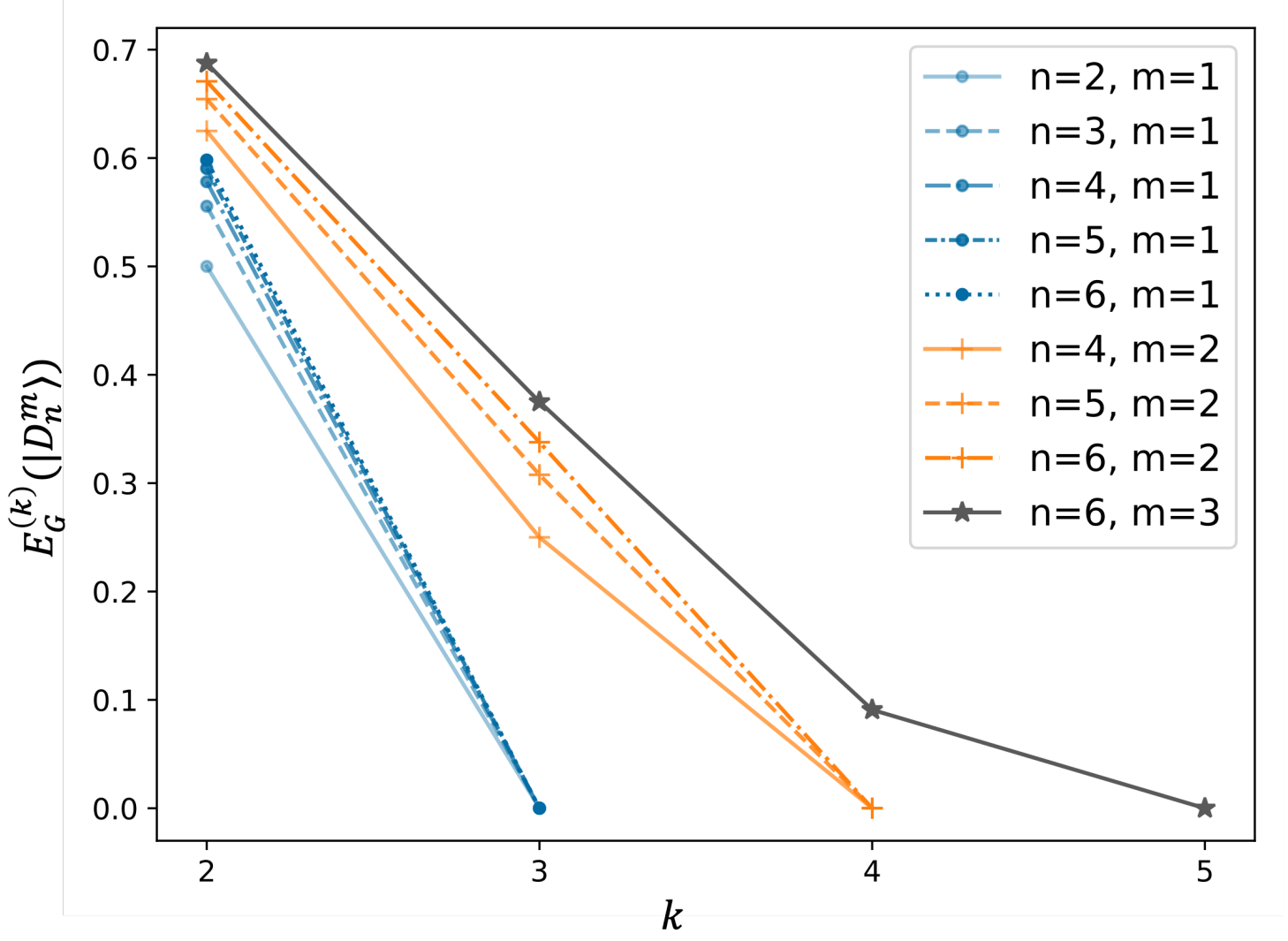

Figure 6.3: Results of $k$-GME for Dicke states $|D_n^m\rangle$ with different $n$ and $m$. The transition of $E_G^{(k)}$ at $k = m + 1$ and $k = m + 2$ implies the border rank of Dicke states.

We compute the $k$-GME $E_G^{(k)}(|D_n^m\rangle)$ for various values of $n$ and $m$, as shown in Fig. 6.3. For the case of $k = 2$, we compare the numerical results with the analytical results given in Eq. (6.29), demonstrating that the error is nearly at the level of machine precision. Furthermore, the analytical form in Eq. (6.28) can also be recovered during the optimization process. In general, identifying the closest state is a difficult task unless the given state possesses specific symmetries [132].

Our methodology provides a powerful tool for exploring these closest states, even within the set of states with bounded tensor rank. Through numerical investigations, we observe distinct transitions from non-zero to zero for different Dicke states, consistent with the postulated border rank of the given $|D_n^m\rangle$.

Border rank and tensor rank are also closely connected to *algebraic complexity theory*. One of the most significant open problems in computer science is understanding the computational complexity of *matrix multiplication*. Specifically, the question is how many multiplication operations are required to multiply $n \times n$ matrices. The naive approach requires $n^3$ multiplications, but Strassen discovered an algorithm that only needs seven multiplications for $2 \times 2$ matrices [175].

The minimum number of multiplications required for matrix multiplication is referred to as the *rank* of matrix multiplication. Meanwhile, the concept of *border rank* arises when certain matrix multiplications can be approximated to arbitrary precision by simpler multiplications with fewer operations [176, 177]. These approximations often lead to faster matrix multiplication algorithms in practice.

Interestingly, the rank of $n \times n$ matrix multiplication is equivalent to the rank of the following tripartite (unnormalized) state [180, 181]:

$$|\Phi_n\rangle_{ABC} =$$

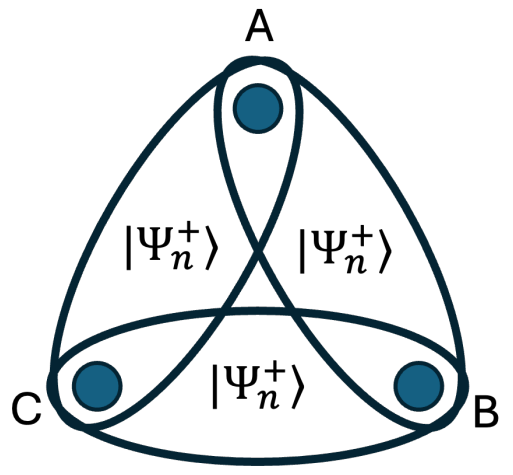

$$= \sum_{i,j,k=0}^{n-1} |ij\rangle_A |ik\rangle_B |jk\rangle_C, \tag{6.30}$$

where any two parties share a $n$-dimensional maximally entangled state $|\Psi_n^+\rangle$ of dimension $n$.

By computing $E_G^{(k)}(|\Phi_2\rangle_{ABC})$ for various $k$, we find that $E_G^{(7)}$ is very close to $\frac{1}{8}$, while $E_G^{(6)}$ is approximately $10^{-14}$. This indicates that $\mathrm{BR}(|\Phi_2\rangle_{ABC}) = 7$. This numerical result aligns with previous findings [182], validating the effectiveness of our approach in analyzing border rank.

## 6.2 Subspace Entanglement

Beyond the analysis of pure state entanglement, a fundamental question arises in the study of quantum systems: determining whether a given *subspace* exhibits entanglement and, if so, identifying its *entanglement dimensionality* [49, 67, 173, 183–190].

In the bipartite setting, the entanglement dimensionality of a pure state is determined by its Schmidt rank, which can be efficiently computed using the Schmidt decomposition [191]. However, assessing the entanglement properties of a *subspace* is a significantly more challenging problem [192]. This task is of critical importance in various applications, including the generation of entangled mixed states [98], the construction of entanglement witnesses [109, 193], quantum error correction [194, 195], and the validation of quantum protocols such as superdense coding [196].

Considerable progress has been made in certifying the presence of entanglement within subspaces. One prominent approach involves deriving *lower bounds* on the entanglement of a subspace through mathematical formulations and analyzing the conditions under which these bounds yield positive values. This line of research has a rich history [197–203], with its origins tracing back to Ref. [204].

Another widely adopted approach focuses on bounding the minimum entanglement using semidefinite programming (SDP) techniques, such as the positive partial transpose (PPT) relaxation [31, 188] and symmetric extension methods [27, 28]. However, mathematical approaches often require intricate or non-intuitive proofs and may lack general applicability. Similarly, SDP-based methods face limitations due to their computational complexity and inability to address certain scenarios, such as bound entanglement [87] and high-dimensional entanglement [205].

More recently, a hierarchical method based on linear systems has been proposed [187]. While this method is capable of certifying high-dimensional entanglement within subspaces, it remains computationally demanding. Furthermore, it does not directly provide explicit entangled states, leaving certain practical challenges unresolved.

### 6.2.1 Bipartite subspaces

Consider a subspace $\mathcal{S}$ spanned by some bipartite pure states $\{|\psi_1\rangle, |\psi_2\rangle, \ldots, |\psi_m\rangle\}$ in the Hilbert space $\mathcal{H}_A \otimes \mathcal{H}_B$. If no product state $|\phi\rangle$ exists within this subspace, we refer to the subspace as *entangled*. The entanglement dimensionality of a subspace can further be characterized by the *minimal rank*, as defined below:

**Definition 19 (Entanglement Dimensionality of Bipartite Subspace)** Given a subspace $\mathcal{S}$ spanned by a set of bipartite states $\{|\psi_1\rangle, |\psi_2\rangle, \ldots, |\psi_m\rangle\}$, the minimal rank of $\mathcal{S}$ is defined as the minimal Schmidt rank of any state within $\mathcal{S}$, *i.e.*,

$$r(\mathcal{S}) = \min_{|\psi\rangle \in \mathcal{S}} \mathrm{SR}(|\psi\rangle), \tag{6.31}$$

where $\mathrm{SR}(|\psi\rangle)$ represents the Schmidt rank of $|\psi\rangle$, and $|\psi\rangle \in \mathcal{S}$ implies that $|\psi\rangle$ can be expressed as a linear combination of the states spanning $\mathcal{S}$, *i.e.*,

$$|\psi\rangle = \sum_{i=1}^{m} c_i |\psi_i\rangle, \tag{6.32}$$

where the coefficients $c_i$ are complex numbers subject to the normalization condition.

A natural idea is to define the $k$-GME for a given subspace as follows:

**Definition 20 ($k$-GME for Bipartite Subspaces)** Given a subspace $\mathcal{S}$ spanned by a set of bipartite states $\{|\psi_1\rangle, |\psi_2\rangle, \ldots, |\psi_m\rangle\}$, the minimal geometric measure of $k$-bounded Schmidt rank is defined as

$$E_G^{(k)}(\mathcal{S}) = \min_{|\psi\rangle \in \mathcal{S}} E_G^{(k)}(|\psi\rangle), \tag{6.33}$$

where $E_G^{(k)}(|\psi\rangle)$ for a bipartite state $|\psi\rangle$ is defined in Definition 17.

Furthermore, starting from the definition, we can derive:

$$
\begin{aligned}
E_G^{(k)}(\mathcal{S}) &= \min_{|\psi\rangle\in\mathcal{S}} E_G^{(k)}(|\psi\rangle) \\
&= \min_{|\psi\rangle\in\mathcal{S}} \left(1 - \max_{\mathrm{SR}(|\phi\rangle)<k} |\langle\phi|\psi\rangle|^2\right) \\
&= 1 - \max_{\mathrm{SR}(|\phi\rangle)<k} \max_{|\psi\rangle\in\mathcal{S}} |\langle\phi|\psi\rangle|^2 \\
&= 1 - \max_{\mathrm{SR}(|\phi\rangle)<k} \left\langle \phi \middle| \mathcal{P}_{\mathcal{S}} \middle| \phi \right\rangle \\
&= \min_{\mathrm{SR}(|\phi\rangle)<k} \left\langle \phi \middle| \mathcal{P}_{\mathcal{S}}^{\perp} \middle| \phi \right\rangle ,
\end{aligned}
\tag{6.34}
$$

where $\mathcal{P}_{\mathcal{S}}$ is the projector onto $\mathcal{S}$, and $\mathcal{P}_{\mathcal{S}}^{\perp}$ is the projector onto the orthogonal complementary subspace $\mathcal{S}^{\perp}$. It is clear that $r(\mathcal{S}) = k$ if and only if $E_G^{(k)}(\mathcal{S}) > 0$ and $E_G^{(k+1)}(\mathcal{S}) = 0$.

The key step in the derivation is the transition from the third line to the fourth line, which follows from the fact that for a given $|\phi\rangle$, the vector from $\mathcal{S}$ that maximizes the quantity $|\langle\phi|\psi\rangle|^2$ is the projection of $|\phi\rangle$ onto $\mathcal{S}$.

In practice, the projector $\mathcal{P}_{\mathcal{S}}^{\perp}$ can be expressed as

$$
\mathcal{P}_{\mathcal{S}}^{\perp} = I - \mathcal{P}_{\mathcal{S}} = I - \sum_{i=1}^{d_{\mathcal{S}}} |e_i\rangle\langle e_i|,
\tag{6.35}
$$

where $\{|e_i\rangle\}$ represents the orthonormal basis of the subspace $\mathcal{S}$, constructed using the *Gram-Schmidt process* applied to $\{|\psi_1\rangle, |\psi_2\rangle, \ldots, |\psi_m\rangle\}$. Here, $d_{\mathcal{S}}$ denotes the dimension of the subspace $\mathcal{S}$.

In addition to certifying the dimensionality of entanglement, $k$-GME also quantifies the *robustness* of entangled subspaces under unitary perturbations, as described in the following theorem (a related discussion can be found in [197]):

**Theorem 6.2.1 (Robustness Analysis for Entangled Subspace)** Let $\mathcal{S} \subset \mathcal{H}_A \otimes \mathcal{H}_B$ be a subspace such that $r(\mathcal{S}) \geq k$. For any perturbation $U = e^{-iH}$, where the operator norm of $H$ is less than $\sqrt{E_G^{(k)}(\mathcal{S})}$, the perturbed subspace $\mathcal{S}'$ will retain a minimal rank $r(\mathcal{S}') \geq k$.

*Proof.* Suppose $\mathcal{S}$ is spanned by $\{|\psi\rangle_1, |\psi\rangle_2, \ldots, |\psi\rangle_m\}$, the subspace after some perturbation $U$ is defined as $\mathcal{S}'$, spanned by $\{U|\psi\rangle_1, U|\psi_2\rangle, \ldots, U|\psi_m\rangle\}$. It is easy to prove that

$$
\mathcal{P}_{\mathcal{S}'}^{\perp} = U\mathcal{P}_{\mathcal{S}}^{\perp}U^{\dagger}.
$$

For convenience, we denote $\sqrt{\boldsymbol{x}^{\dagger} \cdot A \cdot \boldsymbol{x}}$ as $\|\boldsymbol{x}\|_A$. We can verify the following inequalities for

$\|\boldsymbol{x}\|_{\mathcal{P}_{\mathcal{S}}^{\perp}}$:

$$\|\boldsymbol{x}\|_{\mathcal{P}_{\mathcal{S}}^{\perp}} + \|\boldsymbol{y}\|_{\mathcal{P}_{\mathcal{S}}^{\perp}} \geq \|\boldsymbol{x}+\boldsymbol{y}\|_{\mathcal{P}_{\mathcal{S}}^{\perp}} \geq \|\boldsymbol{x}\|_{\mathcal{P}_{\mathcal{S}}^{\perp}} - \|\boldsymbol{y}\|_{\mathcal{P}_{\mathcal{S}}^{\perp}},$$
$$\|\sum_i a_i \boldsymbol{x_i}\|_{\mathcal{P}_{\mathcal{S}}^{\perp}} \leq \sum_i |a_i| \cdot \|\boldsymbol{x_i}\|_{\mathcal{P}_{\mathcal{S}}^{\perp}},$$

where $\boldsymbol{x}, \boldsymbol{y}, \boldsymbol{x_i}$ represent complex vectors and $a_i$'s are complex numbers. Thus, the square root of the geometric measure of $k$-bounded rank for $\mathcal{S}'$

$$\begin{aligned}
E_G^{(k)}(\mathcal{S}')^{\frac{1}{2}} =& \min_{\mathrm{SR}(|\phi\rangle)<k} \||\phi\rangle\|_{\mathcal{P}_{\mathcal{S}'}^{\perp}} \\
=& \min_{\mathrm{SR}(|\phi\rangle)<k} \||\phi\rangle\|_{U\mathcal{P}_{\mathcal{S}}^{\perp}U^\dagger} = \min_{\mathrm{SR}(|\phi\rangle)<k} \|U^\dagger|\phi\rangle\|_{\mathcal{P}_{\mathcal{S}}^{\perp}} \\
\geq& \min_{\mathrm{SR}(|\phi\rangle)<k} \left[\||\phi\rangle\|_{\mathcal{P}_{\mathcal{S}}^{\perp}} - \|(I-U^\dagger)|\phi\rangle\|_{\mathcal{P}_{\mathcal{S}}^{\perp}}\right] \\
\geq& E_G^{(k)}(\mathcal{S})^{\frac{1}{2}} - \max_{\mathrm{SR}(|\phi\rangle)<k} \|(I-U^\dagger)|\phi\rangle\|_{\mathcal{P}_{\mathcal{S}}^{\perp}}.
\end{aligned}$$

Because we have the following inequalities:

$$\begin{aligned}
&\max_{\mathrm{SR}(|\phi\rangle)<k} \|(I-U^\dagger)|\phi\rangle\|_{\mathcal{P}_{\mathcal{S}}^{\perp}} \\
&\quad \leq \max_{|\phi\rangle} \|(I-U^\dagger)|\phi\rangle\|_{\mathcal{P}_{\mathcal{S}}^{\perp}} \\
&\quad \leq \max_{|\phi\rangle} \|(I-U^\dagger)|\phi\rangle\| \\
&\quad = \|(I-U^\dagger)\|_{\mathrm{op}} = \|(I-e^{iH})\|_{\mathrm{op}} \leq \|H\|_{\mathrm{op}},
\end{aligned}$$

where $\|H\|_{\mathrm{op}}$ is the operator norm of $H$. So that if $\|H\|_{\mathrm{op}} < \sqrt{E_G^{(k)}(\mathcal{S})}$, then $r(\mathcal{S}') \geq k$. □

A simple geometric interpretation of this theorem can be provided: observe that $E_G^{(k)}(\mathcal{S})^{\frac{1}{2}}$ represents the sine of the *smallest angle* $\theta^\star$ between states with $k$-bounded Schmidt rank and states within the subspace $\mathcal{S}$. Specifically, $E_G^{(k)}(\mathcal{S})^{\frac{1}{2}} = \sin(\theta^\star) \leq \theta^\star$. A perturbation $U = e^{-iH}$ can be interpreted as a rotation operator within the high-dimensional state space, where the operator norm of $H$ determines its maximal rotation angle. To preserve the entanglement dimensionality, it is necessary to constrain the rotation angle $\theta$ such that $\theta \leq \|H\|_{\mathrm{op}} < E_G^{(k)}(\mathcal{S})^{\frac{1}{2}} \leq \theta^\star$.

Due to the inherent difficulties in characterizing states with bounded Schmidt rank and the non-convex nature of the associated optimization problem, prior research avoided directly computing $k$-GME. Instead, it concentrated on deriving a *lower bound* for $E_G^{(k)}$ through mathematical analysis or by relaxing the problem to enable solutions using the semidefinite programming

(SDP) approach, such as the positive partial transpose (PPT) relaxation:

$$\begin{aligned} E_G^{(2)}(\mathcal{S}) &= \min_{|\phi_{\text{prod}}\rangle} \left\langle \phi_{\text{prod}} \left| \mathcal{P}_{\mathcal{S}}^{\perp} \right| \phi_{\text{prod}} \right\rangle \\ &= \min_{|\phi_{\text{prod}}\rangle} \operatorname{Tr}\left[ \mathcal{P}_{\mathcal{S}}^{\perp} \left|\phi_{\text{prod}}\right\rangle \left\langle \phi_{\text{prod}} \right| \right] \\ &\geqslant \min_{\rho \in \text{PPT}} \operatorname{Tr}\left[ \mathcal{P}_{\mathcal{S}}^{\perp} \rho \right], \end{aligned} \tag{6.36}$$

where PPT denotes the set of quantum states with positive partial transposes. This relaxation allows the computation of $E_G^{(k)}$ to be reformulated as an SDP:

$$\begin{aligned} \min_{\rho} \quad & \operatorname{Tr}[\mathcal{P}_{\mathcal{S}}^{\perp} \rho] \\ \text{subject to} \quad & \operatorname{Tr}[\rho] = 1, \\ & \rho^{\mathrm{T}_A} \succeq 0, \quad \rho \succeq 0. \end{aligned} \tag{6.37}$$

However, the PPT relaxation is only applicable for obtaining a lower bound when $k = 2$. For the quantification of higher-dimensional entanglement, the constraints can be further relaxed using $k$-positive maps. A commonly used $k$-positive map is the generalized reduction map [60, 206]:

$$R_p(X) = \operatorname{Tr}[X] \cdot I_d - p \cdot X, \tag{6.38}$$

where $X \in \mathbb{C}^{d \times d}$. It has been shown that $R_p$ is $k$-positive but not $(k+1)$-positive if $p \in \left(\frac{1}{k+1}, \frac{1}{k}\right]$. For any pure state $|\phi\rangle$ with $\text{SR}(|\phi\rangle) < k$, it holds that $I_{d_A} \otimes R_{\frac{1}{k-1}}(|\phi\rangle\langle\phi|) \succeq 0$. Consequently, the computation of lower bounds for $k$-GME can be reformulated as:

$$\begin{aligned} \min_{\rho} \quad & \operatorname{Tr}[\mathcal{P}_{\mathcal{S}}^{\perp} \rho] \\ \text{subject to} \quad & \operatorname{Tr}[\rho] = 1, \\ & I_{d_A} \otimes R_{\frac{1}{k-1}}(\rho) \succeq 0, \\ & \rho \succeq 0. \end{aligned} \tag{6.39}$$

In this subspace scenario, gradient-based optimization can also be utilized by following the same procedure as described in Sec. 6.1.2. Using similar trivialization maps, the unnormalized states with $k$-bounded Schmidt rank can be parameterized in the Euclidean space. Consequently, the computation in Eq. (6.34) can be reformulated as:

$$\begin{aligned} E_G^{(k)}(\mathcal{S}) &= \min_{\mathrm{SR}(|\phi\rangle)<k} \langle\phi|\mathcal{P}_{\mathcal{S}}^{\perp}|\phi\rangle \\ &= \min_{\boldsymbol{\theta}\in\mathbb{R}^D} \frac{\langle\tilde{\phi}(\boldsymbol{\theta})|\mathcal{P}_{\mathcal{S}}^{\perp}|\tilde{\phi}(\boldsymbol{\theta})\rangle}{\langle\tilde{\phi}(\boldsymbol{\theta})|\tilde{\phi}(\boldsymbol{\theta})\rangle}, \end{aligned} \tag{6.40}$$

where the dimension of the parameter space is $D = (2d_A + 2d_B + 1)(k-1)$. Additionally, the objective function can be expressed in terms of a tensor network as:

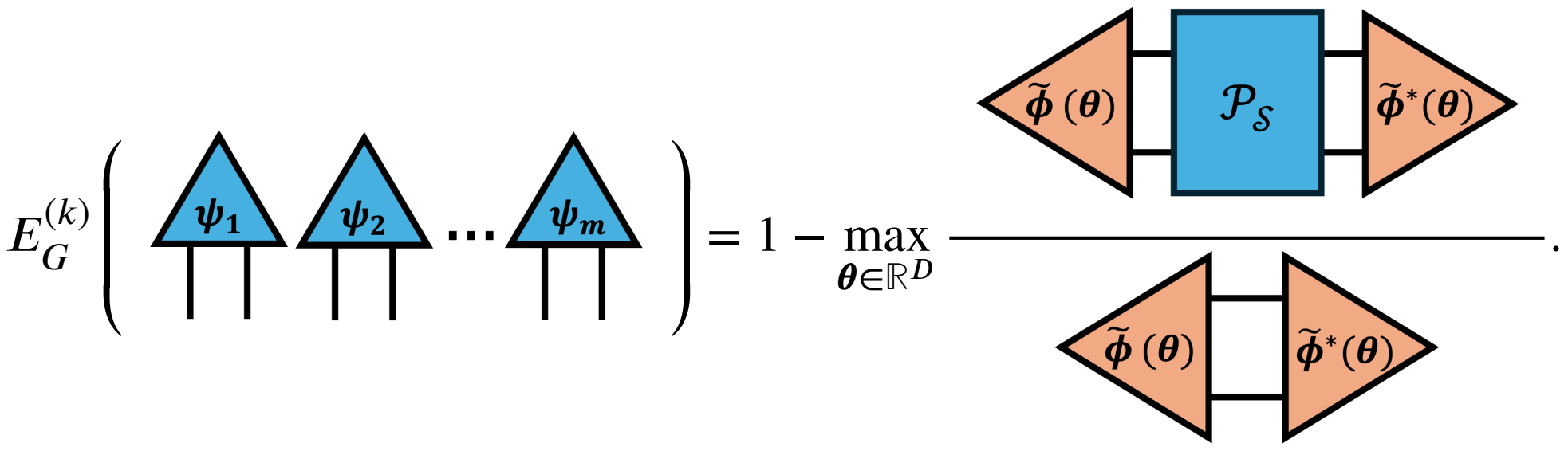

(6.41)

Let us start with a subspace for which the results are analytically known. The subspace $\mathcal{S}_{2\otimes d}^{\theta} \subset \mathcal{H}_A \otimes \mathcal{H}_B$, with $d_A = 2$ and $d_B = d$, is defined as the span of the following set of vectors [188]:

$$|\psi_i\rangle_{AB} = a|0\rangle_A |i\rangle_B + b|1\rangle_A |i+1\rangle_B, \tag{6.42}$$

where $i = 0, 1, \cdots, d-2$, with $a = \cos(\theta/2)$ and $b = \exp(i\xi)\sin(\theta/2)$, $\theta \in (0, \pi)$, and $\xi \in [0, 2\pi)$.

The dimension of $\mathcal{S}_{2\otimes d}^{\theta}$ is $d-1$, which corresponds to the maximal possible dimension of an entangled subspace in this scenario. Furthermore, it has been shown that

$$E_G^{(2)}\left(\mathcal{S}_{2\otimes d}^{\theta}\right) = \frac{1}{2}\left[1 - \sqrt{1 - \sin^2\theta \sin^2\left(\frac{\pi}{d}\right)}\right]. \tag{6.43}$$

In Fig. 6.4(a), we present $E_G^{(2)}(\mathcal{S}_{2\otimes d}^{\theta})$ as a function of $\theta$ for different dimensions, comparing the analytical results with those obtained via the gradient descent (GD) and PPT relaxation methods. For simplicity, we set $\xi = 0$. As shown, for these low-dimensional entangled subspaces, both methods yield accurate results.

The robustness of various entangled subspaces is further investigated in Fig. 6.4(b). Three different subspaces ($\theta = \frac{\pi}{2}, \frac{\pi}{4}, \frac{\pi}{6}$) with $d = 3$ are selected for comparison. For each operator norm of $H$, we randomly generate 1000 unitary perturbations $U = e^{-iH}$. Using the gradient descent strategy, we compute the minimum geometric measure $E_G^{(2)}$ after the perturbations for different operator norms. A nonzero value of $E_G^{(2)}$ signifies that the subspace remains entangled despite the perturbations. As established by Theorem 6.2.1, entangled subspaces with larger $E_G^{(2)}$ values exhibit greater robustness against perturbations.

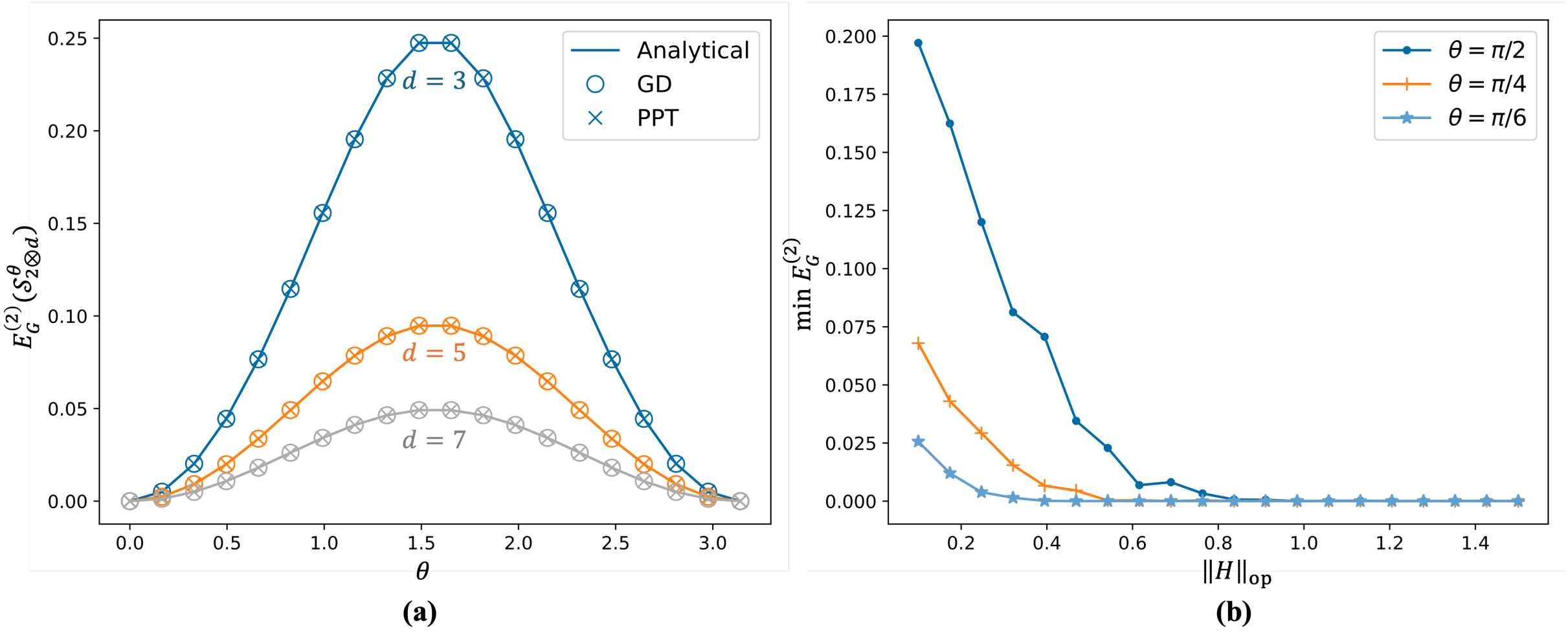

Figure 6.4: Results for $2 \otimes d$ entangled subspace $\mathcal{S}_{2\otimes d}^{\theta}$: (a) Numerical and analytical results for $E_G^{(2)}(\mathcal{S}_{2\otimes d}^{\theta})$ with respect to $\theta$. Here, GD denotes the method based on the gradient descent and PPT represents the results from the PPT relaxation. (b) Minimum values of $E_G^{(2)}$ for the subspace $\mathcal{S}_{2\otimes 3}^{\theta}$ after random unitary perturbation $U = e^{-iH}$ with different operator norms of $H$. A nonzero value indicates the subspace after perturbation maintains its entanglement.

Although the PPT relaxation can give accurate $E_G^{(2)}$ for most subspaces in practice, it fails when faced with bound entanglement and cannot estimate the higher-dimensional entanglement,*i.e.*, $E_G^{(k)}$ with $k > 2$. A well-known method for constructing bound entangled states involves the concept of an *unextendible product basis* (UPB) [75]. Such states can be expressed as:

$$\rho = \frac{1}{d_A d_B - d_{\mathcal{S}}} \mathcal{P}_{\mathcal{S}}^{\perp}, \tag{6.44}$$

where $d_A$ and $d_B$ denote the local dimensions of the bipartite system, $\mathcal{S}$ represents the subspace spanned by the UPB, and $d_{\mathcal{S}}$ is the dimension of this subspace. For instance, consider the five-state tiles UPB introduced in [207]:

$$\begin{aligned}
|\psi_1\rangle &= \frac{1}{\sqrt{2}}|0\rangle \otimes (|0\rangle - |1\rangle), \\
|\psi_2\rangle &= \frac{1}{\sqrt{2}}|2\rangle \otimes (|1\rangle - |2\rangle), \\
|\psi_3\rangle &= \frac{1}{\sqrt{2}}(|0\rangle - |1\rangle) \otimes |2\rangle, \\
|\psi_4\rangle &= \frac{1}{\sqrt{2}}(|1\rangle - |2\rangle) \otimes |0\rangle, \\
|\psi_5\rangle &= \frac{1}{3}(|0\rangle + |1\rangle + |2\rangle) \otimes (|0\rangle + |1\rangle + |2\rangle).
\end{aligned} \tag{6.45}$$

| | PPT | Reduction | GD |
|---|---|---|---|
| $E_G^{(2)}$ | 0.38237 | 0.25 | 0.38237 |
| $E_G^{(3)}$ | - | $10^{-9}$ | 0.06558 |
| $E_G^{(4)}$ | - | $10^{-9}$ | $10^{-13}$ |

Table 6.1: Numerical results for the high-dimensional entangled subspace. Here, "PPT" represents the PPT relaxation method, "Reduction" refers to the relaxation method based on the generalized reduction map, and "GD" denotes the gradient descent approach. A "-" indicates that the method cannot be applied in that scenario.

A UPB refers to a set of orthogonal product bases and no additional product state can be added to this set while maintaining orthogonality with all the existing states in the set. In simpler terms, a UPB is a collection of product states that "block" any other product state from being orthogonal to all of them, yet they leave a subspace of the Hilbert space unspanned. This leftover subspace, known as the orthogonal complementary subspace, contains only entangled states. Thus, for the subspace $\mathcal{S}$ spanned by the states in Eq. (6.45), its orthogonal complementary subspace $\mathcal{S}_\perp$ is entangled. We can utilize the PPT relaxation to estimate the lower bound of $E_G^{(2)}(\mathcal{S}_\perp)$, which gives around $10^{-12}$ close to 0 while gradient descent gives around 0.0284. That means in this case, the PPT relaxation cannot detect the entanglement within $\mathcal{S}_\perp$ while the gradient descent does.

As for high-dimensional entangled subspaces, let us consider the subspace $\mathcal{S}$ with $r(\mathcal{S}) = 3$, spanned by the following states in a $4 \otimes 4$ system [187]:

$$\begin{aligned} |\psi_1\rangle &= \frac{1}{2}(|0\rangle \otimes |0\rangle + |1\rangle \otimes |1\rangle + |2\rangle \otimes |2\rangle + |3\rangle \otimes |3\rangle), \\ |\psi_2\rangle &= \frac{1}{2}(|0\rangle \otimes |1\rangle + |1\rangle \otimes |2\rangle + |2\rangle \otimes |3\rangle + |3\rangle \otimes |0\rangle), \\ |\psi_3\rangle &= \frac{1}{2}(|0\rangle \otimes |2\rangle + |1\rangle \otimes |3\rangle + |2\rangle \otimes |0\rangle - |3\rangle \otimes |1\rangle). \end{aligned} \tag{6.46}$$

In Table 6.1, we summarize the results obtained using the PPT relaxation [Eq. (6.37)], the generalized reduction map [Eq. (6.39)], and the gradient descent method [Eq. (6.40)]. As shown, the PPT relaxation method can certify the entanglement of the subspace but cannot provide information about its entanglement dimensionality. The generalized reduction map successfully detects the entanglement of the subspace, though the lower bound it provides is weaker compared to that from the PPT relaxation method. In principle, the generalized reduction map can be applied to compute higher-dimensional entanglement, but in this case, it fails to provide a nontrivial lower bound for $k = 3$. In contrast, the gradient descent method not only gives the accurate value for $E_G^{(2)}$ but also certifies the high-dimensional entanglement of the subspace by demonstrating that $E_G^{(3)} > 0$.

### 6.2.2 Multipartite subspaces

In the context of multipartite subspaces, the primary emphasis is placed on completely or genuinely entangled subspaces. Completely entangled subspaces are defined as those that lack any *fully product states* [75], whereas genuinely entangled subspaces are devoid of product states across any *bipartition* [208, 209]. The significance of completely entangled subspaces lies in their ability to facilitate the local discrimination of pure quantum states [184, 210]. On the other hand, genuinely entangled subspaces have demonstrated their utility in the realm of quantum cryptography [211].

The identification of genuinely entangled subspaces can be achieved using techniques developed for bipartite settings, as genuine entanglement is defined with respect to bipartitions in multipartite systems. For completely entangled subspaces, we can define $E_G^{(2)}$ for a multipartite subspace $\mathcal{S}$ as follows:

$$E_G^{(2)}(\mathcal{S}) = \min_{\mathrm{TR}(|\phi\rangle)=1} \langle\phi|\mathcal{P}_{\mathcal{S}}^{\perp}|\phi\rangle, \tag{6.47}$$

where $\mathrm{TR}(|\phi\rangle) = 1$ signifies that the state $|\phi\rangle$ is fully product. Clearly, a subspace $\mathcal{S}$ is completely entangled if and only if $E_G^{(2)}(\mathcal{S}) > 0$. A lower bound for $E_G^{(2)}$ can be estimated via a generalized PPT relaxation, expressed as:

$$\begin{aligned} \min_{\rho} \quad & \mathrm{Tr}[\mathcal{P}_{\mathcal{S}}^{\perp}\rho] \\ \text{subject to} \quad & \mathrm{Tr}[\rho] = 1, \\ & \rho \succeq 0, \\ & \rho^{\mathrm{T}_i} \succeq 0 \quad \forall i. \end{aligned} \tag{6.48}$$

Here, $\rho^{\mathrm{T}_i}$ represents the partial transpose of $\rho$ with respect to the $i$th local Hilbert space $\mathcal{H}_i$.

For the gradient descent approach, the optimization becomes straightforward by parameterizing a fully product state $|\phi\rangle$, which is associated with the product manifolds of sphere manifolds introduced in Sec. 5.2. Consequently, the optimization problem in Eq. (6.47) can be reformulated as:

$$E_G^{(2)}(\mathcal{S}) = \min_{\boldsymbol{\theta}\in\mathbb{R}^D} \langle\phi(\boldsymbol{\theta})|\mathcal{P}_{\mathcal{S}}^{\perp}|\phi(\boldsymbol{\theta})\rangle, \tag{6.49}$$

where the dimension of the parameter space is $D = 2\sum_i d_i$, and $d_i$ is the dimension of the $i$th

Table 6.2: Computational time and numerical results for certifying completely entangled subspaces via different methods. Here, the numerical methods include the hierarchical method (Hier), generalized PPT relaxation (PPT), and gradient descent method (GD). “-” means it is not available in acceptable time.

| $(d_1, d_2, d_3)$ | Time(s) | | | $E_G^{(2)}$ | |
|---|---|---|---|---|---|
| | **Hier** | **PPT** | **GD** | **PPT** | **GD** |
| $(2,2,2)$ | 0.01 | 0.09 | 0.07 | $2.00 \times 10^{-1}$ | $2.50 \times 10^{-1}$ |
| $(2,2,6)$ | 0.30 | 0.33 | 0.10 | $1.23 \times 10^{-2}$ | $1.23 \times 10^{-2}$ |
| $(2,3,4)$ | 180.42 | 0.32 | 0.08 | $1.41 \times 10^{-2}$ | $1.41 \times 10^{-2}$ |
| $(2,3,6)$ | - | 1.07 | 0.15 | $2.86 \times 10^{-3}$ | $2.86 \times 10^{-3}$ |
| $(3,3,6)$ | - | 2.71 | 0.18 | $7.20 \times 10^{-4}$ | $7.20 \times 10^{-4}$ |
| $(3,4,7)$ | - | 18.27 | 0.40 | $7.98 \times 10^{-5}$ | $7.98 \times 10^{-5}$ |
| $(4,4,7)$ | - | 189.31 | 0.55 | $2.02 \times 10^{-5}$ | $2.02 \times 10^{-5}$ |
| $(4,5,10)$ | - | - | 2.92 | - | $3.54 \times 10^{-7}$ |

local Hilbert space $\mathcal{H}_i$.

We begin with a three-qubit subspace $\mathcal{S}$ in $\mathcal{H}_A \otimes \mathcal{H}_B \otimes \mathcal{H}_C$, spanned by the following states [212]:

$$\begin{aligned} |\psi_1\rangle &= |0\rangle_A |0\rangle_B |0\rangle_C, \\ |\psi_2\rangle &= |1\rangle_A |+\rangle_B |-\rangle_C, \\ |\psi_3\rangle &= |-\rangle_A |1\rangle_B |+\rangle_C, \\ |\psi_4\rangle &= |+\rangle_A |-\rangle_B |1\rangle_C, \end{aligned} \tag{6.50}$$

where $|-\rangle = \frac{1}{\sqrt{2}}(|0\rangle - |1\rangle)$ and $|+\rangle = \frac{1}{\sqrt{2}}(|0\rangle + |1\rangle)$.

These states form a UPB, meaning that no fully product state exists in the orthogonal complement of the subspace $\mathcal{S}$ spanned by them. In other words, $\mathcal{S}^{\perp}$ is a completely entangled subspace. However, similar to the bipartite UPB case, the PPT relaxation method fails to detect its entanglement, as the lower bound of $E_G^{(2)}$ obtained is approximately $10^{-14}$. In contrast, the gradient descent method yields a value of around 0.08144, which is close to the analytical value $1 - \frac{3\sqrt{6}}{8}$.

Another class of completely entangled subspaces in the $d_1 \otimes d_2 \otimes d_3$ system is described in [213] as follows:

$$\begin{aligned} \mathcal{S} = \operatorname{span}\{ &|i_1\rangle \otimes |i_2\rangle \otimes |i_3\rangle - |j_1\rangle \otimes |j_2\rangle \otimes |j_3\rangle : \\ &\sum_k i_k = \sum_k j_k,\ 0 \le i_k \le d_k - 1,\ \text{for } 1 \le k \le 3\}, \end{aligned} \tag{6.51}$$

which achieves the largest possible dimension $d_1 d_2 d_3 - d_1 - d_2 - d_3 + 2$ in tripartite systems.

In Table 6.2, we compare various numerical methods for certifying the completely entangled subspaces defined in Eq. (6.51) across different dimensions. The results are obtained using the hierarchical method [187], the generalized PPT relaxation [Eq. (6.48)], and the gradient descent approach [Eq. (6.49)].

As shown, the hierarchical method is the most constrained by the size of the dimensions. While the PPT method is capable of detecting most cases successfully and provides accurate lower bounds for $E_G^{(2)}$ in relatively small dimensions, its computational time increases significantly because it requires a quantum state $\rho$ with a dimension of $d_1 d_2 d_3 \times d_1 d_2 d_3$. In contrast, the gradient descent approach can handle much larger dimensions efficiently, as it only requires $2(d_1 + d_2 + d_3)$ real parameters.

## 6.3 Mixed State Entanglement

In practice, calculating an entanglement measure $E(\rho)$ for a given mixed state $\rho$ can often be facilitated by determining its upper and lower bounds. Several approaches have been developed for this purpose. For instance, convex hull approximations [214, 215] can be employed to approximate the set of separable states from the "inside," thereby providing an upper bound. Conversely, semidefinite programming (SDP) techniques can be utilized to derive a lower bound by optimizing over the set of states with positive partial transposes (PPT) or $k$-symmetric extendible states [216], which approximate separable states from the "outside". Despite significant progress in this area, the computation of entanglement measures remains a challenging problem, formally proven to be NP-hard.

### 6.3.1 Bipartite mixed states

In this section, we primarily focus on bipartite mixed states. Extending the definition of $k$-GME for bipartite pure states, we can define $k$-GME for bipartite mixed states using a convex roof construction, as introduced in Sec. 4.2.2:

**Definition 21 ($k$-GME for Bipartite Mixed States)** Consider a bipartite mixed state $\rho$ in the composite Hilbert space $\mathcal{H}_A \otimes \mathcal{H}_B$. The geometric measure of $k$-bounded Schmidt number is defined as:

$$E_G^{(k)}(\rho) = \min_{\{p_i, |\psi_i\rangle\}} \sum_i p_i E_G^{(k)}(|\psi_i\rangle), \tag{6.52}$$

where the minimization is taken over all possible pure-state decompositions $\{p_i, |\psi_i\rangle\}$ satisfying $\rho = \sum_i p_i |\psi_i\rangle\langle\psi_i|$.

Here, $E_G^{(k)}$ is referred to as the geometric measure of $k$-bounded Schmidt number for a mixed state $\rho$, as $k$-GME constructed via the convex roof extension is also a distance-based measure [217]. It quantifies the geometric "distance" between the given state $\rho$ and the set of states with a $k$-bounded Schmidt number:

**Theorem 6.3.1 (Distance-Based Alternative of $k$-GME for Mixed States)** For a mixed state $\rho$ in the composite Hilbert space $\mathcal{H}_A \otimes \mathcal{H}_B$, the $k$-GME expression defined in Eq. (6.52) is equivalent to:

$$E_G^{(k)}(\rho) = \min_{\sigma \in S_{k-1}} D(\rho, \sigma), \tag{6.53}$$

where $D(\rho, \sigma) = 1 - F(\rho, \sigma)$ is a distance measure derived from Uhlmann's fidelity $F(\rho, \sigma) = \|\sqrt{\rho}\sqrt{\sigma}\|_1^2$, and $S_{k-1}$ denotes the set of states with at most $k-1$ Schmidt number.

*Proof.* See Appendix A, Theorem 2, in Ref. [217]. □

Clearly, $\mathrm{SN}(\rho) = k$ if and only if $E_G^{(k)}(\rho) > 0$ and $E_G^{(k+1)}(\rho) = 0$. Furthermore, from Carathéodory's theorem [98], it follows that any state $\sigma$ in $S_{k-1}$ can be represented as a convex combination of at most $n \leq (d_A d_B)^2$ pure states, since $S_{k-1}$ is a convex subset of the entire quantum state space with dimension $d_A d_B$. This leads to the following theorem:

**Theorem 6.3.2** For any state $\rho$ in $\mathcal{H}_A \otimes \mathcal{H}_B$ with dimensions $d_A, d_B$, there always exists an optimal decomposition $\{p_i, |\psi_i\rangle\}_{i=1}^n$ for computing $E_G^{(k)}(\rho)$ such that $n \leq (d_A d_B)^2$.

Furthermore, by definition, $k$-GME is an entanglement monotone [114] that fulfills a series of desirable properties for quantifying high-dimensional entanglement:

**Theorem 6.3.3** The geometric measure of $k$-bounded Schmidt number $E_G^{(k)}$ satisfies the following properties:

- $E_G^{(k)}(\rho) \geq 0$ and equality holds if and only if $\mathrm{SN}(\rho) < k$.
- $E_G^{(k)}$ is local unitary invariant, *i.e.*, $E_G^{(k)}(U\rho U^\dagger) = E_G^{(k)}(\rho)$, where $U = U_A \otimes U_B$.
- $E_G^{(k)}$ is convex under the discarding of information, *i.e.*, $\sum_i p_i E_G^{(k)}(\rho_i) \geq E_G^{(k)}(\sum_i p_i \rho_i)$.
- $E_G^{(k)}(\rho) \leq E_G^{(k')}(\rho)$ for any quantum state $\rho$ if $k > k'$.
- $E_G^{(k)}$ cannot increase under local operations and classical communication (LOCC), *i.e.*, $E_G^{(k)}(\Lambda(\rho)) \leq E_G^{(k)}(\rho)$ for any LOCC map $\Lambda$.

*Proof.* We can prove the first four properties easily from the definition of $E_G^{(k)}(\rho)$ in Eq. (6.52). For the monotonicity under LOCC, we can prove it from the distance expression of $E_G^{(k)}$ in

Theorem 6.3.1:

$$\begin{aligned} E_G^{(k)}(\Lambda(\rho)) &= \min_{\sigma \in S_{k-1}} D(\Lambda(\rho), \sigma) \\ &\leq \min_{\sigma \in S_{k-1}} D(\Lambda(\rho), \Lambda(\sigma)) \\ &\leq \min_{\sigma \in S_{k-1}} D(\rho, \sigma) = E_G^{(k)}(\rho). \end{aligned}$$

The second line follows that the Schmidt number of $\sigma$ will not increase under LOCC [60]. The third line is because $F(\mathcal{E}(\rho), \mathcal{E}(\sigma)) \geq F(\rho, \sigma)$ for any trace-preserving completely positive map $\mathcal{E}$ [218]. □

Similar to the robustness analysis for entangled subspaces, in the case of bipartite mixed states, the $k$-GME quantifies the robustness of a quantum state under unitary perturbations, as expressed below:

**Theorem 6.3.4 (Robustness Analysis for Entangled States)** Let $\rho \in \mathcal{H}_A \otimes \mathcal{H}_B$ be a quantum state with $\mathrm{SN}(\rho) \geq k$, *i.e.*, $E_G^{(k)}(\rho) > 0$. For any unitary perturbation $U = e^{-iH}$ applied to $\rho$, if the half spectral width satisfies $\frac{\Delta\lambda(H)}{2} < \sqrt{E_G^{(k)}(\rho)}$, the resulting perturbed state $\rho'$ will also satisfy $\mathrm{SN}(\rho') \geq k$.

*Proof.* According to Ref. [78], we know that Bures angle $\theta(\rho, \sigma) = \arcsin \sqrt{D(\rho, \sigma)}$ satisfies the triangle inequality, *i.e.*, $\theta(\rho, \sigma) + \theta(\sigma, \tau) \geq \theta(\rho, \tau)$. And obviously, $E_G^{(k)}(\rho) > 0$ is equivalent to $\min_{\sigma \in S_{k-1}} \theta(\rho, \sigma) > 0$. Then we require

$$\min_{\sigma \in S_{k-1}} \theta(\rho, \sigma) - \theta(\rho, U\rho U^\dagger) > 0,$$

so that from the triangle inequality, we have

$$\begin{aligned} \min_{\sigma \in S_{k-1}} \theta(U\rho U^\dagger, \sigma) &\geq \min_{\sigma \in S_{k-1}} [\theta(\rho, \sigma) - \theta(\rho, U\rho U^\dagger)] \\ &= \min_{\sigma \in S_{k-1}} \theta(\rho, \sigma) - \theta(\rho, U\rho U^\dagger) > 0. \end{aligned}$$

In order to ensure $\min_{\sigma \in S_{k-1}} \theta(\rho, \sigma) - \theta(\rho, U\rho U^\dagger) > 0$, we require that

$$D(\rho, U\rho U^\dagger) < \min_{\sigma \in S_{k-1}} D(\rho, \sigma) = E_G^{(k)}(\rho)$$

For $D(\rho, U\rho U^\dagger)$, we have

$$\begin{aligned}
D(\rho, U\rho U^\dagger) =&1 - \left(\sqrt{F(\rho, U\rho U^\dagger)}\right)^2 \\
=&1 - \left(\sqrt{F(\sum_i p_i|\psi_i\rangle\langle\psi_i|, \sum_i p_i U|\psi_i\rangle\langle\psi_i|U^\dagger)}\right)^2 \\
\leq&1 - \left(\sum_i p_i \sqrt{F(|\psi_i\rangle, U|\psi_i\rangle)}\right)^2 \\
=&1 - \left(\sum_i p_i |\langle\psi_i|U|\psi_i\rangle|\right)^2 \\
\leq&1 - \min_\psi |\langle\psi|e^{-iH}|\psi\rangle|^2
\end{aligned}$$

where $\rho = \sum_i p_i|\psi_i\rangle\langle\psi_i|$ is the pure-state decomposition of $\rho$, and the third line uses the strong concavity of $\sqrt{F(\rho,\sigma)}$ [78], *i.e.*,

$$\sqrt{F(\sum_i p_i\rho_i, \sum_i q_i\sigma_i)} \geq \sum_i \sqrt{p_i q_i}\sqrt{F(\rho_i, \sigma_i)}.$$

Now, suppose $U = \sum_j e^{-ih_j}|j\rangle\langle j|$ and $\Delta\lambda(H) = h_{\max} - h_{\min} \in [0, \pi]$, where $h_{\max}$ and $h_{\min}$ represent the maximum and minimum eigenvalues of $H$. For $|\psi\rangle = c_i|i\rangle$, we have

$$\langle\psi|e^{-iH}|\psi\rangle = \sum_j |c_j|^2 e^{-ih_j} = \sum_j p_j e^{-ih_j},$$

where $p_j = |c_j|^2$ and $\sum_j p_j = 1$. So all possible expectation values are a convex combination of the eigenvalues $e^{-ih_j}$. As $|\psi\rangle$ runs over all pure states, we have

$$\min_\psi |\langle\psi|e^{-iH}|\psi\rangle| = \cos\left(\frac{\Delta\lambda(H)}{2}\right),$$

which gives $D(\rho, U\rho U^\dagger) \leq \sin^2(\frac{\Delta\lambda(H)}{2}) \leq (\frac{\Delta\lambda(H)}{2})^2$. Thus we obtain that if $\frac{\Delta\lambda(H)}{2} < E_G^{(k)}(\rho)^{\frac{1}{2}}$, then $E_G^{(k)}(U\rho U^\dagger)$ will be nonzero after the perturbation $U = e^{-iH}$. □

In the following, we present the numerical approaches relevant to computing the $k$-GME in Eq. (6.52). Specifically, the upper bound can be derived using the gradient descent method, while the lower bound is obtained via the SDP approach.

Consider a quantum state $\rho$ defined on the Hilbert space $\mathcal{H}_A \otimes \mathcal{H}_B$ with dimensions $d_A$ and $d_B$. The eigenvalue decomposition of $\rho$ is given by $\rho = \sum_{i=1}^r |\tilde{\lambda}_i\rangle\langle\tilde{\lambda}_i|$, where $\langle\tilde{\lambda}_i|\tilde{\lambda}_j\rangle = \lambda_i\delta_{ij}$,

and $r$ is the rank of $\rho$. It can be verified that for any $n$-entry pure-state decomposition $\rho = \sum_{i=1}^{n} p_i |\psi_i\rangle\langle\psi_i|$, the decomposition can be expressed using a set of auxiliary states $\{|\tilde{\psi}_i\rangle\}_{i=1}^{n}$ satisfying $p_i = \langle\tilde{\psi}_i|\tilde{\psi}_i\rangle$ and $|\psi_i\rangle = \frac{1}{\sqrt{p_i}}|\tilde{\psi}_i\rangle$. These auxiliary states take the following form:

$$|\tilde{\psi}_i\rangle = \sum_{j=1}^{r} X_{ij}|\tilde{\lambda}_j\rangle, \tag{6.54}$$

where $X$ is a $n \times r$ complex matrix referred to as the *Stiefel matrix*, satisfying the orthonormality condition $X^\dagger X = I$. The collection of all $n \times r$ Stiefel matrices forms a special manifold known as the *Stiefel manifold*, denoted by $\mathbb{St}(n,r)$. Consequently, the optimization for a rank-$r$ state over the $n$-entry pure-state decomposition is equivalent to optimization over the Stiefel manifold $\mathbb{St}(n,r)$, *i.e.*,

$$\min_{\{p_i,|\psi_i\rangle\}} \sum_{i=1}^{n} p_i E_G^{(k)}(|\psi_i\rangle) = \min_{X\in\mathbb{St}(n,r)} \sum_{i=1}^{n} p_i(X) E_G^{(k)}(|\psi_i(X)\rangle). \tag{6.55}$$

For $E_G^{(k)}(|\psi\rangle)$, we have $E_G^{(k)}(|\psi\rangle) = 1 - \max_{\mathrm{SR}(|\phi\rangle)<k} |\langle\phi|\psi\rangle|^2$. Thus,

$$\begin{aligned}
\min_{\{p_i,|\psi_i\rangle\}} \sum_{i=1}^{n} p_i E_G^{(k)}(|\psi_i\rangle) &= 1 - \max_{X\in\mathbb{St}(n,r),\mathrm{SR}(|\phi_i\rangle)<k} \sum_{i=1}^{n} p_i(X)|\langle\phi_i|\psi_i(X)\rangle|^2 \\
&= 1 - \max_{X\in\mathbb{St}(n,r),\mathrm{SR}(|\phi_i\rangle)<k} \sum_{i=1}^{n} |\langle\phi_i|\tilde{\psi}_i(X)\rangle|^2 \\
&= 1 - \max_{X\in\mathbb{St}(n,r),\mathrm{SR}(|\phi_i\rangle)<k} \sum_{i=1}^{n} |\langle\phi_i| \sum_{j=1}^{r} X_{ij}|\tilde{\lambda}_j\rangle|^2.
\end{aligned} \tag{6.56}$$

Equivalently, the computation can be represented in the tensor network, as follows

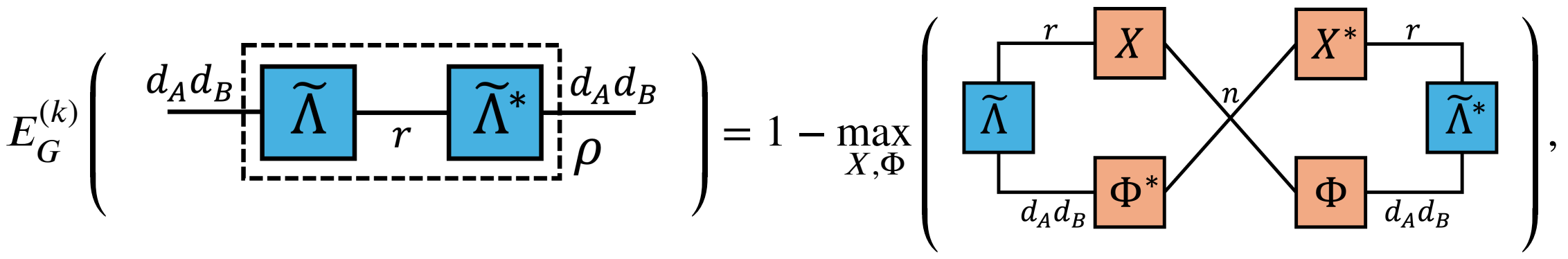

(6.57)

where $\tilde{\Lambda} = (|\tilde{\lambda}_1\rangle, |\tilde{\lambda}_2\rangle, \ldots, |\tilde{\lambda}_r\rangle)$ and $\Phi = (|\phi_1\rangle, |\phi_2\rangle, \ldots, |\phi_n\rangle)$.

For states with $k$-bounded Schmidt rank in the tensor $\Phi$, the same procedure described in Sec. 6.1.2 can be applied to parameterize them, with the addition of a normalization operation. Regarding the Stiefel manifold, an effective trivialization map is the *polar projection*, defined as

$$\phi(A) = A(A^\dagger A)^{-\frac{1}{2}} : \mathbb{C}^{n\times r} \to \mathbb{St}(n,r), \tag{6.58}$$

where $2nr$ real parameters are used to parameterize an $n \times r$ full-rank complex matrix $A$, while the dimension of the Stiefel manifold $\mathbb{St}(n, r)$ is $2nr - r^2$. Alternatively, the polar projection can also be expressed as

$$A = U\Sigma V^\dagger \to \phi(A) = UV^\dagger, \tag{6.59}$$

where $U$, $\Sigma$, and $V$ are obtained via the singular value decomposition (SVD) of $A$. This transformation is continuous and differentiable over the full-rank domain, which is critical for gradient-based optimization. Additionally, the polar projection satisfies the following geometric property:

**Theorem 6.3.5** Let $A$ be a full-rank $n \times r$ complex matrix. The polar projection $\phi : \mathbb{C}^{n\times r} \to \mathbb{St}(n, r)$ satisfies

$$\phi(A) = \arg \min_{X \in \mathbb{St}(n,r)} \|X - A\|_F, \tag{6.60}$$

where $\| \cdot \|_F$ represents the Frobenius norm.

In other words, the polar projection $\phi$ of a full-rank matrix $A$ onto the Stiefel manifold is the closest point on the Stiefel manifold to $A$ in terms of the *Euclidean norm*. This property is particularly useful for optimization over the Stiefel manifold, as it ensures that the smallest possible change is made to the original matrix $A$. This minimal perturbation preserves the original information in $A$ as much as possible, such as the gradient information, which is crucial for optimization algorithms that depend on the structure of $A$. Furthermore, projecting in a way that minimizes the distance at each iteration helps prevent the accumulation of errors, leading to more stable convergence.

Next, we turn our attention to the SDP approach. Similar to the procedure for deriving lower bounds for entangled subspaces, we can use $(k-1)$-positive maps to characterize the set of states with a $k$-bounded Schmidt number. Moreover, as introduced in Sec. 5.1.2, the fidelity function between two quantum states $\rho$ and $\sigma$ can be computed using SDP as follows:

$$\begin{aligned} \sqrt{F(\rho, \sigma)} = \max_{Y,Z} \quad & \mathrm{Tr}[Y] \\ \text{subject to} \quad & \begin{pmatrix} \rho & Y + iZ \\ Y - iZ & \sigma \end{pmatrix} \geq 0. \end{aligned} \tag{6.61}$$

Since Eq. (6.52) is formalized as a distance-based measure (Theorem 6.3.1), for a quantum state $\rho$ in a $d_A \otimes d_B$ system, the upper bound of $\max_{\sigma \in S_{k-1}} \sqrt{F(\rho, \sigma)}$ can be expressed as:

$$\begin{aligned}
\max_{Y,Z} \quad & \mathrm{Tr}[Y] \\
\text{subject to} \quad & \mathrm{Tr}[\sigma] = 1, \\
& \begin{pmatrix} \rho & Y + iZ \\ Y - iZ & \sigma \end{pmatrix} \geq 0, \\
& I_{d_A} \otimes R_{\frac{1}{k-1}}(\sigma) \geq 0, \quad \sigma \geq 0.
\end{aligned} \tag{6.62}$$

Here, we adopt the generalized reduction map as our $(k-1)$-positive map, which is defined as:

$$R_{\frac{1}{k-1}}(X) = \mathrm{Tr}[X] \cdot I - \frac{1}{k-1} X. \tag{6.63}$$

Thus, we can finally compute the lower bound for $E_G^{(k)}(\rho)$ as:

$$E_G^{(k)}(\rho) = 1 - \left( \max_{\sigma \in S_{k-1}} \sqrt{F(\rho, \sigma)} \right)^2. \tag{6.64}$$

Specifically, for $k = 2$, the PPT relaxation, as shown in Eq. (5.17), can provide a better lower bound than the generalized reduction map.

We begin with the $d \otimes d$ Werner states, which are parameterized by $\alpha \in [-1, 1]$, and defined as:

$$\rho_W(\alpha) = \frac{1}{d^2 - d\alpha} I_{d^2} - \frac{\alpha}{d^2 - d\alpha} \sum_{ij} |i, j\rangle\langle i, j|, \tag{6.65}$$

where $I_{d^2}$ denotes the $d^2$-by-$d^2$ identity matrix. For $\alpha \in [-1, 1/d]$, this state is known to be separable [130], while for $\alpha \in [1/d, 1]$, it has been shown that [49]:

$$E_G^{(2)}(\rho_W(\alpha)) = \frac{1}{2} \left( 1 - \sqrt{1 - \left( \frac{d\alpha - 1}{\alpha - d} \right)^2} \right). \tag{6.66}$$

Additionally, it has been proven that the Schmidt number of Werner states is at most two [61]. In other words, $E_G^{(k)}(\rho_W(\alpha)) = 0$ for $k > 2$.

In Fig. 6.5, we calculate the $k$-GME for $4 \otimes 4$ Werner states using the gradient descent method. For $k = 2$, the numerical results align with the analytical expression given in Eq. (6.66). The sharp transition of $E_G^{(2)}$ clearly demonstrates the boundary between separable states (with Schmidt number one) and entangled states (with Schmidt number two). For $k > 2$, as expected, the numerical results are approximately zero.

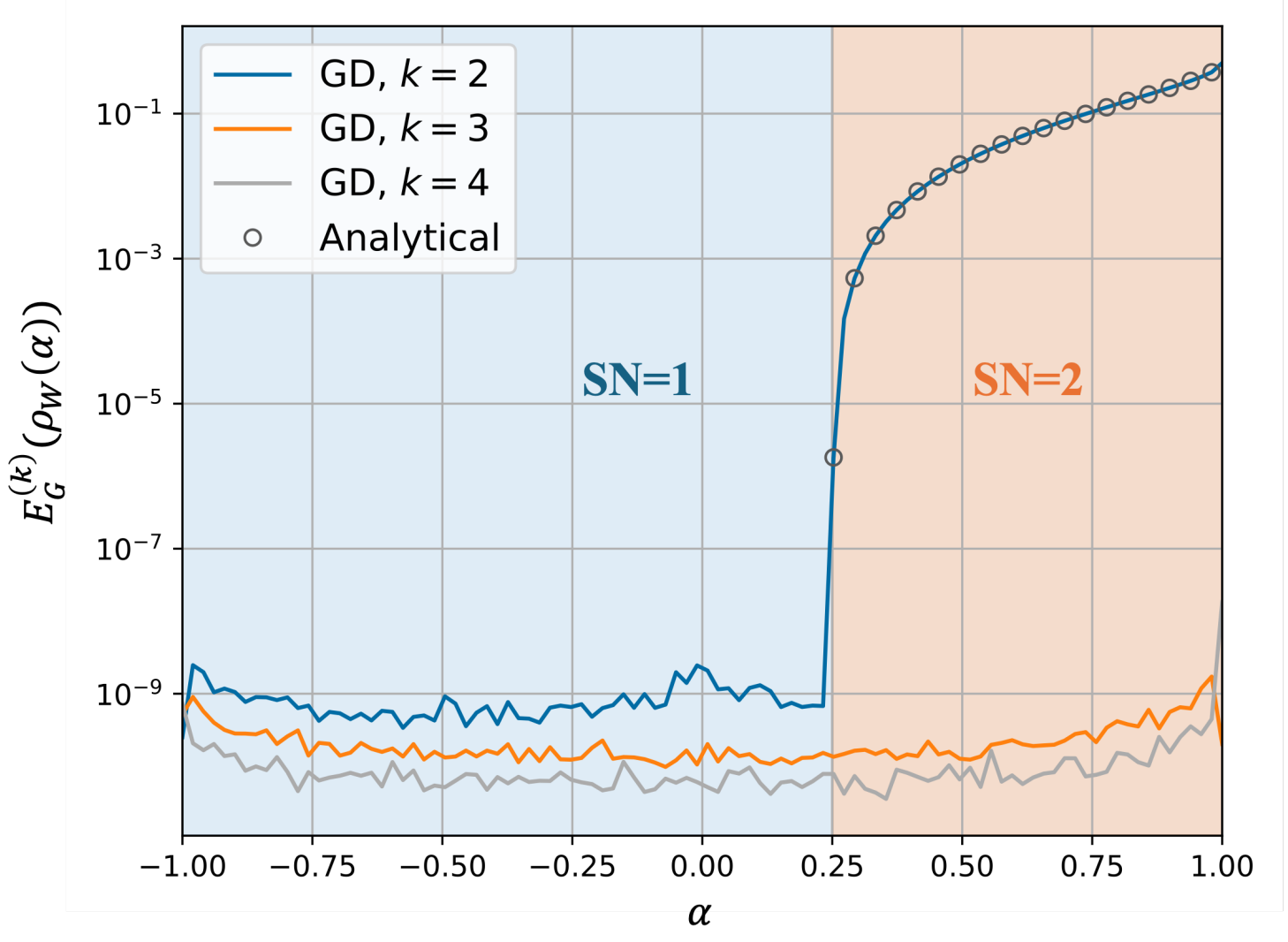

Figure 6.5: The $k$-GME results for $4 \otimes 4$ Werner states. The solid lines represent numerical results obtained via gradient descent (GD), while the dots indicate analytical results. Werner states are categorized into two regions based on their Schmidt number.

For the $d \otimes d$ isotropic states, parameterized by $F \in [0, 1]$, the states are expressed as:

$$\rho_I(F) = \frac{1-F}{d^2-1}\left(I_{d^2} - |\Psi_d^+\rangle\langle\Psi_d^+|\right) + F|\Psi_d^+\rangle\langle\Psi_d^+|, \tag{6.67}$$

where $|\Psi_d^+\rangle = \frac{1}{\sqrt{d}}\sum_{i=0}^{d-1}|ii\rangle$ represents the maximally entangled state. For $F \in [0, 1/d]$, this state is known to be separable [106], while for $F \geq 1/d$, it has been shown that [49]:

$$E_G^{(2)}(\rho_I(F)) = 1 - \frac{1}{d}\left[\sqrt{F} + \sqrt{(1-F)(d-1)}\right]^2. \tag{6.68}$$

Unlike the Werner states, which have a Schmidt number of at most two, isotropic states are mixtures of the maximally entangled state $|\Psi_d^+\rangle$ and the maximally mixed state $\frac{1}{d^2}I_{d^2}$. This implies that the Schmidt number of $\rho_I(F)$ can range from 1 to $d$ as the parameter $F$ changes. Here, we extend the analytical results for the case $k = 2$ to cases where $k > 2$, as follows:

**Theorem 6.3.6 ($k$-GME for Isotropic States)** For a $d \otimes d$ isotropic state $\rho_I(F)$, we have

$$E_G^{(k)}(\rho_I(F)) = \begin{cases} 0 & 0 \leq F \leq \frac{k-1}{d}, \\ 1 - \frac{1}{d}\left[\sqrt{F(k-1)} + \sqrt{(1-F)(d-k+1)}\right]^2 & \frac{k-1}{d} < F \leq 1, \end{cases} \tag{6.69}$$

where $k = 2, 3, \ldots, d$.

*Proof.* One important property of isotropic states is their invariance under the operation $U \otimes U^*$

for any unitary transformation $U$. We introduce a twirling superoperator $S^{U\otimes U^*}$ as follows:

$$S^{U\otimes U^*}(\rho) = \frac{1}{\mathrm{Vol}(U)} \int dU\, U \otimes U^* \rho U^\dagger \otimes U^{*\dagger}.$$

Let us consider an arbitrary pure state under the Schmidt decomposition $|\psi\rangle = \sum_{i=1}^d \mu_i |a_i, b_i\rangle$, where $\sum_i \mu_i^2 = 1$ and $\mu_i \geq 0$. Applying the twirling operation, we can rewrite $|\psi\rangle$ as $|\psi\rangle = U_A \otimes U_B \sum_i \mu_i |i,i\rangle$. From [129], we know:

$$S^{U\otimes U^*}(|\psi\rangle\langle\psi|) = \frac{|\sum_i \nu_i|^2}{d} |\Psi_d^+\rangle\langle\Psi_d^+| + \frac{1 - |\sum_i \nu_i|^2/d}{d^2-1}\left(\mathbb{I}_{d^2} - |\Psi_d^+\rangle\langle\Psi_d^+|\right) = \rho_I\left(\frac{|\sum_i \nu_i|^2}{d}\right),$$

where $\nu_i = \mu_i V_{ii} = \mu_i \langle i|U_A^T U_B|i\rangle$. For general $V$, we observe:

$$|\sum_i \nu_i|^2 \leq \left(\sum_i |\nu_i|\right)^2 \leq \left(\sum_i \mu_i\right)^2, \tag{6.70}$$

since $|V_{ii}| \leq 1$ for all $i$. Thus, given the fixed $\boldsymbol{\mu}$, the largest value for $F$ can be obtained from $|\psi\rangle = \sum_{i=1}^d \mu_i |ii\rangle$.

Assume the optimal pure-state decomposition of $\rho_I(F)$ is $\{p_i, |\psi_i(V^i, \boldsymbol{\mu}^i)\rangle\}$ for computing $E_G^{(k)}$, where $V^i$ is determined by the corresponding local operations on $|\psi_i\rangle$ and $\boldsymbol{\mu}^i$ denotes the Schmidt vector of $|\psi_i\rangle$. By twirling both sides of $\rho_I(F) = \sum_i p_i |\psi_i\rangle\langle\psi_i|$, we find $\rho_I(F) = \sum_i p_i \rho_I(F_i) = \rho_I\left(\sum_i p_i F_i\right)$, with:

$$F_i(V^i, \boldsymbol{\mu}^i) = \frac{|\sum_j V_{jj}^i \mu_j^i|^2}{d}.$$

Thus, we have:

$$E_G^{(k)}\left(\rho_I(\sum_i p_i F_i)\right) = \sum_i p_i E_G^{(k)}(|\psi_i\rangle) = \sum_i p_i \left|1 - \sum_{j=1}^{k-1} (\mu_j^i)^2\right|,$$

where $F_i = |\sum_j V_{jj}^i \mu_j^i|^2/d$ and $\boldsymbol{\mu}^i = (\mu_1^i, \mu_2^i, \ldots, \mu_d^i)$ is the Schmidt vector of $|\psi_i\rangle$.

Now, consider the function:

$$R_V^{(k)}(F) = \min_{\boldsymbol{\mu}} \left\{ 1 - \sum_{i=1}^{k-1} \mu_i^2 \,\middle|\, F = \frac{|\sum_{i=1}^d V_{ii}\mu_i|^2}{d}, \sum_i \mu_i^2 = 1, \mu_i \geq 0 \right\},$$

which leads to the fact that $E_G^{(k)}(\rho_I(F)) = \mathrm{conv}(R_V^{(k)}(F))$. Here, $\mathrm{conv}(\cdot)$ means the convex hull of all functions. Furthermore, we can consider another function that does not rely on $V$:

$$R^{(k)}(F) = \min_{\boldsymbol{\mu}} \left\{ 1 - \sum_{i=1}^{k-1} \mu_i^2 \,\middle|\, F = \frac{\left(\sum_{i=1}^{d} \mu_i\right)^2}{d}, \sum_i \mu_i^2 = 1, \mu_i \geq 0 \right\}. \tag{6.71}$$

From Eq. (6.70), it can be shown that for any $V$, there exists an $F' \geq F$ such that $R_V^{(k)}(F) = R^{(k)}(F')$, so $\mathrm{conv}(R_V^{(k)}(F)) = \mathrm{conv}(R_+^{(k)}(F))$, where $R_+^{(k)}(F) = \min_x \{R^{(k)}(x) \,|\, x \geq F\}$.

We now determine the function $R^{(k)}(F)$ defined in Eq. (6.71). Without loss of generality, due to the symmetry of $\mu_i$, assume $\mu_i = a$ for $i = 1, \ldots, k-1$ and $\mu_i = b$ for $i = k, \ldots, d$ such that $a \geq b \geq 0$. From the constraints, we have:

$$\begin{cases} (k-1)a + (d-k+1)b = \sqrt{Fd}, \\ (k-1)a^2 + (d-k+1)b^2 = 1. \end{cases}$$

Solving for $a$ and $b$, we find:

$$\begin{cases} a = \sqrt{\frac{F}{d}} + \sqrt{\frac{(d-k+1)(1-F)}{d(k-1)}}, \\ b = \sqrt{\frac{F}{d}} - \sqrt{\frac{(k-1)(1-F)}{d(d-k+1)}}. \end{cases}$$

Since we require $b \geq 0$, so that $\frac{k-1}{d} \leq F \leq 1$, and

$$R^{(k)}(F) = 1 - (k-1)a^2 = 1 - \frac{1}{d}\left[\sqrt{F(k-1)} + \sqrt{(1-F)(d-k+1)}\right]^2.$$

While for $0 \leq F < \frac{k-1}{d}$, it is easy to see $R^{(k)}(F) = 0$. Then we are going to prove that $R^{(k)}(F)$ is a monotonically increasing function of $F$ in $[\frac{k-1}{d}, 1]$. Firstly, we have

$$\frac{\mathrm{d}}{\mathrm{d}F} R^{(k)}(F) = -\frac{2}{d} L(F) L'(F),$$

where $L(F) = \sqrt{F(k-1)} + \sqrt{(1-F)(d-k+1)}$ and

$$L'(F) = \frac{1}{2}\left(\frac{\sqrt{k-1}}{\sqrt{F}} - \frac{\sqrt{d-k+1}}{\sqrt{1-F}}\right).$$

For $F \in [\frac{k-1}{d}, 1]$, $L'(F) \leq 0$, so that $\frac{\mathrm{d}}{\mathrm{d}F} R^{(k)}(F) \geq 0$, *i.e.*, $R^{(k)}(F)$ is monotonically increasing. From the monotonicity, we know that $R_+^{(k)}(F) = R^{(k)}(F)$.

Similarly, we can calculate the second derivative of $R^{(k)}(F)$:

$$\frac{\mathrm{d}^2}{\mathrm{d}F^2}R^{(k)}(F) = -\frac{2}{d}\left(L''(F)L(F) + [L'(F)]^2\right),$$

where

$$L''(F) = -\frac{1}{4}\left(\frac{\sqrt{k-1}}{F^{\frac{3}{2}}} + \frac{\sqrt{d-k+1}}{(1-F)^{\frac{3}{2}}}\right).$$

For simplicity, we denote $A = k-1$ and $B = d-k+1$, then we have

$$[L'(F)]^2 = \frac{1}{4}\left(\frac{\sqrt{A}}{\sqrt{F}} - \frac{\sqrt{B}}{\sqrt{1-F}}\right)^2 = \frac{1}{4}\left(\frac{A}{F} - 2\frac{\sqrt{AB}}{\sqrt{F(1-F)}} + \frac{B}{1-F}\right),$$

$$L''(F)L(F) = -\frac{1}{4}\left(\frac{A}{F} + \frac{B}{1-F} + \frac{\sqrt{AB}\sqrt{F}}{(1-F)^{\frac{3}{2}}} + \frac{\sqrt{AB}\sqrt{1-F}}{F^{\frac{3}{2}}}\right),$$

So that

$$L''(F)L(F) + [L'(F)]^2 = -\frac{\sqrt{AB}}{4}\left(\frac{2}{\sqrt{F(1-F)}} + \frac{\sqrt{F}}{(1-F)^{\frac{3}{2}}} + \frac{\sqrt{1-F}}{F^{\frac{3}{2}}}\right) \leq 0,$$

which implies $R^{(k)}(F)$ is a convex function.

Due to its convexity, we finally obtain that $E_G^{(k)}(\rho_I(F)) = \mathrm{conv}(R_V^{(k)}(F)) = R^{(k)}(F)$ □

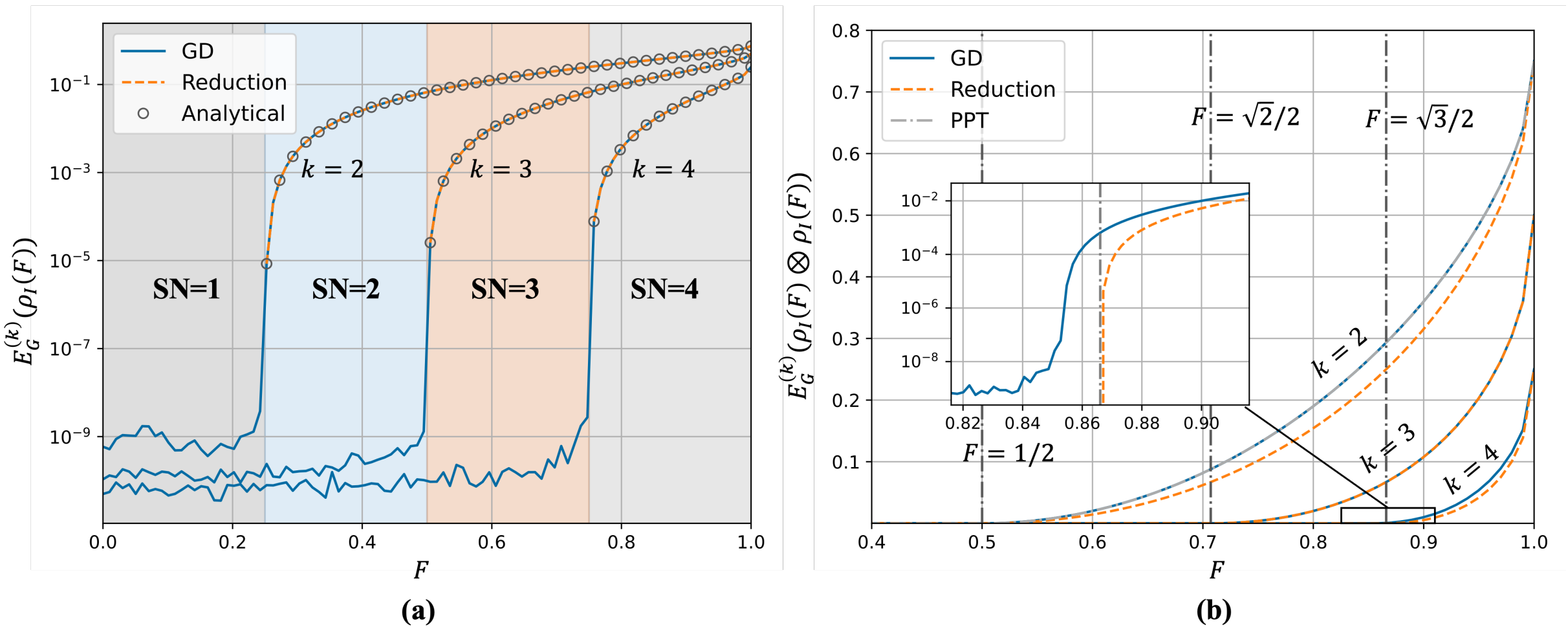

Figure 6.6: The $k$-GME results for isotropic states: (a) Results for $4\otimes 4$ istropic state $\rho_I(F)$. (b) Results for two copies of $2\otimes 2$ isotropic states $\rho_I(F)\otimes\rho_I(F)$ in $4\otimes 4$ system. Here, “GD” represents the gradient descent method, “Reduction” means the generalized reduction maps, and “PPT” denotes the PPT relaxation method.

In Fig. 6.6(a), we present a comparison of the numerical results obtained via gradient descent (GD) and the generalized reduction map for $4\otimes 4$ isotropic states. The results demonstrate con-sistency between the numerical data and the analytical findings derived in Theorem 6.3.6. The sharp transitions of $E_G^{(k)}$ for various $k$ effectively delineate the boundaries of different Schmidt number regions. In Fig. 6.6(b), we further study the case of two copies of isotropic states, $\rho_I(F) \otimes \rho_I(F)$. Notable gaps are observed between the results obtained from gradient descent and those from the reduction maps. Additionally, we include results from the PPT relaxation method, which provides improved lower bounds for $E_G^{(2)}$.

In [60], it was asserted that $F = \frac{1}{2}, \frac{\sqrt{2}}{2}, \frac{\sqrt{3}}{2}$ represent the tight boundaries of different Schmidt number regions for two copies, though the tightness of $F = \frac{\sqrt{3}}{2}$ was not proven. However, our numerical results from gradient descent strongly suggest that $F = \frac{\sqrt{3}}{2}$ is not a tight boundary between Schmidt numbers three and four, as illustrated in the inset of this subfigure. Specifically, at $F = \frac{\sqrt{3}}{2}$, the gradient descent method gives $E_G^{(4)} \approx 6.45 \times 10^{-4}$.

Next, we consider a more intricate scenario known as bound entanglement, in which no pure entanglement can be distilled, even with an infinite number of copies. In [98], Horodecki introduced a family of bound entangled states, referred to as the Horodecki states, defined as follows:

$$\rho_H(a) = \frac{1}{8a+1}\begin{pmatrix} a & 0 & 0 & 0 & a & 0 & 0 & 0 & a \\ 0 & a & 0 & 0 & 0 & 0 & 0 & 0 & 0 \\ 0 & 0 & a & 0 & 0 & 0 & 0 & 0 & 0 \\ 0 & 0 & 0 & a & 0 & 0 & 0 & 0 & 0 \\ a & 0 & 0 & 0 & a & 0 & 0 & 0 & a \\ 0 & 0 & 0 & 0 & 0 & a & 0 & 0 & 0 \\ 0 & 0 & 0 & 0 & 0 & 0 & \frac{1+a}{2} & 0 & \frac{\sqrt{1-a^2}}{2} \\ 0 & 0 & 0 & 0 & 0 & 0 & 0 & a & 0 \\ a & 0 & 0 & 0 & a & 0 & \frac{\sqrt{1-a^2}}{2} & 0 & \frac{1+a}{2} \end{pmatrix}, \tag{6.72}$$

which can be shown to be PPT entangled for $a \in (0, 1)$. Thus, its entanglement cannot be detected using the PPT criterion. However, we can still compute the geometric measure of entanglement $E_G^{(2)}$ for the Horodecki states using the gradient descent method, as depicted in Fig. 6.7. The numerical results clearly confirm the existence of bound entanglement in the range $a \in (0, 1)$.

It was initially believed that bound entangled states exhibit weak entanglement and are unsuitable for quantum information tasks [219]. However, a family of PPT states with logarithmically increasing Schmidt numbers relative to the local dimension was first discovered in [93]. Subsequently, PPT states with linearly scaling Schmidt numbers were proposed in [94, 95, 220]. While various mathematical techniques have been developed to construct PPT entangled states with high Schmidt numbers, prior studies lack effective methods for determining the Schmidt number, let alone quantifying the corresponding entanglement. In the following, we demonstrate how this challenge is addressed in this context.

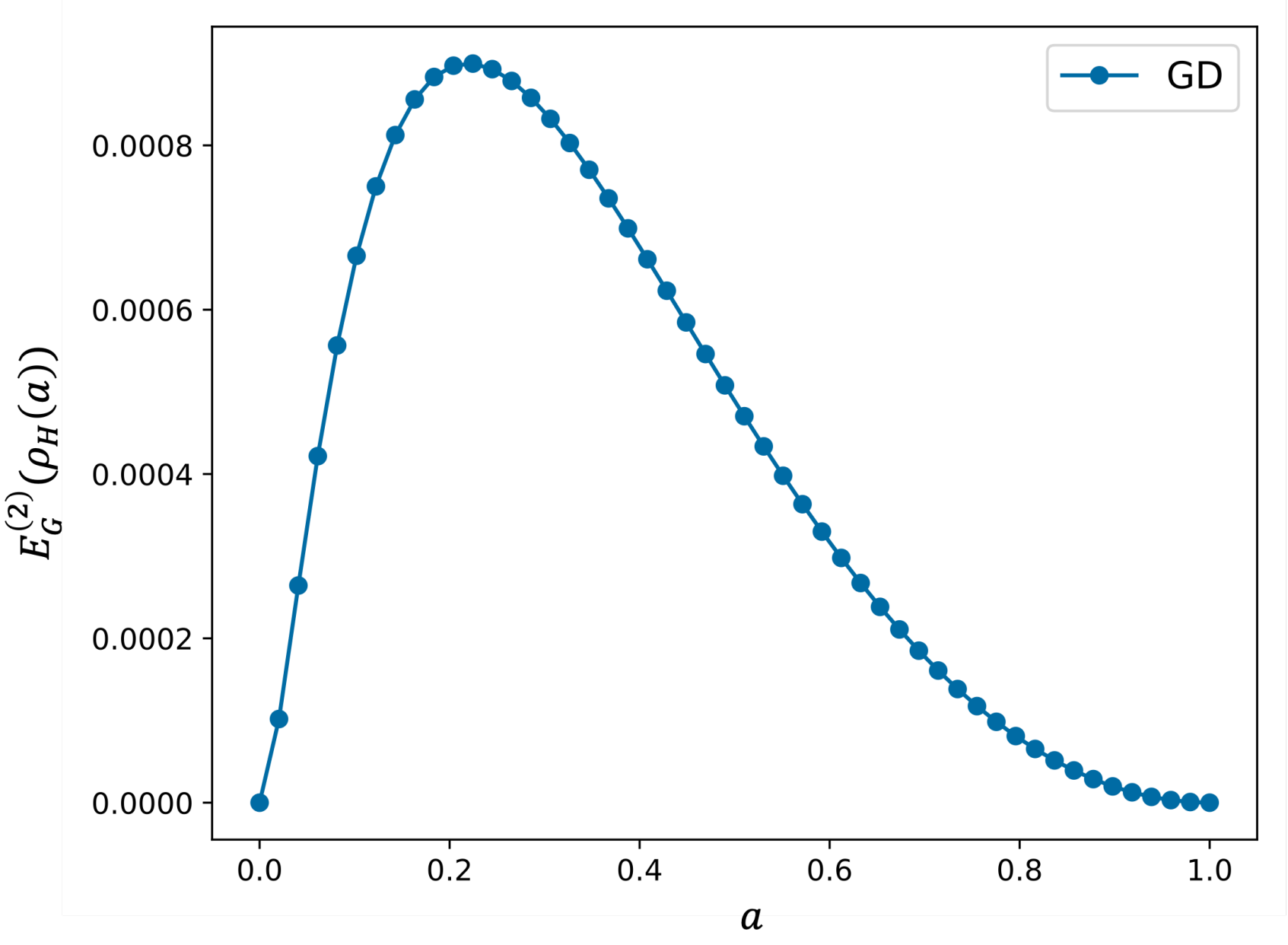

Figure 6.7: Geometric measure of entanglement for Horodecki states $\rho_H(a)$. The numerical results obtained from gradient descent (GD) reveal the bound entanglement for $a \in (0, 1)$.

Consider the following $d \otimes d$ (unnormalized) state in the $(A_1A_2)(B_1B_2)$ system, where $d$ is an even integer satisfying $d \geq 4$:

$$\begin{aligned} R_{AB} =& (I_4 - |\Psi_2^+\rangle\langle\Psi_2^+|)_{A_1B_1} \otimes (I_{d^2/4} - |\Psi_{d/2}^+\rangle\langle\Psi_{d/2}^+|)_{A_2B_2} \\ &+ \left(\frac{d}{2}+1\right) |\Psi_4^+\rangle\langle\Psi_4^+|_{A_1B_1} \otimes |\Psi_{d/2}^+\rangle\langle\Psi_{d/2}^+|_{A_2B_2}, \end{aligned} \tag{6.73}$$

where $|\Psi_d^+\rangle = \frac{1}{\sqrt{d}}\sum_{i=0}^{d-1}|ii\rangle$ represents the maximally entangled state in the $A_1B_1$ or $A_2B_2$ systems, with $d_{A_1} = d_{B_1} = 2$ and $d_{A_2} = d_{B_2} = \frac{d}{2}$. It has been proven that $\rho_{AB} = R_{AB}/\operatorname{Tr}[R_{AB}]$ is PPT and possesses a Schmidt number that is at least $\left\lceil \frac{d}{4} \right\rceil$ [94].

| Dimension ($d$) | $E_G^{(2)}$ | $E_G^{(3)}$ | $E_G^{(4)}$ |
|---|---|---|---|
| 4 | $8.68 \times 10^{-9}$ | $1.86 \times 10^{-9}$ | $4.938 \times 10^{-9}$ |
| 6 | **0.0476** | $3.5069 \times 10^{-8}$ | $9.99 \times 10^{-10}$ |
| 8 | **0.05** | $3.08 \times 10^{-8}$ | $2.238 \times 10^{-8}$ |
| 10 | **0.046** | **0.01538** | $2.6798 \times 10^{-8}$ |

Table 6.3: The $k$-GME results for high-dimensional bound entanglement. An obvious non-zero value indicates the emergence of entanglement with certain Schmidt number.

In the following, we utilize the gradient descent (GD) to detect and quantify such high-dimensional entanglement within these PPT entangled states, as shown in Table 6.3. It clearly shows the feasibility of GD for determining Schmidt numbers and quantifying corresponding entanglement in the bound entanglement region. This task is infeasible for other previous methods like the generalized reduction map [60], covariance matrix criterion [221], and trace norm of correlations [222].

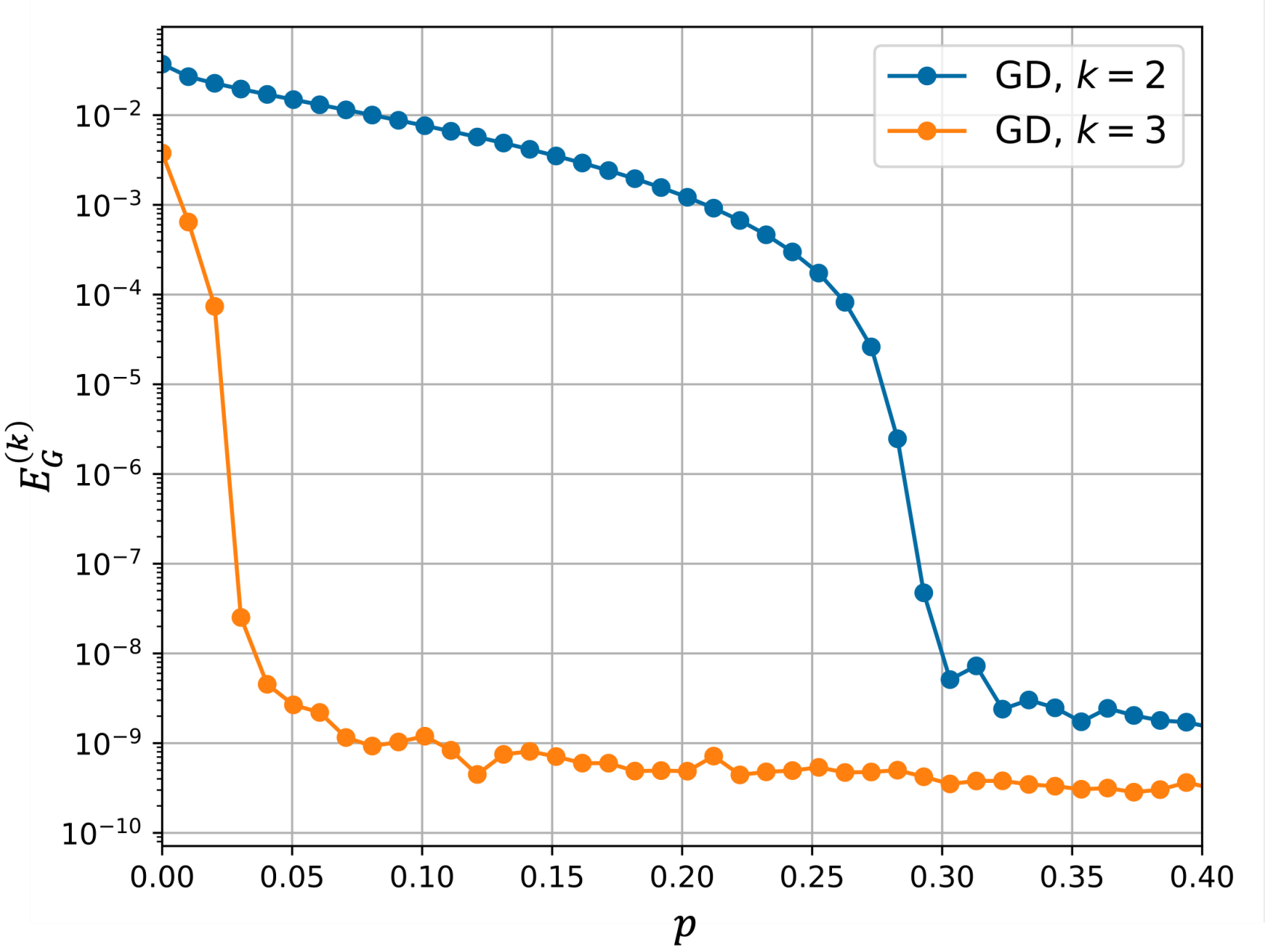

Figure 6.8: The $k$-GME results for the edge state under a depolarizing channel. The noise level of the depolarizing channel is controlled by $p$. A sharp transition indicates the disappearance of entanglement with a certain Schmidt number.

We can also investigate the robustness of a high-dimensional bound entangled state. For example, here, we consider the depolarizing channel for $d \otimes d$ states as follows

$$\mathcal{N}(\rho) = (1-p)\rho + \frac{p}{d^2} I_{d^2}, \tag{6.74}$$

where $p \in [0, 1]$ characterizes the noise level. In Fig. 6.8, we study the influence of the depolarizing channel for an edge state proposed in [220], which is the simplest example of PPT entangled states with Schmidt number 3. As we can expect, the high-dimensional entanglement (SN = 3) within the state can be much more vulnerable to depolarizing noise compared to the low-dimensional entanglement (SN = 2).

## 6.3.2 Multipartite mixed states

This section focuses on the detection and quantification of entanglement in completely entangled and genuinely entangled multipartite mixed states.

For a given bipartition, the geometric measure of entanglement (GME) in the bipartite setting can be utilized to determine whether a multipartite mixed state is biseparable with respect to the specified bipartition. In the context of completely entangled states, the $E_G^{(2)}$ for a multipartite mixed state is defined as:

$$E_G^{(2)}(\rho) = \min_{\{p_i, |\psi_i\rangle\}} \sum_i p_i E_G^{(2)}(|\psi_i\rangle), \tag{6.75}$$

where $\{p_i, |\psi_i\rangle\}$ denotes a pure-state decomposition of $\rho$, satisfying $\rho = \sum_i p_i |\psi_i\rangle\langle\psi_i|$. For a multipartite pure state $|\psi\rangle$, the geometric measure of entanglement is expressed as:

$$E_G^{(2)}(|\psi\rangle) = 1 - \max_{|\phi\rangle = \otimes_i |\phi^{(i)}\rangle} |\langle\phi|\psi\rangle|^2, \tag{6.76}$$

where $|\phi\rangle$ represents a fully product state in the multipartite quantum system. It is evident that a multipartite mixed state $\rho$ is completely entangled if and only if $E_G^{(2)}(\rho) > 0$.

We begin with the multipartite states introduced in Sec. 3.2.2, which are biseparable with respect to any bipartition but not fully separable. A notable example is the three-qubit mixed state presented in [75], defined as:

$$\rho_{ABC} = \frac{1}{4}\left(I - \sum_{i=1}^{n} |\psi_i\rangle\langle\psi_i|\right), \tag{6.77}$$

where the set $\{|\psi_i\rangle\}$ forms an unextendible product basis (UPB) in the $2 \otimes 2 \otimes 2$ system:

$$\begin{cases} |\psi_1\rangle = \frac{1}{\sqrt{2}}|0\rangle \otimes |1\rangle \otimes (|0\rangle + |1\rangle), \\ |\psi_2\rangle = \frac{1}{\sqrt{2}}|1\rangle \otimes (|0\rangle + |1\rangle) \otimes |0\rangle, \\ |\psi_3\rangle = \frac{1}{\sqrt{2}}(|0\rangle + |1\rangle) \otimes |0\rangle \otimes |1\rangle, \\ |\psi_4\rangle = \frac{1}{2\sqrt{2}}(|0\rangle - |1\rangle) \otimes (|0\rangle - |1\rangle) \otimes (|0\rangle - |1\rangle). \end{cases} \tag{6.78}$$

| | $A\|B\|C$ | $A\|BC$ | $B\|AC$ | $C\|AB$ |
|---|---|---|---|---|
| $E_G^{(2)}$ | **0.08144** | $\approx 10^{-14}$ | $\approx 10^{-14}$ | $\approx 10^{-14}$ |

Table 6.4: The numerical results of GME for three-qubit state $\rho_{ABC}$ with respect to different partitions. An obvious non-zero value indicates the entanglement under the given partition.

In Table 6.4, we analyze the GME of the given state for various partitions. For the bipartitions $A|BC$, $B|AC$, and $C|AB$, the GME values are found to be nearly zero, indicating that the state

is biseparable with respect to each bipartition. However, for the partition $A|B|C$, a distinctly non-zero GME value reveals that the state is fully entangled.

Next, we examine certain multipartite mixed states exhibiting high symmetry. For instance, consider a mixture of two distinct Dicke states, $|D_n^{k_1}\rangle$ and $|D_n^{k_2}\rangle$, represented in the form given by Eq. (6.27):

$$\rho_{(n,k_1,k_2)}(r) = r|D_n^{k_1}\rangle\langle D_n^{k_1}| + (1-r)|D_n^{k_2}\rangle\langle D_n^{k_2}|. \tag{6.79}$$

Since $|D_n^{k_1}\rangle$ and $|D_n^{k_2}\rangle$ are orthogonal when $k_1 \neq k_2$, the range of $\rho_{(n,k_1,k_2)}(r)$ is spanned solely by these two states. Consequently, for any pure-state decomposition $\{p_i, |\psi_i\rangle\}$ satisfying $\rho_{(n,k_1,k_2)}(r) = \sum_i p_i|\psi_i\rangle\langle\psi_i|$, each pure state $|\psi_i\rangle$ must be a linear combination of $|D_n^{k_1}\rangle$ and $|D_n^{k_2}\rangle$, *i.e.*,

$$|\psi_i\rangle = |\psi(r_i,\phi_i)\rangle = \sqrt{r_i}|D_n^{k_1}\rangle + e^{i\phi_i}\sqrt{1-r_i}|D_n^{k_2}\rangle, \tag{6.80}$$

where $r_i \in [0,1]$ and $\sum_i p_i r_i = r$. We can apply a series of local unitary operations $U^{\otimes n}$ on $|\psi(r,\phi)\rangle$, where

$$U = \begin{pmatrix} 1 & 0 \\ 0 & e^{i\frac{\phi}{k_1-k_2}} \end{pmatrix}, \tag{6.81}$$

resulting in $U^{\otimes n}|\psi(r,\phi)\rangle = e^{i\frac{k_1}{k_1-k_2}\phi}|\psi(r,0)\rangle$. Since local unitary operations do not alter the GME, we have

$$E_G^{(2)}(|\psi(r,\phi)\rangle) = E_G^{(2)}(U^{\otimes n}|\psi(r,\phi)\rangle) = E_G^{(2)}(|\psi(r,0)\rangle). \tag{6.82}$$

Thus, it follows that

$$\begin{aligned} E_G^{(2)}\left(\rho_{(n,k_1,k_2)}(r)\right) &= \min_{\{p_i,|\psi_i\rangle\}} \sum_i p_i E_G^{(2)}(|\psi_i\rangle) \\ &= \min_{\{\sum_i p_i r_i = r, \phi_i\}} \sum_i p_i E_G^{(2)}(|\psi(r_i,\phi_i)\rangle) \\ &= \min_{\sum_i p_i r_i = r} \sum_i p_i E_G^{(2)}(|\psi(r_i,0)\rangle), \end{aligned} \tag{6.83}$$

implying that $E_G^{(2)}(\rho_{(n,k_1,k_2)}(r))$ can be determined from the convex hull of $E_G^{(2)}(|\psi(r,0)\rangle)$.

For the computation of $E_G^{(2)}(|\psi(r,0)\rangle)$, we assume that the closest fully product state has the following form (see Theorem 4.2.1):

$$|\phi\rangle = (\cos\theta|0\rangle + \sin\theta|1\rangle)^{\otimes n}, \tag{6.84}$$

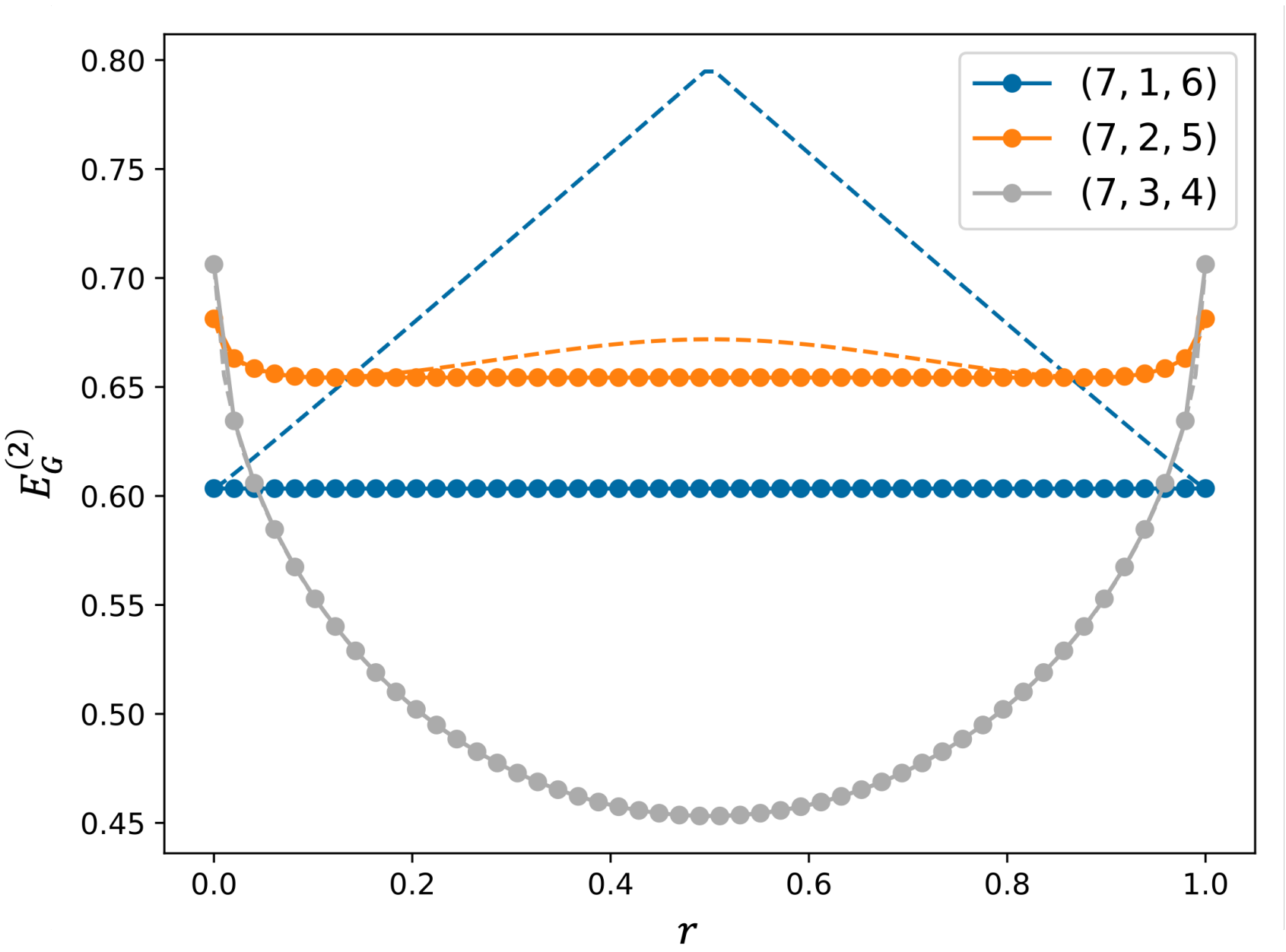

Figure 6.9: The geometric measure of entanglement (GME) for the mixed state $\rho_{(n,k_1,k_2)}(r)$. Here, $(n, k_1, k_2)$ indicates a mixture of two distinct Dicke states, $|D_n^{k_1}\rangle$ and $|D_n^{k_2}\rangle$. The dashed lines represent the numerical results for the GME of corresponding pure states $|\psi(r, 0)\rangle$. Theoretically, $E_G^{(2)}(\rho_{(n,k_1,k_2)}(r))$ is the convex hull of $E_G^{(2)}(|\psi(r, 0)\rangle)$.

where $\theta \in [0, \frac{\pi}{2}]$. The inner product with a Dicke state is given by

$$\langle \phi | D_n^k \rangle = \sqrt{\binom{n}{k}} (\cos\theta)^{n-k} (\sin\theta)^k. \tag{6.85}$$

For the target pure state $|\psi(r, 0)\rangle = \sqrt{r}|D_n^{k_1}\rangle + \sqrt{1-r}|D_n^{k_2}\rangle$, the inner product becomes

$$\langle \phi | \psi(r, 0) \rangle = \sqrt{r\binom{n}{k_1}} (\cos\theta)^{n-k_1} (\sin\theta)^{k_1} + \sqrt{(1-r)\binom{n}{k_2}} (\cos\theta)^{n-k_2} (\sin\theta)^{k_2}. \tag{6.86}$$

As a result, the expression for $E_G^{(2)}(|\psi(r, 0)\rangle)$ is given by a maximization over $\theta$:

$$\begin{aligned} E_G^{(2)}(|\psi(r, 0)\rangle) =& 1 - \Bigg\{ \max_{\theta\in[0,\frac{\pi}{2}]} \Bigg( \sqrt{r\binom{n}{k_1}} (\cos\theta)^{n-k_1} (\sin\theta)^{k_1} \\ &+ \sqrt{(1-r)\binom{n}{k_2}} (\cos\theta)^{n-k_2} (\sin\theta)^{k_2} \Bigg) \Bigg\}^2. \end{aligned} \tag{6.87}$$

However, due to the complexity of the maximization, obtaining a closed-form expression for $E_G^{(2)}(|\psi(r, 0)\rangle)$ is generally not feasible.

In Fig. 6.9, we compute the GME for the mixed states $\rho_{(n,k_1,k_2)}(r)$ with $n = 7$ and their cor-responding pure states $|\psi(r,0)\rangle$. From the numerical results, it is evident that $E_G^{(2)}(\rho_{(n,k_1,k_2)}(r))$ is indeed the convex hull of $E_G^{(2)}(|\psi(r,0)\rangle)$.

## 6.4 Summary and Discussion

In this section, we provide a concise summary of the geometric measure of entanglement (GME) across various quantum scenarios. As elaborated in the preceding sections, $k$-GME offers a unified and robust framework for identifying the dimensionality of entanglement and quantifying its magnitude. For clarity, we focus on bipartite systems in the following discussion, with the key results summarized in Table 6.5.

| **Quantum Scenario** | **Dimensionality** | **$k$-GME Definition** |
|---|---|---|
| Pure state $\vert\psi\rangle$ | Schmidt rank $\mathrm{SR}(\vert\psi\rangle)$ | $E_G^{(k)}(\vert\psi\rangle) = 1 - \max_{\mathrm{SR}(\vert\phi\rangle)<k} \vert\langle\phi\vert\psi\rangle\vert^2$ |
| Subspace $\mathcal{S}$ | Minimal Schmidt rank $r(\mathcal{S})$ | $E_G^{(k)}(\mathcal{S}) = \min_{\vert\psi\rangle\in\mathcal{S}} E_G^{(k)}(\vert\psi\rangle)$ |
| Mixed state $\rho$ | Schmidt number $\mathrm{SN}(\rho)$ | $E_G^{(k)}(\rho) = \min_{\{p_i,\vert\psi_i\rangle\}} \sum_i p_i E_G^{(k)}(\vert\psi_i\rangle)$ |

Table 6.5: Definitions of $k$-GME in bipartite systems and their corresponding interpretations of entanglement dimensionality.

The definitions of $k$-GME across these quantum scenarios are deeply interconnected. For example, the following theorem establishes a fundamental relationship between them:

**Theorem 6.4.1** For a mixed state $\rho$, its $k$-GME is bounded below by the $k$-GME of its range space Range($\rho$). Specifically,

$$E_G^{(k)}(\rho) \geq E_G^{(k)}[\mathrm{Range}(\rho)]. \tag{6.88}$$

*Proof.* Let the optimal pure-state decomposition of $\rho$ for computing $E_G^{(k)}(\rho)$ be $\{p_i, |\psi_i\rangle\}$. Since the range space of $\rho$ is spanned by $\{|\psi_i\rangle\}$, we have

$$E_G^{(k)}(\rho) = \sum_i p_i E_G^{(k)}(|\psi_i\rangle) \geq \min_i E_G^{(k)}(|\psi_i\rangle) \geq \min_{\mathbf{c}} E_G^{(k)}\left(\sum_i c_i|\psi_i\rangle\right) = E_G^{(k)}[\mathrm{Range}(\rho)].$$

□

This result implies that the $k$-GME of the range space of $\rho$ serves as a nontrivial lower bound for the $k$-GME of $\rho$. Conceptually, this theorem generalizes the range criterion discussed in Sec. 4.1.3, which asserts that a mixed state is entangled if its range space contains no product states. Furthermore, the theorem establishes that $\mathrm{SN}(\rho) \geq k$ if the minimal Schmidt rank of Range($\rho$) satisfies $r[\text{Range}(\rho)] \geq k$, meaning the range space excludes all states with Schmidt rank less than $k$.

Additionally, the robustness of a mixed state $\rho$ is closely related to the robustness of its corresponding range space Range($\rho$), as demonstrated in the following theorem:

**Theorem 6.4.2** For a mixed state $\rho$ with $\text{SN}(\rho) \geq k$ (i.e., $E_G^{(k)}(\rho) > 0$), consider any unitary perturbation $U = e^{-iH}$ applied to $\rho$. If the operator norm of $H$ satisfies $\|H\|_{\text{op}} < \sqrt{E_G^{(k)}[\text{Range}(\rho)]}$, the perturbed state $\rho'$ will retain $\text{SN}(\rho') \geq k$.

*Proof.* According to Theorem 6.3.4, retaining the Schmidt number of $\rho$ after perturbation requires that

$$\frac{\Delta\lambda(H)}{2} < \sqrt{E_G^{(k)}(\rho)}.$$

Using the following inequalities:

$$\frac{\Delta\lambda(H)}{2} \leq \|H\|_{\text{op}},$$
$$E_G^{(k)}[\text{Range}(\rho)] \leq E_G^{(k)}(\rho),$$

we can conclude that

$$\frac{\Delta\lambda(H)}{2} \leq \|H\|_{\text{op}} < \sqrt{E_G^{(k)}[\text{Range}(\rho)]} \leq \sqrt{E_G^{(k)}(\rho)}.$$

□

Last but not least, although the previous sections primarily focused on the 2-GME for multipartite subspaces and mixed states, the definitions of $k$-GME can naturally be extended to cases where $k > 2$ by building on the definition of $k$-GME for multipartite pure states:

**Definition 22 ($k$-GME for Multipartite Mixed States)** Consider a multipartite mixed state $\rho$ in the composite Hilbert space $\mathcal{H}_1 \otimes \mathcal{H}_2 \otimes \cdots \otimes \mathcal{H}_n$. The geometric measure is defined as:

$$E_G^{(k)}(\rho) = \min_{\{p_i, |\psi_i\rangle\}} \sum_i p_i E_G^{(k)}(|\psi_i\rangle), \tag{6.89}$$

where the minimization is taken over all possible decompositions of $\rho$ into multipartite pure states, *i.e.*, $\rho = \sum_i p_i |\psi_i\rangle\langle\psi_i|$, with $\{p_i, |\psi_i\rangle\}$.

The Schmidt number, initially defined for bipartite systems, can be generalized to multipartite settings as follows [223]:

**Definition 23 (Border Number of Multipartite Mixed State)** The border number of a multipartite density matrix $\rho$, denoted as $\mathrm{BN}(\rho)$, is defined by the following conditions: (i) For any pure-state decomposition of $\rho$, *i.e.*, $\rho = \sum_i p_i |\psi_i\rangle\langle\psi_i|$ with $\{p_i, |\psi_i\rangle\}$, at least one pure state in the decomposition must have a border rank $\mathrm{BR} \geq k$. (ii) There exists at least one decomposition of $\rho$ such that all pure states $|\psi_i\rangle$ in the decomposition have border rank $\mathrm{BR} \leq k$. Formally, $\mathrm{BN}(\rho) = k$ if and only if both conditions are satisfied.

The convex hull of multipartite pure states with bounded border rank, *i.e.*, the set of density matrices with bounded border number, is *compact* in the Euclidean topology [224, 225]. By definition, we have $\mathrm{BN}(\rho) = k$ if and only if $E_G^{(k)}(\rho) > 0$ and $E_G^{(k+1)}(\rho) = 0$.

# REFERENCES

[1]M. Planck, “On the law of distribution of energy in the normal spectrum”, Annalen der physik **4**, 1 (1901) (cit. on p. 1).

[2]A. Einstein, “On a heuristic point of view concerning the production and transformation of light”, Annalen der Physik **17**, 1–16 (1905) (cit. on p. 1).

[3]N. Bohr, *The quantum postulate and the recent development of atomic theory* (Good Press, 2025) (cit. on p. 1).

[4]E. Schrödinger, “Quantisierung als eigenwertproblem”, Annalen der physik **386**, 109–139 (1926) (cit. on p. 1).

[5]W. Heisenberg, “Über den anschaulichen inhalt der quantentheoretischen kinematik und mechanik”, Zeitschrift für Physik **43**, 172–198 (1927) (cit. on p. 1).

[6]P. A. M. Dirac, *The principles of quantum mechanics*, 27 (Oxford university press, 1981) (cit. on p. 1).

[7]M. Born, “Quantum mechanics of collision processes”, Uspekhi Fizich **1926**, 456 (1926) (cit. on p. 1).

[8]E. Schrödinger, “Discussion of probability relations between separated systems”, in Mathematical proceedings of the cambridge philosophical society, Vol. 31, 4 (Cambridge University Press, 1935), pp. 555–563 (cit. on p. 1).

[9]A. Einstein et al., “Can quantum-mechanical description of physical reality be considered complete?”, Phys. Rev. **47**, 777–780 (1935) (cit. on p. 1).

[10]A. Aspect et al., “Experimental test of bell’s inequalities using time-varying analyzers”, Phys. Rev. Lett. **49**, 1804–1807 (1982) (cit. on p. 1).

[11]A. K. Ekert, “Quantum cryptography based on bell’s theorem”, Phys. Rev. Lett. **67**, 661–663 (1991) (cit. on p. 1).

[12]J. Yin et al., “Entanglement-based secure quantum cryptography over 1,120 kilometres”, Nature **582**, 501–505 (2020) (cit. on p. 1).

[13]T. Jennewein et al., “Quantum cryptography with entangled photons”, Phys. Rev. Lett. **84**, 4729–4732 (2000) (cit. on p. 1).

[14]R. Ursin et al., “Entanglement-based quantum communication over 144 km”, Nature Phys **3**, 481–486 (2007) (cit. on p. 1).

[15]N. Zou, “Quantum entanglement and its application in quantum communication”, in J. Phys.: Conf. Ser. Vol. 1827, 1 (IOP Publishing, 2021), p. 012120 (cit. on p. 1).

[16]G. Alber et al., “Mixed-state entanglement and quantum communication”, Quantum information: An introduction to basic theoretical concepts and experiments, 151–195 (2001) (cit. on p. 1).

[17]R. Raussendorf et al., “A one-way quantum computer”, Phys. Rev. Lett. **86**, 5188–5191 (2001) (cit. on p. 1).

[18]P. Walther et al., “Experimental one-way quantum computing”, Nature **434**, 169–176 (2005) (cit. on p. 1).

[19]V. Giovannetti et al., “Advances in quantum metrology”, Nature Photon **5**, 222–229 (2011) (cit. on p. 1).

[20]G. Tóth et al., “Quantum metrology from a quantum information science perspective”, J. Phys. A: Math. Theor. **47**, 424006 (2014) (cit. on p. 1).

[21] V. Giovannetti et al., "Quantum metrology", Phys. Rev. Lett. **96**, 010401 (2006) (cit. on p. 1).

[22] L. Gurvits, "Classical deterministic complexity of edmonds' problem and quantum entanglement", in Proceedings of the thirty-fifth annual acm symposium on theory of computing, STOC '03 (2003), pp. 10–19 (cit. on pp. 1, 31).

[23] A. Peres, "Separability criterion for density matrices", Phys. Rev. Lett. **77**, 1413–1415 (1996) (cit. on pp. 1, 31, 32).

[24] B. M. Terhal, "Bell inequalities and the separability criterion", Phys. Lett. A **271**, 319–326 (2000) (cit. on pp. 2, 36).

[25] S. Boyd et al., *Convex optimization* (Cambridge University Press, 2004) (cit. on pp. 2, 50, 51, 65).

[26] P. Skrzypczyk et al., *Semidefinite programming in quantum information science* (IOP Publishing, 2023) (cit. on pp. 2, 53, 60).

[27] A. C. Doherty et al., "Distinguishing separable and entangled states", Phys. Rev. Lett. **88**, 187904 (2002) (cit. on pp. 2, 89).

[28] A. C. Doherty et al., "Complete family of separability criteria", Phys. Rev. A **69**, 022308 (2004) (cit. on pp. 2, 89).

[29] S. Gerke et al., "Numerical construction of multipartite entanglement witnesses", Phys. Rev. X **8**, 031047 (2018) (cit. on p. 2).

[30] H. Fawzi et al., "Semidefinite programming lower bounds on the squashed entanglement", arXiv preprint arXiv:2203.03394 (2022) (cit. on p. 2).

[31] Z. Zhang et al., "Numerical and analytical results for geometric measure of coherence and geometric measure of entanglement", Sci. Rep. **10**, 12122 (2020) (cit. on pp. 2, 89).

[32] M. I. Jordan et al., "Machine learning: trends, perspectives, and prospects", Science **349**, 255–260 (2015) (cit. on pp. 2, 61).

[33] S. J. Orfanidis, *Introduction to signal processing* (Prentice-Hall, Inc., 1995) (cit. on p. 2).

[34] M. Hayashi et al., *Introduction to quantum information science* (Springer, 2014) (cit. on p. 2).

[35] A. Lucas, "Ising formulations of many NP problems", Front. Phys. **2**, 5 (2014) (cit. on p. 2).

[36] Y. LeCun et al., "Deep learning", Nature **521**, 436–444 (2015) (cit. on p. 2).

[37] S. Kirkpatrick et al., "Optimization by simulated annealing", Science **220**, 671–680 (1983) (cit. on p. 2).

[38] D. C. Liu et al., "On the limited memory BFGS method for large scale optimization", Mathematical programming **45**, 503–528 (1989) (cit. on pp. 2, 63).

[39] D. P. Kingma et al., "Adam: a method for stochastic optimization", arXiv preprint arXiv:1412.6980 (2014) (cit. on p. 2).

[40] A. G. Nikolaev et al., "Simulated annealing", Handbook of metaheuristics, 1–39 (2010) (cit. on p. 2).

[41] O. Kramer et al., *Genetic algorithms* (Springer, 2017) (cit. on p. 2).

[42] C. Szepesvári, *Algorithms for reinforcement learning* (Springer nature, 2022) (cit. on p. 2).

[43] C. Finn et al., "Model-agnostic meta-learning for fast adaptation of deep networks", in International conference on machine learning (PMLR, 2017), pp. 1126–1135 (cit. on p. 2).

[44] N. Boumal, *An introduction to optimization on smooth manifolds* (Cambridge University Press, 2023) (cit. on pp. 3, 71).

[45] J. Hu et al., "A brief introduction to manifold optimization", J. Oper. Res. Soc. China **8**, 199–248 (2020) (cit. on p. 3).

[46]M. Lezcano Casado, "Trivializations for gradient-based optimization on manifolds", Advances in Neural Information Processing Systems **32** (2019) (cit. on pp. 3, 73).

[47]A. Shimony, "Degree of entanglement a", Annals of the New York Academy of Sciences **755**, 675–679 (1995) (cit. on pp. 3, 44, 78).

[48]H. Barnum et al., "Monotones and invariants for multi-particle quantum states", J. Phys. A: Math. Gen. **34**, 6787 (2001) (cit. on pp. 3, 44, 78).

[49]T.-C. Wei et al., "Geometric measure of entanglement and applications to bipartite and multipartite quantum states", Phys. Rev. A **68**, 042307 (2003) (cit. on pp. 3, 44, 45, 78, 87, 89, 105, 106).

[50]M. Erhard et al., "Advances in high-dimensional quantum entanglement", Nat Rev Phys **2**, 365–381 (2020) (cit. on pp. 3, 78, 85).

[51]L.-J. Kong et al., "High-dimensional entanglement-enabled holography", Phys. Rev. Lett. **130**, 053602 (2023) (cit. on pp. 3, 78, 85).

[52]D. Cozzolino et al., "High-dimensional quantum communication: benefits, progress, and future challenges", Advanced Quantum Technologies **2**, 1900038 (2019) (cit. on pp. 3, 78, 85).

[53]P. A. M. Dirac, "A new notation for quantum mechanics", in Mathematical proceedings of the cambridge philosophical society, Vol. 35, 3 (Cambridge University Press, 1939), pp. 416–418 (cit. on pp. 5, 85).

[54]R. Penrose et al., "Applications of negative dimensional tensors", Combinatorial mathematics and its applications **1**, 3 (1971) (cit. on p. 12).

[55]A. Peres, *Quantum theory: concepts and methods* (Springer, 2002) (cit. on p. 17).

[56]R. F. Werner, "Quantum states with einstein-podolsky-rosen correlations admitting a hidden-variable model", Phys. Rev. A **40**, 4277–4281 (1989) (cit. on pp. 17, 21).

[57]C. Carathéodory, "Über den variabilitätsbereich der fourier'schen konstanten von positiven harmonischen funktionen", Rendiconti del Circolo Matematico di Palermo **32**, 193–217 (1911) (cit. on p. 19).

[58]W. K. Wootters, "Entanglement of formation of an arbitrary state of two qubits", Phys. Rev. Lett. **80**, 2245–2248 (1998) (cit. on pp. 19, 41, 43).

[59]A. Sanpera et al., "Quantum inseparability as local pseudomixture", arXiv preprint quant-ph/9801024 (1998) (cit. on p. 19).

[60]B. M. Terhal et al., "Schmidt number for density matrices", Phys. Rev. A **61**, 040301 (2000) (cit. on pp. 19, 36, 93, 101, 110, 112).

[61]S.-H. Kye, "Positive maps in quantum information theory", Lecture Notes, Seoul National University, Seoul (2023) (cit. on pp. 21, 105).

[62]G. Svetlichny, "Distinguishing three-body from two-body nonseparability by a bell-type inequality", Phys. Rev. D **35**, 3066–3069 (1987) (cit. on p. 21).

[63]D. M. Greenberger et al., "Going beyond bell's theorem", in *Bell's theorem, quantum theory and conceptions of the universe* (Springer Netherlands, Dordrecht, 1989), pp. 69–72 (cit. on p. 21).

[64]A. Zeilinger et al., "Higher-order quantum entanglement", in Nasa. goddard space flight center, workshop on squeezed states and uncertainty relations (1992) (cit. on p. 21).

[65]W. Dür et al., "Three qubits can be entangled in two inequivalent ways", Phys. Rev. A **62**, 062314 (2000) (cit. on pp. 21, 22).

[66]V. Scarani et al., "Spectral decomposition of bell's operators for qubits", J. Phys. A: Math. Gen. **34**, 6043 (2001) (cit. on p. 22).

[67]J. Eisert et al., "Schmidt measure as a tool for quantifying multiparticle entanglement", Phys. Rev. A **64**, 022306 (2001) (cit. on pp. 22, 89).

[68]H. A. Carteret et al., "Multipartite generalization of the schmidt decomposition", J. Math. Phys. **41**, 7932–7939 (2000) (cit. on p. 22).

[69]J. Håstad, "Tensor rank is np-complete", J. Algorithms **11**, 644–654 (1990) (cit. on pp. 22, 85).

[70]C. J. Hillar et al., "Most tensor problems are np-hard", J. ACM **60**, 1–39 (2013) (cit. on pp. 22, 85).

[71]T. Lickteig, "A note on border rank", Inform. Process Lett. **18**, 173–178 (1984) (cit. on pp. 22, 85).

[72]J. M. Landsberg et al., "On the ranks and border ranks of symmetric tensors", Found Comput Math **10**, 339–366 (2009) (cit. on pp. 22, 85).

[73]A. K. Jensen, "Tensors and the entanglement of pure quantum states", PhD thesis (University of Copenhagen, Faculty of Science, Department of Mathematical …, 2019) (cit. on p. 22).

[74]G. Tóth et al., "Optimal spin squeezing inequalities detect bound entanglement in spin models", Phys. Rev. Lett. **99**, 250405 (2007) (cit. on p. 24).

[75]C. H. Bennett et al., "Unextendible product bases and bound entanglement", Phys. Rev. Lett. **82**, 5385–5388 (1999) (cit. on pp. 24, 32, 95, 97, 113).

[76]L. A. Lessa et al., "Mixed-state quantum anomaly and multipartite entanglement", Phys. Rev. X **15**, 011069 (2025) (cit. on p. 24).

[77]H.-K. Lo et al., "Concentrating entanglement by local actions: beyond mean values", Phys. Rev. A **63**, 022301 (2001) (cit. on pp. 25, 80).

[78]M. A. Nielsen et al., *Quantum computation and quantum information* (Cambridge university press, 2010) (cit. on pp. 25, 55, 101, 102).

[79]M. A. Nielsen, "Conditions for a class of entanglement transformations", Phys. Rev. Lett. **83**, 436–439 (1999) (cit. on pp. 26, 27, 80).

[80]R. Bhatia, *Matrix analysis*, Vol. 169 (Springer Science & Business Media, 2013) (cit. on p. 26).

[81]G. Vidal, "Entanglement of pure states for a single copy", Phys. Rev. Lett. **83**, 1046–1049 (1999) (cit. on p. 27).

[82]D. Jonathan et al., "Entanglement-assisted local manipulation of pure quantum states", Phys. Rev. Lett. **83**, 3566–3569 (1999) (cit. on pp. 28, 84).

[83]S. Lloyd, "Capacity of the noisy quantum channel", Phys. Rev. A **55**, 1613–1622 (1997) (cit. on p. 28).

[84]V. Scarani et al., "The security of practical quantum key distribution", Rev. Mod. Phys. **81**, 1301–1350 (2009) (cit. on p. 29).

[85]C. H. Bennett et al., "Concentrating partial entanglement by local operations", Phys. Rev. A **53**, 2046–2052 (1996) (cit. on pp. 29, 30).

[86]C. H. Bennett et al., "Mixed-state entanglement and quantum error correction", Phys. Rev. A **54**, 3824–3851 (1996) (cit. on pp. 29, 39, 41, 42).

[87]M. Horodecki et al., "Mixed-state entanglement and distillation: is there a "bound" entanglement in nature?", Phys. Rev. Lett. **80**, 5239–5242 (1998) (cit. on pp. 30, 32, 90).

[88]A. K. Ekert, "Quantum cryptography based on bell's theorem", Phys. Rev. Lett. **67**, 661–663 (1991) (cit. on p. 30).

[89]K. Horodecki et al., "Secure key from bound entanglement", Phys. Rev. Lett. **94**, 160502 (2005) (cit. on p. 30).

[90]K. Horodecki et al., "Low-dimensional bound entanglement with one-way distillable cryptographic key", IEEE Transactions on Information Theory **54**, 2621–2625 (2008) (cit. on p. 30).

[91]T. Moroder et al., "Steering bound entangled states: a counterexample to the stronger peres conjecture", Phys. Rev. Lett. **113**, 050404 (2014) (cit. on p. 30).

[92]T. Vértesi et al., “Disproving the peres conjecture by showing bell nonlocality from bound entanglement”, Nat Commun **5**, 5297 (2014) (cit. on p. 30).

[93]L. Chen et al., “Schmidt number of bipartite and multipartite states under local projections”, Quantum Inf Process **16**, 75 (2017) (cit. on pp. 30, 110).

[94]M. Huber et al., “High-dimensional entanglement in states with positive partial transposition”, Phys. Rev. Lett. **121**, 200503 (2018) (cit. on pp. 30, 110, 111).

[95]K. F. Pál et al., “Class of genuinely high-dimensionally-entangled states with a positive partial transpose”, Phys. Rev. A **100**, 012310 (2019) (cit. on pp. 30, 110).

[96]R. Horodecki et al., “Quantum entanglement”, Rev. Mod. Phys. **81**, 865–942 (2009) (cit. on pp. 31, 42).

[97]M. Horodecki et al., “Separability of mixed states: necessary and sufficient conditions”, Phys. Lett. A **223**, 1–8 (1996) (cit. on pp. 32, 35).

[98]P. Horodecki, “Separability criterion and inseparable mixed states with positive partial transposition”, Phys. Lett. A **232**, 333–339 (1997) (cit. on pp. 32, 34, 89, 100, 110).

[99]D. Bruß et al., “Construction of quantum states with bound entanglement”, Phys. Rev. A **61**, 030301 (2000) (cit. on p. 32).

[100]W. Dür et al., “Distillability and partial transposition in bipartite systems”, Phys. Rev. A **61**, 062313 (2000) (cit. on p. 33).

[101]D. P. DiVincenzo et al., “Evidence for bound entangled states with negative partial transpose”, Phys. Rev. A **61**, 062312 (2000) (cit. on p. 33).

[102]Ł. Pankowski et al., “A few steps more towards npt bound entanglement”, IEEE Transactions on Information Theory **56**, 4085–4100 (2010) (cit. on p. 33).

[103]P. W. Shor et al., “Nonadditivity of bipartite distillable entanglement follows from a conjecture on bound entangled werner states”, Phys. Rev. Lett. **86**, 2681–2684 (2001) (cit. on p. 33).

[104]K. G. H. Vollbrecht et al., “Activating distillation with an infinitesimal amount of bound entanglement”, Phys. Rev. Lett. **88**, 247901 (2002) (cit. on p. 33).

[105]R. Duan, “Super-activation of zero-error capacity of noisy quantum channels”, arXiv preprint arXiv:0906.2527 (2009) (cit. on p. 33).

[106]M. Horodecki et al., “Reduction criterion of separability and limits for a class of distillation protocols”, Phys. Rev. A **59**, 4206–4216 (1999) (cit. on pp. 33, 34, 106).

[107]E. Størmer, *Positive linear maps of operator algebras* (Springer, 2013) (cit. on p. 35).

[108]M.-D. Choi, “Completely positive linear maps on complex matrices”, Linear Alg. Appl. **10**, 285–290 (1975) (cit. on p. 35).

[109]D. Chruściński et al., “Entanglement witnesses: construction, analysis and classification”, J. Phys. A: Math. Theor. **47**, 483001 (2014) (cit. on pp. 36, 89).

[110]M. Lewenstein et al., “Optimization of entanglement witnesses”, Phys. Rev. A **62**, 052310 (2000) (cit. on p. 36).

[111]M. Reed et al., *Methods of modern mathematical physics: functional analysis*, Vol. 1 (Gulf Professional Publishing, 1980) (cit. on p. 37).

[112]B. M. Terhal, “A family of indecomposable positive linear maps based on entangled quantum states”, Linear Alg. Appl. **323**, 61–73 (2001) (cit. on p. 38).

[113]V. Vedral et al., “Quantifying entanglement”, Phys. Rev. Lett. **78**, 2275–2279 (1997) (cit. on pp. 39, 48).

[114]G. V. and, "Entanglement monotones", Journal of Modern Optics **47**, 355–376 (2000) (cit. on pp. 40, 41, 100).

[115]A. Uhlmann, "Entropy and optimal decompositions of states relative to a maximal commutative subalgebra", Open Systems & Information Dynamics **5**, 209–228 (1998) (cit. on p. 41).

[116]M. Horodecki et al., "Classical capacity of a noiseless quantum channel assisted by noisy entanglement", Quantum Info. Comput. **1**, 70–78 (2001) (cit. on p. 41).

[117]D. Bruß, "Characterizing entanglement", J. Math. Phys. **43**, 4237–4251 (2002) (cit. on p. 41).

[118]S. A. Hill et al., "Entanglement of a pair of quantum bits", Phys. Rev. Lett. **78**, 5022–5025 (1997) (cit. on pp. 41, 43).

[119]W. K. Wootters, "Entanglement of formation and concurrence", Quantum Info. Comput. **1**, 27–44 (2001) (cit. on p. 41).

[120]A. Uhlmann, "Fidelity and concurrence of conjugated states", Phys. Rev. A **62**, 032307 (2000) (cit. on p. 41).

[121]P. Rungta et al., "Universal state inversion and concurrence in arbitrary dimensions", Phys. Rev. A **64**, 042315 (2001) (cit. on pp. 41, 43).

[122]A. Osterloh et al., "Scaling of entanglement close to a quantum phase transition", Nature **416**, 608–610 (2002) (cit. on p. 41).

[123]L.-A. Wu et al., "Quantum phase transitions and bipartite entanglement", Phys. Rev. Lett. **93**, 250404 (2004) (cit. on p. 41).

[124]S. Ghosh et al., "Entangled quantum state of magnetic dipoles", Nature **425**, 48–51 (2003) (cit. on p. 41).

[125]S. P. Walborn et al., "Experimental determination of entanglement with a single measurement", Nature **440**, 1022–1024 (2006) (cit. on p. 41).

[126]C. E. Shannon, "A mathematical theory of communication", The Bell system technical journal **27**, 379–423 (1948) (cit. on p. 42).

[127]G. Tóth et al., "Evaluating convex roof entanglement measures", Phys. Rev. Lett. **114**, 160501 (2015) (cit. on p. 43).

[128]F. Buscemi et al., "Linear entropy as an entanglement measure in two-fermion systems", Phys. Rev. A **75**, 032301 (2007) (cit. on p. 43).

[129]B. M. Terhal et al., "Entanglement of formation for isotropic states", Phys. Rev. Lett. **85**, 2625–2628 (2000) (cit. on pp. 44, 107).

[130]K. G. H. Vollbrecht et al., "Entanglement measures under symmetry", Phys. Rev. A **64**, 062307 (2001) (cit. on pp. 44, 48, 105).

[131]T.-C. Wei et al., "Matrix permanent and quantum entanglement of permutation invariant states", J. Math. Phys. **51**, 092203 (2010) (cit. on pp. 45, 87).

[132]R. Hübener et al., "Geometric measure of entanglement for symmetric states", Phys. Rev. A **80**, 032324 (2009) (cit. on pp. 46, 47, 88).

[133]T. M. Cover, *Elements of information theory* (John Wiley & Sons, 1999) (cit. on p. 48).

[134]G. Lindblad, "Expectations and entropy inequalities for finite quantum systems", Commun. Math. Phys. **39**, 111–119 (1974) (cit. on p. 48).

[135]V. Vedral et al., "Entanglement measures and purification procedures", Phys. Rev. A **57**, 1619–1633 (1998) (cit. on p. 48).

[136]V. Vedral, "The role of relative entropy in quantum information theory", Rev. Mod. Phys. **74**, 197–234 (2002) (cit. on p. 48).

[137]E. M. Rains, "Bound on distillable entanglement", Phys. Rev. A **60**, 179–184 (1999) (cit. on p. 48).

[138]P. M. Hayden et al., "The asymptotic entanglement cost of preparing a quantum state", J. Phys. A: Math. Gen. **34**, 6891–6898 (2001) (cit. on p. 49).

[139]M. Wright, "The interior-point revolution in optimization: history, recent developments, and lasting consequences", Bull. Amer. Math. Soc. **42**, 39–56 (2005) (cit. on p. 51).

[140]Y. Nesterov et al., *Interior-point polynomial algorithms in convex programming* (SIAM, 1994) (cit. on p. 51).

[141]A. V. Fiacco et al., *Nonlinear programming: sequential unconstrained minimization techniques* (SIAM, 1990) (cit. on p. 52).

[142]I. Bengtsson et al., *Geometry of quantum states: an introduction to quantum entanglement* (Cambridge university press, 2017) (cit. on p. 53).

[143]J. Watrous, "Simpler semidefinite programs for completely bounded norms", arXiv preprint arXiv:1207.5726 (2012) (cit. on p. 55).

[144]C. Lemaréchal, "Cauchy and the gradient method", Doc Math Extra **251**, 10 (2012) (cit. on p. 61).

[145]N. Qian, "On the momentum term in gradient descent learning algorithms", Neural networks **12**, 145–151 (1999) (cit. on p. 63).

[146]J. Nocedal et al., *Numerical optimization* (Springer, 1999) (cit. on p. 63).

[147]G. H. Golub et al., *Matrix computations* (JHU press, 2013) (cit. on p. 64).

[148]M. E. Jerrell, "Automatic differentiation and interval arithmetic for estimation of disequilibrium models", Computational Economics **10**, 295–316 (1997) (cit. on p. 65).

[149]G. F. Corliss, "Applications of differentiation arithmetic", in *Reliability in computing* (Elsevier, 1988), pp. 127–148 (cit. on p. 66).

[150]A. G. Baydin et al., "Automatic differentiation in machine learning: a survey", Journal of machine learning research **18**, 1–43 (2018) (cit. on p. 66).

[151]A. Paszke et al., "Automatic differentiation in pytorch", (2017) (cit. on p. 66).

[152]M. Abadi et al., "TensorFlow: a system for Large-Scale machine learning", in 12th usenix symposium on operating systems design and implementation (osdi 16) (2016), pp. 265–283 (cit. on p. 66).

[153]R. E. Wengert, "A simple automatic derivative evaluation program", Communications of the ACM **7**, 463–464 (1964) (cit. on p. 66).

[154]S. Linnainmaa, "Taylor expansion of the accumulated rounding error", BIT **16**, 146–160 (1976) (cit. on p. 66).

[155]A. Edelman et al., "The geometry of algorithms with orthogonality constraints", SIAM journal on Matrix Analysis and Applications **20**, 303–353 (1998) (cit. on p. 72).

[156]P.-A. Absil et al., "Optimization algorithms on matrix manifolds", in *Optimization algorithms on matrix manifolds* (Princeton University Press, 2009) (cit. on p. 73).

[157]J. Biamonte et al., "Quantum machine learning", Nature **549**, 195–202 (2017) (cit. on p. 73).

[158]M. Cerezo et al., "Variational quantum algorithms", Nat Rev Phys **3**, 625–644 (2021) (cit. on p. 73).

[159]H. Zheng et al., "Improving deep neural networks using softplus units", in 2015 international joint conference on neural networks (IJCNN) (IEEE, 2015), pp. 1–4 (cit. on p. 74).

[160]V. Nair et al., "Rectified linear units improve restricted boltzmann machines", in Proceedings of the 27th international conference on international conference on machine learning, ICML'10 (2010), pp. 807–814 (cit. on p. 74).

[161]D. H. Ackley et al., “A learning algorithm for boltzmann machines”, Cognitive science **9**, 147–169 (1985) (cit. on p. 75).

[162]C. M. Bishop et al., *Pattern recognition and machine learning*, Vol. 4, 4 (Springer, 2006) (cit. on p. 75).

[163]M. Blasone et al., “Hierarchies of geometric entanglement”, Phys. Rev. A **77**, 062304 (2008) (cit. on p. 78).

[164]J. J. Duistermaat et al., *Lie groups* (Springer Science & Business Media, 2012) (cit. on p. 80).

[165]P. Hayden et al., “Aspects of generic entanglement”, Commun. Math. Phys. **265**, 95–117 (2006) (cit. on p. 80).

[166]D. N. Page, “Average entropy of a subsystem”, Phys. Rev. Lett. **71**, 1291–1294 (1993) (cit. on p. 80).

[167]S. Lloyd et al., “Complexity as thermodynamic depth”, Annals of Physics **188**, 186–213 (1988) (cit. on p. 80).

[168]K. Zyczkowski et al., “Induced measures in the space of mixed quantum states”, J. Phys. A: Math. Gen. **34**, 7111 (2001) (cit. on p. 80).

[169]B. Sharmila et al., “Exact eigenvalue order statistics for the reduced density matrix of a bipartite system”, Annals of Physics **446**, 169107 (2022) (cit. on pp. 80, 81).

[170]S. N. Majumdar et al., “Exact minimum eigenvalue distribution of an entangled random pure state”, J Stat Phys **131**, 33–49 (2008) (cit. on p. 81).

[171]P. Vivo, “Largest schmidt eigenvalue of random pure states and conductance distribution in chaotic cavities”, J. Stat. Mech. **2011**, P01022 (2011) (cit. on p. 81).

[172]C. Nadal et al., “Statistical distribution of quantum entanglement for a random bipartite state”, J Stat Phys **142**, 403–438 (2011) (cit. on p. 81).

[173]W. Bruzda et al., “Rank of a tensor and quantum entanglement”, Linear Multilinear Algebra, 1–64 (2023) (cit. on pp. 85, 89).

[174]S. Etcheverry et al., “Quantum key distribution session with 16-dimensional photonic states”, Sci Rep **3**, 2316 (2013) (cit. on p. 85).

[175]V. Strassen et al., “Gaussian elimination is not optimal”, Numer. Math. **13**, 354–356 (1969) (cit. on pp. 85, 88).

[176]D. Bini et al., “O (n2. 7799) complexity for n× n approximate matrix multiplication”, Inf. Process. Lett. **8**, 234–235 (1979) (cit. on pp. 85, 88).

[177]D. Bini et al., “Approximate solutions for the bilinear form computational problem”, SIAM J. Comput. **9**, 692–697 (1980) (cit. on pp. 85, 88).

[178]V. De Silva et al., “Tensor rank and the ill-posedness of the best low-rank approximation problem”, SIAM J. Matrix Anal. Appl. **30**, 1084–1127 (2008) (cit. on p. 86).

[179]M. Gharahi et al., “Algebraic-geometric characterization of tripartite entanglement”, Phys. Rev. A **104**, 042402 (2021) (cit. on p. 87).

[180]E. Chitambar et al., “Tripartite entanglement transformations and tensor rank”, Phys. Rev. Lett. **101**, 140502 (2008) (cit. on p. 89).

[181]M. Christandl et al., “The resource theory of tensor networks”, arXiv preprint arXiv:2307.07394 (2023) (cit. on p. 89).

[182]J. Landsberg, “The border rank of the multiplication of 2× 2 matrices is seven”, J. Am. Math. Soc. **19**, 447–459 (2006) (cit. on p. 89).

[183]K. R. Parthasarathy, “On the maximal dimension of a completely entangled subspace for finite level quantum systems”, Proc. Math. Sci. **114**, 365–374 (2004) (cit. on p. 89).

[184] J. Walgate et al., "Generic local distinguishability and completely entangled subspaces", J. Phys. A: Math. Theor. **41**, 375305 (2008) (cit. on pp. 89, 97).

[185] M. Demianowicz, "Universal construction of genuinely entangled subspaces of any size", Quantum **6**, 854 (2022) (cit. on p. 89).

[186] M. Demianowicz et al., "Simple sufficient condition for subspace to be completely or genuinely entangled", New J. Phys. **23**, 103016 (2021) (cit. on p. 89).

[187] N. Johnston et al., "Complete hierarchy of linear systems for certifying quantum entanglement of subspaces", Phys. Rev. A **106**, 062443 (2022) (cit. on pp. 89, 90, 96, 99).

[188] M. Demianowicz et al., "Entanglement of genuinely entangled subspaces and states: exact, approximate, and numerical results", Phys. Rev. A **100**, 062318 (2019) (cit. on pp. 89, 94).

[189] S. Liu et al., "Characterizing entanglement dimensionality from randomized measurements", PRX Quantum **4**, 020324 (2023) (cit. on p. 89).

[190] I. Nape et al., "Measuring dimensionality and purity of high-dimensional entangled states", Nat Commun **12**, 5159 (2021) (cit. on p. 89).

[191] A. Ekert et al., "Entangled quantum systems and the schmidt decomposition", Am. J. Phys. **63**, 415–423 (1995) (cit. on p. 89).

[192] J. F. Buss et al., "The computational complexity of some problems of linear algebra", J. Comput. Syst. Sci. **58**, 572–596 (1999) (cit. on p. 89).

[193] R. Augusiak et al., "A note on the optimality of decomposable entanglement witnesses and completely entangled subspaces", J. Phys. A: Math. Theor. **44**, 212001 (2011) (cit. on p. 89).

[194] G. Gour et al., "Entanglement of subspaces and error-correcting codes", Phys. Rev. A **76**, 042309 (2007) (cit. on p. 89).

[195] F. Huber et al., "Quantum codes of maximal distance and highly entangled subspaces", Quantum **4**, 284 (2020) (cit. on p. 89).

[196] M. Horodecki et al., "On quantum advantage in dense coding", J. Phys. A: Math. Theor. **45**, 105306 (2012) (cit. on p. 89).

[197] D. Cavalcanti, "Connecting the generalized robustness and the geometric measure of entanglement", Phys. Rev. A **73**, 044302 (2006) (cit. on pp. 89, 91).

[198] Y.-C. Ou et al., "Bounds on negativity of superpositions", Phys. Rev. A **76**, 022320 (2007) (cit. on p. 89).

[199] J. Niset et al., "Tight bounds on the concurrence of quantum superpositions", Phys. Rev. A **76**, 042328 (2007) (cit. on p. 89).

[200] W. Song et al., "Bounds on the multipartite entanglement of superpositions", Phys. Rev. A **76**, 054303 (2007) (cit. on p. 89).

[201] Y. Xiang et al., "The bound of entanglement of superpositions with more than two components", Eur. Phys. J. D **47**, 257–260 (2008) (cit. on p. 89).

[202] S. J. Akhtarshenas, "Concurrence of superpositions of many states", Phys. Rev. A **83**, 042306 (2011) (cit. on p. 89).

[203] Z. Ma et al., "Improved bounds on negativity of superpositions", Quantum Inf. Comput. **12**, 983–988 (2012) (cit. on p. 89).

[204] N. Linden et al., "Entanglement of superpositions", Phys. Rev. Lett. **97**, 100502 (2006) (cit. on p. 89).

[205] T. Cubitt et al., "On the dimension of subspaces with bounded schmidt rank", J. Math. Phys. **49**, 022107 (2008) (cit. on p. 90).

[206] J. Tomiyama, "On the geometry of positive maps in matrix algebras. II", Linear algebra and its applications **69**, 169–177 (1985) (cit. on p. 93).

[207] D. P. DiVincenzo et al., "Unextendible product bases, uncompletable product bases and bound entanglement", Commun. Math. Phys. **238**, 379–410 (2003) (cit. on p. 95).

[208] M. Demianowicz et al., "From unextendible product bases to genuinely entangled subspaces", Phys. Rev. A **98**, 012313 (2018) (cit. on p. 97).

[209] S. Agrawal et al., "Genuinely entangled subspace with all-encompassing distillable entanglement across every bipartition", Phys. Rev. A **99**, 032335 (2019) (cit. on p. 97).

[210] B. Lovitz et al., "Entangled subspaces and generic local state discrimination with pre-shared entanglement", Quantum **6**, 760 (2022) (cit. on p. 97).

[211] A. H. Shenoy et al., "Maximally nonlocal subspaces", J. Phys. A: Math. Theor. **52**, 095302 (2019) (cit. on p. 97).

[212] C. Branciard et al., "Evaluation of two different entanglement measures on a bound entangled state", Phys. Rev. A **82**, 012327 (2010) (cit. on p. 98).

[213] B. R. Bhat, "A completely entangled subspace of maximal dimension", Int. J. Quantum Inf. **4**, 325–330 (2006) (cit. on p. 98).

[214] S. Lu et al., "Separability-entanglement classifier via machine learning", Phys. Rev. A **98**, 012315 (2018) (cit. on p. 99).

[215] S.-Y. Hou et al., "Upper bounds for relative entropy of entanglement based on active learning", Quantum Sci. Technol. **5**, 045019 (2020) (cit. on p. 99).

[216] A. C. Doherty et al., "Distinguishing separable and entangled states", Phys. Rev. Lett. **88**, 187904 (2002) (cit. on p. 99).

[217] A. Streltsov et al., "Linking a distance measure of entanglement to its convex roof", New J. Phys. **12**, 123004 (2010) (cit. on p. 100).

[218] M. A. Nielsen, "The entanglement fidelity and quantum error correction", arXiv preprint quant-ph/9606012 (1996) (cit. on p. 101).

[219] A. Peres, "All the bell inequalities", Foundations of Physics **29**, 589–614 (1999) (cit. on p. 110).

[220] R. Krebs et al., "High schmidt number concentration in quantum bound entangled states", Phys. Rev. Lett. **132**, 220203 (2024) (cit. on pp. 110, 112).

[221] S. Liu et al., "Bounding entanglement dimensionality from the covariance matrix", Quantum **8**, 1236 (2024) (cit. on p. 112).

[222] A. Tavakoli et al., "Enhanced schmidt-number criteria based on correlation trace norms", Phys. Rev. A **110**, 062417 (2024) (cit. on p. 112).

[223] F. Shahandeh et al., "Structural quantification of entanglement", Phys. Rev. Lett. **113**, 260502 (2014) (cit. on p. 117).

[224] J. M. Landsberg, *Tensors: geometry and applications*, Vol. 128 (American Mathematical Soc., 2011) (cit. on p. 118).

[225] R. Hartshorne, *Algebraic geometry*, Vol. 52 (Springer Science & Business Media, 2013) (cit. on p. 118).